\pdfoutput=1
\documentclass[11pt,twoside,a4paper,cmspaper,final,collab]{cms-tdr}
\def\svnVersion{dfcec1c}\def\svnDate{2025/01/25}\def\cmsCernNoTag{CERN-EP-2024-106}\def\cmsCernDate{\today}\def\cmsMessage{Published in Physics Reports as \href{http://dx.doi.org/10.1016/j.physrep.2024.09.013}{\doi{10.1016/j.physrep.2024.09.013}.}}
\begin{document}\cmsNoteHeader{EXO-23-005}

\providecommand{\cmsTable}[1]{\resizebox{\textwidth}{!}{#1}}
\newcommand\numberthis{\addtocounter{equation}{1}\tag{\theequation}}
\newcommand\cmsParagraph[1]{\paragraph{#1}}
\providecommand{\cmsLeft}{left\xspace}
\providecommand{\cmsRight}{right\xspace}
\renewcommand{\pmb}[1]{#1}
\setcounter{secnumdepth}{4}

\newcommand{\dark}{\ensuremath{\text{dark}}\xspace}
\newcommand{\MADANALYSIS} {\textsc{MadAnalysis}\xspace}
\newcommand{\MADDM} {\textsc{MadDM}\xspace}
\newcommand{\Vjets}{\ensuremath{\PV}\text{+jets}\xspace}
\newcommand{\Zvv}{\ensuremath{\PZ(\PGn\PAGn)}\xspace}
\newcommand{\Zvvjets}{\ensuremath{\PZ(\PGn\PAGn)}\text{+jets}\xspace}
\newcommand{\Wlvjets}{\ensuremath{\PW(\Pell\PGn)}\text{+jets}\xspace}
\newcommand{\phojets}{\ensuremath{\PGg}\text{+jets}\xspace}
\newcommand{\phojet}{\ensuremath{\PGg}\text{+jet}\xspace}
\newcommand{\hinv}{\ensuremath{\PH \to \text{inv}}\xspace}
\newcommand{\brhinv}{\ensuremath{{\mathcal{B}(\hinv)}}\xspace}
\newcommand{\jet}{\ensuremath{j}}
\newcommand{\detajj}{\ensuremath{\abs{\Delta\eta_{\jet\jet}}}\xspace}
\newcommand{\mjj}{\ensuremath{m_{\jet\jet}}\xspace}
\newcommand{\ggH}{\ensuremath{\Pg\Pg\PH}\xspace}
\newcommand{\vh}{\ensuremath{\PV\PH}\xspace}
\newcommand{\vbf}{VBF\xspace}
\newcommand{\mZ}{\ensuremath{m_{\PZ}}\xspace}
\newcommand{\PS}{\ensuremath{\text{S}}\xspace}
\newcommand{\mS}{\ensuremath{m_{\PS}}\xspace}
\newcommand{\mA}{\ensuremath{m_{\PA}}\xspace}
\newcommand{\PSS}{\ensuremath{\widetilde{\text{S}}}\xspace}
\newcommand{\PHone}{\ensuremath{\Ph_{1}}\xspace}
\newcommand{\PHtwo}{\ensuremath{\Ph_{2}}\xspace}
\newcommand{\PP}{\ensuremath{\text{P}}\xspace}
\newcommand{\PA}{\ensuremath{\text{A}}\xspace}
\newcommand{\PAprime}{\ensuremath{\PA^{\prime}}\xspace}
\newcommand{\mAprime}{\ensuremath{m_{\PAprime}}\xspace}
\newcommand{\PGgdark}{\ensuremath{\PGg_{\text{D}}}\xspace}
\newcommand{\mDM}{\ensuremath{m_{\text{DM}}}\xspace}
\newcommand{\mMed}{\ensuremath{m_{\text{med}}}\xspace}
\newcommand{\PDM}{\ensuremath{\chi}\xspace}
\newcommand{\PADM}{\ensuremath{\overline{\chi}}\xspace}
\newcommand{\mZprime}{\ensuremath{m_{\PZpr}}\xspace}
\newcommand{\mDark}{\ensuremath{m_{\dark}}\xspace}
\newcommand{\rinv}{\ensuremath{r_{\text{inv}}}\xspace}
\newcommand{\Pqdark}{\ensuremath{\PQq_{\dark}}\xspace}
\newcommand{\mqdark}{\ensuremath{m_{\Pqdark}}\xspace}
\newcommand{\Paqdark}{\ensuremath{\PAQq_{\dark}}\xspace}
\newcommand{\gq}{\ensuremath{g_{\PQq}}\xspace}
\newcommand{\Ih}{\ensuremath{I_{\text{h}}}\xspace}
\newcommand{\msd}{\ensuremath{m_{\mathrm{SD}}}\xspace}

\newcommand{\gqdark}{\ensuremath{g_{\Pqdark}}\xspace}
\newcommand{\gDM}{\ensuremath{g_{\text{DM}}}\xspace}
\newcommand{\gell}{\ensuremath{g_{\Pell}}\xspace}
\newcommand{\yq}{\ensuremath{y_{\PQq}}\xspace}
\newcommand{\yDM}{\ensuremath{y_{\text{DM}}}\xspace}
\newcommand{\ZprimeToDark}{\ensuremath{\PZpr \to \Pqdark\Paqdark}\xspace}
\newcommand{\ZprimeToQQ}{\ensuremath{\PZpr \to \qqbar}\xspace}
\newcommand{\Nc}{\ensuremath{N_{\text{c}}^{\dark}}\xspace}
\newcommand{\Nf}{\ensuremath{N_{\text{f}}^{\dark}}\xspace}
\newcommand{\ptg}{\ensuremath{p^{\PGg}_\mathrm{T}}\xspace}
\newcommand{\dphifull}{\ensuremath{\Delta\phi(\ptvec^{\,\text{jet}},\ptvecmiss)}\xspace}
\newcommand{\mindphi}{\ensuremath{\Delta\phi_{\text{min}}}\xspace}
\newcommand{\nbjets}{\ensuremath{N_{{\PQb}\text{-jet}}}\xspace}
\newcommand{\Ppidark}{\ensuremath{\PGp_{\text{dark}}}\xspace}
\newcommand{\Prhodark}{\ensuremath{\PGr_{\dark}}\xspace}
\newcommand{\Petadark}{\ensuremath{\PGh_{\dark}}\xspace}
\newcommand{\Pomegadark}{\ensuremath{\omega_{\dark}}\xspace}
\newcommand{\metadark}{\ensuremath{m_{\Petadark}}\xspace}
\newcommand{\momegadark}{\ensuremath{m_{\Pomegadark}}\xspace}
\newcommand{\aDark}{\ensuremath{\alpha_{\dark}}\xspace}
\newcommand{\lamDark}{\ensuremath{\Lambda_{\dark}}\xspace}
\newcommand{\aDarkPeak}{\ensuremath{\aDark^{\text{peak}}}\xspace}
\newcommand{\lamDarkPeak}{\ensuremath{\lamDark^{\text{peak}}}\xspace}
\newcommand{\tdark}{\ensuremath{\tau_{\dark}}\xspace}
\newcommand{\tdarkmax}{\ensuremath{\tdark^{\text{max}}}\xspace}
\newcommand{\Nstable}{\ensuremath{N_{\text{stable}}}\xspace}
\newcommand{\Nunstable}{\ensuremath{N_{\text{unstable}}}\xspace}
\newcommand{\mq}{\ensuremath{m_{\PQq}}\xspace}
\newcommand{\mdown}{\ensuremath{m_{\PQd}}\xspace}
\newcommand{\Qdark}{\ensuremath{Q_\text{dark}}\xspace}
\newcommand{\zjets}{\ensuremath{{\PZ}}+\text{jets}\xspace}
\newcommand{\wjets}{\ensuremath{{\PW}}+\text{jets}\xspace}
\newcommand{\Pbifun}{\ensuremath{\Phi}\xspace}
\newcommand{\dcoupl}{\ensuremath{\kappa_{\alpha i}}\xspace}
\newcommand{\enc}[1]{\ensuremath{\left( #1 \right)}}
\newcommand{\fpidark}{\ensuremath{f_{\Ppidark}}\xspace}
\newcommand{\Paell}{\ensuremath{{\overline{\Pell}}}\xspace}
\newcommand{\gammamu}{\ensuremath{\gamma^{\mu}}\xspace}
\newcommand{\thetaH}{\ensuremath{\theta_{\text{h}}}\xspace}
\newcommand{\thetaW}{\ensuremath{\theta_{\PW}}\xspace}
\newcommand{\thetaa}{\ensuremath{\theta_{a}}\xspace}
\newcommand{\DeltaZ}{\ensuremath{\Delta_{\PZ}}\xspace}
\newcommand{\HTmiss}{\ensuremath{\HT^\text{miss}}\xspace}
\newcommand{\ptsub}[1]{\ensuremath{p_{\text{T},#1}}\xspace}
\newcommand{\etsub}[1]{\ensuremath{E_{\text{T},#1}}\xspace}
\newcommand{\ptvecsub}[1]{\ensuremath{\vec{p}_{\text{T},#1}}\xspace}
\newcommand{\phijjmiss}{\ensuremath{\phi_{jj,\text{miss}}}\xspace}
\newcommand\Tstrut{\rule{0pt}{2.6ex}}
\newcommand\Bstrut{\rule[-0.9ex]{0pt}{0pt}}
\newcommand{\ffbar}{\ensuremath{f\overline{f}}\xspace}
\newcommand{\dVV}{\ensuremath{d_{\mathrm{VV}}}\xspace}
\newcommand{\ttbarh}{\ensuremath{\PQt\PAQt\PH}\xspace}
\newcommand{\SYY}{SY$\overline{\text{Y}}$\xspace}
\newcommand{\pipi}{\ensuremath{\pi^{+}\pi^{-}}\xspace}
\newcommand{\BR}{\ensuremath{\mathcal{B}}\xspace}
\newcommand{\TD}{\ensuremath{T_{\dark}}\xspace}
\newcommand{\CP}{\ensuremath{\text{CP}}\xspace}
\newcommand{\ctau}{\ensuremath{c\tau_0}\xspace}
\newcommand{\Lint}{\ensuremath{\mathcal{L}_{\mathrm{int}}\xspace}}
\newcommand{\Hdark}{\ensuremath{\PH_{\mathrm{D}}}\xspace}
\newcommand{\mHdark}{\ensuremath{m_{\Hdark}}\xspace}
\newcommand{\PQtdark}{\ensuremath{\PQt_{\text{dark}}}\xspace}
\newcommand{\mtdark}{\ensuremath{m_{\PQtdark}}\xspace}
\newcommand{\mzero}{\ensuremath{m_{0}}\xspace}
\newcommand{\dEdx}{\ddinline{E}{x}\xspace}
\newcommand{\nLow}{\ensuremath{N_{\text{hits}}^{\text{low \dEdx}}}\xspace}
\newcommand{\zinv}{\ensuremath{\PZ \to \text{inv}}\xspace}
\newcommand{\zll}{\ensuremath{\PZ \to \Pell\Pell}\xspace}

\newcommand{\zlljets}{\ensuremath{\PZ(\Pell\Pell)}+ \text{jets}\xspace}

\newcommand{\llost}{\ensuremath{\Pell_{\text{lost}}}\xspace}

\newcommand{\ttbarjets}{\ensuremath{\ttbar+\text{jets}}\xspace}
\newlength\cmsTabSkip\setlength{\cmsTabSkip}{1ex}
\newcommand{\sigmatilde}{\ensuremath{\widetilde{\sigma}}\xspace}
\newcommand{\wi}{\ensuremath{\Gamma_{i}}\xspace}
\newcommand{\wf}{\ensuremath{\Gamma_{f}}\xspace}
\newcommand{\wtot}{\ensuremath{\Gamma_{\text{tot}}}\xspace}
\newcommand{\mtrim}{\ensuremath{m_\text{trim}}\xspace}
\newcommand{\FiPixels}{\ensuremath{F_{i}^{\text{Pixels}}}\xspace}
\newcommand{\GiStrip}{\ensuremath{G_{i}^{\text{Strip}}}\xspace}
\newcommand{\ddbar}{\ensuremath{\PQd\PAQd}\xspace}

\newlength{\myfigurewidth}
\setlength{\myfigurewidth}{1.0\textwidth}
\newlength{\myfigureskipeqh}
\setlength{\myfigureskipeqh}{0.1cm}
\newlength{\myfigureheight}
\setlength{\myfigureheight}{1cm}
\newcommand{\threefigeqh}[3]{\resizebox{\myfigurewidth}{!}{\includegraphics[height=\myfigureheight]{#1}\hspace{\myfigureskipeqh}\includegraphics[height=\myfigureheight]{#2}\hspace{\myfigureskipeqh}\includegraphics[height=\myfigureheight]{#3}}}

\cmsNoteHeader{EXO-23-005}

\title{Dark sector searches with the CMS experiment}

\date{\today}

\abstract{
   Astrophysical observations provide compelling evidence for gravitationally interacting dark matter in the universe that cannot be explained by the standard model of particle physics. The extraordinary amount of data from the CERN LHC presents a unique opportunity to shed light on the nature of dark matter at unprecedented collision energies. This Report comprehensively reviews the most recent searches with the CMS experiment for particles and interactions belonging to a dark sector and for dark-sector mediators. Models with invisible massive particles are probed by searches for signatures of missing transverse momentum recoiling against visible standard model particles. Searches for mediators are also conducted via fully visible final states. The results of these searches are compared with those obtained from direct-detection experiments. Searches for alternative scenarios predicting more complex dark sectors with multiple new particles and new forces are also presented. Many of these models include long-lived particles, which could manifest themselves with striking unconventional signatures with relatively small amounts of background. Searches for such particles are discussed and their impact on dark-sector scenarios is evaluated. Many results and interpretations have been newly obtained for this Report.
}

\hypersetup{
pdfauthor={CMS Collaboration},
pdftitle={Dark sector searches with the CMS experiment},
pdfsubject={CMS},
pdfkeywords={CMS, Dark Matter, Dark Sectors, BSM}}

\maketitle 

\setcounter{tocdepth}{3}
\makeatletter
\ifthenelse{\boolean{cms@external}}{}{\renewcommand{\l@subsubsection}{\@dottedtocline{3}{7.3em}{3.5em}}}
\makeatother
\tableofcontents

\section{Introduction}\label{sec:intro}
The \emph{dark sector} (DS) is a collection of yet-unobserved quantum fields and their corresponding new particles. The DS particles interact weakly with standard model (SM) particles, and these interactions are typically mediated through other new particles, such as the dark photon, sterile neutrino, and axion. The primary motivation for a DS is to explain the source of dark matter (DM). The DS may also solve other fundamental issues in the SM, such as providing a natural value for the Higgs vacuum expectation value and a mechanism for baryogenesis~\cite{Foot:2014uba}. The CERN LHC provides the highest center-of-mass energy collisions at an accelerator laboratory in the world and provides a unique facility for probing a wide variety of DS models. This Report comprehensively reviews the most recent searches with the CMS experiment for particles and interactions belonging to a DS and for DS mediators.

There is strong astrophysical and cosmological evidence that DM exists and makes up approximately 26\% of the total mass-energy budget of the universe~\cite{Planck:2018vyg,ARBEY2021103865}. This evidence is based on numerous observations of its gravitational interaction on galactic scales. The rotation curves of most galaxies do not match the expected behavior from visible matter~\cite{Rubin:1980zd,Persic:1995ru}. Recently, several galaxies have been observed whose rotation curves do match the expectation~\cite{vanDokkum:2018vup,PinaMancera:2021wpc}, suggesting DM is unevenly distributed. Strong lensing observations of galaxy cluster collisions~\cite{Clowe:2006eq} and weak gravitational lensing from large-scale structures~\cite{Chang:2017kmv} both indicate the presence of DM at super-galactic scales. Accurate modeling of the cosmic microwave background power spectrum~\cite{Planck:2018vyg} and the matter power spectrum of the universe~\cite{Dodelson:2011qv,Planck:2018nkj} requires the presence of DM. Various scenarios beyond the standard model (BSM) that contain DM particle candidates may also resolve discrepancies in the SM, such as the predictions for light-element abundances from Big Bang nucleosynthesis~\cite{Pospelov:2010hj}.

Several complementary approaches~\cite{Bauer:2013ihz} including direct-detection (DD) experiments, indirect-detection (ID) experiments, and proton-proton ($\Pp\Pp$) and heavy ion (HI) collisions in experiments such as CMS at the LHC can study potential interactions of DM particles with the SM. While DD and ID explore DM coming from astrophysical sources, the LHC provides a laboratory environment in which we can control every aspect of the collisions and therefore, hopefully, of DM production. DD experiments directly probe DM scattering from ordinary matter, usually nuclei. The search for such scattering is the basis of experiments such as XENON~\cite{XENON:2023cxc}, LUX-ZEPLIN~\cite{LZ:2022lsv}, PandaX~\cite{PandaX-4T:2021bab}, PADME~\cite{PADME:2022xly}, and others (a review can be found in Ref.~\cite{Antel:2023hkf}). This approach is very sensitive to low values of the scattering cross section, down to the zeptobarn scale, but may face difficulties detecting DM-lepton interactions or light DM particles ($\lesssim$1\GeV in mass). These difficulties arise from the fact that the liquid xenon and liquid argon energy resolutions are poor for low-energy recoils. To probe low recoil energies,  different technologies are needed. Conversely, the ID approach looks for signals of DM-DM annihilation into SM particles, which are being searched for by experiments such as AMS-02~\cite{AMS:2021nhj}, EGRET~\cite{EGRET:1999szp}, Fermi-LAT~\cite{Fermi-LAT:2015xeq}, and IceCube~\cite{IceCube:2006tjp}. This approach is sensitive to the coupling of DM to SM particles, while also probing the nature of the DM-DM annihilation process that plays a fundamental role in the observed thermal-relic density. The main difficulty is the need for accurate modeling of the astrophysical background sources and of the DM density profile in the region of interest. There also exist beam-dump experiments that could potentially produce DM~\cite{Beacham:2019nyx}, which are beyond the scope of this Report.

While DD and ID experiments could determine the mass and nuclear cross section of DM and its abundance, the LHC experiments could determine the properties of the DM particles. In particular, the CMS experiment can probe many BSM scenarios that predict the existence of a DS. At particle colliders, searches for DM often involve the production of a pair of DM candidates, leading to a signature of missing transverse momentum (\ptmiss) recoiling against an SM particle. Simplified benchmark models have been put forward by the community to guide these searches~\cite{Abercrombie:2015wmb}, together with recommendations on the presentation of experimental results~\cite{Boveia:2016mrp} and guidelines for the comparison between the collider and DD/ID experiments~\cite{Albert:2017onk}. These benchmark models have a DM candidate and a mediator particle, which may also be a BSM state. Collider searches generally present their results in terms of the masses and spins of both of these particles. As will be shown in this Report, the collider approach can provide sensitivity that is complementary to those of the DD and ID experiments. In the particular case of simplified models, certain assumptions on the mediator couplings to both SM and DM particles allow us to compare collider and DD searches. The collider probes a different phase space leading to different assumptions about the assumed particle nature and interactions. Under these assumptions, lighter DM particles (masses down to a few {\GeVns}) and models where the nuclear interaction is spin-dependent can lead to strong bounds on the dark sector coupling. Compared to DD, collider searches also have the benefit of not being dependent on the local DM density and the assumption that the local DM present in the current universe is the same particle used to produce the DM thermal-relic density. However, it is important to note that if a DM candidate is found at a collider, the stability of the particle once it has exited the detector is not guaranteed, and so the interpretation of any positive result should be treated with care.

Going beyond the simplified-model picture entails the construction of an extended DS of particles, based on concepts such as weak-scale supersymmetry (SUSY)~\cite{Baer:2020kwz},  extra dimensions~\cite{Rappoccio:2018qxp}, or extended scalar sectors~\cite{Ilnicka:2018def}. An alternative approach is to hypothesize that these new particles are neutral under all the SM charges: electric, weak, and color. This new DS can have rich dynamics with previously unexplored signatures~\cite{Alexander:2016aln} that are now the target of dedicated searches by the CMS Collaboration. In this Report, we review CMS DS searches, using the Run~2 $\Pp\Pp$ and HI collision data sets collected by the CMS detector from 2016--2018, or, in some cases, using data sets from Run~1 or Run~3, collected in 2010--2012 and 2022, respectively.

Several types of signatures are explored in this Report. Direct DM signatures, in their simplest form, consist of the production of the mediator particle, which subsequently decays into DM. Final states from such processes feature the presence of \ptmiss because the DM particles interact sufficiently weakly to be invisible in the detector; here, we call the DM particle a weakly interacting massive particle (WIMP)~\cite{Jungman:1995df}. To be detectable, the DM particle must be accompanied by at least one visible object, such as a jet (which is a collimated spray of energetic particles produced by the hadronization of a high-energy quark or gluon), lepton, photon, or the decay products of a heavy SM boson, such as the Higgs (\PH), \PW, or \PZ boson. These characteristic signatures are the mainstay of ``mono-X'' searches, where X denotes the visible, radiated object that recoils off the system that directly produces the DM. In this Report, the DM particle is generally assumed to be a Dirac fermion, unless otherwise stated; however, this does not preclude sensitivity to other DM spin states. Moreover, we focus on DM with masses ranging from the {\MeVns} to the {\TeVns} range and avoid consideration of DM when it is below this range (often referred to as ultra-light DM) or above this range (often called ultra-heavy DM).

Any mediator between the DS and the SM that is produced at colliders by the interaction of SM particles must also be able to decay back to those SM particles, such as the process $\cPq\cPaq \to \PZpr \to \cPq^{\prime}\cPaq^{\prime}$. Correspondingly, we can also search for the DM indirectly via fully visible resonances arising from the mediator production. This approach is only sensitive to the SM interactions of the mediator and therefore makes no additional assumptions about the portal. However, accessing different resonant mass ranges for evidence of DS particles may require different search strategies at colliders, as discussed in subsequent sections.

Different signatures appear from rich DS dynamics that can produce more mediators, additional unstable particles, or new interactions. These extended DM models give rise to a number of signatures that can be probed at the LHC. Moreover, these added signatures enhance the sensitivity of the LHC to the DS with additional visible particles and energy in the final state, compared to mono-X searches.

One such signature is particles with long lifetimes, which often appear in BSM scenarios, notably in models that describe the elementary particle nature of DM. Long-lived particles (LLPs) can be produced as a result of small couplings, little available phase space for the particle’s decay, or high mass scales~\cite{Curtin:2018mvb}. These mechanisms generically appear in BSM scenarios, including those in the DS. For example, small portal couplings between the SM and the DS give rise to LLPs in hidden valley models~\cite{Strassler:2006im,Strassler:2006ri}. In addition, phase space restrictions are often predicted in models where WIMPs coannihilate with an additional particle in the early universe; thus, LLPs appear in coannihilation models such as the one discussed in Ref.~\cite{Guo:2021vpb}. In inelastic dark matter (IDM) models, the small mass splitting creates a small kinematic phase space available for the decay and can produce LLPs~\cite{Izaguirre:2015zva}. In freeze-in scenarios such as those discussed in Refs.~\cite{Hall:2009bx,Co:2015pka}, the decay length required for the observed DM abundance leads to LLPs and displaced signals at colliders. Lastly, decays suppressed by high mass scales naturally arise in asymmetric DM scenarios~\cite{Kaplan:2009ag}.

The CMS DS search program has grown in coverage and sophistication over the course of the recent and ongoing LHC runs. Mono-X and fully visible searches have incorporated more data and additional final states, aided by the development of boosted object reconstruction and advanced analysis methodologies. New searches targeting LLPs and other signatures of extended DSs have been enabled by advances in triggering, such as data scouting, dedicated displaced reconstruction techniques, and machine learning methodologies. These developments have facilitated greater utilization of the detector for sensitivity to more complex final states. The dedicated LHC searches for DM and DSs have significantly expanded our exploration and knowledge of the available parameter space of masses, couplings, and lifetimes. The resulting exclusions from these searches significantly narrow the allowed parameter space where new physics related to DM and DS might lie and thus will guide future theoretical and experimental advances.

As indicated by the above discussion, a given theoretical model can produce multiple distinct signatures, and a given signature can be produced by multiple theoretical models. The collider effort to search for DM is organized in terms of observable final states, while the results are interpreted in terms of theoretical models. Therefore, we consider both the theoretical and observable perspectives in this Report.

We begin by presenting the theoretical framework of the DM models used for CMS DM analyses in Section~\ref{sec:theoreticalFramework}. Subsequently, we discuss the experimental apparatus in Section~\ref{sec:detector}. We then present the results and their interpretations in the context of the theoretical framework in Section~\ref{sec:reinterpretationAndResults}. Then, we present the experimental methods that are typical in these searches in Section~\ref{sec:commonExperimentalChallenges}. The data and simulation used are described in Section~\ref{sec:datasetAndSignalSimulation}, and the final-state signatures probed by each CMS DM analysis are detailed in Section~\ref{sec:signatures}. Finally, we summarize this Report in Section~\ref{sec:conclusion}.

\subsection{Summary of searches, models, and interpretations}

Table~\ref{tab:bigSummaryTable} lists the analyses presented in this Report, which are detailed in Section~\ref{sec:signatures}, the models in which they are interpreted, which are detailed in Section~\ref{sec:theoreticalFramework}, and the figures in which their results are presented, which are detailed in Section~\ref{sec:reinterpretationAndResults}.

\newcommand{\clineInner}{}
\newcolumntype{F}{>{\raggedright\arraybackslash}p{1.2cm}}
\begin{table}[htbp]
\centering
\renewcommand{\arraystretch}{1.2}
\topcaption{The mapping of the theoretical models in Section~\ref{sec:theoreticalFramework} (columns) and the signature-based experimental searches in Section~\ref{sec:signatures} (rows) to the interpretations in Section~\ref{sec:reinterpretationAndResults} (figure numbers).}
\cmsTable{
\begin{tabular}{p{2cm}p{2cm}>{\hangindent=0.2cm}p{6.3cm}FFFFFFF}
\multicolumn{1}{c}{Signature}&\multicolumn{1}{c}{Type}&\multicolumn{1}{c}{Search for}&\hyperref[sec:spinoneportal]{Spin-1 portal}&\hyperref[sec:scalarportal]{Spin-0 portal}&\hyperref[sec:bifunportal]{Fermion portal}&\hyperref[sec:2HDM]{2HDM +a scenario}&\hyperref[sec:HAHM]{HAHM} and \hyperref[sec:theory_susy]{SUSY}&\hyperref[sec:iDM]{Inelastic DM}&\hyperref[sec:darkqcd]{Hidden valleys}\\
\hline\multirow{7}{2cm}[-3\baselineskip]{\hyperref[sec:signatures_invis]{Invisible final states}}&\multirow{5}{2cm}[-1.5\baselineskip]{\hyperref[sec:monoX]{Mono-X searches}}&\hyperref[sec:EXO-20-004]{Monojet and hadronically-decaying mono-V dark matter}&\ref{fig:summary_vector_axial}, \ref{fig:summary_spindependent_dependent}, \ref{fig:summary_gq}, \ref{fig:eps_darkpho}&\ref{fig:summary_scalar}, \ref{fig:summary_pseudo}&\ref{fig:monojet_fermionportal}&\ref{fig:summary_2hdma_mamA}&\NA&\NA&\ref{fig:svj_reinterp}, \ref{fig:svj_gq_limit}\\
\clineInner&&\hyperref[sec:EXO-19-003]{New physics in leptonically decaying \PZ}&\ref{fig:summary_vector_axial}&\ref{fig:summary_scalar}, \ref{fig:summary_pseudo}&\ref{fig:monojet_fermionportal}&\ref{fig:summary_2hdma_mamA}&\NA&\NA&\NA\\
\clineInner&&\hyperref[sec:EXO-16-053]{Monophoton events}&\ref{fig:summary_vector_axial}&\NA&\NA&\NA&\NA&\NA&\NA\\
\clineInner&&\hyperref[sec:EXO-18-011]{Dark matter in Higgs boson associated production}&\NA&\NA&\NA&\ref{fig:summary_2hdma_mamA}&\NA&\NA&\NA\\
\clineInner&&\hyperref[sec:EXO-21-012]{Dark matter in dark Higgs+\ptmiss}&\NA&\ref{fig:darkhiggsww}&\NA&\NA&\NA&\NA&\NA\\
\cline{2-10}&\multirow{1}{2cm}{\hyperref[sec:higgs_invisible]{Higgs to invisible}}&\hyperref[para:HIG-21-007]{Higgs boson decay into invisible final states}&\NA&\ref{fig:darkhiggsthetaH}, \ref{fig:higgs_portal_comb}, \ref{fig:spin_independent_DM_cross_section}&\NA&\ref{fig:summary_h-aa-2hdma}&\NA&\NA&\ref{fig:higgsllp_high}, \ref{fig:higgsllp_low}, \ref{fig:higgsllp_vlow}\\
\cline{2-10}&\hyperref[sec:svj]{Hidden valley models}&\hyperref[sec:EXO-19-020]{Semivisible jets}&\NA&\NA&\NA&\NA&\NA&\NA&\ref{fig:svj_reinterp}, \ref{fig:svj_gq_limit}\\
\hline\multirow{10}{2cm}[-5\baselineskip]{\hyperref[sec:signatures_vis]{Fully visible and prompt signatures}}&\multirow{4}{2cm}[-3\baselineskip]{\hyperref[sec:lowMass]{Low-mass resonance searches}}&\hyperref[par:EXO-18-012]{Low-mass vector resonances decaying into quark-antiquark pairs}&\ref{fig:summary_vector_axial}, \ref{fig:summary_gq}&\NA&\NA&\NA&\NA&\NA&\NA\\
\clineInner&&\hyperref[par:EXO-17-027]{Low-mass quark-antiquark resonances in combination with a photon}&\ref{fig:summary_gq}&\NA&\NA&\NA&\NA&\NA&\NA\\
\clineInner&&\hyperref[paragraph:EXO-19-018]{A prompt dark photon resonance decaying into two muons including data scouting}&\ref{fig:darkphotonscouting}&\NA&\NA&\NA&\NA&\NA&\NA\\
\clineInner&&\hyperref[paragraph:EXO-21-005]{Prompt dimuon resonances with data scouting}&\ref{fig:darkphotonscouting}&\NA&\NA&\NA&\NA&\NA&\NA\\
\cline{2-10}&\multirow{3}{2cm}[-1.5\baselineskip]{\hyperref[sec:highMass]{High-mass resonance searches}}&\hyperref[par:EXO-19-004]{Dijet resonances using events with three jets}&\ref{fig:summary_vector_axial}, \ref{fig:summary_gq}&\NA&\NA&\NA&\NA&\NA&\NA\\
\clineInner&&\hyperref[sec:EXO-19-012]{High-mass dijet resonances}&\ref{fig:summary_vector_axial}, \ref{fig:summary_vector_axial_gl0p1}, \ref{fig:summary_gq}&\NA&\NA&\NA&\NA&\NA&\ref{fig:svj_reinterp}, \ref{fig:svj_gq_limit}\\
\clineInner&&\hyperref[sec:EXO-19-019]{New physics in high-mass dilepton final state}&\ref{fig:summary_vector_axial_gl0p1}&\NA&\NA&\NA&\NA&\NA&\NA\\
\cline{2-10}&\multirow{3}{2cm}[-0.5\baselineskip]{\hyperref[sec:otherSig]{Other signatures}}&\hyperref[sec:SUEPsearch]{Soft unclustered energy patterns}&\NA&\NA&\NA&\NA&\NA&\NA&\ref{fig:3D_limits}\\
\clineInner&&\hyperref[sec:signatures_stealth]{Stealth top squarks}&\NA&\NA&\NA&\NA&\ref{fig:stealth}, \ref{fig:stealth_syy_shh_2D}&\NA&\NA\\
\clineInner&&\hyperref[sec:FSQ-16-012]{Axion-like particles in ultraperipheral PbPb collisions}&\NA&\ref{fig:ALPresults}&\NA&\NA&\NA&\NA&\NA\\
\hline\multirow{11}{2cm}[-5\baselineskip]{\hyperref[sec:signatures_llp]{Searches for long-lived particles}}&\multirow{4}{2cm}[-1.5\baselineskip]{\hyperref[sec:displacedLeptons]{Displaced leptons}}&\hyperref[sec:EXO-18-003]{Leptons with large impact parameters in $\Pe\pmb{\Pgm}$, $\Pe\Pe$, and $\pmb{\Pgm}\pmb{\Pgm}$ final states}&\NA&\NA&\NA&\NA&\NA&\NA&\ref{fig:higgsllp_high}, \ref{fig:higgsllp_low}\\
\clineInner&&\hyperref[sec:EXO-21-006]{Muon pairs from a displaced vertex}&\NA&\NA&\NA&\NA&\ref{fig:CMS_Longlived_DarSector}&\NA&\NA\\
\clineInner&&\hyperref[paragraph:HIG-18-003]{Prompt and displaced dimuons in final states with $4\pmb{\Pgm}$+X}&\NA&\NA&\NA&\NA&\ref{fig:CMS_Longlived_DarSector}&\NA&\NA\\
\clineInner&&\hyperref[paragraph:EXO-20-014]{Displaced dimuon resonances with data scouting}&\NA&\NA&\NA&\NA&\ref{fig:CMS_Longlived_DarSector}&\NA&\ref{fig:higgsllp_high}, \ref{fig:higgsllp_low}, \ref{fig:higgsllp_vlow}\\
\cline{2-10}&\multirow{3}{2cm}[-1.5\baselineskip]{\hyperref[sec:hadronicLLPdecays]{Hadronic LLP decays}}&\hyperref[parag:disp_jets]{LLPs decaying into displaced jets}&\NA&\NA&\NA&\NA&\ref{fig:stealth_syy_shh_2D}&\NA&\ref{fig:higgsllp_high}, \ref{fig:higgsllp_low}, \ref{fig:zprimeLL_4b}, \ref{fig:zprimeLL_2b2nu}, \ref{fig:higgsLLP_4b}, \ref{fig:higgsLLP_2b2nu}\\
\clineInner&&\hyperref[sec:EXO-19-013]{LLPs decaying to jets with displaced vertices}&\NA&\NA&\NA&\NA&\ref{fig:stealth_syy_shh_2D}&\NA&\ref{fig:higgsllp_high}, \ref{fig:higgsllp_low}, \ref{fig:zprimeLL_4b}, \ref{fig:higgsLLP_4b}\\
\clineInner&&\hyperref[subsec:emj_run1]{Emerging jets}&\NA&\NA&\NA&\NA&\NA&\NA&\ref{fig:emj_reinterp}\\
\cline{2-10}&\multirow{4}{2cm}[-2\baselineskip]{\hyperref[sec:LLPsAndMET]{Signatures with LLPs and \ptmiss}}&\hyperref[sec:msclusters]{Neutral LLPs decaying in the muon system}&\NA&\NA&\NA&\NA&\NA&\NA&\ref{fig:stealth_syy_shh_2D}, \ref{fig:emj_reinterp}, \ref{fig:clustersdarkshowers}, \ref{fig:higgsllp_high}, \ref{fig:higgsllp_low}, \ref{fig:higgsllp_vlow}, \ref{fig:higgsLLP_4b}\\
\clineInner&&\hyperref[parag:inelDMEXO20010]{Inelastic dark matter}&\NA&\NA&\NA&\NA&\NA&\ref{fig:iDM_limits}&\NA\\
\clineInner&&\hyperref[sec:EXO-19-001]{New physics with delayed jets}&\NA&\NA&\NA&\NA&\NA&\NA&\ref{fig:zprimeLL_4b}, \ref{fig:zprimeLL_2b2nu}\\
\clineInner&&\hyperref[sec:EXO-21-014]{LLPs with trackless and out-of-time jets and \ptmiss}&\NA&\NA&\NA&\NA&\ref{fig:stealth_syy_shh_2D}&\NA&\ref{fig:zprimeLL_4b}, \ref{fig:zprimeLL_2b2nu}, \ref{fig:higgsLLP_4b}, \ref{fig:higgsLLP_2b2nu}\\
\end{tabular}
}
\label{tab:bigSummaryTable}
\end{table}

\section{Theoretical framework}
\label{sec:theoreticalFramework}

There are numerous proposed models accessible in high-energy collisions that include new particles satisfying the cosmological and astrophysical constraints for a DM candidate. Reviews of such models can be found in Refs.~\cite{ARBEY2021103865,Arcadi:2017kky,Kahlhoefer:2017dnp}. Dark matter searches at the LHC, therefore, are characterized by final states that include a DM particle or are otherwise consistent with a BSM scenario that can produce DM candidates.

In addition to the DM particle, every model includes an additional sector, called a ``portal'', that couples SM particles to DM particles. In most DM models, this portal consists of a new mediator particle. However, in the models considered in this Report, the portal can also include a \PZ or \PH boson with couplings modified to include the possibility of decays to DM. We do not consider models in which the portal is the \PW boson, as this Report does not cover scenarios in which the DM carries an SM charge.

\begin{figure}[htb]
    \centering
    \includegraphics[width=0.95\textwidth]{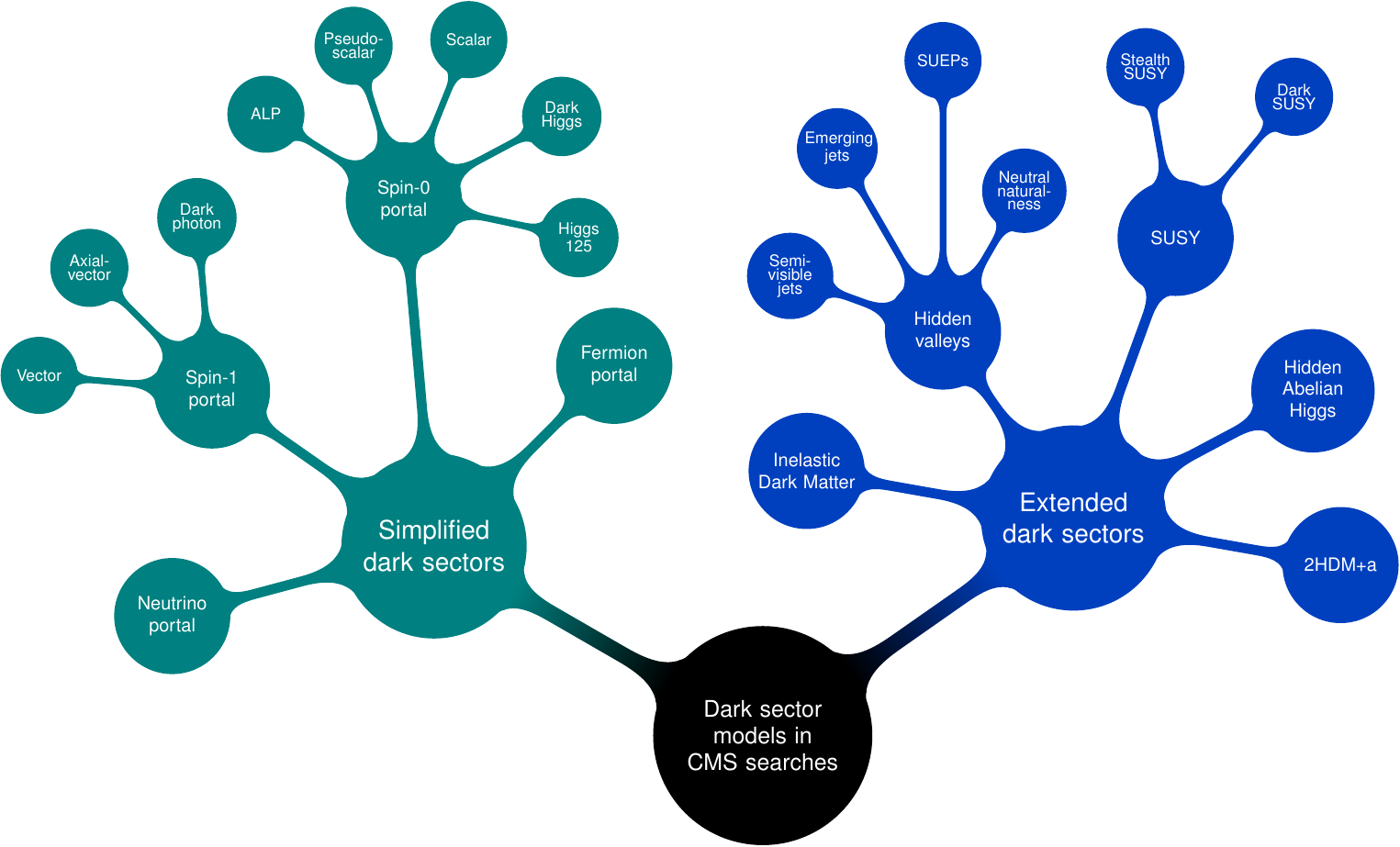}
\caption{Map of the models probed in CMS searches for dark sectors. Nodes are separated by emphasis of the interpretation and may overlap in physics origin.}
    \label{fig:nomological}
\end{figure}

\begin{figure}[htbp!]
    \centering
    \includegraphics[width=0.9\textwidth]{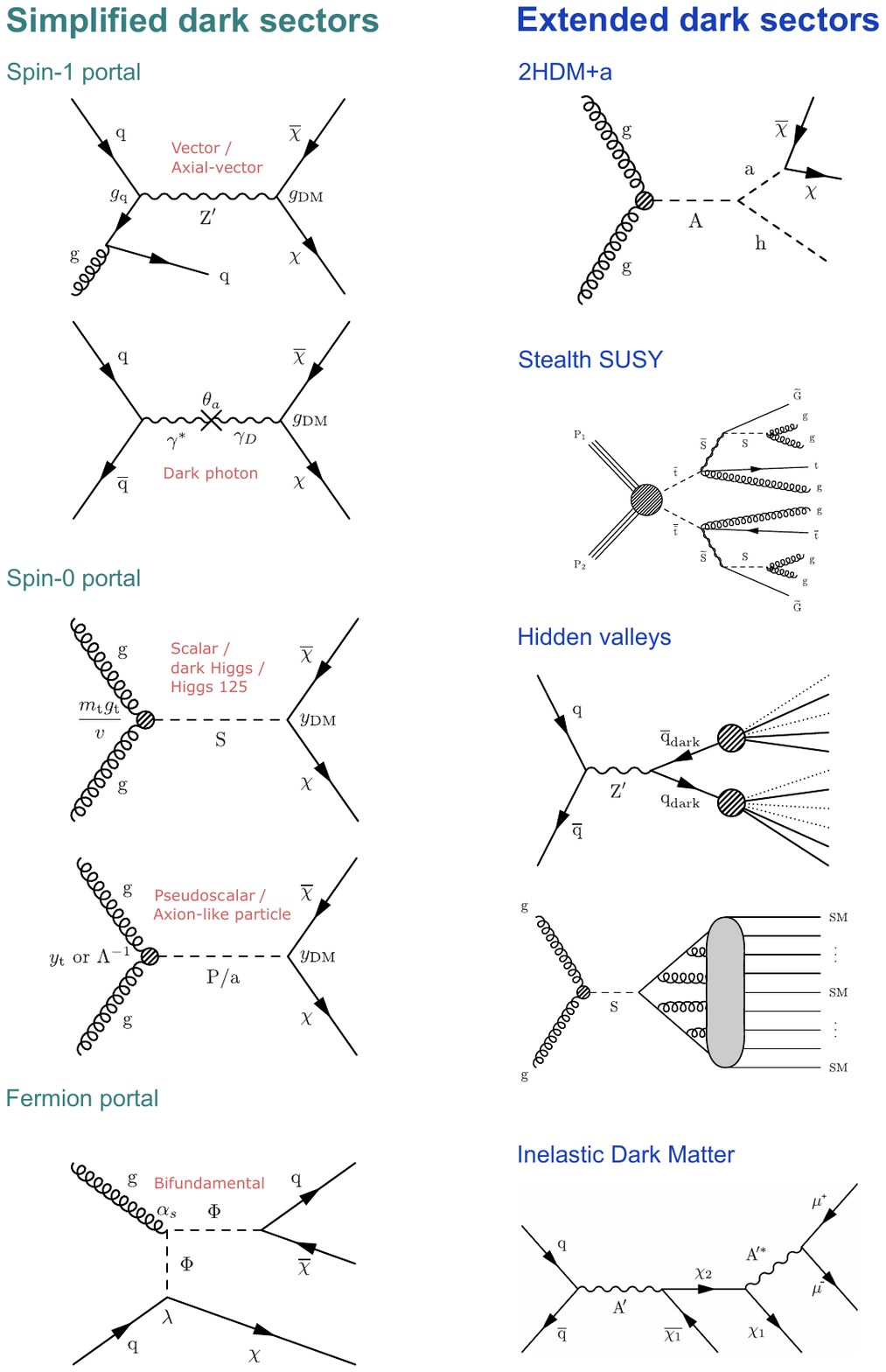}
\caption{Example Feynman diagrams in the taxonomy of dark sector models.}
    \label{fig:diagramtable}
\end{figure}

In this section, we present the scope of DS models probed with the CMS experiment. We classify the models into two categories: models that consist of a single mediator particle and DM are denoted \emph{simplified DSs} and are discussed in Section~\ref{sec:portals}, and models with more complicated DS dynamics are denoted \emph{extended DSs} and are discussed in Section~\ref{sec:models}. In simplified DSs, the DM particle is a WIMP~\cite{Jungman:1995df}, whereas in extended DSs, the nature of the DM particle or particles may vary. We note that simplified DS models are not complete models, but proxies for more complete models~\cite{Morgante:2018tiq,Kahlhoefer:2015bea,Abdallah:2015ter}. Figures~\ref{fig:nomological} and \ref{fig:diagramtable} give an overview of the models probed in CMS searches, which are explained in the following.

\subsection{Simplified dark sectors}\label{sec:portals}

Initially, the exploration of the DS proceeded using an effective field theory (EFT) approach, with a single parameter $\Lambda$~\cite{Rajaraman:2011wf,Beltran:2010ww,Bai:2010hh,Goodman:2010ku}. This parameter defines either the coupling strength or the interaction scale, which cannot be disentangled. Therefore, bounds on the DM production cross section are presented in terms of $\Lambda$, using a prescribed fixed coupling, when compared to noncollider experimental results. However, the higher energies at the LHC allow for exploring more physical features that are not captured by EFT models because they are valid only for momentum transfers much smaller than the scale of the interaction. Therefore, they have largely been superseded by two classes of DM models: simplified models and hidden sector portal models, the latter of which is also known as ``feebly interacting particle'' (FIP) models~\cite{Buchmueller:2013dya,Lebedev:2014bba,Fairbairn:2014aqa,Buchmueller:2014yoa,Malik:2014ggr,Buckley:2014fba,Harris:2014hga,Xiang:2015lfa,Chala:2015ama,Abdallah:2015ter,Abercrombie:2015wmb,Harris:2015kda,Choudhury:2015lha,DeSimone:2016fbz,Albert:2016osu,Kahlhoefer:2017dnp,Albert:2017onk,Arcadi:2017kky}. The difference between the FIP and simplified models is minimal and  mostly related with the mixing with the SM \PZ boson, but it cannot be neglected. However, because of this, one can easily translate results between those two models. Results interpreted with EFTs in place of a mediator are not discussed further in this Report. However, effective couplings connecting a mediator to a higher scale are considered in a few instances.

The simplified models were developed explicitly to compare LHC results with those from DD and ID searches, while the DS models were developed to facilitate comparisons with beam dump experiments targeting light DSs. These two classes largely overlap and methods exist to interpret results from one class for the other class.

For both the simplified models and the DS models, there is a framework that connects the DS with the visible sector through a mediator. The existence of a mediator mitigates the limitations of present in EFTs, which can yield unphysical distributions because of the lack of a mediator at a collision scale comparable to the EFT scale. The mediator enables resonant production and a physical production mechanism but also adds complexity because several other parameters need to be scanned to produce interpretable results. Moreover, aspects such as renormalizability and ultraviolet completion are typically not taken into account. Despite these shortcomings, established and reliable schemes exist to present the results, and established models exist that aim to cover a variety of mediators and DM interactions\cite{Abercrombie:2015wmb,Boveia:2016mrp}. 

In this Section, we present both classes, namely the simplified models and DS models, together, highlighting differences when needed~\cite{Boveia:2022jox,Antel:2023hkf}. 
To ensure broad coverage, four separate categories of portals are commonly utilized. These models are classified by the spin and the properties of the portal:
\begin{itemize}
\item {\textbf{Spin-1 portal:}} This category of models (Section~\ref{sec:spinoneportal}) has a spin-1 mediator that couples to the SM with couplings that are uniform across flavors but deviates by particle type (leptons and quarks can have different couplings)~\cite{Buchmueller:2013dya,Lebedev:2014bba,Fairbairn:2014aqa,Buchmueller:2014yoa,Malik:2014ggr,Harris:2014hga,Xiang:2015lfa,Chala:2015ama,Abdallah:2015ter,Abercrombie:2015wmb,Harris:2015kda,Choudhury:2015lha,DeSimone:2016fbz,Albert:2016osu,Kahlhoefer:2017dnp,Albert:2017onk,Arcadi:2017kky,Boveia:2016mrp,Alexander:2016aln,Beacham:2019nyx,Gori:2022vri}. With simplified models, a minimal model with only quark couplings is taken as the baseline, although other couplings are possible, and both a pure vector and pure axial-vector coupling are allowed. In FIP models, the spin-1 mediator is assumed to mix with the \PZ boson and photon, yielding a dark-photon model. Where critical, interference with Drell--Yan (DY) processes are also considered~\cite{Albert:2017onk}.
\item {\textbf{Spin-0 portal:}} This category of models (Section~\ref{sec:scalarportal}) has a scalar or pseudoscalar particle as the mediator. The simplified model assumes the scalar particle does not mix with the Higgs boson. In the FIP models, the scalar portal mediator mixes with the Higgs boson and is often referred to as the dark-Higgs or the Higgs portal mediator (\Hdark). The FIP version of the mediator of the axion (\Pa) portal is often referred to as an axion-like particle (ALP), which has the same coupling structure as the pseudoscalar mediator in the simplified model~\cite{Buckley:2014fba,Harris:2014hga,Abercrombie:2015wmb,Harris:2015kda,Choudhury:2015lha,DeSimone:2016fbz,Albert:2016osu,Kahlhoefer:2017dnp,Albert:2017onk,Arcadi:2017kky,Boveia:2016mrp,Alexander:2016aln,Beacham:2019nyx,Gori:2022vri,Haisch:2015ioa}.
\item {\textbf{Neutrino portal:}} This category of models (Section~\ref{sec:neutrinoportal}) includes a heavy neutral lepton (HNL), which often takes the form of a right-handed neutrino~\cite{Canetti:2012vf,Drewes:2013gca,Abdullahi:2022jlv,Boyarsky:2018tvu,Drewes:2016upu,Canetti:2012zc}. 
\item {\textbf{Fermion portal:}} This category of models (Section~\ref{sec:bifunportal}) includes a scalar mediator \Pbifun with a Yukawa coupling between DM and SM fermions, which allows $t$-channel interactions~\cite{Papucci:2014iwa,Arina:2020udz,Arina:2023msd}.
\end{itemize}

\begin{figure}
    \centering
    \includegraphics[width=0.33\textwidth]{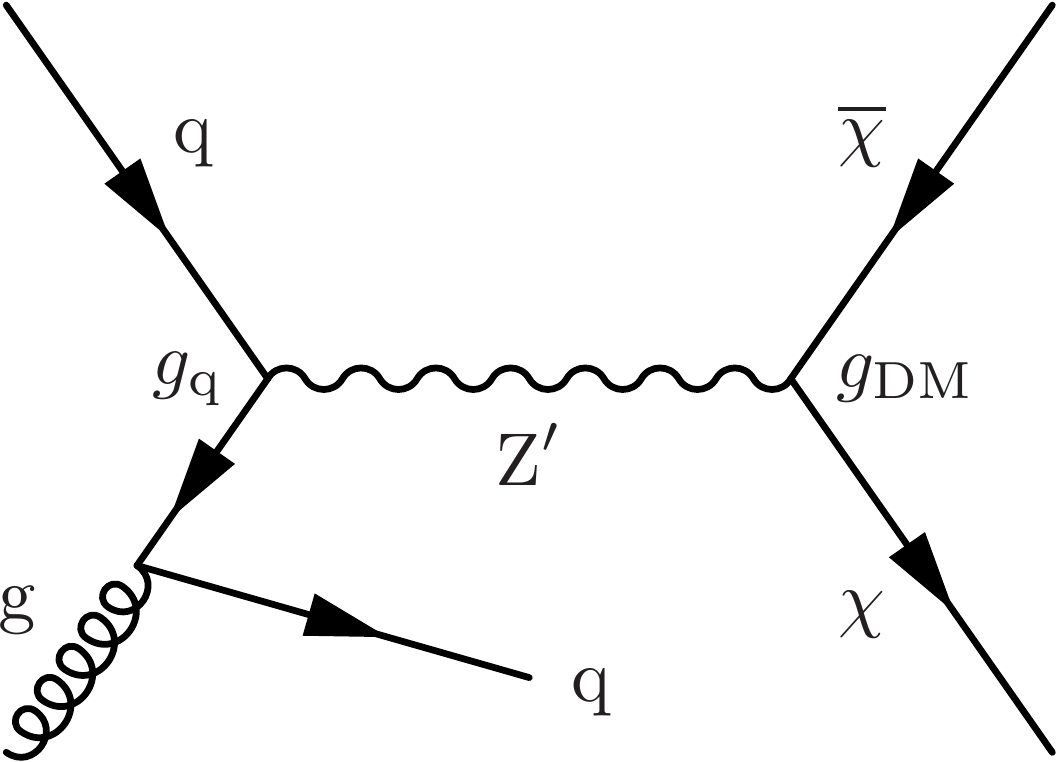}\hspace{2em}
    \includegraphics[width=0.33\textwidth]{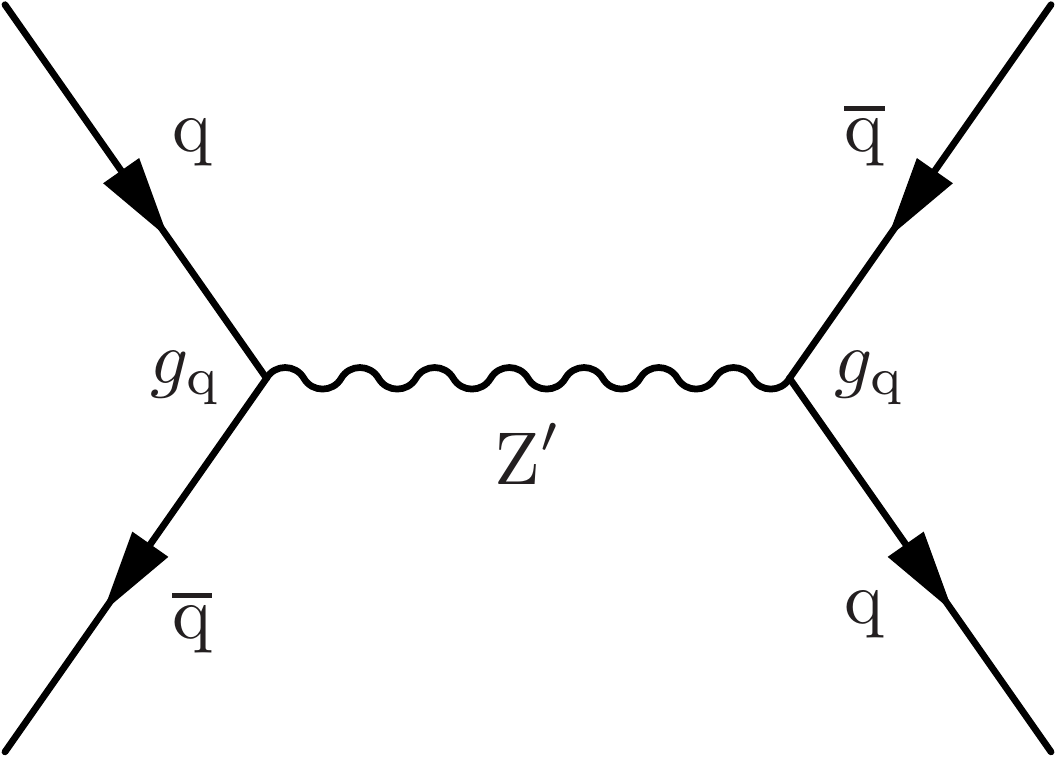}\\[2ex]
    \includegraphics[width=0.33\textwidth]{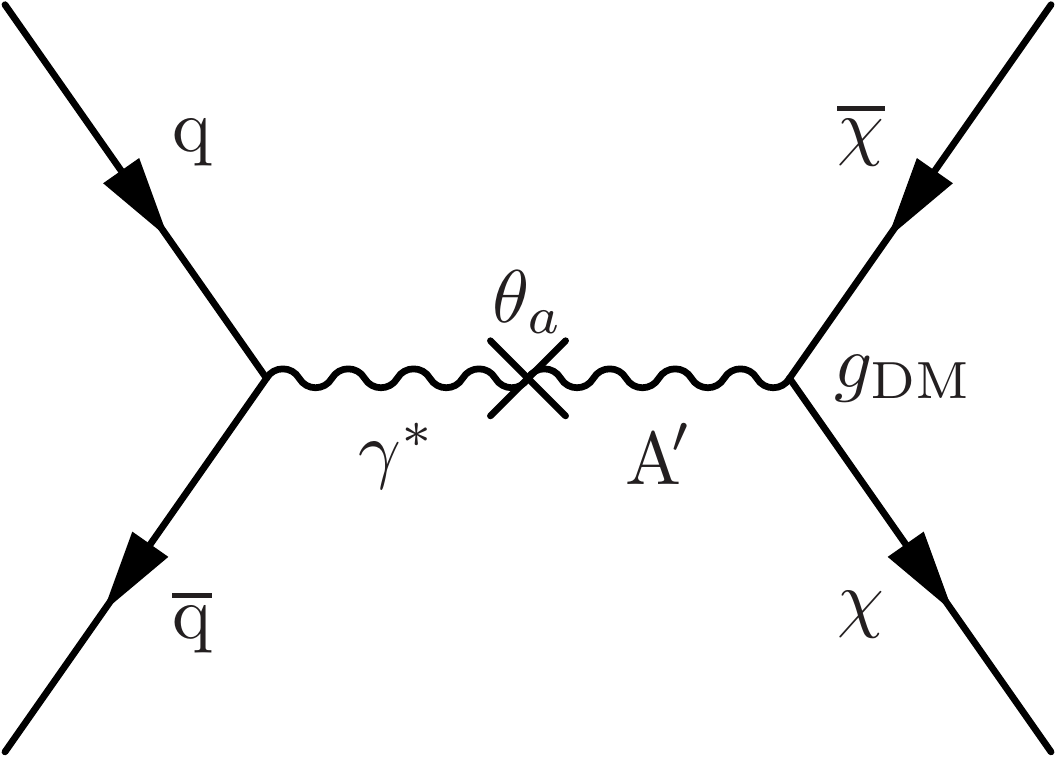}\\[2ex]
    \includegraphics[width=0.33\textwidth]{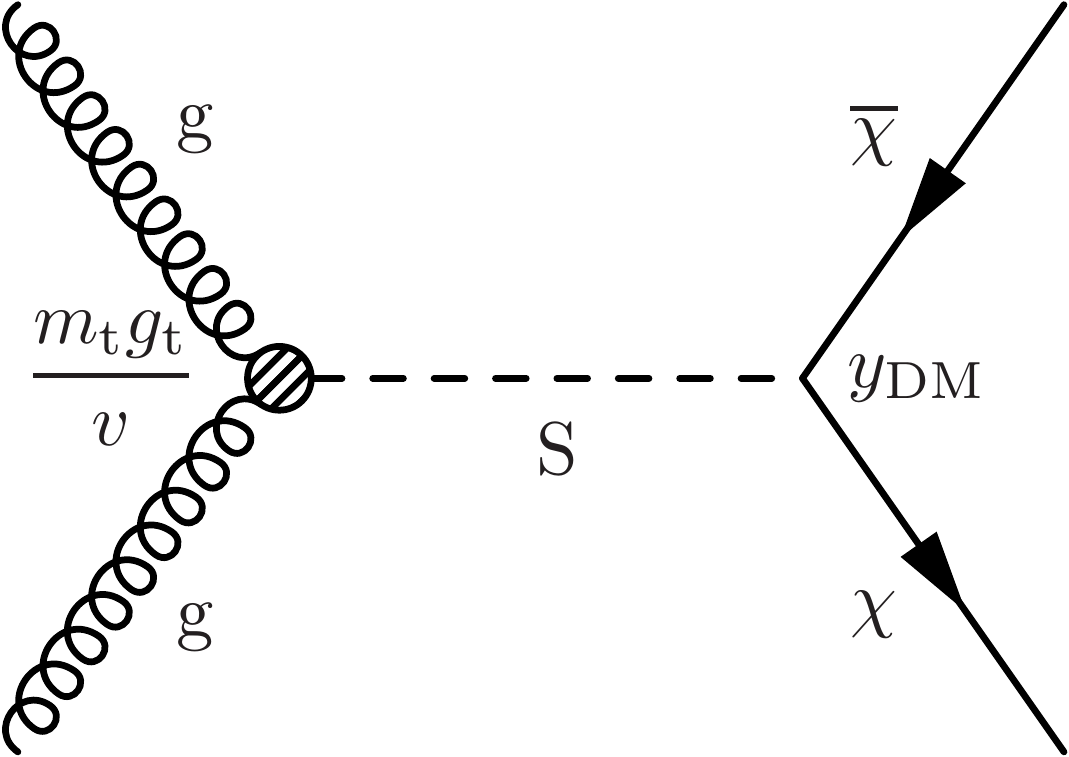}\hspace{2em}
    \includegraphics[width=0.33\textwidth]{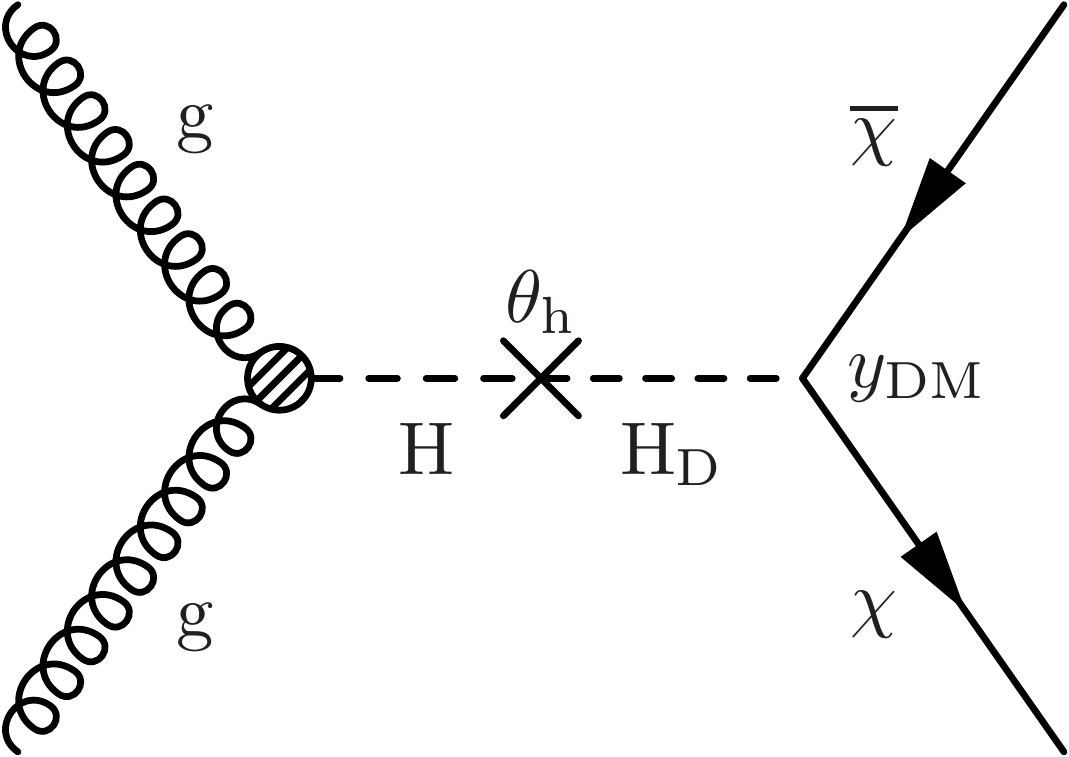}\\[2ex]
    \includegraphics[width=0.33\textwidth]{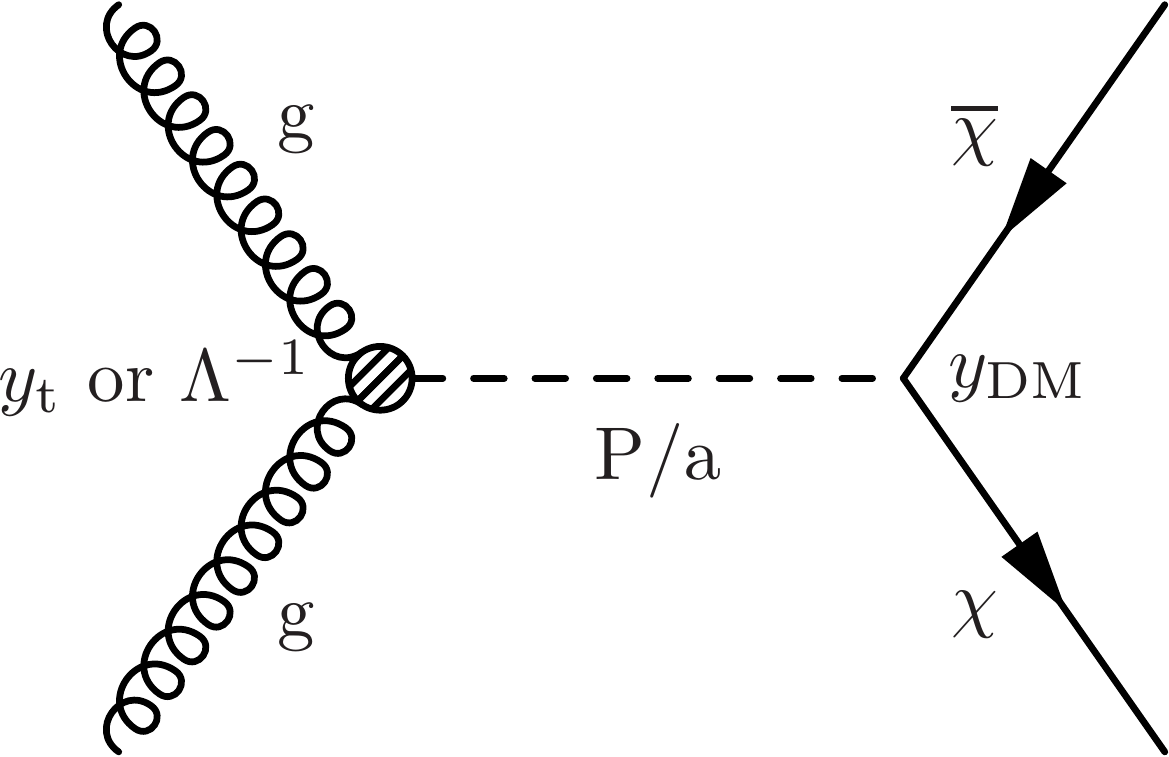}\hspace{2em}
    \includegraphics[width=0.33\textwidth]{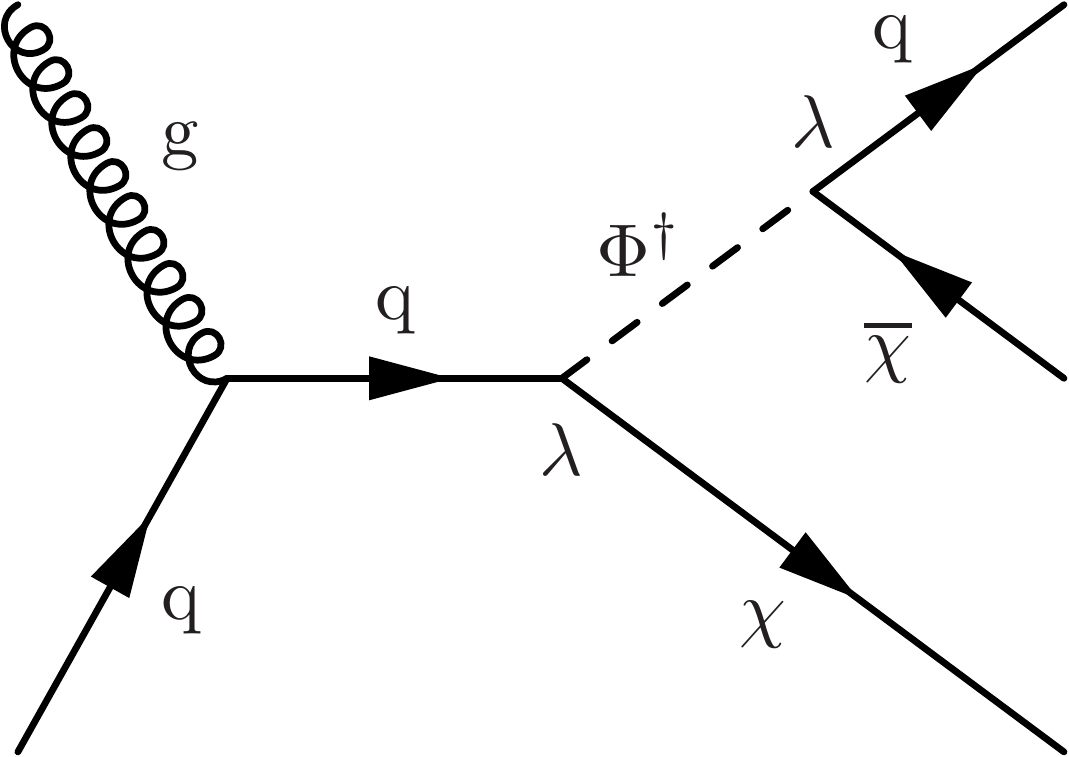}\\[2ex]
\caption{Representative Feynman diagrams for simplified model processes of DM pair production via different mediators.
First row \cmsLeft: \PZpr mediator, with \gq and \gDM couplings to the quarks and the DM candidate $\PDM$, respectively, discussed in Section~\ref{sec:zprimeportal}. In this diagram, we also show the initial-state radiation that is regularly used as an additional component in the searches.
First row \cmsRight: \PZpr mediator, with \gq couplings to the quarks, also discussed in Section~\ref{sec:zprimeportal}.
Second row: dark-photon mediator \PAprime, via mixing with the SM photon, discussed in Section~\ref{sec:darkphotonportal}.
Third row \cmsLeft: generic scalar mediator \PS, with Yukawa couplings $y_\PQq = m_\PQq g_\PQq/v$, and $y_\textrm{DM}$ and gluon coupling induced primarily via the top quark loop, 
discussed in Section~\ref{subsec:scalarportal}.
Third row \cmsRight: dark Higgs mediator \Hdark, produced via mixing \thetaH with the SM Higgs boson, discussed in Section~\ref{sec:darkhiggs}. As discussed in Section~\ref{sec:higgsportal}, the Higgs portal scenario can be seen as a subcase of the dark-Higgs portal.
Fourth row \cmsLeft: pseudoscalar and ALP (\PP/\Pa) mediators, either with Yukawa-like coupling $y_\PQt$ or effective coupling $\Lambda^{-1}$, as described in Sections~\ref{sec:pseudoscalarmodel} and \ref{subsec:ALPs}, respectively. 
Fourth row \cmsRight: the fermion portal via the bifundamental mediator $\Phi$, discussed in Section~\ref{sec:bifunportal}.}
    \label{fig:theoryDiagrams}
\end{figure}

In the following subsections, we present each model from the above list. Where required, we discuss the differences between the simplified and FIP versions of the models and how to reinterpret the bounds on these models. Figure~\ref{fig:theoryDiagrams} shows representative diagrams for each theoretical model addressed in this Report. Note that there are no diagrams shown for HNL models, as these models are the subject of their own Report~\cite{EXO-23-006}.

In order to provide constraints that are applicable to a wide range of scenarios, the analyses discussed in this Report are often interpreted using additional simplified models in which the branching fractions to exotic particles, LLP lifetimes, and final states are fixed independently of any theoretical or experimental constraints. This allows the results of these searches to be reinterpreted using both the models discussed below and extended DS models.

\subsubsection{Spin-1 portal}\label{sec:spinoneportal} 

This section discusses both commonly used spin-1 portal models, the \PZpr portal and the minimal dark photon models. In addition to presenting both models, we discuss how results can be re-interpreted between the two. In both cases, the couplings are assumed to be uniform with respect to flavor. Despite that, flavor-specific spin-1 mediators do exist in the literature. These include models that motivate an explanation for the observed deviations in the anomalous magnetic moment of the muon~\cite{Muong-2:2023cdq}. These models are typically reinterpreted from the flavor symmetric bounds and are not extensively discussed further (more details can be found in Refs.~\cite{HFLAV:2019otj,Bernlochner:2021vlv,BaBar:2013mob,LHCb:2017rln,Belle:2019rba,Altmannshofer:2019zhy,Greljo:2021npi,AlonsoAlvarez:2021ktn,Cen:2021iwv,Biswas:2021dan,Greljo:2022dwn}). 

\cmsParagraph{Vector and axial-vector portal\label{sec:zprimeportal}} A vector mediator arises from a broken $U(1)$ symmetry with couplings to both the SM and the DS. These couplings can be strictly vector or axial-vector in nature, and they are typically assumed to be universal for each type of matter particle. The interaction terms in the Lagrangian for a vector \PZpr boson are given by:
\begin{equation}
\mathcal{L}_{\text{vector}} \supset -\gDM \PZpr_{\mu} \PADM \gammamu \PDM - \gq \sum_{\PQq}{\PZpr_{\mu} \PAQq \gammamu \PQq} - \gell \sum_{\Pell}{\PZpr_{\mu} \Paell \gammamu \Pell},
\end{equation}
where \Pell are the leptons; $\PDM$ is the DM field; and \gDM, \gq, and \gell are the couplings of the \PZpr boson to DM, quarks, and leptons, respectively. The axial-vector mediator has the same terms with $\gammamu$ replaced by $\gammamu\gamma_{5}$.
Dilepton resonance searches (e.g.,~\cite{ATLAS:2019erb,CMS:EXO19019}) heavily constrain \PZpr mediators with lepton couplings, so the coupling \gell is often set to 0, leading to a leptophobic \PZpr boson. The LHC DM Working Group has established benchmark scenarios~\cite{Albert:2017onk} that were intended for the presentation of DM results in early LHC Run~2 and are used for the interpretations in this Report. In the leptophobic case, the benchmark quark coupling value is $\gq = 0.25$. This value is close to the SM quark coupling values for the \PZ boson, such that the \PZpr boson in this universal coupling model has a production rate and width similar to the \PZpr boson in the sequential SM~\cite{Altarelli:1989ff}. In the nonleptophobic case, the benchmark quark coupling is $\gq = 0.1$ and the lepton coupling may take values of 0.1 or 0.01. The former value for the lepton coupling is chosen to give similar sensitivity in dijet and dilepton final states, while the latter value is an example of the case with $\gell \ll \gq$, which can occur for a pure vector mediator. The axial-vector mediator requires an appreciable \gell for anomaly cancellation. In all cases discussed in this Report, the DM coupling is set to $\gDM = 1.0$.

\cmsParagraph{Dark-photon portal\label{sec:darkphotonportal}} A dark photon (\PAprime) is a spin-1 mediator with a pure vector coupling that mixes with the SM \PZ boson. The simplest form of the Lagrangian is written in terms of a mixing angle $\thetaa$. The mixing terms are:
\begin{equation}
\mathcal{L} \supset \gDM \cos{\thetaa}\PZ^{\prime}_{\mu}\PADM\gamma^{\mu}\PDM + \gDM \sin{\thetaa}\PZ_{\mu}\PADM\gamma^{\mu}\PDM.
\end{equation}
In this equation, the dark photon is represented as \PZpr to indicate that the mixing occurs after electroweak symmetry breaking, such that only massive eigenstates can mix.

The mixing gives rise to the production of invisible particles through DY decays into DM. This mixing is resonantly enhanced at the mass of the dark photon (\mAprime). Dark-sector particles may also be produced in the decays of vector mesons, and Dalitz decays of light mesons. A massless dark photon cannot be observed unless additional weakly charged DS particles, so-called millicharged particles, exist, because the effects of a massless dark photon would be indistinguishable from a redefinition of the electromagnetic fields~\cite{Berlin:2022hmt}. 

Dark sector bounds are often presented in terms of the mixing parameter $\epsilon$, defined by this approximate relationship:
\begin{equation}
\quad\tan{\thetaa}\approx\epsilon\frac{\tan{\thetaW}}{\DeltaZ-1},\label{eq:eps_mixing}
\end{equation}
where $\Delta_{\PZ} = (\mAprime/\mZ)^{2}$ and \thetaW is the weak mixing angle. A variety of SM couplings are assumed for the dark-photon model, and these are presented in Ref.~\cite{Ilten:2018crw}. These models have different lepton and baryon couplings, leading to different production modes. In all cases, to ensure sufficient DM annihilation in the early universe to achieve the observed thermal-relic density, a large DM coupling of $\aDark=\frac{\gDM^2}{4\pi}=0.5$ is assumed~\cite{CarrilloGonzalez:2021lxm}. This ensures a branching fraction to DM of nearly 100\% when the DM is lighter than $\mAprime/2$ and coupling-dependent SM decay modes when the DM is heavier. The signatures for \PZpr boson searches and dark photon searches largely mirror each other. Resonant mediator search bounds are driven by dilepton and dijet resonances, with dilepton searches typically being substantially more sensitive when lepton couplings are present. Additionally, when the DM is sufficiently light, the same mono-X searches used in \PZpr simplified model searches can be applied to dark photons.

When the mixing, $\epsilon$, is very weak, and the DM is sufficiently heavy, it is possible for the dark photon to be long lived, leading to displaced signatures. The proper decay length of the DS mediator can be written as:
\begin{equation}
  \ctau=\frac{1}{\Gamma} = \frac{3}{N_{\text{eff}} \mAprime\alpha\epsilon^{2}}\,,
\end{equation}
where $\alpha=e^2/4\pi$ and $N_{\text{eff}}$ is the effective number of particle species into which the dark photon can decay~\cite{Ilten:2015hya,Boveia:2016mrp}. The $N_{\text{eff}}$ varies with the mass of the dark photon, as more decays are kinematically allowed. This proper decay length equates to 80\micron for a dark-photon mass of 100\MeV, with $\epsilon=10^{-4}$. As a consequence, LLP searches are capable of excluding extremely small values of $\epsilon$ because their unique signature is associated with a small amount of SM background.

Finally, dark-photon models are often presented in the context of pseudo-Dirac DM particles~\cite{Hsieh:2007wq}. In pseudo-Dirac models, there is a mass splitting between the DM particles that renders DM DD searches insensitive by preventing elastic scattering~\cite{Tucker-Smith:2001myb,DeSimone:2010tf}. Furthermore, the sensitivity of ID searches is also suppressed to the point that they are found to be less sensitive than collider searches~\cite{Lanfranchi:2020crw}. Hence, these models are often presented solely in the context of collider searches and beam dump experiments. This type of model, also called inelastic DM, is discussed further in Section~\ref{sec:iDM}.

\cmsParagraph{Connecting dark photons and \texorpdfstring{\PZpr}{Z'} bosons} The bounds on dark-photon models and \PZpr models can be connected by noting that the signatures are the same, excluding the mixing with the SM $\PGg^{*}/\PZ^{*}$. As a result, we can approximately express $\epsilon$ in terms of \gq, starting from Eq.~\eqref{eq:eps_mixing} and using $\DeltaZ=(\mZprime/\mZ)^2$, where $\DeltaZ$ is small~\cite{Curtin:2014cca}:
\begin{equation}
  \gq = \epsilon e  \frac{1}{\cos^{2}\thetaW} \frac{1}{1-\DeltaZ}\left(Q\cos^{2}\thetaW+\Delta_{z}Y\right)\label{eq:eps_gq}
\end{equation}
where $Q$ is the charge operator yielding the quark charge and $Y$ is the hypercharge operator applied to the quarks. This formula can be estimated numerically in $\Pp\Pp$ collisions by summing the respective charges and hypercharges over the left and right helicities: 
\begin{equation}
  \gq \simeq \epsilon e \frac{1}{\cos^{2}\thetaW} \frac{1}{1-\DeltaZ}\left(\sqrt{\langle Q^2\rangle}\cos^{2}\thetaW+\Delta_{z}\sqrt{\langle Y^2\rangle}\right),
\end{equation}
where $\langle Q^2\rangle = 0.3$ and $\langle Y^2\rangle \approx 0.7$.
This approximation facilitates the recasting of simplified model results with roughly 20\% accuracy. This formula breaks down when the \PZpr boson mass is equal to the mass of the \PZ boson, in which case we refer to the bounds on the \PZ boson to invisible decay. The connection in Eq.~\eqref{eq:eps_gq} is valid only for the simplest dark-photon model. Other models may lead to slightly modified signatures, for instance, a $B-L$ model with couplings of neutrinos to the dark photon~\cite{Choi:2020dec}, where $B$ is baryon number and $L$ is lepton number. In that case, a connection is still possible but would require a more careful interpretation.

\subsubsection{Spin-0 portal}\label{sec:scalarportal}

The spin-0 portal consists of a scalar or pseudoscalar mediator that couples to DM. Like the spin-1 portal, both FIP models and simplified models can have a scalar mediator, and the FIP model deviates from the simplified model by assuming the scalar mixes with the Higgs boson. The addition of this mixing gives rise to the Higgs boson decay into invisible final states (\hinv) signature that is the cornerstone of many DM searches. The FIP terminology for this portal is ``dark Higgs'', whereas it is simply called ``scalar portal'' in simplified models.
Similarly, both FIP models and simplified models can have a pseudoscalar mediator, which is an ALP in the FIP case and directly related to the simplified pseudoscalar model as discussed below.

\cmsParagraph{Scalar portal\label{subsec:scalarportal}} The scalar portal assumes mass-dependent Yukawa couplings between the mediator \PS and the SM particles, in analogy with the SM. As a result, heavy-flavor-induced processes drive the production of the scalar mediator at the LHC. We can write the scalar Lagrangian as:
\begin{equation}
\mathcal{L} \supset \gq\frac{\PS}{\sqrt{2}}\sum_{\PQq}\yq\qqbar + \yDM \PS \PADM\PDM + \ldots,
\end{equation}
where \yDM is the DM Yukawa coupling, $\yq=\mq/v$ is proportional to the mass of the coupled particle, $v$ is the Higgs vacuum expectation value, and an additional scale factor, denoted here as \gq, is applied to allow for variations in the cross section. As a result of the Yukawa couplings, the production modes largely follow those of the Higgs boson. Gluon fusion through a top quark loop is the dominant production mode, followed by top quark-antiquark (\ttbar) associated production of the scalar. No vector boson couplings are assumed in this model, and therefore vector boson fusion and vector boson associated production are not possible. As with the spin-1 simplified model, no mixing is assumed.

Scalar models retain the same fermionic couplings and branching fractions as a fermiophilic Higgs boson. Moreover, more exotic variations of the scalar model exist that couple exclusively to a specific generation, such as the third or second generation, although we do not focus on these variations in this Report. When the DM is heavier than $\mS/2$, we find that top quark final states form the dominant decay mode for heavy scalars, with lighter scalar decays dominated by the heaviest fermion that is lighter than $\mS/2$, or photon decays through a heavy quark loop for very light scalars.

Bounds on scalar models from modern DD experiments probe a very different phase space. When recast into bounds on scalar models, the DD experiments tend to be comparable in sensitivity to collider bounds in models with similar coupling strength, as shown in Section~\ref{sec:reinterpretationAndResults}. Thermal-relic density constraints on scalar models are particularly restrictive because only models with very large couplings avoid an overabundance of DM. As a result, the remaining parameter space is found to not be compatible with thermal-relic constraints~\cite{Krnjaic:2015mbs}.

\cmsParagraph{Dark-Higgs boson portal\label{sec:darkhiggs}}  Dark-Higgs boson portal models include an additional scalar DM mediator that has the same properties as the Higgs boson except for the mass. This scalar mediator, the dark Higgs mediator (\Hdark), acts as the portal to the DS and mixes with the SM Higgs boson.

This mixing is introduced to the Lagrangian of the dark-Higgs model through a coupling between the dark Higgs boson and the SM Higgs boson. The extra terms in the Lagrangian are given by:
\begin{equation}
\mathcal{L} \supset -\yDM \Hdark\PADM\PDM + (\mu \Hdark + \lambda \Hdark^2)\PH^{\dagger}\PH.
\end{equation}
The $\mu$ interaction term yields dark matter signatures from decays of mixtures of the \PH and \Hdark fields. The SM Higgs boson can therefore decay into DS particles, and the dark Higgs boson can decay into SM particles. 

The mixing of the dark Higgs boson with the SM Higgs boson is typically written in terms of the eigenstates \PHone, \PHtwo and the angle \thetaH between them:
\begin{equation}
\begin{aligned}
&\mathcal{L} \supset &&\yDM \left(  \PHone \sin \thetaH +  \PHtwo \cos \thetaH \right)\PADM\PDM~+ \\
& &&\left(\PHone \cos \thetaH - \PHtwo \sin \thetaH \right) \left(2\frac{m_{\PW}^2}{v} \PW^{+}_{\mu}\PW^{-\mu} + \frac{\mZ^{2}}{v}\PZ_{\mu}\PZ^{\mu} - \sum_{\Pf} \frac{m_{\Pf}}{v} \overline{\Pf}\Pf\right) ,
\end{aligned}
\end{equation}
where \PHone is the SM Higgs boson, \PHtwo is aligned with the dark Higgs boson, and \Pf is an SM fermion. 

The mixing implies that the SM couplings of the Higgs boson deviate from their SM values by $\cos \thetaH$. Furthermore, it also introduces the possibility of decays into invisible particles, with a contribution to the Higgs boson width given by:
\begin{equation}
    \Gamma(\PHone\to\PADM\PDM) = \frac{\yDM^{2}m_{\PHone}\sin^{2}\thetaH}{8\pi} \left(1-\frac{4\mDM^{2}}{m^{2}_{\PHone}}\right)^{3/2}.
\end{equation}

Additionally, there is the possibility to produce the dark Higgs boson \Hdark through the same SM production modes as the Higgs boson. However, this production is suppressed by a factor of $\sin^{2} \thetaH$. Furthermore, in the case of a large DM coupling, the dark Higgs boson decays immediately into invisible particles, leading to a further increase in the overall measured cross section for \hinv. 

Constraints on the dark-Higgs model primarily come from constraints on the Higgs boson couplings and the bounds on \hinv. Higgs boson couplings are modified by the mixing angle $\cos \thetaH$. The \hinv bound is driven by the \hinv branching fraction $\propto\Gamma(\PHone)$, which is proportional to $\sin^{2}\thetaH$~\cite{FERBER2024104105}. This model has also been referred to as the singlet mixing model. When the dark Higgs boson is light, $\mHdark < 10\GeV$, it is possible for its production to exceed that of the SM Higgs boson, yielding additional constraints on this model from direct production~\cite{Albert:2016osu}. The final state for direct dark-Higgs boson production is the same as for \hinv. Lastly, in some extended models, the dark Higgs boson gives rise to the mass of the dark photon~\cite{PhysRevD.103.035027}. This mass generation mechanism leads to additional couplings with the possibility that the SM Higgs boson decays into dark photons through the mixing with the dark Higgs boson.

As with the scalar simplified model, DM DD and collider searches provide similar bounds on dark-Higgs boson models. Colliders are more sensitive than DD because of the additional vector boson fusion production modes. Thermal-relic density constraints are also very restrictive and the remaining parameter space from minimal dark-Higgs models is found to not be consistent with the thermal-relic density.

In addition to simplified models that introduce a single portal, it is possible to introduce multiple mediators. One example has both a spin-1 \PZpr boson and a dark Higgs boson, which we denote as the $\PZpr\Hdark$ model, with the dark Higgs providing a mechanism to give mass to the \PZpr boson~\cite{Autran:2015mfa,Duerr:2017uap}. These models largely follow the behavior of the simplified models, with searches possible for either of the mediators. Additionally, there is the possibility of new portal-portal interactions, such as dark-Higgs boson radiation.

\begin{figure}[htbp!]
\centering
\includegraphics[width=0.34\linewidth]{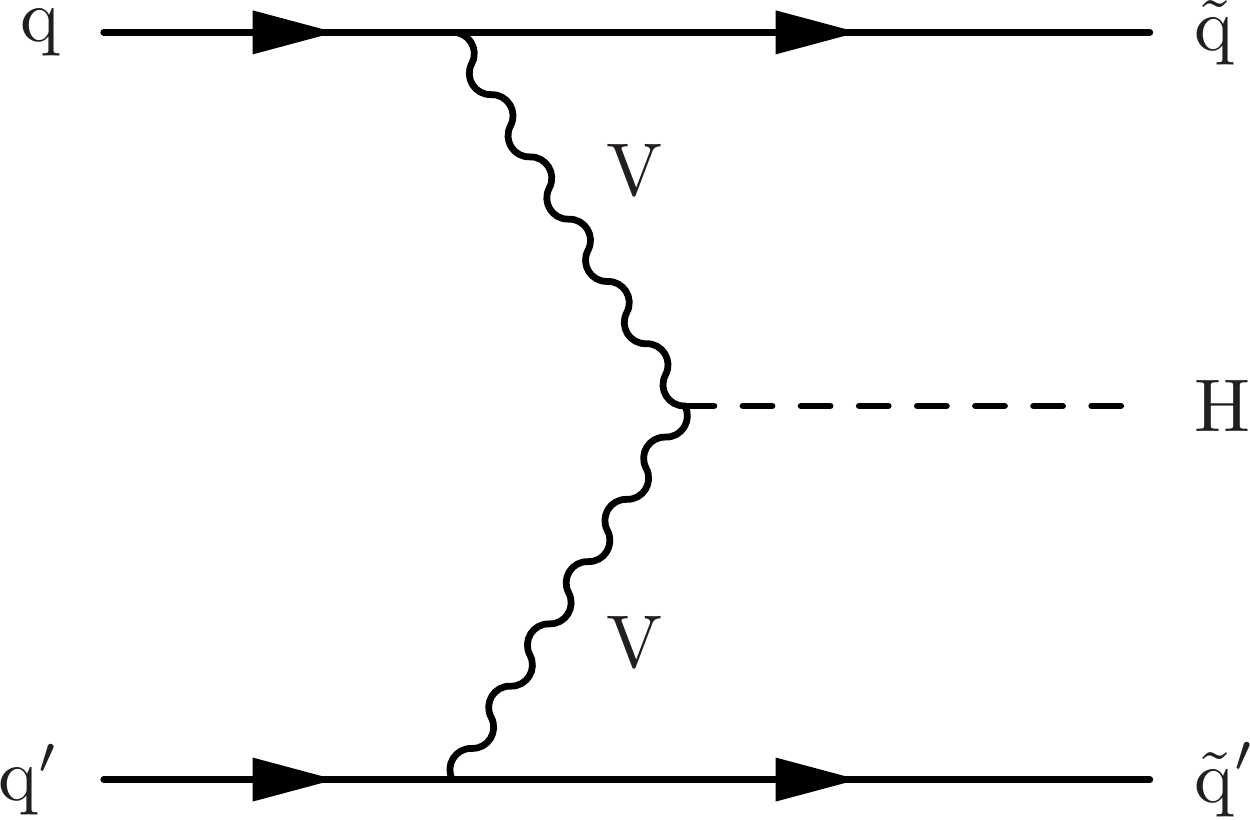}\hspace{2em}
\includegraphics[width=0.34\linewidth]{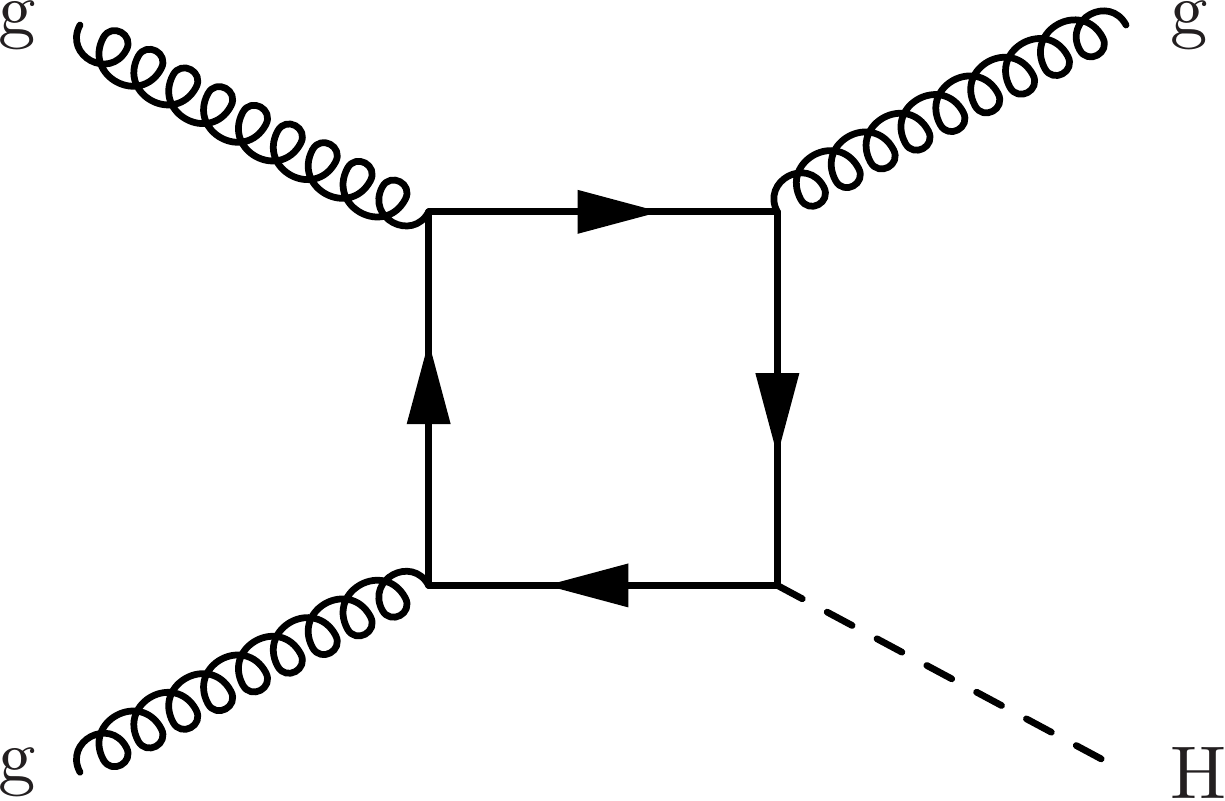}
\\[2ex]
\includegraphics[width=0.34\linewidth]{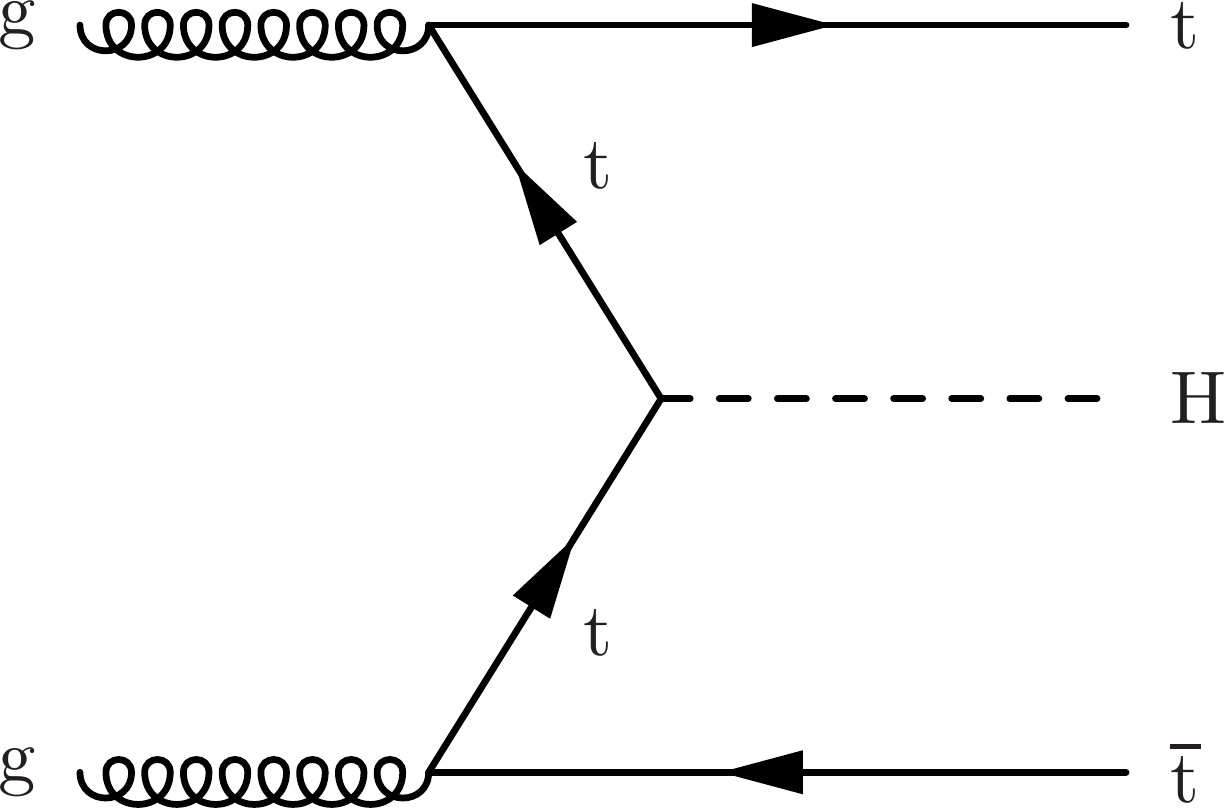}\hspace{2em}
\includegraphics[width=0.34\linewidth]{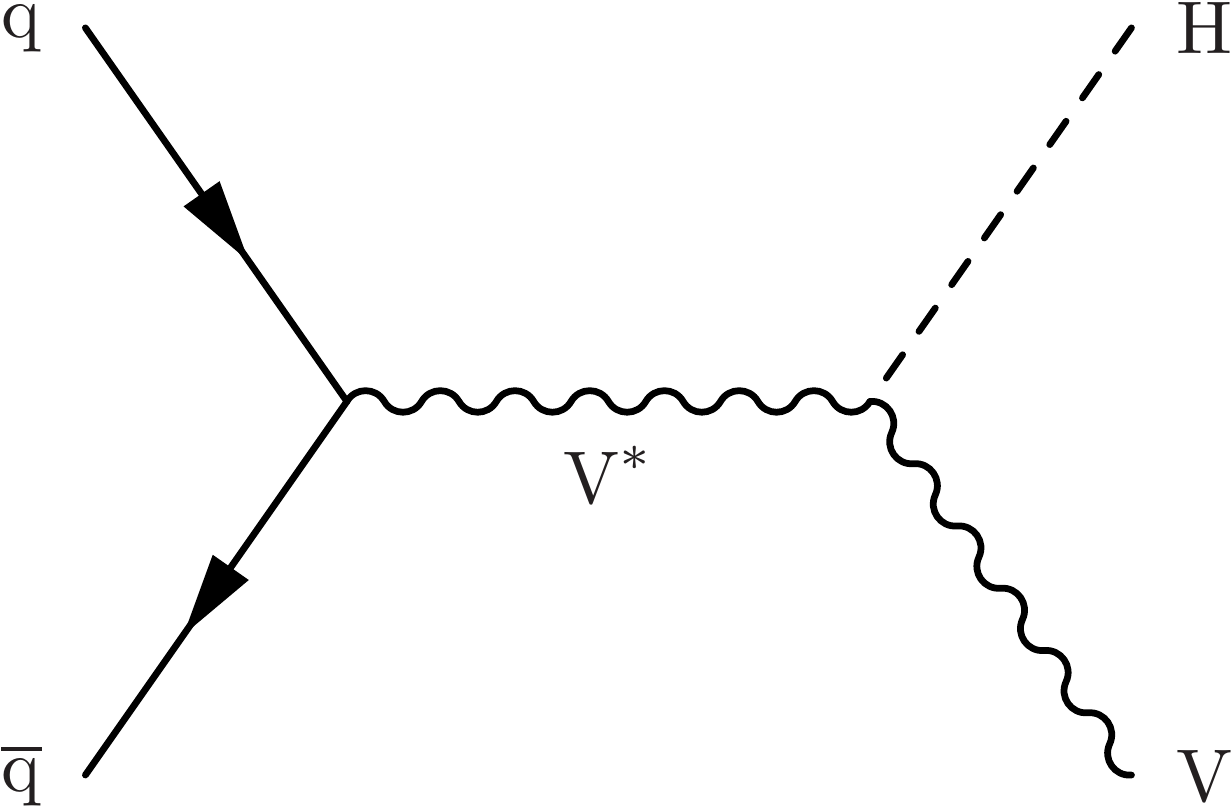}
\caption{Feynman diagrams of the \vbf, \ggH, \ttbarh, and \vh Higgs boson production modes analyzed in the \hinv searches.}
\label{fig:higgshinvsignatures}
\end{figure}

\cmsParagraph{Higgs boson portal\label{sec:higgsportal}} The Higgs boson portal is effectively a less parameterized version of the dark-Higgs boson model aimed at probing the \hinv final state. In other words, the Higgs boson portal model has fewer parameters than the Dark Higgs version because it does not assume the existence of a new mediator particle with its own new parameters. The SM Higgs boson branching fraction to invisible final states, \brhinv, is only about 0.1\%~\cite{Dittmaier:2011ti}, from the decay of the Higgs boson via $\PZ\PZ^* \to 4\PGn$. Several BSM scenarios predict much higher values of \brhinv~\cite{Belanger:2001am,Datta:2004jg,Dominici:2009pq,SHROCK1982250,Argyropoulos:2021sav}. In particular, in Higgs portal models, the Higgs boson acts as the mediator between SM particles and DM~\cite{Djouadi:2011aa,Baek:2012se,Djouadi:2012zc,Beniwal:2015sdl}, strongly enhancing \brhinv. Currently, the combination of Higgs boson precision measurements still allows for a branching fraction as large as 16\% for decays of the Higgs boson to undetectable particles~\cite{CMS:2022dwd}. This motivates the search for exotic decays of the Higgs boson to invisible or semivisible states, leading to signatures of imbalanced visible momenta. Searches for \hinv using Run~2 CMS data have been performed targeting various Higgs boson production modes: vector boson fusion (\vbf)~\cite{CMS:2016dhk,CMS:2022qva}, gluon-gluon fusion (\ggH)~\cite{CMS:2016dhk,CMS:2021far}, and in association with either a \ttbar quark pair (\ttbarh)~\cite{CMS:2019ysk,CMS:2020pyk,CMS:2021eha,CMS:2023sdw}, or a vector boson (\vh, where V stands for either a \PW or a \PZ boson)~\cite{CMS:2016dhk,CMS:2020ulv,CMS:2021far,CMS:2023sdw}, including both leptonic and hadronic final states of the latter two modes. The signatures for these Higgs boson production modes are illustrated in Fig.~\ref{fig:higgshinvsignatures}.
A semivisible decay of the Higgs boson could arise if a new unbroken $U(1)$ symmetry in the DS leads to an effective coupling of the Higgs boson to an SM photon and a stable, massless, dark photon~\cite{Curtin:2013fra,Djouadi:1997gw,Petersson:2012dp,Gabrielli:2013jka,Gabrielli:2014oya,Biswas:2016jsh,Biswas:2017anm}. As mentioned in Section~\ref{sec:darkphotonportal}, such semivisible decays would imply the existence of millicharged particles.

\cmsParagraph{Pseudoscalar portal\label{sec:pseudoscalarmodel}}  The pseudoscalar simplified model adds these terms to the Lagrangian:
\begin{equation}
\mathcal{L} \supset -i \gq\frac{\PP}{\sqrt{2}}\sum_{\PQq}{\yq\PQq\gamma^{5}\PAQq} -i \yDM \PP\PADM\gamma^{5}\PDM
\end{equation}
where \PP is the pseudoscalar mediator. The experimental searches cover a variety of final states and include both gluon fusion and $\ttbar\PP$ production modes, which tend to be the most important for spin-0 DM production. In one case, the search identifies a monojet plus \ptmiss signal, targeting the process where the mediator is radiated from top-quark loops. In the other case, the search relies on detecting the top-quark decay products that arise from the tree-level reaction \ttbar+\ptmiss. The coupling structure, combined with the initial state, leads to a larger cross section for pseudoscalar production via gluon fusion, compared to that of scalar production~\cite{Buckley:2014fba,Harris:2014hga}. On the other hand, the production cross section of \ttbar\PS is enhanced when compared to that of pseudoscalar production. Vector boson couplings to the scalar and pseudoscalar are also possible and most commonly added in dark-Higgs models, where a new scalar is introduced that mixes with the Higgs boson. Further details are discussed in Section~\ref{sec:signatures}.

Despite the pseudoscalar and scalar having similar sensitivities at the LHC, the sensitivity to these models at other experiments differs by a large amount. On the one hand, the DD of pseudoscalar mediator depends on the square of the velocity of the DM, which significantly suppresses sensitivity, making the predicted cross sections unobtainable with DD technology (though some studies suggest a cancellation of the velocity suppression through cross section enhancements at the nucleon level~\cite{Dienes:2013xya}). On the other hand, ID bounds have the opposite behavior, with sensitivity to pseudoscalar-mediated DM annihilation being enhanced.

Finally, the pseudoscalar portal also occurs in an extended DS scenario, denoted the 2HDM+a framework. This is an ultraviolet-complete, renormalizable DM model, which is discussed further in Section~\ref{sec:2HDM}.

\cmsParagraph{Axion-like particle portal}\label{subsec:ALPs} The ALP portal is of particular interest as ALPs are motivated by and could play a role in the strong charge conjugation parity (\CP) puzzle from quantum chromodynamics (QCD)~\cite{Agrawal:2017ksf,Hook:2019qoh,Gherghetta:2020keg}, but they are more generally motivated in DSs as pseudo-Goldstone bosons and so are naturally light and feebly coupled. The ALP portal differs from the pseudoscalar simplified model by assuming an effective coupling from a Yukawa interaction. This replaces the top quark loop interactions in the pseudoscalar simplified model, allowing all kinds of new physics, including heavy particles, in the loop. The Lagrangian terms for photon and gluon interactions with the ALP portal are~\cite{Jaeckel:2010ni,Izaguirre:2016dfi,Aloni:2018vki,Beacham:2019nyx}:
\begin{align}
\mathcal{L}_{\PGg} &\supset \frac{c_{\PGg}}{4\Lambda} \Pa F_{\mu\nu}\widetilde{F}^{\mu\nu}, \\
\mathcal{L}_{\Pg} &\supset \frac{4\pi\alpS c_{\Pg}}{\Lambda} \Pa G_{i,\mu\nu}\widetilde{G}_{i}^{\mu\nu},
\end{align}
where $F_{\mu\nu}$ and $G_{i,\mu\nu}$ are the photon and strong force field strength tensors; $c_{\PGg}$ and $c_{\Pg}$ are the dimensionless vertex coupling constants between the ALP and the EW and strong fields, respectively; $\Lambda$ is the cutoff scale; and $\alpS$ is the strong coupling constant, following the conventions of Ref.~\cite{Beacham:2019nyx} for the photon coupling and those of Ref.~\cite{Aloni:2018vki} for the gluon coupling. As a consequence, the ALP (\Pa) interaction is related to the pseudoscalar coupling \gq~\cite{Boveia:2022jox,Chetyrkin:1998mw} as:
\begin{equation}
\frac{c_{\Pg}}{\Lambda} = \gq/(32 \pi^2 v).
\end{equation}
Moreover, the translation between the pseudoscalar portal and the ALP portal is direct under these assumptions, since there are no additional interference terms or other changes in the behavior of the ALP particle, beyond the couplings that may be turned on or off. The ALP models are often well motivated because they offer a solution to the strong \CP problem~\cite{Peccei:1988ci}. There are no strong cosmological constraints on the existence of heavy ALPs that could mediate DS interactions. 

\subsubsection{Fermion portal} \label{sec:bifunportal}
An alternative mediator \Pbifun possesses couplings between SM fermions and DS particles~\cite{Bai:2013iqa,DiFranzo:2013vra}:
\begin{equation}
\mathcal{L}_{\Pbifun} = \sum_{i, j}{\lambda_{ij} \PADM_{i} \Pbifun_{ij} \Pf_{j}} ,
\end{equation}
where $i$ is the DM flavor index, $j$ is the SM flavor index, and $\lambda_{ij}$ are the corresponding Yukawa couplings. The mediator is called leptophobic if all leptonic couplings are set to zero, as in the \PZpr boson case (Section~\ref{sec:zprimeportal}). In order to conserve the quantum numbers of the interacting DM and SM particles, the mediator must be part of the fundamental representation for all gauge groups under which they are charged. Hence, assuming at least one gauge group for each type of particle and that \Pbifun carries its own non-Abelian gauge charges, \Pbifun may be described as bifundamental~\cite{Cohen:2017pzm}. It is assumed to be a complex scalar particle.

\begin{figure}
    \centering
    \includegraphics[width=0.33\linewidth]{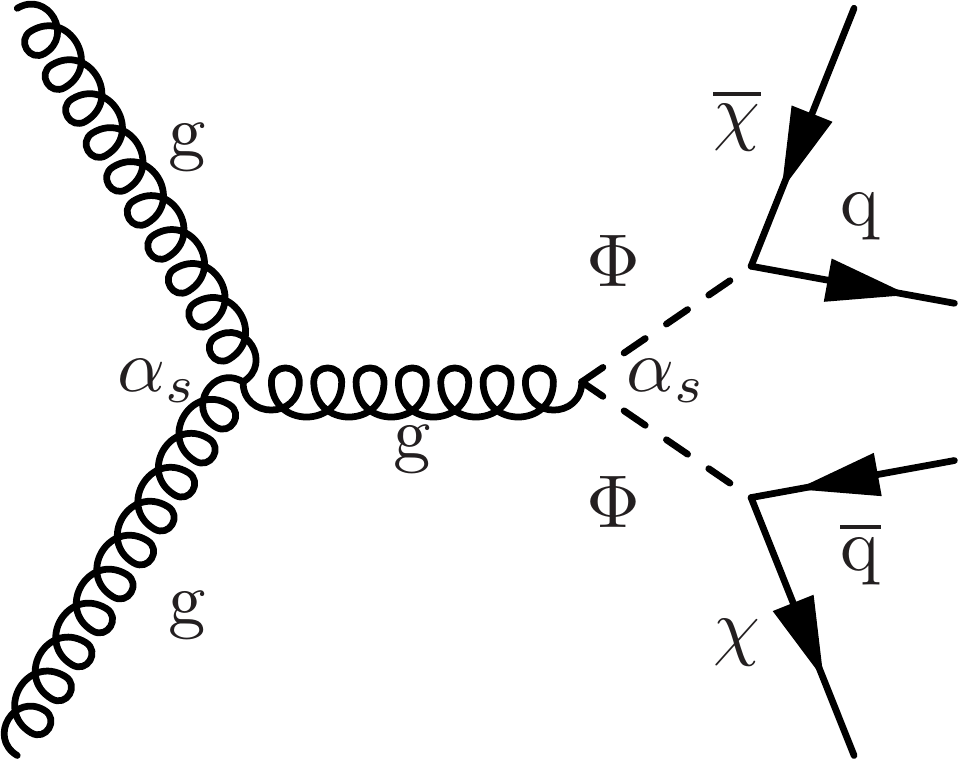}\hspace{2em}
    \includegraphics[width=0.33\linewidth]{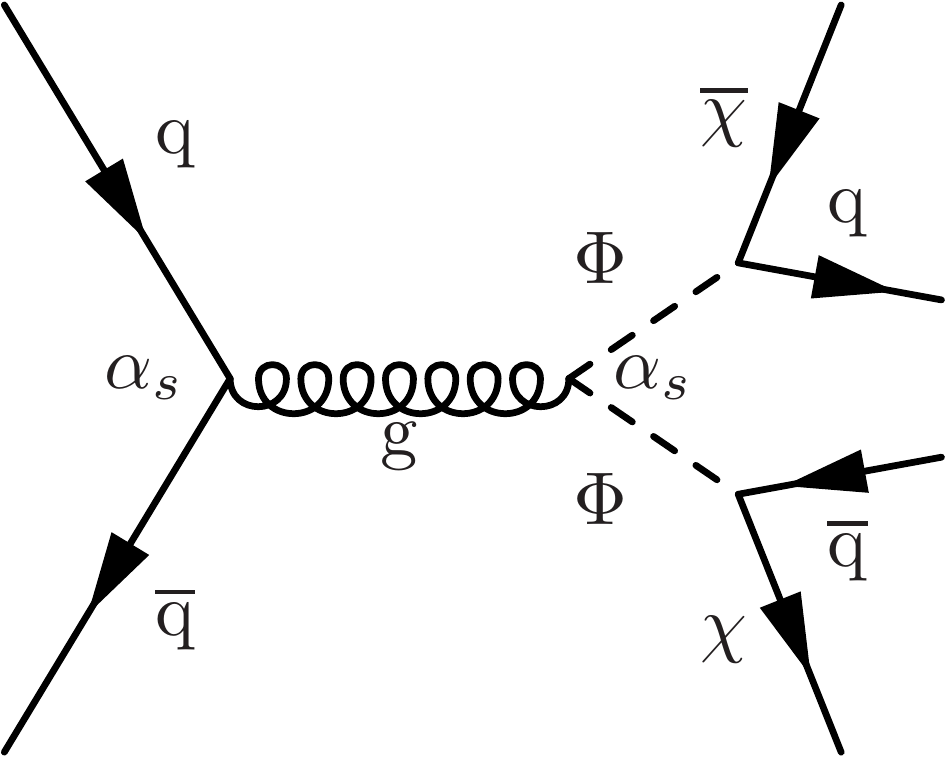}\\
    \vspace{2ex}
    \includegraphics[width=0.33\linewidth]{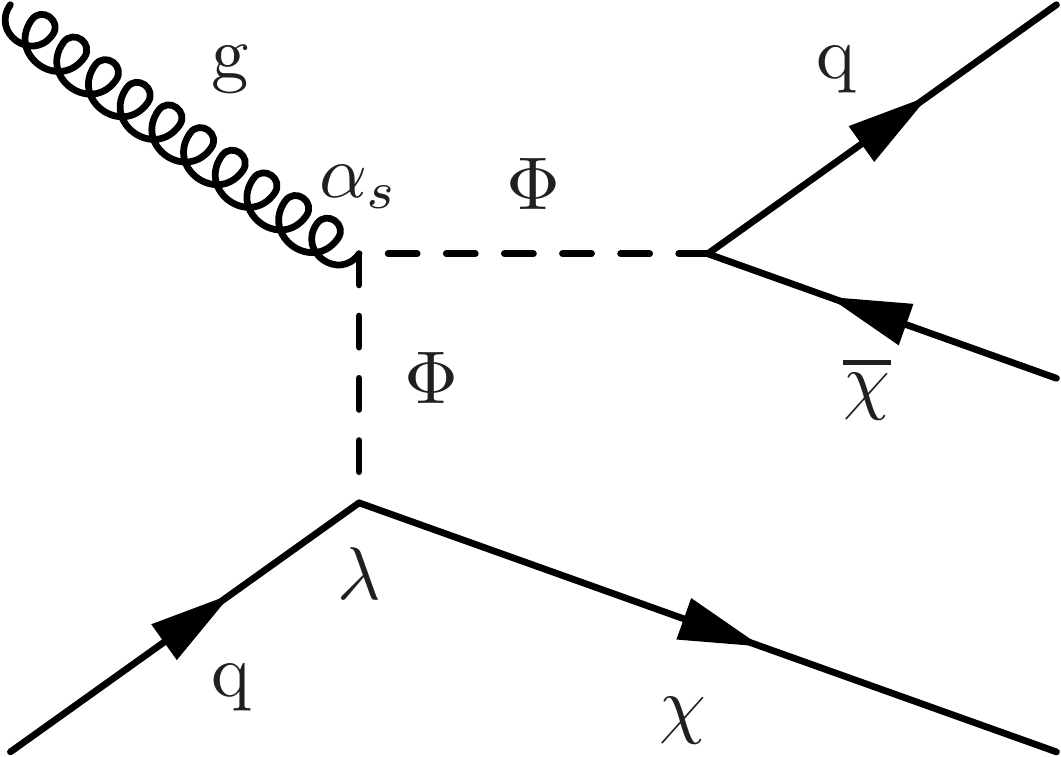}\hspace{2em}
    \includegraphics[width=0.33\linewidth]{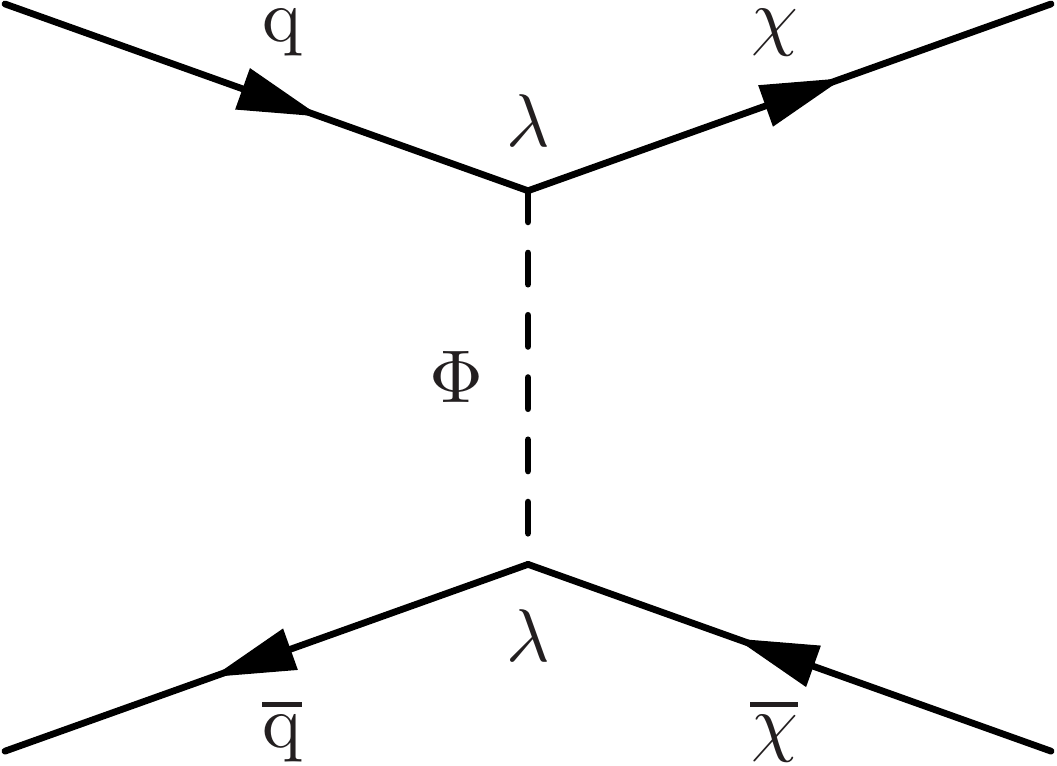}
    \caption{Feynman diagrams for production channels involving the bifundamental mediator \Pbifun: pair production via gluon-gluon fusion (upper \cmsLeft), pair production via quark-antiquark annihilation (upper \cmsRight), single production in association with a DM particle \PDM (lower \cmsLeft), and $t$-channel nonresonant DM production (lower \cmsRight).}
    \label{fig:tchan_diagrams}
\end{figure}

When \Pbifun has non-zero quark couplings, it is strongly charged and therefore may be produced in pairs via gluon-gluon fusion or quark-antiquark annihilation. Other possible processes involving this mediator include single production in association with a DM particle via quark-gluon scattering or virtual mediation of $t$-channel nonresonant DM production. The representative diagrams for these processes are shown in Fig.~\ref{fig:tchan_diagrams}. The pair production cross section is dominated by the gluon-gluon fusion process, which is independent of $\lambda_{ij}$.

\subsubsection{Neutrino portal}\label{sec:neutrinoportal}

The neutrino portal, often referred to as the HNL model~\cite{Abdullahi:2022jlv,Dasgupta:2021ies}, consists of the addition of right-handed neutrinos, either paired with the three existing neutrino flavors or with a fourth, sterile neutrino. The HNL model is capable of producing the observed DM thermal-relic density when the HNL masses and flavors are adjusted appropriately. While HNLs play a role as one of the DS portals, we do not further discuss HNL models in this Report. A detailed review of the CMS searches for such models is provided in Ref.~\cite{EXO-23-006}.

\subsection{Extended dark sectors}\label{sec:models}

Many models with complex dynamics in the DS have been theorized. They potentially communicate with the SM through any of the portals described above. Extended models of DM typically incorporate more than a single particle species in the DS, in contrast to, for example, minimal models that feature WIMPs. The additional states can give rise to enriched dynamics in the DS, with potential relevant experimental footprints in $\Pp\Pp$ collision events. Specific cases motivating CMS searches are highlighted in the following sections. Some of the models are selected based on theoretical interest, in order to explore the implications of well-motivated extensions to simplified DSs. Other models are selected primarily on the basis of their novel phenomenology, such that they are not covered by mono-X searches targeting simplified DSs.

\subsubsection{A 2HDM-type complete model: 2HDM+a}\label{sec:2HDM}

A class of ultraviolet-complete models has been developed with a focus on DM. One of those models is an extension of the existing two-Higgs-doublet models (2HDM)~\cite{Branco:2011iw,Gunion:2002zf}, which adds an additional spin-0 (pseudoscalar) mediator along with a DM particle candidate. Thus, it is described as the 2HDM plus a pseudoscalar (2HDM+a). The Lagrangian of such a model is described in Ref.~\cite{Bauer:2017ota}.

The interaction between the DM candidates and the SM particles is achieved by incorporating interaction terms between the 2HDM Higgs doublet fields ($\Ph_{1,2}$) and the newly introduced pseudoscalar mediator field (\PP). This interaction generates a mixing between the \CP-odd pseudoscalar mediator and the particles present in the 2HDM, which in turn allows for SM interactions. This yields a nondiagonal mass matrix, of which one mass eigenstate corresponds to the mediator (\Pa), while the other eigenstates correspond to the \CP-odd Higgs boson (\PA) and the other 2HDM fields (\PH, \Ph, $\PH^{\pm}$). The latter fields also acquire couplings of different kinds with the mediator via the trilinear and quartic couplings introduced in the scalar potential. We follow the convention that the heaviest neutral Higgs boson in a particular model is represented by \PH and other neutral scalar bosons, if any, are represented by \Ph. The DM particle nature is characterized by the Dirac fermion field \PDM, which couples to the two \CP-odd states and whose respective coupling strengths are controlled by the mixing angle of the \CP-odd sector. The Yukawa sector is taken to be the same as in the usual 2HDMs, where the structure is selected to avoid the appearance of flavor changing neutral currents. This often results in four possible configurations in terms of scalar and fermions couplings, labeled as scenarios of type: I, II, III, and IV. For the purpose of this publication, we focus our attention on the type-II scenario, where there is a differentiated interaction between the scalars and fermions for up-type and down-type quarks~\cite{Branco:2011iw}.

\begin{figure}
    \centering
    \includegraphics[width=0.3\linewidth]{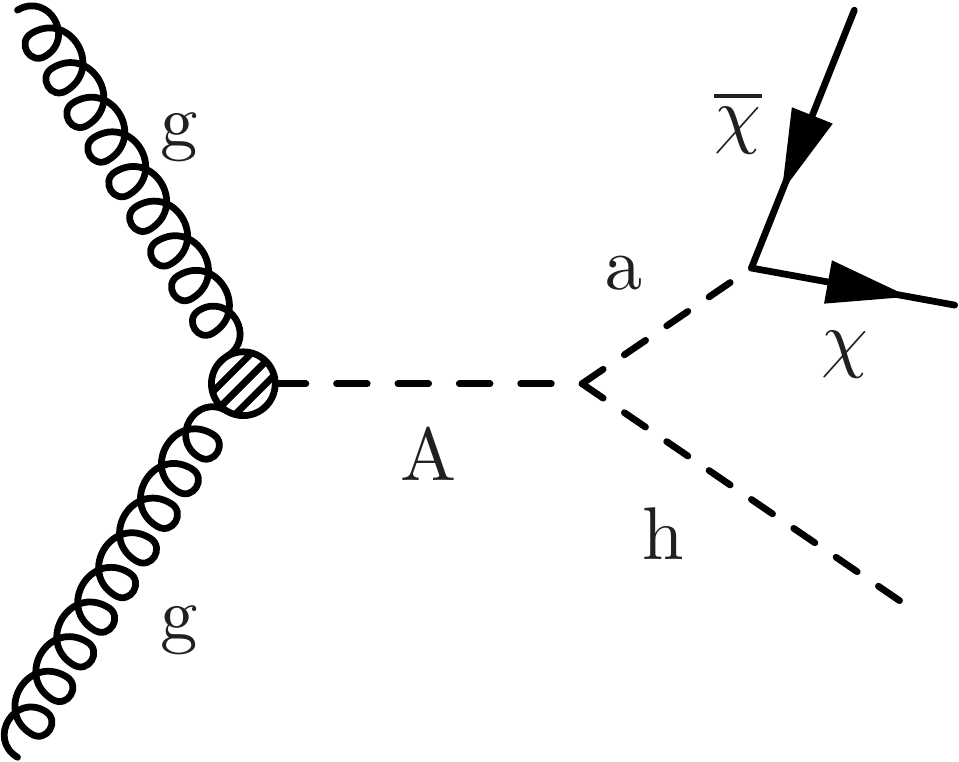}\hspace{2em}
    \includegraphics[width=0.28\linewidth]{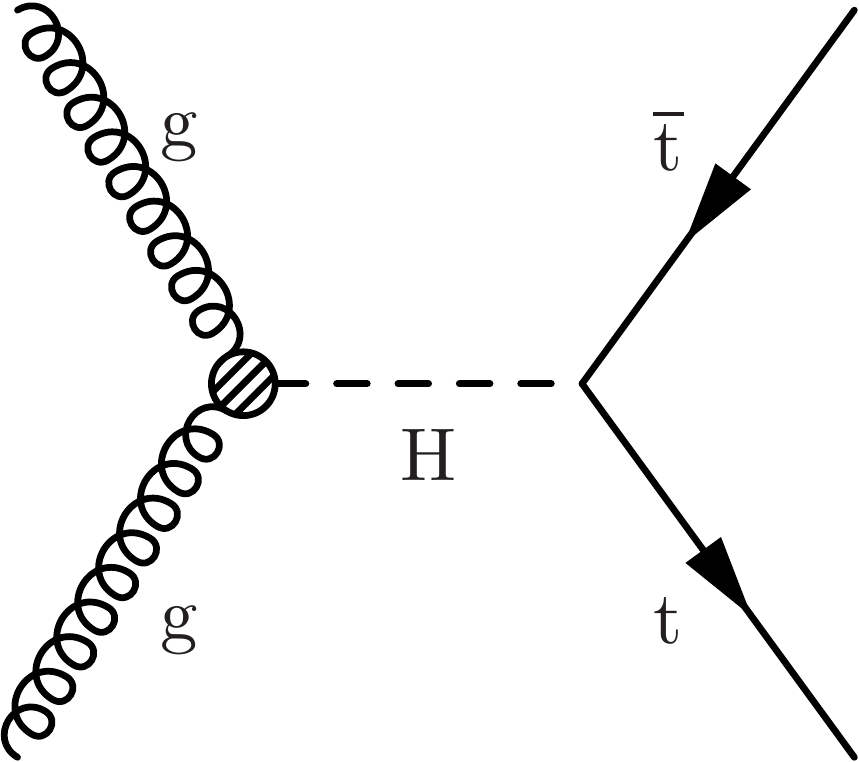}\hspace{2em}
    \includegraphics[width=0.3\linewidth]{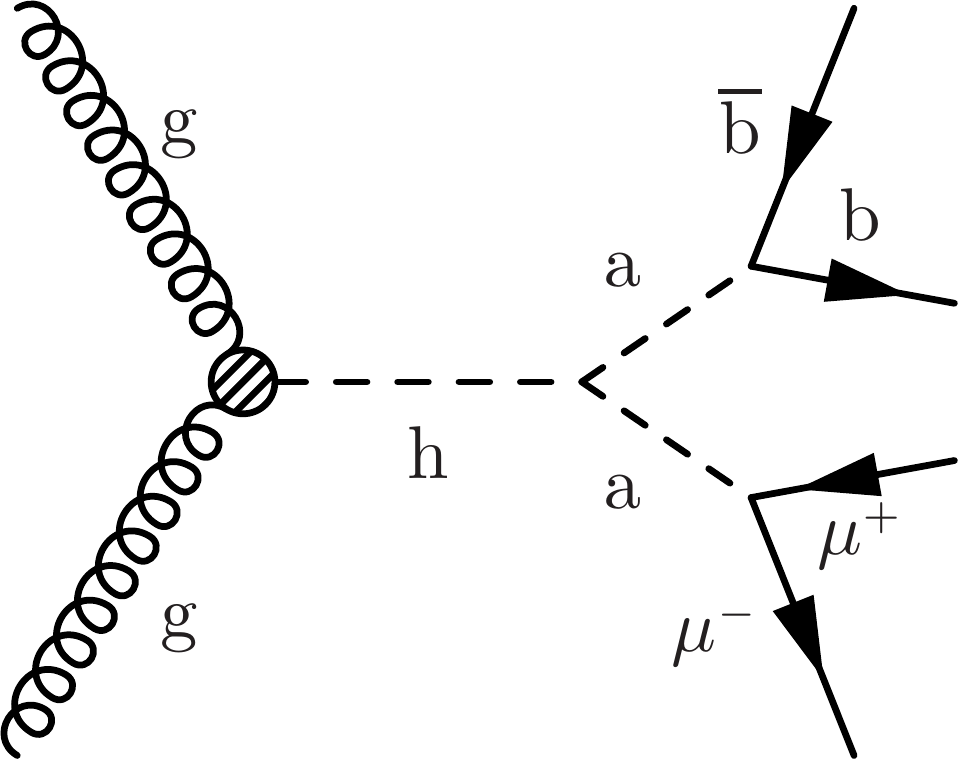}
    \caption{Feynman diagrams for 2HDM+a signatures.
    Left: a mono-\PH signature, mediated by the heavy pseudoscalar $\PA$.
    Center: $\ttbar$ resonant production, mediated by the heavy scalar $\PH$.
    Similar processes involve $\PH^\pm$ particles, \eg $\PH^\pm \to \PQt\PQb$.
    Right: exotic decay of the SM-like Higgs boson \Ph.
    }
    \label{fig:2hdma_diagrams}
\end{figure}

After electroweak symmetry breaking, the dynamics is determined by 14 parameters: $v$, $m_{\Ph}$, $m_{\PH}$, $m_{\PA}$, $m_{\PH^{\pm}}$, $m_{\Pa}$, \mDM, $\cos(\beta-\alpha)$, $\tan\beta$, $\sin\theta$, \yDM, $\lambda_{3}$, $\lambda_{\PP_{1}}$, $\lambda_{\PP_{2}}$. This number is typically reduced when the existing theoretical and experimental constraints are imposed on the model. The usual theoretical constraints resulting from the unitarity and perturbativity considerations apply. Among the experimental constraints are the measurements of the properties of the SM Higgs boson, and EW precision and flavor physics observables. A more detailed description of the list of the constraints restricting the parameter phase space is given in Ref.~\cite{LHCDarkMatterWorkingGroup:2018ufk}, where the following benchmark parameter choices are motivated, which will be used for most results in this Report:
\begin{equation}
\begin{split}
    & m_{\PH} = m_{\PA} = m_{\PH^{\pm}}, \hspace{0.2cm} \mDM=10\GeV, \\
    & \cos(\beta-\alpha)=0, \hspace{0.2cm} \tan\beta = 1, \hspace{0.2cm} \sin\theta = 0.35, \\
    & \lambda_{3}=\lambda_{\PP_{1}}=\lambda_{\PP_{2}}=3, \hspace{0.2cm} \yDM = 1.
\end{split}
\label{Sec2:Eq:BenchmarkChoice}
\end{equation}
\begin{sloppypar}Given the above-mentioned items and the natural complexity presented by the 2HDM+a framework, there exists very rich phenomenology in both the Higgs and dark sectors. A large number of signatures can be naturally produced in the 2HDM+a~\cite{LHCDarkMatterWorkingGroup:2018ufk}. In the context of DM, the list includes resonant mono-X production with an SM particle recoiling against DM particles, nonresonant production of SM particles accompanied by DM, and many others. An example Feynman diagram is shown in the \cmsLeft panel of Fig.~\ref{fig:2hdma_diagrams}. Being a 2HDM-type, conventional signatures of heavy resonances decaying into SM particles can be copiously produced in the 2HDM+a context. Decays of the neutral scalar states to a pair of top quarks become important when their mass is above the $\ttbar\ttbar$ threshold, which can produce signatures containing either two or four top quarks if one considers the gluon-gluon fusion and the \ttbar-associated production modes of the resonance, as shown in the middle panel of Fig.~\ref{fig:2hdma_diagrams}. Other cases such as $\PA/\Pa \to \PGtp\PGtm$, though still present, have reduced production rates in this model because of the heavy competition with the dark channel $\PA/\Pa \to \PDM\PADM$. In that sense, signatures involving the decay of the SM-like Higgs boson (\Ph) are more diversified in this model. For very low $m_{\Pa}$, the exotic decay $\Ph \to \Pa\Pa$ is possible, which can involve invisible, semivisible, and visible final states; an example diagram is shown in the \cmsRight panel of Fig.~\ref{fig:2hdma_diagrams}. A more comprehensive discussion of the various decay channels, covering not only the neutral scalar sector but also the charged resonances, is given in Ref.~\cite{Bauer:2017ota}.\end{sloppypar}

\subsubsection{Hidden Abelian Higgs model (HAHM)\label{sec:HAHM}} The hidden Abelian Higgs model (HAHM) is an extension of the SM based on the group $G=SU(3)_{C}\times SU(2)_{L} \times U(1)_{Y} \times U(1)_{\PDM}$. The extra $U(1)_{\PDM}$ gauge group is added to the SM. The new dark Higgs boson \Hdark and the neutral gauge boson (the dark photon \PAprime) are allowed to mix with the corresponding SM fields~\cite{Wells:2008xg}. The mixing of the SM and dark Higgs bosons (\PH--\Hdark) is via a parameter $\kappa$, while the dark photon mixes through the hypercharge portal with the SM photon and the \PZ boson via the kinetic mixing parameter $\epsilon$. When the dark photon mass is ${<}12\GeV$, its dominant coupling mode is to electrically charged SM particles. In the HAHM model, the production of the dark photon is via the decay of the Higgs boson ($\PH \to \PAprime\PAprime$ or $\PH \to \PZ \PAprime$) or via the DY process ($\qqbar \to \PAprime$). The lifetime of the dark photon depends on $\epsilon$~\cite{EXO-21-006}.

\subsubsection{Supersymmetry}\label{sec:theory_susy}

Many SUSY models predict a lightest supersymmetric particle that is a good candidate for DM~\cite{Jungman:1995df,Alimena:2019zri,Liu:2015bma} and that has an abundance that naturally agrees with the observed thermal-relic density~\cite{Jungman:1995df}. These SUSY models include the minimal supersymmetric SM~\cite{Nilles:1983ge,Barbieri:1987xf,Haber:1984rc,Gunion:1988yc}, gauge-mediated SUSY breaking (GMSB)~\cite{Dimopoulos:1996vz,Dimopoulos:1996gy}, and split SUSY~\cite{Giudice:2004tc}. In this Report, we will focus on SUSY models that include a dark sector, as described below.

\cmsParagraph{Dark supersymmetry\label{sec:DarkSUSY}} Dark matter is naturally embedded in extensions of the SM motivated by solving the hierarchy problem, particularly with low-energy SUSY~\cite{Nima_Arkani-Hamed_2008}. A hidden gauge symmetry $U(1)_{\mathrm{D}}$ is broken near the {\GeVns} scale, giving rise to new dark vector bosons. A completely generic prediction is that those new bosons can be produced in cascade decays of the minimal supersymmetric SM superpartners. The lightest {\GeVns}-scale dark Higgs bosons and gauge bosons eventually decay back into light SM states, and dominantly into leptons.  In this scenario, the lightest SUSY particle decays into the lightest particle in the DS, which escapes detection, plus a dark photon (\PAprime) that decays into leptons with a sizeable branching fraction.  The dark-photon decay can occur promptly or after traveling some distance producing a displaced vertex (DV). Regardless of the DVs, the lepton pairs will have a small mass O({\GeVns}), and in typical decays, will come out with small angular separation. Thus, one can produce ``lepton jets'', which are boosted groups of collimated leptons with small masses. The presence of lepton jets dramatically reduces backgrounds and probes direct EW production (\eg, slepton pair production, gaugino pair production) at higher masses. 

\cmsParagraph{Stealth supersymmetry\label{sec:theory_stealth}} Supersymmetric models with $R$-parity conservation often include a neutralino as the lightest SUSY particle, which makes it a candidate for WIMP DM. Searches at the LHC have placed strong constraints on these models, which has prompted interest in scenarios that could have evaded detection. One example of such a new scenario is the extension of the usual minimal supersymmetric SM particle content with a dark ``stealth'' sector~\cite{Fan:2011yu,Fan:2012jf,Fan:2015mxp}, containing in the minimal case a scalar singlet \PS and its fermionic superpartner the singlino \PSS. There are multiple options for communication between the SM and the stealth sector, including the Higgs portal via mixing and a new vector-like $SU(5)$ messenger.

In these models, the portal between the stealth sector and the SUSY breaking sector is suppressed, such that SUSY is approximately conserved and the \PS and \PSS are nearly mass degenerate. These stealth sector particles are not stable. Once produced, the singlino decays into the singlet and a stable DM particle. The stable DM particle is often assumed to be a gravitino (\sGra), but it could also be an axino. In both cases, the stable DM is typically assumed to be light in these models, of order 1\GeV. Depending on the size of the mass splitting and the involved couplings, the singlino can be long lived. If it is long lived on cosmological scales, it can be a viable DM candidate and results in co-decaying DM~\cite{Dror:2016rxc,Dery:2019jwf}, which is a mechanism for thermal DM freeze-out where degenerate particles in complex DSs and out-of-equilibrium decays can both decay to obtain the observed thermal-relic density. The singlet decay depends on the assumed portal between the stealth sector and the SM. In the case of the Higgs portal and singlet masses of order 100\GeV, the decay is predominantly to two bottom quarks, whereas in the case of the vector portal, the decay is predominantly to two gluons. 

\begin{figure*}[htb!]
\centering
\includegraphics[width=0.35\linewidth]{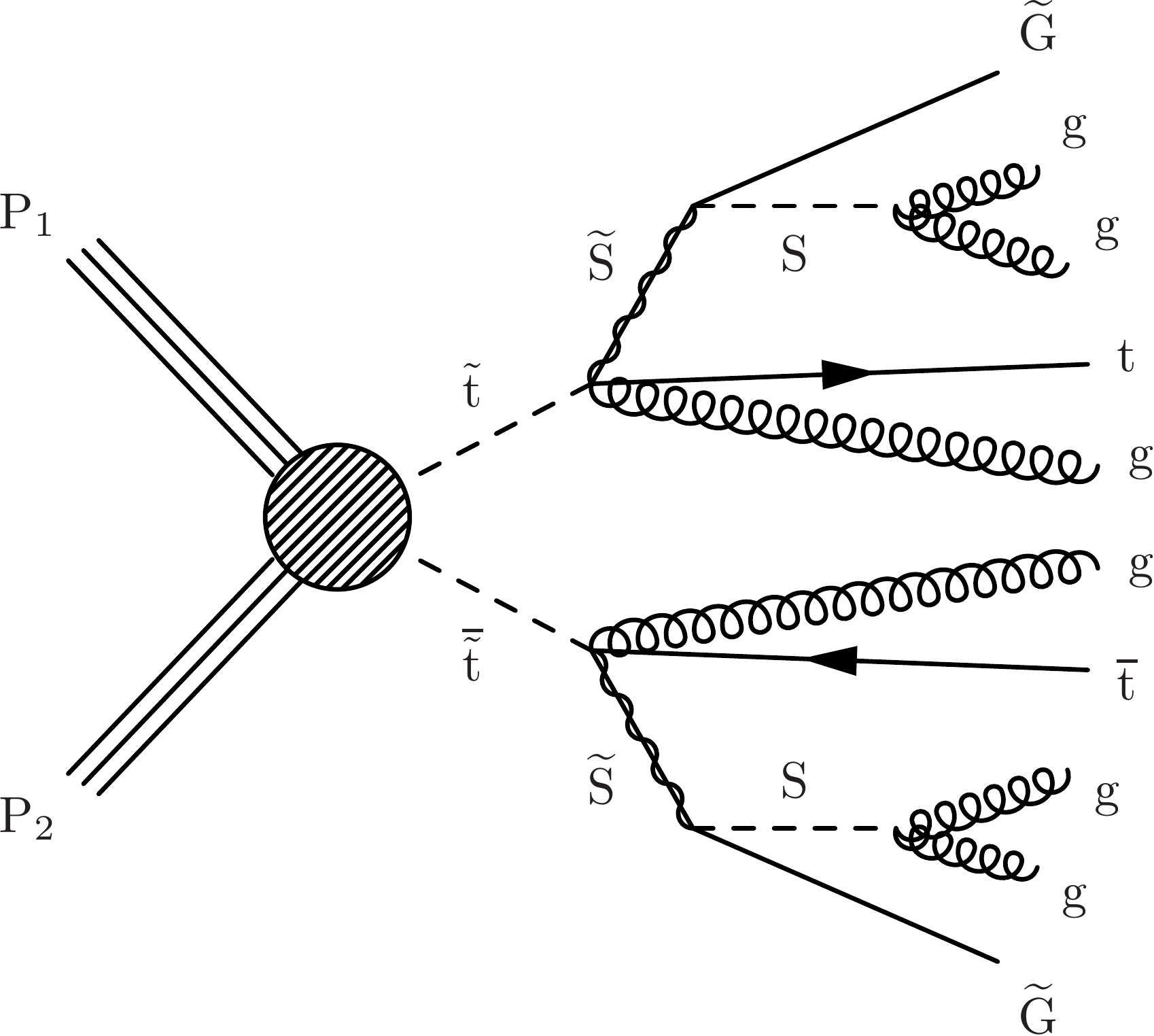}
\hspace{2em}
\includegraphics[width=0.35\linewidth]{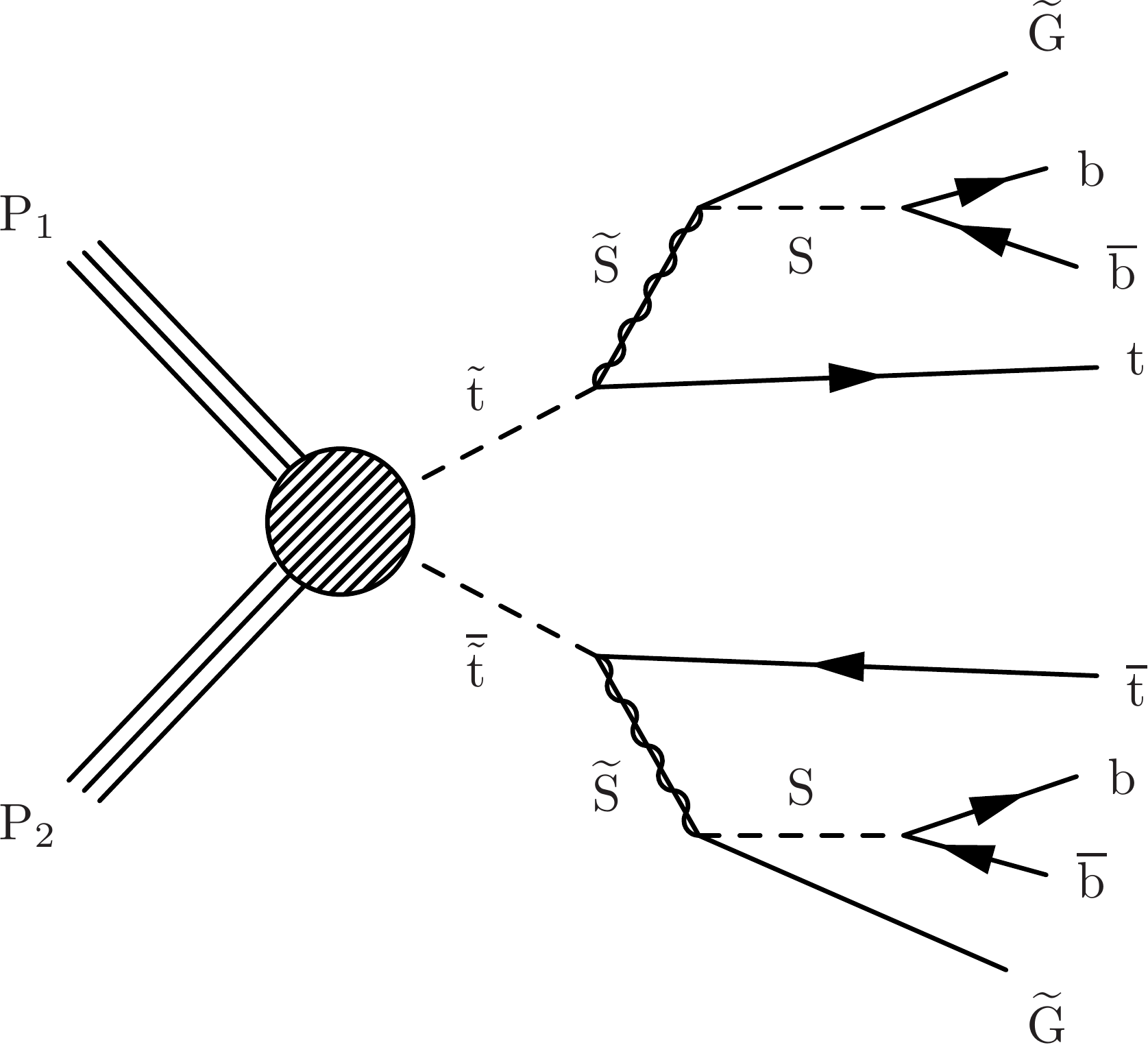}
\caption{Feynman diagrams for pair production of top squarks under the stealth SYY (\cmsLeft) and stealth SHH (\cmsRight) models. In these models, the signature is a pair of SM top quarks, with additional jets originating from gluons (SYY) or \PQb quarks (SHH).\label{fig:stealth_diagrams}}
\end{figure*}

At the LHC, the stealth sector particles are assumed to be produced in the decay of a SUSY particle, such as a squark. Between the many options for the production channel and possibilities for the interaction portal, the phenomenology of these models is varied. Importantly, the small assumed DM mass in combination with the small mass splitting between the singlet and singlino results in a common experimental signature with little to no \ptmiss. Searches for stealth SUSY are therefore highly complementary to traditional high-\ptmiss SUSY searches, which are reviewed in Refs.~\cite{Melzer-Pellmann:2014eta,Canepa:2019hph,Costanzo:2008zz,Pralavorio:2013yta}. Feynman diagrams for two stealth SUSY models are shown in Fig.~\ref{fig:stealth_diagrams}, where depending on the portal, additional gluons (stealth SYY) or \PQb quarks (stealth SHH) are produced in the final state.

\subsubsection{Inelastic dark matter}\label{sec:iDM}

In IDM models~\cite{Tucker-Smith:2001myb,Izaguirre:2015zva,Berlin:2018jbm}, two DS states are predicted with near mass degeneracy. These states can be scalars or fermions, since this degeneracy can be induced in both cases via different mechanisms. For small mass splittings relative to the average mass, the elastic couplings between same-flavor states are suppressed compared to the inelastic ones, leading to the preferred simultaneous production of both states in $\Pp\Pp$ collisions at the LHC. This production is mediated by one of the portal interactions, typically taken to be the dark-photon portal. These models can both evade increasingly stringent DM scattering constraints from DD and ID experiments and predict the correct thermal-relic DM abundance as indicated by cosmological observations.

Focusing on the scenario with fermionic DM, a Dirac fermion can be defined as the bispinor $\psi = ( \eta \; \overline{\xi}  )$. Assuming vector and axial-vector couplings to quarks, the interactions are described by~\cite{Tucker-Smith:2001myb}
\begin{equation}
    \mathcal{L} \supset \overline{\psi} \, \gamma_\mu \, ( g_V^\prime + g_A^\prime \gamma_5 ) \, \psi \, \PAQq \, \gamma^\mu \, ( g_V + g_A \gamma_5 ) \, \PQq.
\end{equation}
If we also add a small Majorana mass term $\frac{\Delta}{2}(\eta\eta + \overline{\xi}\overline{\xi})$ to the Lagrangian, where $\Delta$ is the small mass splitting between states, the fermion mass eigenstates become 

\begin{equation}\begin{aligned}
\PDM_1 & \approx \frac{i}{\sqrt{2}} \left( \eta - \xi \right), \\
\PDM_2 & \approx \frac{1}{\sqrt{2}} \left( \eta + \xi \right).
\end{aligned}\end{equation}

The vector current $\overline{\psi} \, \gamma_\mu \, \psi$ in this scenario has the form
\begin{equation}
    \overline{\psi} \, \gamma_\mu \, \psi \approx i ( \overline{\PDM}_1 \, \overline{\sigma}_\mu \, \PDM_2 - \overline{\PDM}_2 \, \overline{\sigma}_\mu \, \PDM_1 ) + \frac{\Delta}{2m} (\overline{\PDM}_2 \, \overline{\sigma}_\mu \, \PDM_2 - \overline{\PDM}_1 \, \overline{\sigma}_\mu \, \PDM_1 ).
\end{equation}
The elastic couplings in the second term are suppressed by a factor of $\Delta/m$ relative to the inelastic couplings in the first term and are negligible.

\begin{figure}
    \centering
   \includegraphics[width=0.5\textwidth]{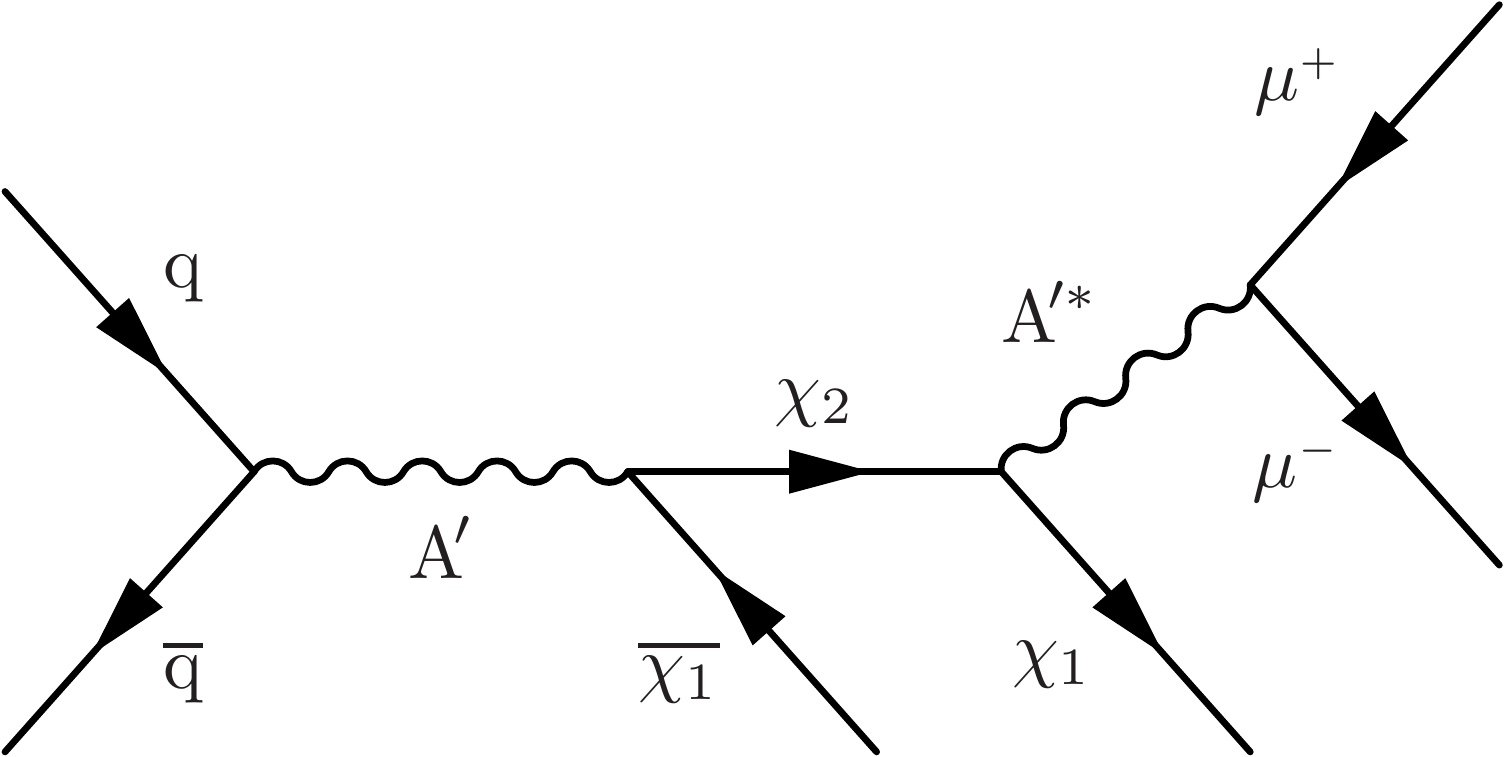}
    \caption{Feynman diagram of inelastic dark matter production and decay processes in $\Pp\Pp$ collisions, for fermionic DM states. The heavier DM state $\PDM_2$ can be long-lived, and decays into $\PDM_1$ and to a muon pair via an off-shell dark photon \PAprime. }\label{fig:inelasticDM_diagram}
\end{figure}

The excited state $\PDM_2$, once produced in tandem with the DM ground state $\PDM_1$ via $\Pp\Pp \to \PAprime \to \PDM_2 \, \PDM_1$, eventually decays into a $\PDM_1$ plus a pair of SM fermions by emission of an off-shell dark photon ($\PDM_2 \to \PDM_1\ffbar)$. The model is efficiently parameterized by the mass splitting $\Delta$, the lighter state mass $m_1 = \mDM$, and the interaction strength $y = \epsilon^2\aDark (\mDM/\mAprime)^4$, where \mAprime is the dark photon mass, $\epsilon$ is the kinetic mixing between the dark photon and the SM hypercharge, and \aDark is the coupling strength of the DS gauge interaction. The small mass splitting between the states leaves only a small kinematic phase space available for the decay, leading both to a small decay width (and hence a large lifetime) of the excited state and to the production of low-energy SM fermions at the end of the decay chain. Additionally, there is near collinearity between the SM fermion pair and between the SM fermions and the $\PDM_1$ states. The displaced and low-energy SM fermion pair in the final state, combined with significant \ptmiss from the $\PDM_1$, presents a unique and compelling experimental signature that can be searched for in $\Pp\Pp$ collision events. Figure~\ref{fig:inelasticDM_diagram} shows a diagram for the displaced $\PGm^+\PGm^-$ + \ptmiss signature.

\subsubsection{Hidden valleys}
\label{sec:darkqcd}

Nonminimal DSs may include multiple new particles and potentially new interactions that are decoupled from the SM. These kinds of models are often referred to as ``hidden valleys'' (HV)~\cite{Strassler:2006im} because the DSs may contain rich dynamics and phenomenology at relatively low energy scales, similar to those of the lightest SM particles, while nevertheless being accessible via collider production only at high energy scales corresponding to the mass of the mediator particle.
Hidden valleys provide DM candidates if they contain new, stable, invisible particles. Generally, in HV models, the SM is supplemented by a DS with a non-Abelian confining $SU(\Nc)$ gauge interaction with \Nc dark colors, gauge coupling \aDark, and massless dark gluons as the carriers of the new force. All SM particles are neutral under $SU(\Nc)$, but there are new light particles that are charged under $SU(\Nc)$ and neutral under the SM gauge groups. The basic particle content in the DS comprises \Nf flavors of dark quarks (\Pqdark) charged under $SU(\Nc)$ with masses \mqdark.

Higher-dimensional operators, induced by a high-mass \PZpr boson or a loop of heavy particles carrying both SM and hidden-sector charges, allow interactions between SM fields and the new light particles of the DS. In a simple HV scenario, adding a broken $U^{\prime}(1)$ gauge group introduces a heavy vector portal mediating between the two sectors. In such a scenario, the kinetic mixing between the DS group $U^{\prime}(1)$ and the SM group $U(1)_Y$ cannot be forbidden, implying the possible existence of an HV dark photon that may communicate with the SM via kinetic mixing. This class of models is sometimes called ``dark QCD'' in analogy with the SM QCD, though not all such models evince QCD-like behavior.

The confinement of this Yang--Mills theory at a scale \lamDark is guaranteed only for $\Nf < 3\Nc$~\cite{Albouy:2022cin}. Confinement and hadronization in the DS result in a spray of composite hidden-sector states, dark hadrons. This process is called a dark shower and produces dark jets. A key feature of dark shower signatures is the evolution of energy within the DS that follows the initial production at the hard process energy scale \Qdark. In QCD, the momentum flow from the hard scattering energy scale $Q$ to the confinement scale is dominated by the soft and collinear singularities and can be described using perturbation theory (parton shower). This feature holds generally for theories that, like QCD, have small 't~Hooft couplings $\lambda = \aDark^2\Nc$. In these theories, the 't~Hooft coupling can become large, but only in a limited energy range near the confinement scale. The small 't~Hooft coupling regime defines a QCD-like parton evolution, where well-established parton shower algorithms allow for good modeling of the partonic component of the hidden-sector evolution~\cite{Knapen:2021eip}.

\begin{figure*}[htb!]
\centering
\includegraphics[width=0.55\linewidth]{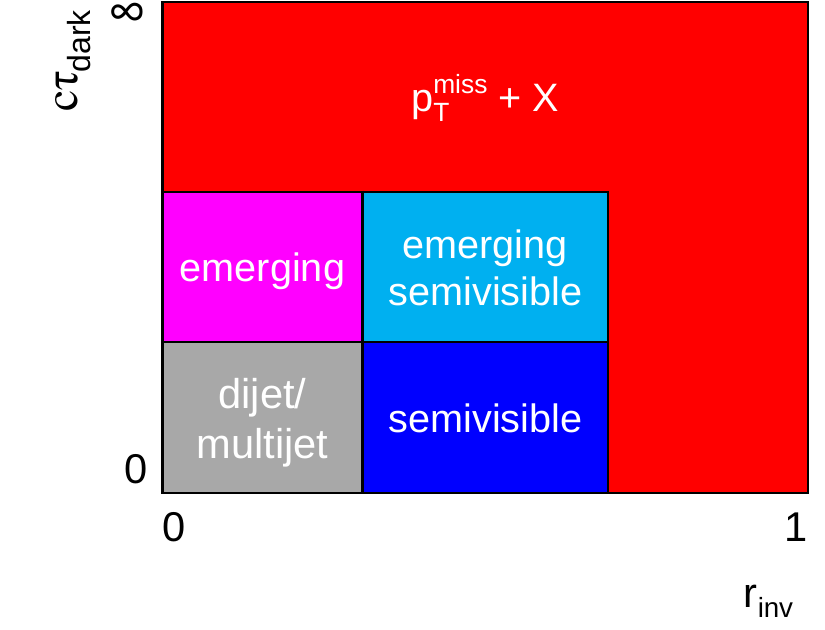}
\caption{A qualitative depiction of the phenomenological behavior of dark QCD models depending on the fraction of invisible particles within a jet \rinv and the proper decay length of dark hadrons c\tdark. The \rinv parameter is defined in Section~\ref{sec:svjtheory}.}
\label{fig:dark_qcd_diagram}
\end{figure*}

\begin{figure*}[htb!]
\centering
\threefigeqh{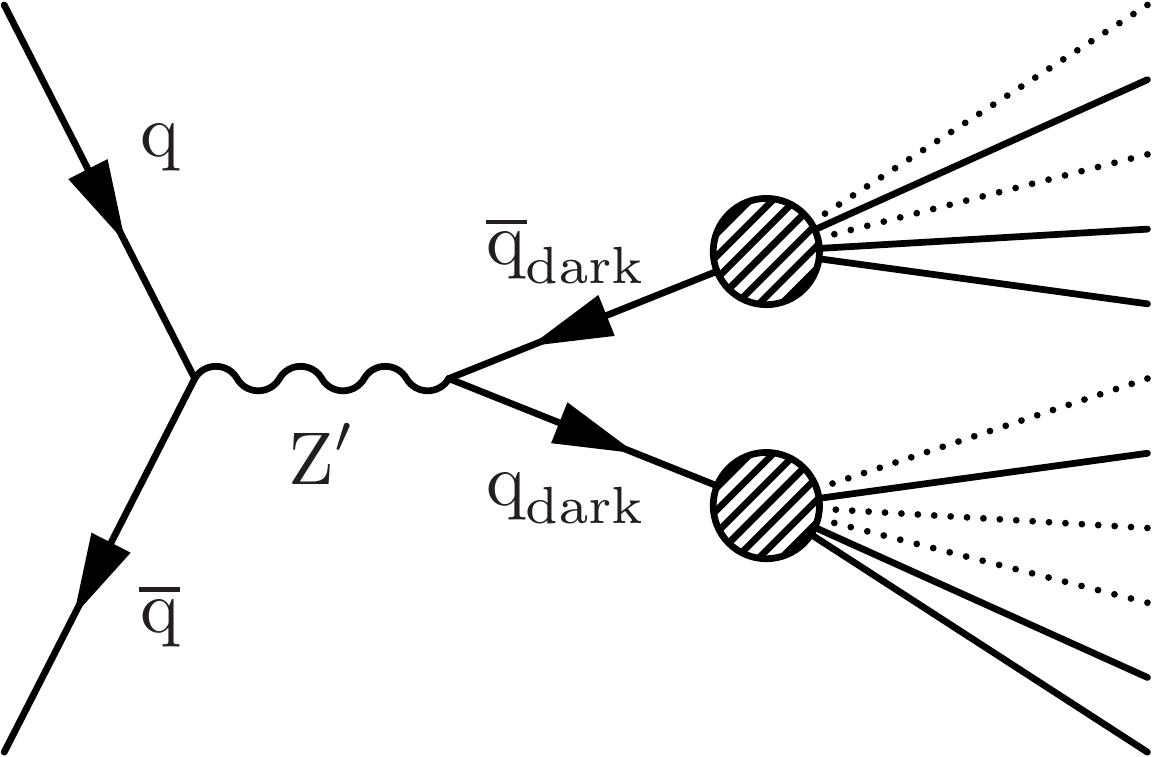}{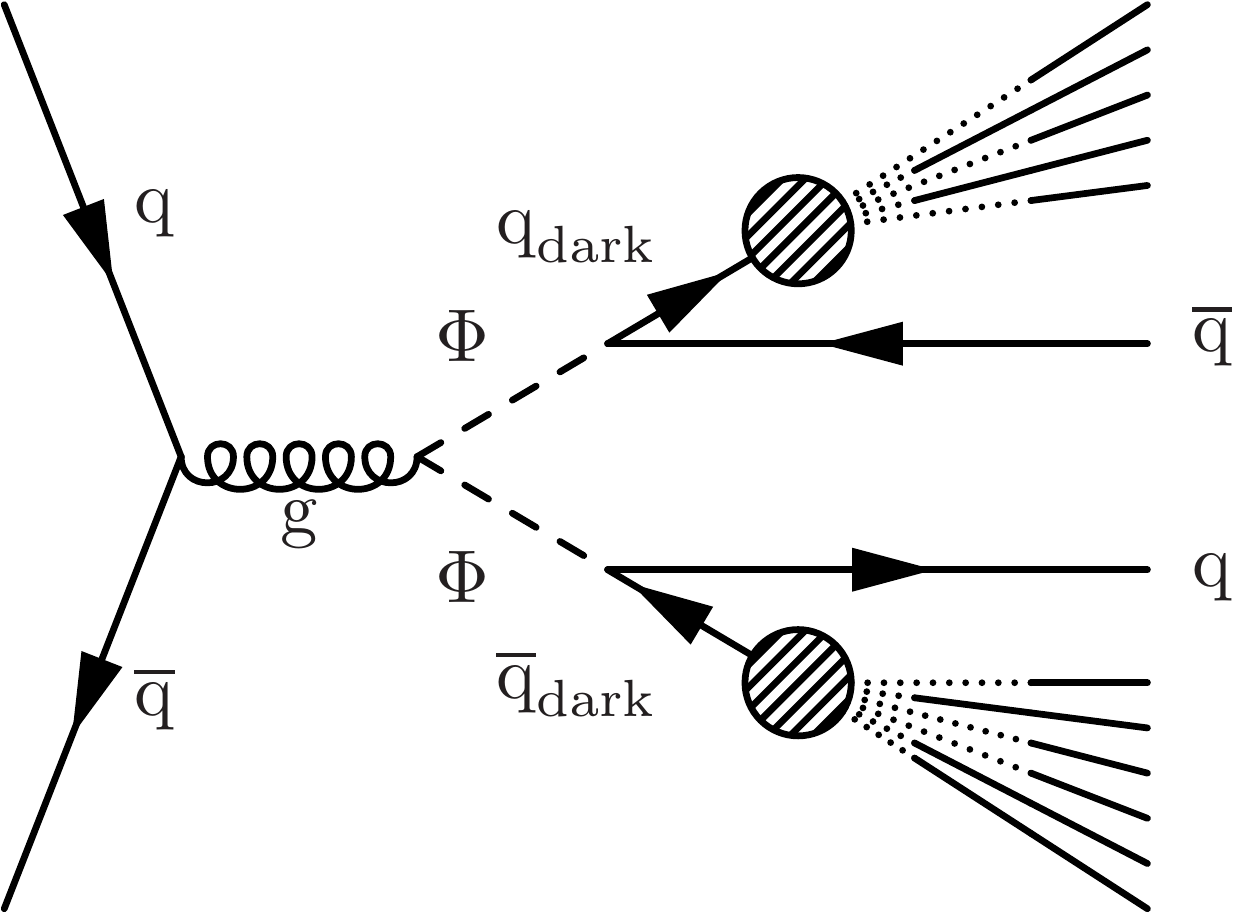}{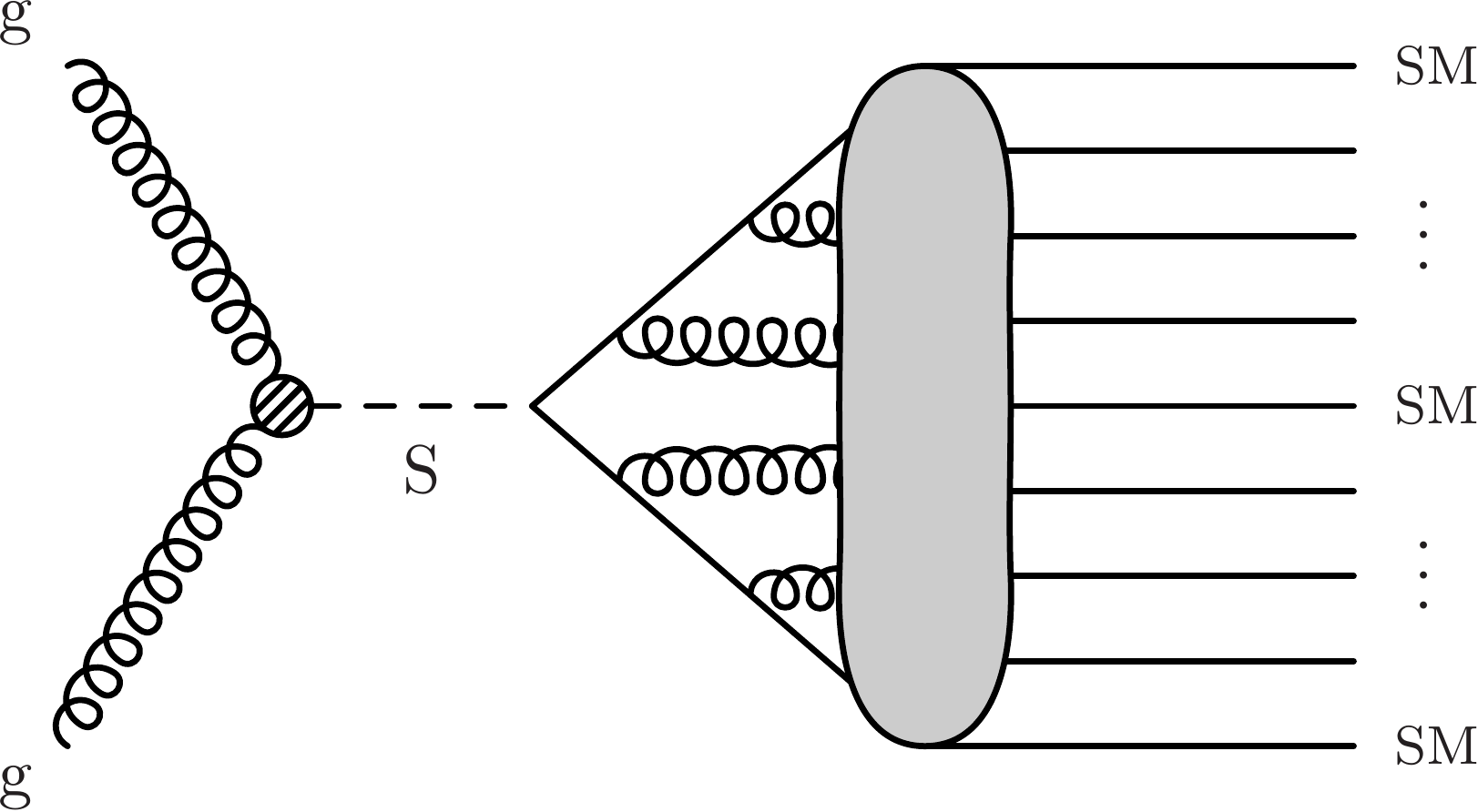}
\caption{Illustrative Feynman diagrams showing example production modes for different hidden valley phenomena: semivisible jets (\cmsLeft), emerging jets (center), and soft unclustered energy patterns (\cmsRight). Dotted lines indicate invisible particles.}
\label{fig:HV_diagrams}
\end{figure*}

The dark mesons produced in the dark shower may or may not be degenerate with mass(es) \mDark and proper decay length(s) c\tdark, while dark baryons are typically neglected, as their masses scale with \Nc and therefore their production is suppressed~\cite{Witten:1979kh}. Alternatively, if $\Nf=0$ or $\mqdark > \lamDark$, dark glueballs form~\cite{Juknevich:2009ji}, along with quirks~\cite{Kang:2008ea} in the latter case. Numerous phenomenological signatures are possible, depending on the values of these parameters that define the dark QCD model. Two major categories in the case of a small 't~Hooft coupling are semivisible jets (SVJs) and emerging jets (EJs), described in Sections~\ref{sec:svjtheory} and \ref{sec:emjtheory}, respectively. The relationships between these two signatures are shown in Fig.~\ref{fig:dark_qcd_diagram} in terms of the novel parameters of the models, which are explained in the following sections. Generically, visible jets resulting from dark showers are expected to be wider than SM jets if there is a mass hierarchy between the dark hadrons and their decay products, such that the latter are produced with significant momenta. Here, we discuss the particular models used to motivate and design CMS searches. Comparisons of these and other models, along with other details, are detailed in Ref.~\cite{Albouy:2022cin}. Alternatively, a large 't~Hooft coupling produces soft unclustered energy patterns (SUEPs), discussed in Section~\ref{sec:SUEPs}. Figure~\ref{fig:HV_diagrams} shows examples of final states including each of the three phenomena. It is generically expected that signals of composite DM are highly suppressed at DD experiments~\cite{Cohen:2017pzm}, complementing other models, such as simplified DSs with vector or scalar portals, where DD may have more sensitivity than collider production.

\cmsParagraph{Semivisible jets\label{sec:svjtheory}} A scheme that produces both stable and unstable dark hadrons in varying proportions will lead to collimated mixtures of visible and invisible particles, or semivisible jets. Dark-hadron stability depends on the conservation of accidental symmetries in the DS. This behavior is captured in an effective parameter called the invisible fraction, defined as $\rinv = \langle \Nstable/(\Nstable+\Nunstable) \rangle$, where \Nstable is the number of stable dark hadrons and \Nunstable is the number of unstable dark hadrons. Semivisible jets occur when $\rinv$ assumes values different from 0 or 1.

References~\cite{Cohen:2015toa,Cohen:2017pzm} introduce a simple strongly coupled DS with $\Nc=2$ and $\Nf=2$, connected to the SM via a leptophobic \PZpr mediator. In this model, the dark baryon number is conserved, so dark baryons cannot decay into SM particles. Similarly, the dark isospin is conserved, so dark vector mesons and pseudoscalars carrying nonzero dark isospin cannot decay. Therefore, combinations of different flavors of dark quarks are stable. The multiplicity of such states is proportional to $\Nf(\Nf-1)$; however, the production of these stable hadrons may be suppressed by a mass splitting between the dark quark flavors, if $\Delta\mqdark^2>\lamDark^2$. Therefore, variations of \Nf and the mass splitting allow \rinv to take any value from 0 to 1. This parameter accordingly incorporates the effects of nonperturbative dynamics in the DS.

Both vector dark mesons \Prhodark and pseudoscalar dark mesons \Ppidark may form, with the former expected to occur with 75\% probability, if the masses for the dark mesons are degenerate~\cite{Cohen:2015toa}. These dark mesons are assumed to have similar mass scales, parameterized as a single value \mDark, and we set the constituent dark quark mass $\mqdark = \mDark/2$. The unstable \Prhodark mesons decay democratically to any pair of SM quarks satisfying $\mDark \geq 2\mq$. The unstable \Ppidark mesons decay via a mass insertion, in analogy with SM pion decay, preferring the most massive species of SM quarks satisfying the above relationship. All decays of unstable dark mesons are assumed to be prompt, in accordance with theoretical predictions for this class of models~\cite{Cohen:2015toa}. The stable dark mesons traverse the detector invisibly and represent DM candidates~\cite{Beauchesne:2018myj,Beauchesne:2019ato,Bernreuther:2023kcg}. The impact of the dark-coupling scale \lamDark depends on \mDark, so its value is set to $\lamDarkPeak = 3.2(\mDark)^{0.8}$, which is empirically found to maximize the number of dark hadrons produced in a typical dark shower~\cite{CMS:2021dzg}. The running coupling of the dark force can then be calculated as $\aDark(\lamDark) = \pi/\left(b_{0}\log\left(\Qdark/\lamDark\right)\right)$, with $b_{0} = (11\Nc - 2\Nf)/6$ and $\Qdark = 1\TeV$; the value at \lamDarkPeak is called \aDarkPeak. The mediator in this model is a leptophobic \PZpr boson with universal couplings to SM quarks \gq and to dark quarks \gqdark, as described in Section~\ref{sec:zprimeportal}). To account for the multiple flavors and colors in the DS, we set $\gqdark = 1.0/\sqrt{\smash[b]{\Nc\Nf}} = 0.5$. This produces a branching fraction to DM of 47\% and a width of 5.6\%, consistent with the LHC DM Working Group benchmark $\gDM = 1.0$ for minimal DM models~\cite{Albert:2017onk}.

\cmsParagraph{Emerging jets\label{sec:emjtheory}} In some strongly coupled DS models, parton showering and fragmentation in the DS create dark mesons on a shorter time scale than that of the dark-meson decay into SM particles. Therefore, these dark mesons travel long distances before decaying into SM particles. This behavior leads to the signature of an emerging jet, which is a jet that encompasses the multiple smaller displaced jets formed by the dark-meson decays. We consider three scenarios with $\Nc=3$, such that the stable dark hadrons are dark baryons: $\Nf=7$, $\Nf=3$, and $\Nf=1$.

References~\cite{EMJ1:Bai:2013xga,EMJ2:Schwaller:2015gea} introduce a strongly coupled DS with $\Nf=7$ fermionic dark quarks. The dark quarks are produced via the decay of a complex scalar mediator \Pbifun (Section~\ref{sec:bifunportal}), which is charged under both QCD and dark QCD. When produced resonantly, the mediator decays into a dark quark and SM quark: $\Pbifun \to \Pqdark\PAQq$. This model assumes all dark quarks are degenerate and coupled through the mediator to SM down-type quarks, as required by the quantum numbers of the mediator, and is therefore described as ``unflavored''. The undetermined model parameters that influence the kinematic behavior include the mediator and dark meson masses and the dark meson lifetimes. The proper decay length can be computed as:
\begin{equation}
    c\tdark = 80\unit{mm} \enc{\frac{1}{\kappa^4}} \enc{\frac{2\GeV}{\fpidark}}^2 \enc{\frac{100\MeV}{\mdown}}^2 \enc{\frac{2\GeV}{\mDark}} \enc{\frac{m_{\Pbifun}}{1\TeV}}^4 , 
\end{equation}
where $\kappa$ is the Yukawa coupling between \Pbifun, \Pqdark, and the SM down quark; \fpidark is the dark pion decay constant; and \mdown is the mass of the SM down quark.

A related model with $\Nf=3$~\cite{Renner:2018fhh}, includes a coupling matrix \dcoupl for the mediator \Pbifun, where $\alpha$ is the dark quark flavor and $i$ is the SM quark flavor. In particular, the ``flavor-aligned'' version of this model is considered, where the matrix is given by $\dcoupl = \kappa_0 \delta_{\alpha i}$, such that each flavor of dark quark couples to a single flavor of down-type SM quarks. Decays into the most massive allowed SM particles are preferred, leading to \PQb quark enriched final states when the dark mesons are sufficiently massive. In this model, the proper decay length for a dark meson composed of dark quarks of flavors $\alpha$ and $\beta$ is:
\begin{equation}
    c\tdark^{\alpha \beta} = \frac{8\pi m_{\Pbifun}^{4}}{N_{\text{c}} \mDark \fpidark^2 \abs{\dcoupl\kappa_{\beta j}^*}^2 \enc{m^2_{i}+m^2_{j}} \sqrt{\enc{1-\frac{(m_{i}+m_{j})^2}{\mDark^2}} \enc{1-\frac{(m_{i}-m_{j})^2}{\mDark^2}}}}.
\end{equation}
These models may also be characterized by the maximum proper decay length of any dark meson species, denoted c\tdarkmax.

Reference~\cite{Knapen:2021eip} introduces a set of models with similar phenomenological behavior: the formation of long-lived dark hadrons that eventually decay into SM particles. These models fix $\Nc=3$ and $\Nf=1$, resulting in a spectrum with a spin-0 dark meson \Petadark and a spin-1 dark meson \Pomegadark. Two benchmark scenarios for the dark hadron masses and dark-QCD scale are considered: $\lamDark = \momegadark = \metadark$ and $\lamDark = \momegadark = 2.5\metadark$. In the first scenario, the \Pomegadark is typically stable and formed during hadronization with 75\% probability (as in Section~\ref{sec:svjtheory}), while in the second scenario, the decay $\Pomegadark \to \Petadark\Petadark$ occurs and \Pomegadark forms with 32\% probability. Typically, the \Petadark is unstable and decays into SM particles, though there are some exceptions. The SM Higgs boson portal described in Section~\ref{sec:higgsportal} is employed, producing a pair of dark quarks. However, after dark hadrons are formed, their decays into SM particles may proceed through different portals, leading to distinct phenomenology:
\begin{itemize}
    \item gluon portal, with the decay $\Petadark \to \Pg\Pg$ producing hadron-rich showers;
    \item photon portal, with the decay $\Petadark \to \PGg\PGg$ producing photon showers;
    \item vector portal (Section~\ref{sec:spinoneportal}), in particular a heavy kinetically-mixed dark photon that allows both leptonic and hadronic decays of the vector \Pomegadark while the \Petadark is stable, producing SVJs with a default $\rinv = 0.25$ (Section~\ref{sec:svjtheory});
    \item Higgs boson portal, with preferred decays $\Petadark \to \bbbar$, $\Petadark \to \ccbar$, and $\Petadark \to \PGtp\PGtm$ producing heavy flavor rich showers; and
    \item dark-photon portal (Section~\ref{sec:darkphotonportal}), less massive than the vector portal to allow the decay $\Petadark \to \PA^{\prime}\PA^{\prime}$, with the \PAprime decaying into quarks and leptons, producing lepton-rich showers.
\end{itemize}
The minimum lifetime of the unstable dark-hadron species depends on which decay portal is used; therefore, this model can be referred to as the ``decay portal'' model. The dark-photon portal leads to a short minimum lifetime; the photon and vector portals lead to intermediate minimum lifetimes; and the gluon and Higgs boson portals lead to very long minimum lifetimes. One or more collimated decays of these particles may be observed in the tracker, calorimeter, and/or muon system of the detector, depending on the lifetime of the dark hadron.

\cmsParagraph{Soft unclustered energy patterns\label{sec:SUEPs}} Dark showers produced in HV models do not necessarily result in collimated jets similar to SM QCD. In particular, SUEPs comprising a large multiplicity of spherically distributed low-momentum charged particles are also possible signatures of HV models. The underlying physics that produces such events can be varied; here, we consider quasi-conformal models in which the dark QCD force has a large 't~Hooft coupling $\lambda \gg 1$ above its confinement scale~\cite{Knapen:2016hky}. When new particles shower with efficient branching over a wider energy range than in SM QCD, the initial parton momenta are not preserved, resulting in soft and isotropic emissions. In this case, the production of dark mesons proceeds similarly to hadron production in high-temperature QCD.

Following Ref.~\cite{Knapen:2016hky}, we focus on a benchmark model with a heavy scalar mediator \PS connecting the SM and DS, produced via gluon fusion. We assume the dark quark masses \mqdark are less than the confinement scale \lamDark, and also that $\lamDark \ll \sqrt{s}$. Therefore, the dark quarks undergo a quasi-conformal showering, forming dark pseudoscalar mesons \Ppidark. The dark-meson transverse momentum (\pt) spectrum follows a Boltzmann distribution that depends on the dark-meson mass \mDark and a temperature $\TD \approx \lamDark$. The pseudoscalar mesons decay into a pair of dark photons \PAprime. The dark photon kinetically mixes with the SM photon and decays promptly to SM particles including electrons, muons, and pions, with branching fractions (\BR) that depend on its mass. Three benchmark \mAprime values are considered, each with corresponding branching fractions: $\mAprime = 0.5\GeV$ ($\PAprime \to \EE, \MM, \pipi$ with $\BR = 40, 40, 20\%$), $\mAprime = 0.7\GeV$ ($\PAprime \to \EE, \MM, \pipi$ with $\BR = 15, 15, 70\%$), and $\mAprime = 1.0\GeV$ ($\PAprime \to \pipi$ with $\BR = 100\%$).

\cmsParagraph{Neutral naturalness\label{sec:neutnat}} 
One of the major puzzles in particle physics is the large difference between the Planck scale and the measured Higgs boson mass. As an elementary scalar, the Higgs boson is not protected from quantum corrections to its mass, which should drive the EW scale to be much higher than what has been observed~\cite{Weisskopf:1939zz,Wilson:1970ag,tHooft:1979rat,Susskind:1978ms}. This is referred to as the EW hierarchy problem. A common solution is to introduce a new symmetry to protect the Higgs boson mass from quantum corrections. A canonical example is SUSY, where the introduction of superpartners cancels the quadratic divergences in the radiative corrections to the Higgs boson mass from SM particles. The most significant of these corrections is associated with the top quark, and therefore the top squark plays an important role. However, as the top squark carries SM color charge, it can be copiously produced at the LHC and thus faces stringent constraints from collider searches~\cite{CMS:2021eha}. 

An alternative approach is to introduce new discrete symmetries, instead of continuous symmetries as in SUSY. In this case, the new particles need not carry SM charges and thus can naturally arise from a DS or HV. Consequently, the top quark partner is not abundantly produced at the LHC and evades the stringent constraints on colored top quark partners. This scenario is referred to as ``neutral naturalness'', realizations of which include the twin Higgs (TH)~\cite{Chacko:2005pe}, folded SUSY (FSUSY)~\cite{Burdman_2007}, and quirky little Higgs~\cite{Cai_2009} models. To address the EW hierarchy problem, the necessary ingredients in the DS include a dark Higgs boson, a dark top quark (uncolored top quark partner), and a dark QCD gauge interaction with $\Nc=3$. The SM Higgs boson provides a portal to the DS because the top quark partner must interact with it to ensure naturalness. In many cases, the lightest dark quark is assumed to be heavier than the dark QCD confinement scale, in order to avoid cosmological constraints from Big Bang nucleosynthesis~\cite{Craig:2015pha,Harigaya:2019shz}. This means that the lightest hadronic states in the DS are usually glueballs~\cite{Juknevich:2009ji,Juknevich:2009gg}. Once the glueballs are produced, they decay back to SM particles through the Higgs portal. The lifetime of the lightest glueball is fixed by the naturalness requirement and can be estimated as~\cite{Curtin:2015fna,deFlorian:2016spz}: 
\begin{equation}
\ctau\approx\left(\frac{\mtdark}{400\GeV}\right)^{4}\left(\frac{20\GeV}{\mzero}\right)^{7}\times\begin{cases} 35\cm & [\mathrm{FSUSY}] \\ 8.8\cm & [\mathrm{TH}] \end{cases},
\label{eqn:glulife}
\end{equation}
where \mtdark is the dark top quark mass and \mzero is the lightest glueball mass. The preferred range for \mzero can be derived using the renormalization group flow argument and is found to be between 10 and 60\GeV~\cite{Craig:2015pha,Curtin:2015fna,deFlorian:2016spz}. Equation~\eqref{eqn:glulife} naturally produces a macroscopic lifetime for a large fraction of the preferred phase space. When $\mzero \geq 2m_{\PQb}$, the glueball predominantly decays to a pair of bottom quarks, motivating displaced-jets signatures arising from exotic Higgs decays. A common benchmark signature is the exotic decay of the 125\GeV Higgs boson to two long-lived scalars, each of which further decays to a pair of displaced jets~\cite{Craig:2015pha,Curtin:2015fna}. As a result of the hadronization in the DS, more complicated decay topologies are also possible, where the Higgs boson can decay to more than two glueballs, or undergo cascade decays because of the additional states in the DS. A long lifetime is a generic feature of these decays and can lead to EJ or SVJ signatures, especially when the Higgs boson is produced with high momentum~\cite{Batz:2023zef}.

In addition to exotic Higgs boson decays, other collider signatures can arise from neutral naturalness. For example, the dark Higgs boson can be directly produced via mixing with the SM Higgs boson, with a smaller production cross section than SM Higgs boson production. The dark Higgs boson may decay to a pair of SM Higgs bosons or gauge bosons. It can also decay to the DS glueballs, leading to signatures with multiple displaced decays~\cite{Kilic:2018sew,Alipour-Fard:2018rbc}.

The cosmological implications of neutral naturalness have also been extensively studied. There are many possible DM candidates, especially in the context of TH models, including the twin tau lepton~\cite{GarciaGarcia:2015fol,Craig:2015xla,Curtin:2021spx}, the twin neutrino~\cite{Cheng:2018vaj}, the twin electron~\cite{Koren:2019iuv}, and twin baryons or twin atoms~\cite{Farina:2015uea,Farina:2016ndq,Terning:2019hgj}. There can also be gravitational-wave signals arising from the DS phase transition~\cite{Schwaller:2015tja,Fujikura:2018duw,Zu:2023olm}, providing complementary probes in cosmological and astrophysical observations.

\section{The CMS detector}
\label{sec:detector}

The central feature of the CMS apparatus is a superconducting solenoid of 6\unit{m} internal diameter, providing a magnetic field of 3.8\unit{T}. Within the solenoid volume are a silicon pixel and strip tracker, a lead tungstate crystal electromagnetic calorimeter (ECAL), and a brass and scintillator hadron calorimeter (HCAL), each composed of a barrel and two endcap sections. The ECAL barrel (endcap) covers the pseudorapidity range $\abs{\eta}<1.479$ ($1.479<\abs{\eta}<3.0$), while the HCAL barrel (endcap) covers the $\abs{\eta}<1.3$ ($1.3<\abs{\eta}<3.0$) range. Forward calorimeters extend the pseudorapidity coverage provided by the barrel and endcap detectors. Muons are measured in gas-ionization detectors embedded in the steel flux-return yoke outside the solenoid. The muon system is composed of three types of chambers: drift tubes (DTs) in the barrel ($\abs{\eta}<1.2$), cathode strip chambers (CSCs) in the endcaps ($0.9<\abs{\eta}<2.4$), and resistive-plate chambers (RPCs) in both the barrel and the endcaps. A more detailed description of the CMS detector, together with a definition of the coordinate system used and the relevant kinematic variables, can be found in Ref.~\cite{CMS:2008xjf}.

Events of interest are selected using a two-tiered trigger system. The first level (``level-1''), composed of custom hardware processors, uses information from the calorimeters and muon detectors to select events at a rate of around 100\unit{kHz} within a fixed latency of about 4\mus~\cite{CMS:2020cmk}. The second level, known as the high-level trigger (HLT), consists of a farm of processors running a version of the full event reconstruction software optimized for fast processing, and reduces the event rate to around 1\unit{kHz} before data storage~\cite{CMS:2016ngn}.

\section{Interpretations}
\label{sec:reinterpretationAndResults}

This section summarizes the results of the DS searches described in this Report. None of the searches produce evidence for the existence of new physics. Accordingly, limits on model parameters are presented in the following. The results also include interpretations of certain analyses in terms of DS models that are presented for the first time. The results are organized in terms of the DS models introduced in Section~\ref{sec:theoreticalFramework}. 

\subsection{Simplified dark sectors}

\label{sec:simpDSResults}
For simplified models of DSs, limits are presented as a function of the essential parameters of such models, which are the masses of the mediator (\ie, the portal) states, of the DM, as well as couplings and, for FIP models, mixing strengths.

\subsubsection{Spin-1 portal}
\begin{figure}[htbp!]
    \centering
    \includegraphics[width=1\textwidth]{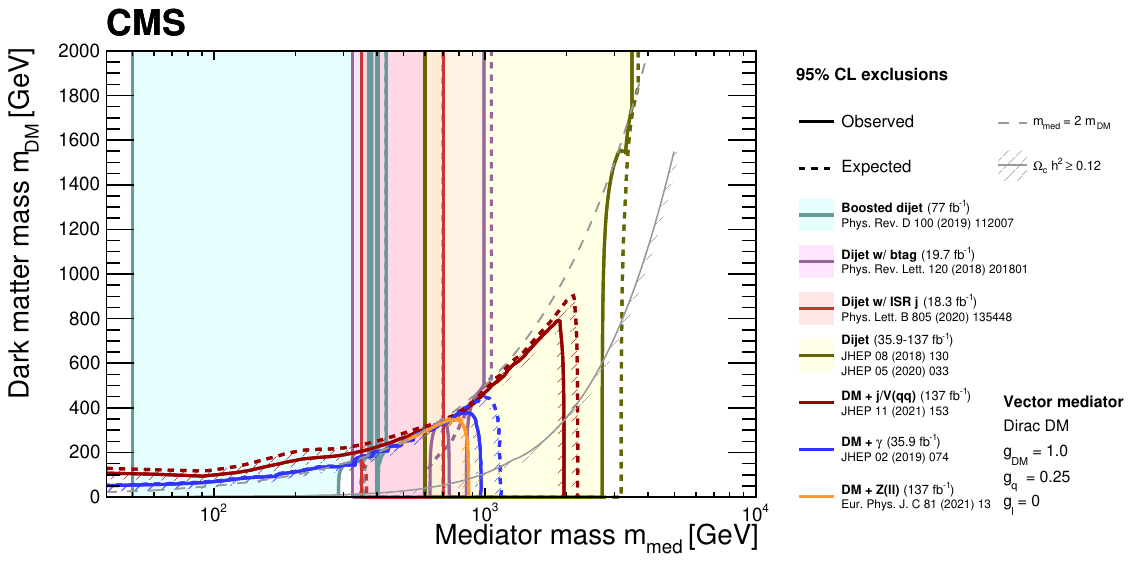}
    \includegraphics[width=1\textwidth]{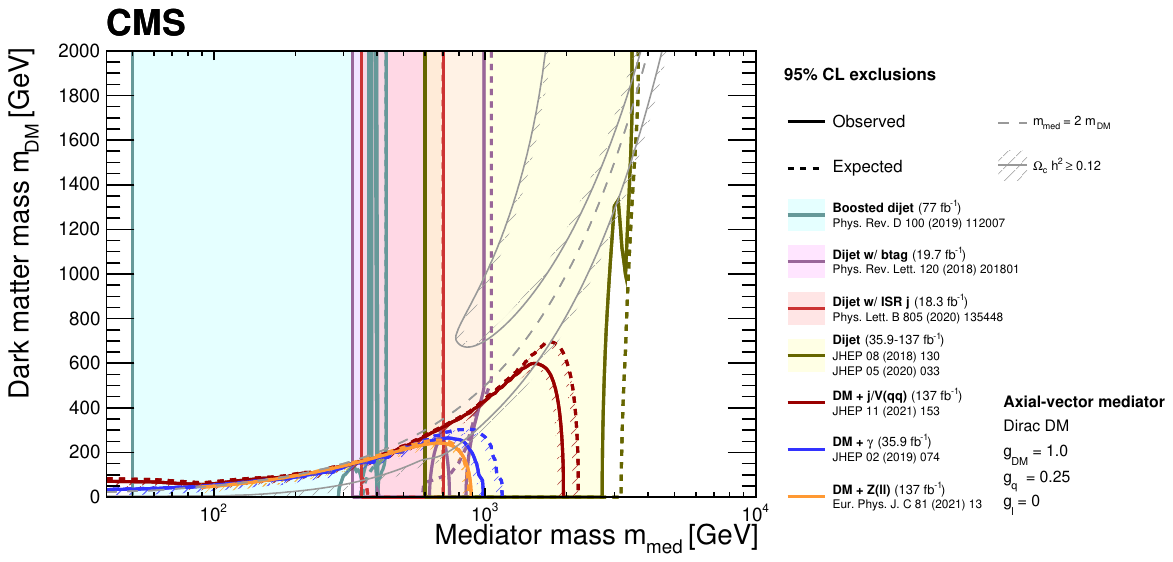}
    \caption{Observed and expected 95\% \CL exclusion regions in the \mMed-\mDM plane for dijet searches~\cite{CMS-PAPERS-EXO-18-012,EXO-16-057,EXO-19-004,CMS:2018mgb,CMS:2019gwf} and different \ptmiss-based DM searches~\cite{CMS:2021far,CMS:2018ffd,CMS:2020ulv} from CMS in the leptophobic vector mediator model (upper) and the axial-vector mediator model (lower). Following the recommendation of the LHC DM Working Group~\cite{Boveia:2016mrp,Albert:2017onk}, the exclusions are computed for a universal quark coupling of $\gq = 0.25$ and for a DM coupling of $\gDM = 1.0$. The resonant annihilation constraint $\mDM = 0.5\mMed$ is plotted as the gray dashed line, while the constraint from the thermal-relic density ($\Omega \mathrm{h}^2>0.12$), obtained from WMAP~\cite{WMAP:2012fli} and Planck~\cite{Planck:2015fie}, is plotted as the gray solid line. It should also be noted that the absolute exclusion of the different searches as well as their relative importance, will strongly depend on the chosen coupling and model scenario. Therefore, the exclusion regions, thermal-relic density contours, and unitarity curve shown in this plot are not applicable to other choices of coupling values or models.
    }
    \label{fig:summary_vector_axial}
\end{figure}

\begin{figure}[htbp!]
    \centering
    \includegraphics[width=1\textwidth]{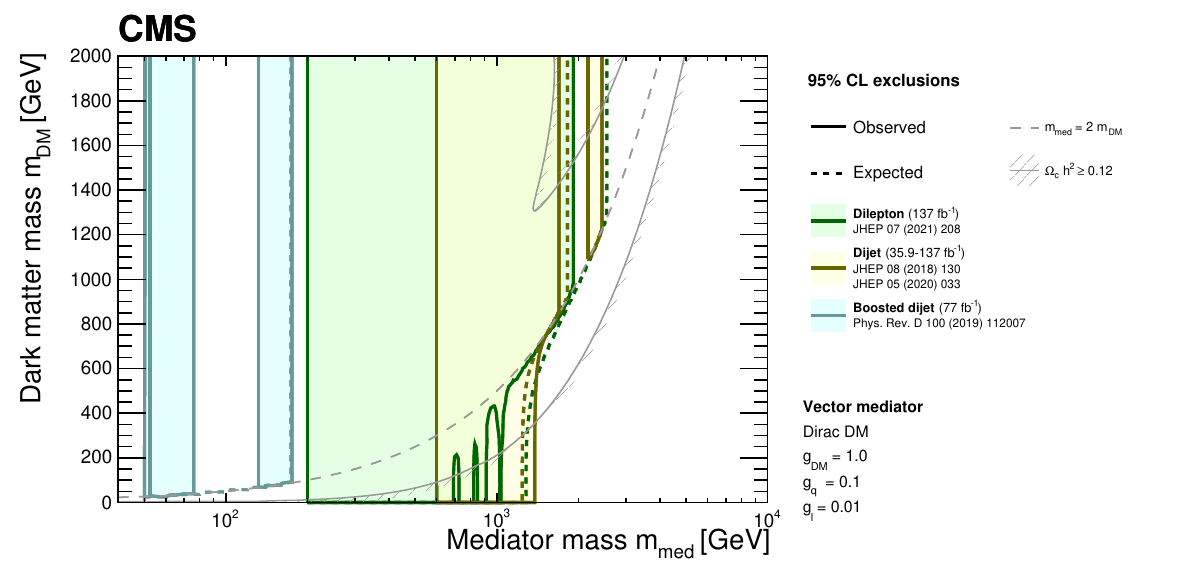}
    \includegraphics[width=1\textwidth]{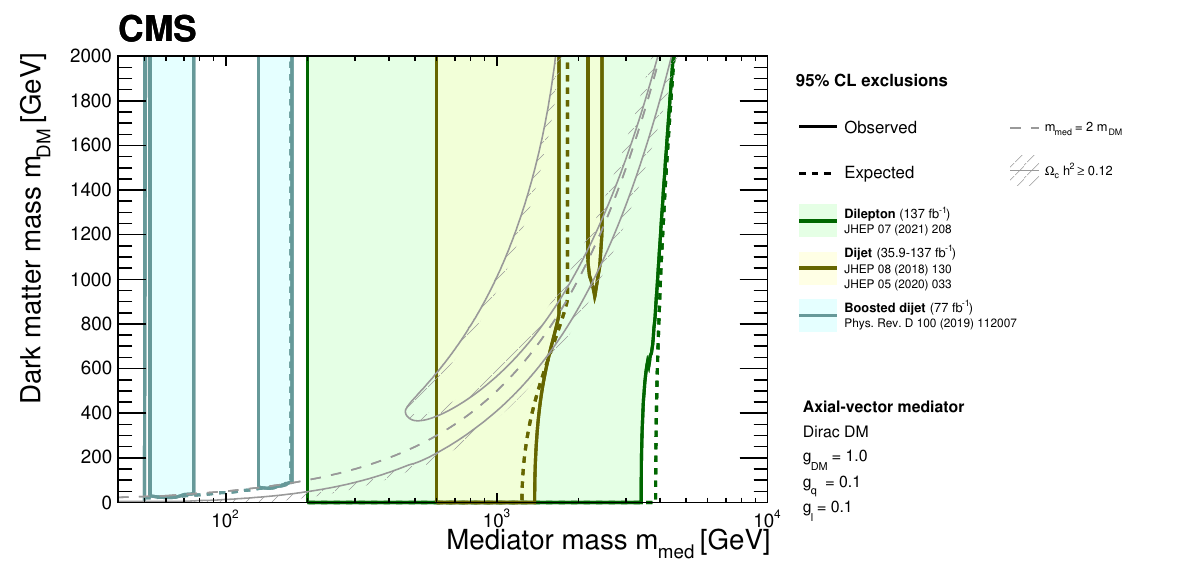}
    \caption{Observed and expected 95\% \CL exclusion regions in the \mMed-\mDM plane for dijet~\cite{CMS:2018mgb,CMS:2019gwf,CMS-PAPERS-EXO-17-001} and dilepton~\cite{CMS:EXO19019} searches from CMS in the vector mediator model (upper) and the axial-vector mediator model (lower). Following the recommendation of the LHC DM Working Group ~\cite{Boveia:2016mrp,Albert:2017onk}, the exclusions are computed for a universal quark coupling of $\gq = 0.1$, lepton coupling $\gell = 0.01$ (upper) and $\gell = 0.1$ (lower), and for a DM coupling of $\gDM = 1.0$. The resonant annihilation constraint $\mDM = 0.5\mMed$ is plotted as the gray dashed line, while the constraint from the thermal-relic density ($\Omega \mathrm{h}^2>0.12$), obtained from WMAP~\cite{WMAP:2012fli} and Planck~\cite{Planck:2015fie}, is plotted as the gray solid line. It should also be noted that the absolute exclusion of the different searches as well as their relative importance, will strongly depend on the chosen coupling and model scenario. Therefore, the exclusion regions, thermal-relic density contours, and unitarity curve shown in this plot are not applicable to other choices of coupling values or models. 
    }
    \label{fig:summary_vector_axial_gl0p1}
\end{figure}

\begin{figure}
    \centering
    \includegraphics[width=0.94\textwidth]{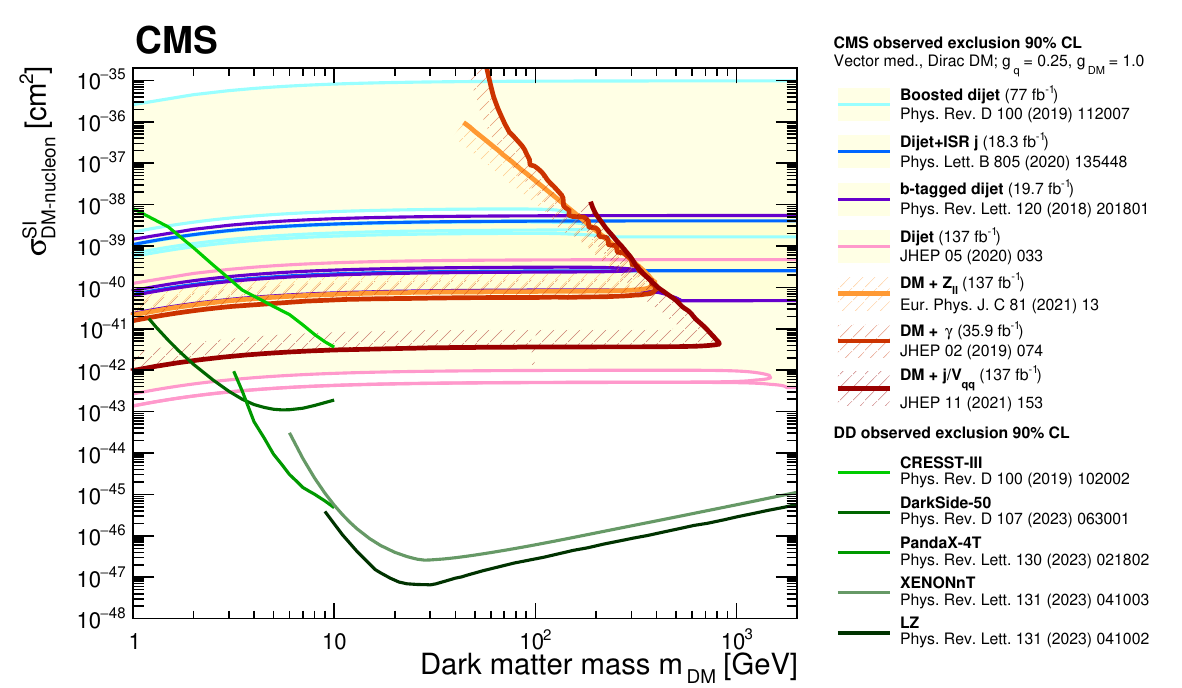}
    \includegraphics[width=0.94\textwidth]{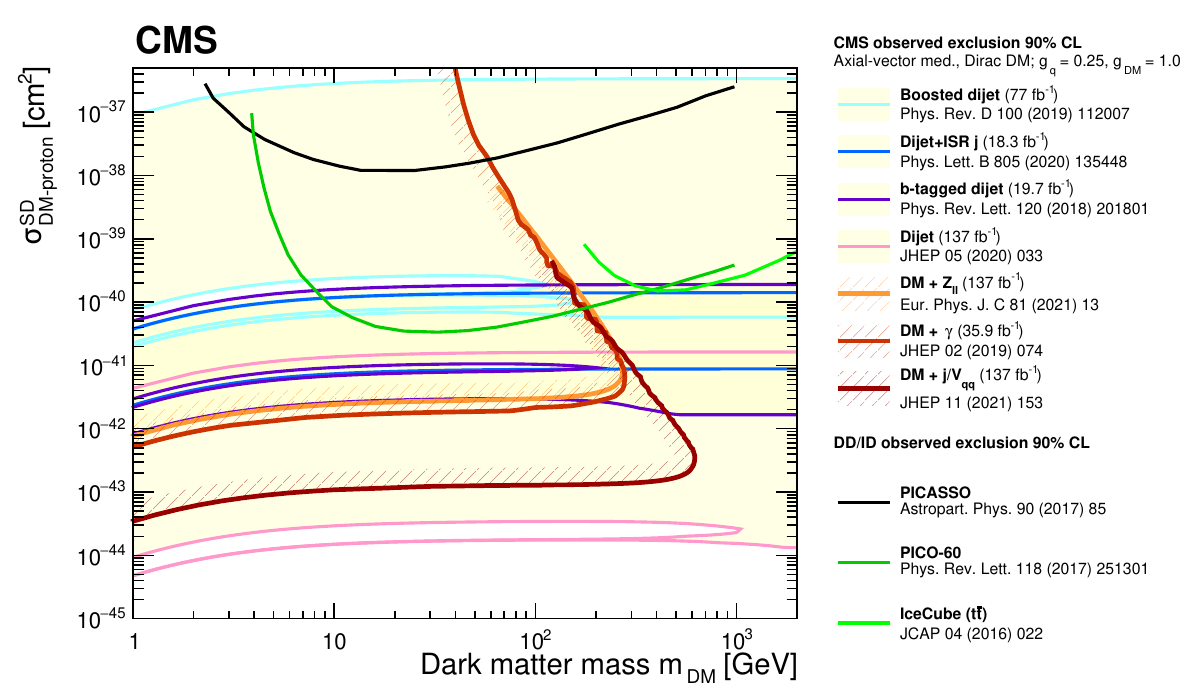}
    \caption{A comparison of CMS exclusions in the \mDM-$\sigma_\mathrm{SI}$ plane (upper), which are derived from exclusion limits on the vector model, and the \mDM-$\sigma_\mathrm{SD}$ plane (lower), which are derived from exclusion limits on the axial-vector model. The exclusions are derived from the model with a vector mediator, Dirac DM, and couplings of $\gq=0.25$ and $\gDM=1.0$. Unlike for the \mDM-\mMed plane, the limits are shown at 90\% \CL. The CMS SI exclusion contour is compared with limits from the CRESST-III~\cite{CRESST:2019jnq}, DarkSide-50~\cite{DarkSide-50:2022qzh}, PandaX-4T~\cite{PandaX:2022aac}, XENONnT~\cite{XENON:2023cxc}, and LZ~\cite{LZ:2022lsv} experiments. The CMS SD exclusion contour is compared with limits from the PICASSO~\cite{Behnke:2016lsk} and PICO~\cite{PICO:2019vsc} experiments, as well as the IceCube limit for the \ttbar annihilation channel~\cite{IceCube:2016dgk,IceCube:2016yoy}. The CMS limits do not include a constraint on the thermal-relic density, and the absolute exclusion of the different CMS searches as well as their relative importance will strongly depend on the chosen coupling and model scenario. Therefore, the shown CMS exclusion regions in this plot are not applicable to other choices of coupling values or models.}
    \label{fig:summary_spindependent_dependent}
\end{figure}

\cmsParagraph{Vector and axial-vector portal\label{par:DMsimp_zprime_result}} Summaries of the 95\% \CL observed exclusion limits in the plane of the mediator mass and the DM mass (the \mMed-\mDM plane) for different \ptmiss-based DM searches in the leptophobic vector and axial-vector models are presented in Table~\ref{tab:summary_vector_av} and Fig.~\ref{fig:summary_vector_axial}. Summaries of the 95\% \CL observed exclusion limits for a nonleptophobic vector and axial-vector mediator are presented in Fig.~\ref{fig:summary_vector_axial_gl0p1}. Looking at the sensitivities of the different mono-X channels, the monojet search is the one with the largest statistical power and thus, in general, provides the most stringent limits in comparison to mono-\PZ and monophoton searches. For mono-\PH and mono-\PQt searches, the recoiling SM object is an integral part of the SM-DM interaction; therefore, they are not directly comparable to the monojet/mono-V/monophoton exclusions. As can be seen in Fig.~\ref{fig:summary_vector_axial_gl0p1}, the highest-mass constraints on a leptophobic mediator are provided by dijet searches. 
In order to compare with various DD experiments, the 90\% \CL observed exclusion limits from the vector (axial-vector) model are converted to upper limits on the spin-independent (-dependent) DM-nucleon scattering cross section~\cite{Boveia:2016mrp} and shown in Fig.~\ref{fig:summary_spindependent_dependent}, where $\sigma_\mathrm{SI}$ ($\sigma_\mathrm{SD}$) is the spin-independent (-dependent) DM-nucleon scattering cross section.

\begin{table}[htbp]
\centering
\topcaption{Summary of 95\% \CL observed exclusion limits on $\mMed = \mZprime$ for \mbox{\ptmiss-based} DM searches in the leptophobic vector and axial-vector model. Following the recommendation of the LHC DM Working Group~\cite{Boveia:2016mrp,Albert:2017onk}, the exclusions are computed for a universal quark coupling of $\gq = 0.25$ and for a DM coupling of $\gDM = 1.0$.}
\renewcommand{\arraystretch}{1.2}
\begin{tabular}{lccrl}
Reference & $\Lint$~[{\fbinv}] & Channel & 95\% \CL & Notes \Bstrut \\
&&&limit [{\TeVns}]&\\
\hline
\cite{CMS:2021far} & 137   & Monojet    &        $\mZprime > 1.95$ & \\
\cite{CMS:2020ulv} &  137  & Mono-\PZ   &        $\mZprime > 0.87$ & Vector coupling\\
                                                   &    &            &        $\mZprime > 0.80$ & Axial coupling\\
\cite{CMS:2018gbj} &  36  & Mono-${\PQt}$    & $\mZprime < 0.20$ & Portal is FCNC \\
 & & & or $\mZprime > 1.75$ & \\
\cite{CMS:2018ffd} &  36  & Monophoton &       $ \mZprime > 0.95$ & \\
\cite{CMS:2018zjv} &  36 & Mono-${\PH}(\bbbar)$ &       $ \mZprime > 1.60$ &  \\
\label{tab:summary_vector_av}
\end{tabular}
\end{table}

\begin{figure}[htbp!]
    \centering
    \includegraphics[width=1\textwidth]{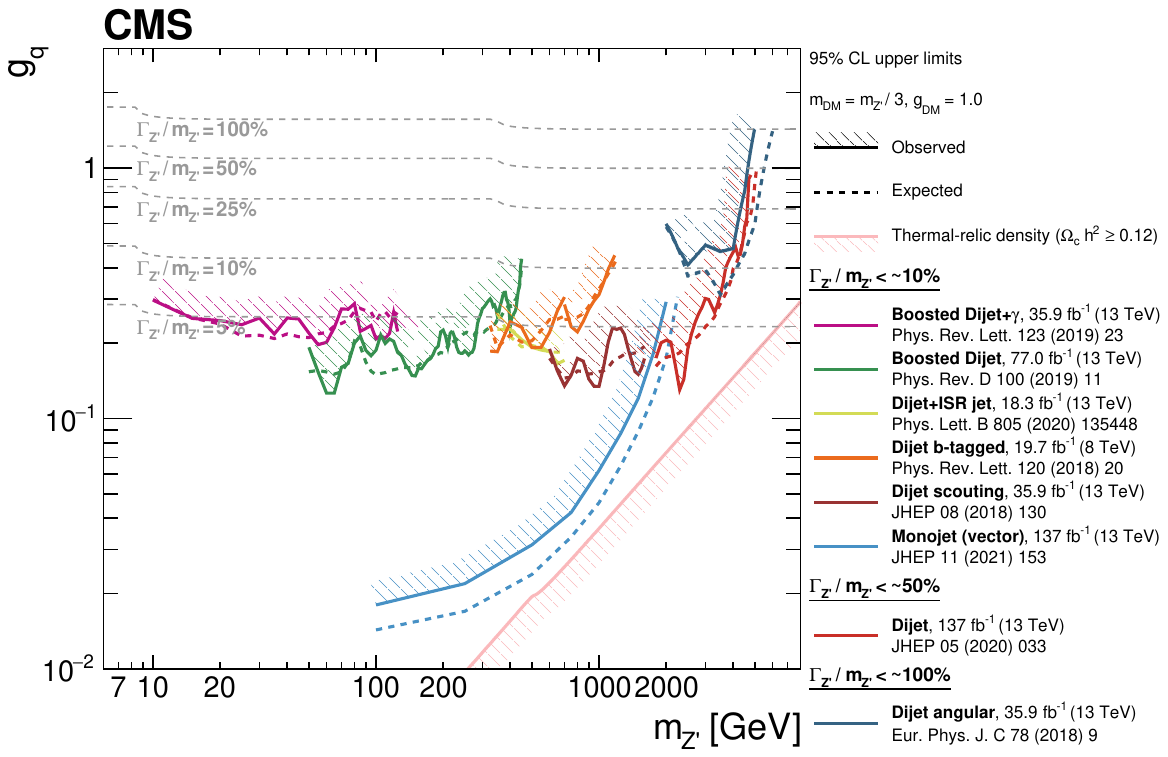}
    \caption{Observed and expected 95\% \CL exclusion regions for the universal quark coupling \gq, assuming a DM coupling $\gDM = 1.0$, for varying \PZpr mediator mass~\cite{CMS-PAPERS-EXO-17-027,CMS-PAPERS-EXO-18-012,EXO-19-004,EXO-16-057,CMS:2018mgb,CMS:2021far,CMS:2019gwf,CMS:2018ucw}. The hashed areas indicate the direction of the excluded area from the observed limits. The gray dashed lines show the \gq values at fixed values of the relative width $\Gamma_{\PZpr}/\mZprime$. Most searches assume that the intrinsic \PZpr width is negligible compared to the experimental resolution and hence are valid for $\Gamma_{\PZpr}/\mZprime \lesssim 10\%$. The dijet search is valid for $\Gamma_{\PZpr}/\mZprime \lesssim 50\%$, and the dijet angular analysis is valid for $\Gamma_{\PZpr}/\mZprime \lesssim 100\%$. The observed DM thermal-relic density is also shown; it drops to $2.17\times10^{-4}$ for $\mZprime = 5\GeV$.}
    \label{fig:summary_gq}
\end{figure}

Cross section exclusions can be converted to limits on \gq assuming the benchmark values for the DM coupling $\gDM = 1.0$ and DM mass $\mDM = \mZprime/3$~\cite{Izaguirre:2015yja}, following the procedure outlined in Ref.~\cite{Albert:2022xla}. Briefly, in the narrow-width approximation, the dependence of the cross section and branching fraction on the couplings is encapsulated in a few factors: $\sigmatilde \approx \wi\wf/\wtot$, where \wi is the partial width of the initial state, \wf is the partial width of the final state, and \wtot is the total width~\cite{Harris:2011bh}. Dividing the cross section limit by the theoretical cross section produces a dimensionless signal strength limit $r$ for the original model, which is taken to be the benchmark model with $\gq = 0.25$ and $\gDM = 1.0$. The signal strength limit can be scaled by the ratio of \sigmatilde factors to obtain the limit for another model with a different coupling value $\gq^{\prime}$: $r^{\prime} = r \sigmatilde / \sigmatilde^{\prime}$. The excluded coupling value is extracted by setting $r^{\prime} = 1$ and solving for $\gq^{\prime}$. The thermal-relic density can also be determined by calculating the predicted thermal-relic density for the desired simplified model, and labeling the region where, under the assumptions of the simplified model, the predicted DM density is above the observed DM density from CMB measurements. The computed thermal-relic density is only valid for the chosen coupling values in the minimal scenario of a single fermionic DM candidate, as prescribed by the simplified model. Figure~\ref{fig:summary_gq} shows the 95\% \CL observed exclusion for the \gq coupling for varying \PZpr mediator mass, including the monojet search with an invisible final state (Section~\ref{sec:EXO-20-004}) and dijet searches with visible final states (Sections~\ref{par:EXO-18-012}, \ref{par:EXO-17-027}, \ref{par:EXO-19-004}, \ref{sec:EXO-19-012}). The dijet search strategy provides the best exclusion at large \mZprime, while the monojet search provides the best exclusion at small \mZprime.

\begin{figure}[htbp!]
    \centering
    \includegraphics[width=0.7\textwidth]{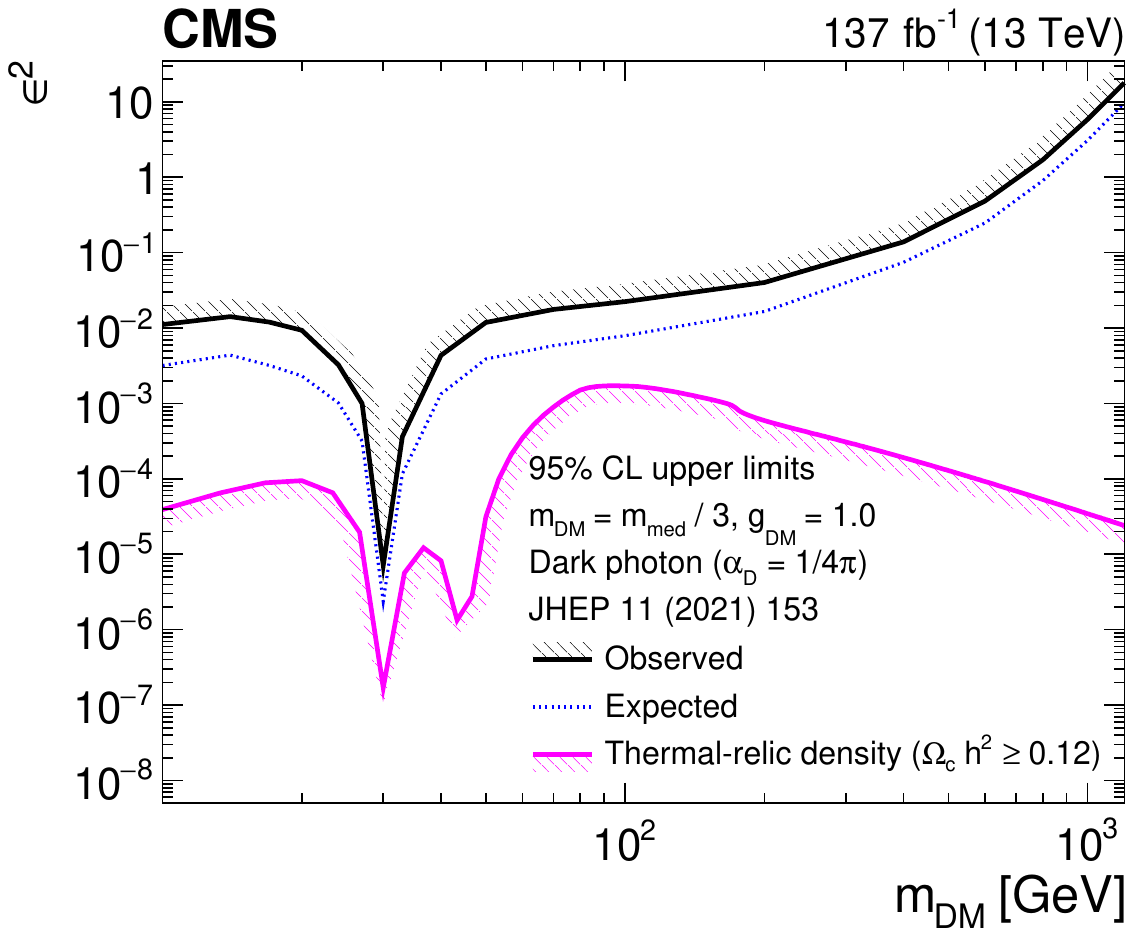}
    \caption{Limits at 95\% \CL from the monojet search~\cite{CMS:2021far} interpreted via \MADANALYSIS~\cite{DVN/IRF7ZL_2021} for a dark-photon model with a DM coupling. The limits are presented in terms of the mixing parameter $\epsilon^{2}$ with $\gDM = 1.0$ and $\aDark = \gDM^{2}/(4\pi)$. The constraint from the thermal-relic density ($\Omega_{c} \mathrm{h}^2 \geq 0.12$), obtained from WMAP~\cite{WMAP:2012fli} and Planck~\cite{Planck:2015fie}, is plotted in magenta.}
    \label{fig:eps_darkpho}
\end{figure}

\cmsParagraph{Dark-photon portal}\label{sec:darkPhotonPortalResults} Figure~\ref{fig:eps_darkpho} presents the 95\% \CL limits from the monojet search for a dark-photon model with a DM coupling. The exclusion is presented in terms of the mixing parameter $\epsilon^{2}$ as a function of DM mass. The interference with the \PZ boson can be observed for \mDM near $\mZ/3$, which leads to a more stringent limit in that region by up to three orders of magnitude. At small \mDM values, ${<}1\GeV$, low-energy experiments are more sensitive~\cite{Antel:2023hkf}.
The thermal-relic density additionally constrains $\epsilon^{2}$ to lower values for \mDM near $\mZ/2$, when thermal \PZ boson production becomes resonant. The thermal-relic density constraint also tightens at large \mDM, which corresponds to large \mMed. This occurs when the dark-photon width is dominated by kinetic mixing decays and is therefore proportional to $\epsilon^{2}$; when the mediator DM and SM couplings are of a similar order, the $\epsilon$ dependence in the DM annihilation cross section nearly cancels~\cite{Izaguirre:2015yja}.

\begin{figure*}[htbp!]
\centering
\includegraphics[width=0.975\linewidth]{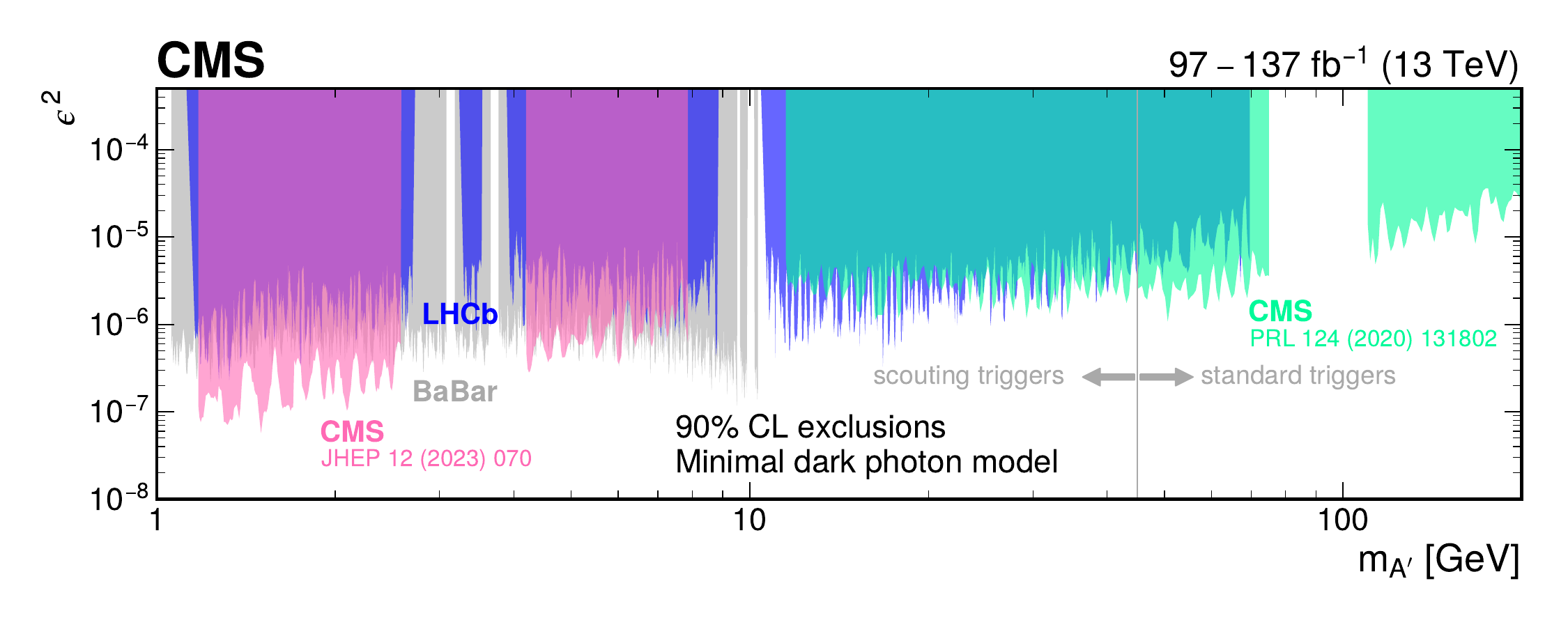}
\caption{Observed upper limits at 90\% \CL on the square of the kinetic mixing coefficient $\epsilon$ in the minimal model of a dark photon from a CMS dimuon search~\cite{CMS-PAS-EXO-21-005} in the mass ranges of 1.1--2.6\GeV and 4.2--7.9\GeV (pink) and from another CMS dimuon search~\cite{CMS:2019buh} at larger masses (green). The limits are compared with the existing limits at 90\% \CL provided by LHCb (blue)~\cite{LHCb:2017trq,LHCb:2019vmc} and BaBar (gray)~\cite{BaBar:2014zli}.}
\label{fig:darkphotonscouting}
\end{figure*}

Figure~\ref{fig:darkphotonscouting} presents the 90\% \CL limits on the squared kinetic mixing coefficient from the prompt dimuon searches with and without data scouting as a function of \mAprime along with the LHCb~\cite{LHCb:2017trq,LHCb:2019vmc} and BaBar~\cite{BaBar:2014zli} limits. Values of the squared kinetic mixing coefficient in the dark-photon model above are excluded over $10^{-6}$ for most of the dark-photon mass range of the search.

For the dark-photon search in Higgs boson production via vector boson fusion and in association with \PZ bosons, the combined observed upper limit at 95\% \CL on the branching fraction for a Higgs boson decaying into such an invisible particle and a photon is 2.9\%. 

\subsubsection{Spin-0 portal}

\begin{figure}[htbp!]
    \centering
    \includegraphics[width=0.7\textwidth]{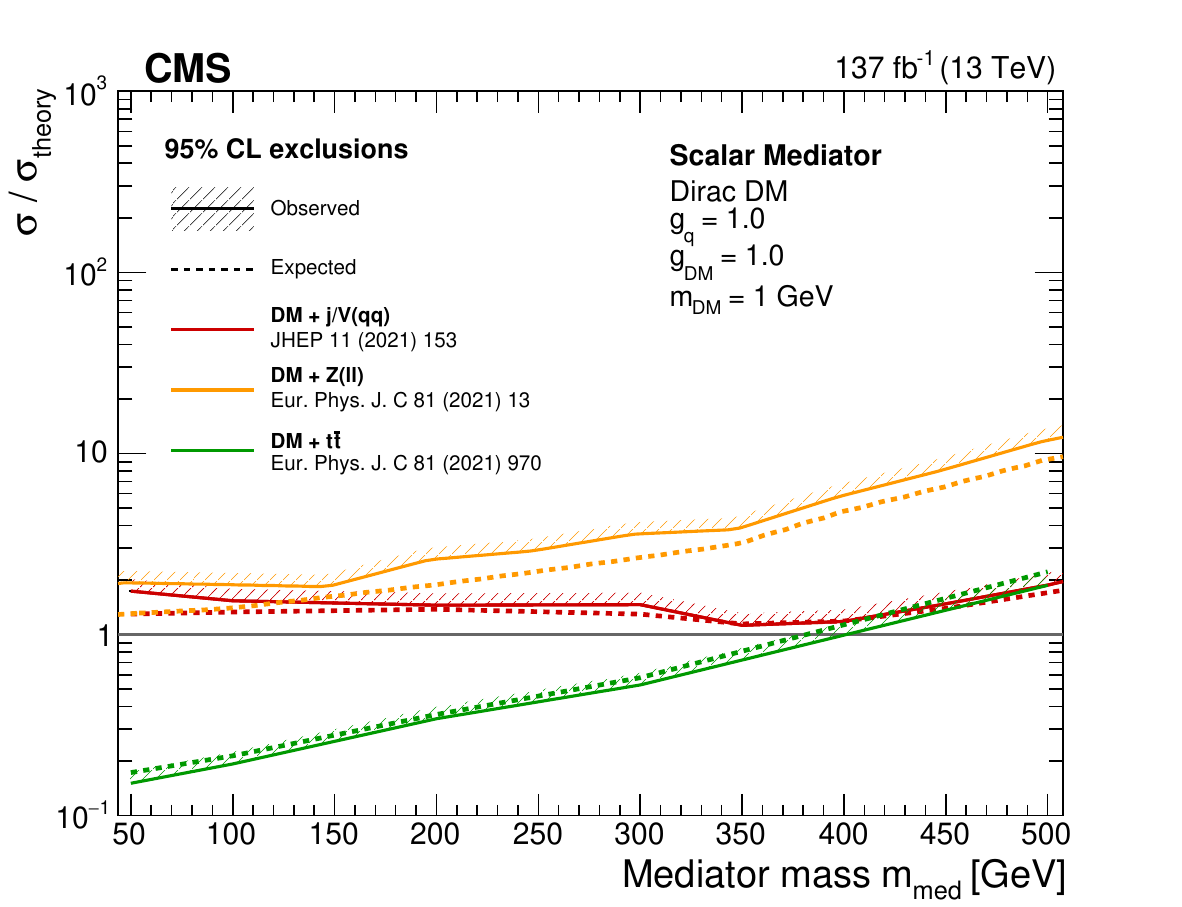}
    \caption{Observed (solid lines) and expected (dashed lines) 95\% \CL exclusion limits for the scalar model as a function of \mMed for different \ptmiss-based DM searches from CMS~\cite{CMS:2021far,CMS:2020ulv,CMS:2021eha}. The hashed areas indicate the direction of the excluded area from the observed limits. Following the recommendation of the LHC DM Working Group~\cite{Boveia:2016mrp,Albert:2017onk}, the exclusions are computed for a universal quark coupling of $\gq = 1.0$ and for a DM coupling of $\gDM = 1.0$. The exclusion away from $\sigma / \sigma_{\text{theory}} =1$ only applies to coupling combinations that yield the same kinematic distributions as the benchmark model considered here.}
    \label{fig:summary_scalar}
\end{figure}

\cmsParagraph{Scalar portal} A summary table and a plot for the 95\% \CL observed exclusion limits on \mMed for different \ptmiss-based DM searches from CMS in the scalar model are presented in Table~\ref{tab:summary_scalar} and Fig.~\ref{fig:summary_scalar}, respectively. From the different analyses interpreting for the scalar model, the search for {\ttbar}+\ptmiss signatures is the most sensitive, excluding scalar masses up to $\mS = 0.40\TeV$ for $\gDM=\gq=1.0$. For the monojet search, the limits show distinctive features around the top quark decay threshold of $\mMed = 2m_\PQt$. As the mediator is produced via a top quark loop, the signal cross section is enhanced as \mMed approaches the threshold. Above the threshold, the decay of the mediator into a pair of top quarks becomes possible, leading to a significant suppression of the branching fraction to DM, and therefore of the effective signal cross section.

\begin{table}[htbp]
\centering
\topcaption{Summary of 95\% \CL observed exclusion limits on $\mMed = \mS$ for~\mbox{\ptmiss-based} DM searches from CMS in the scalar model. Following the recommendation of the LHC DM Working Group~\cite{Boveia:2016mrp,Albert:2017onk}, the exclusions are computed for a universal quark coupling of $\gq = 1.0$ and for a DM coupling of $\gDM = 1.0$. Each search listed here used data corresponding to $\Lint=137\fbinv$.}
\begin{tabular}{lcrl}
Reference&Channel&95\% \CL lower limit& Notes \\
&  & on \mS [{\TeVns}] &  \\
\hline
\cite{CMS:2021far} &  Monojet    &  \NA & Excludes $\sigma/\sigma_{\text{theory}} = 1.12$ for $\mS = 350\GeV$\\
\cite{CMS:2020ulv} &  Mono-\PZ   &  \NA & Excludes $\sigma/\sigma_{\text{theory}} = 1.86$ for $\mS = 150\GeV$\\
\cite{CMS:2021eha} &  {\ttbar}+\ptmiss    &   $0.40$   & \\
\label{tab:summary_scalar}
\end{tabular}
\end{table}

\begin{figure}[thb!]
    \centering
    \includegraphics[width=0.8\textwidth]{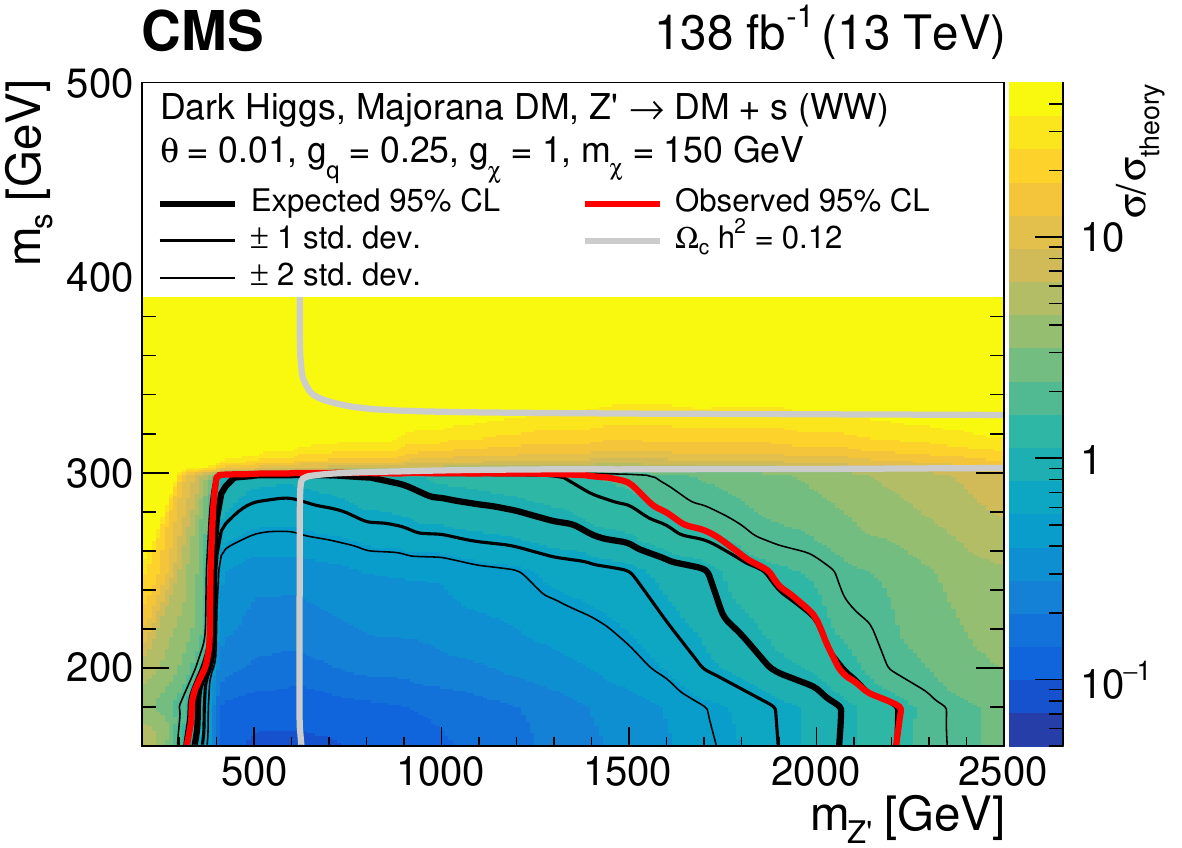}
    \raisebox{0.25em}{\includegraphics[width=0.8\textwidth]{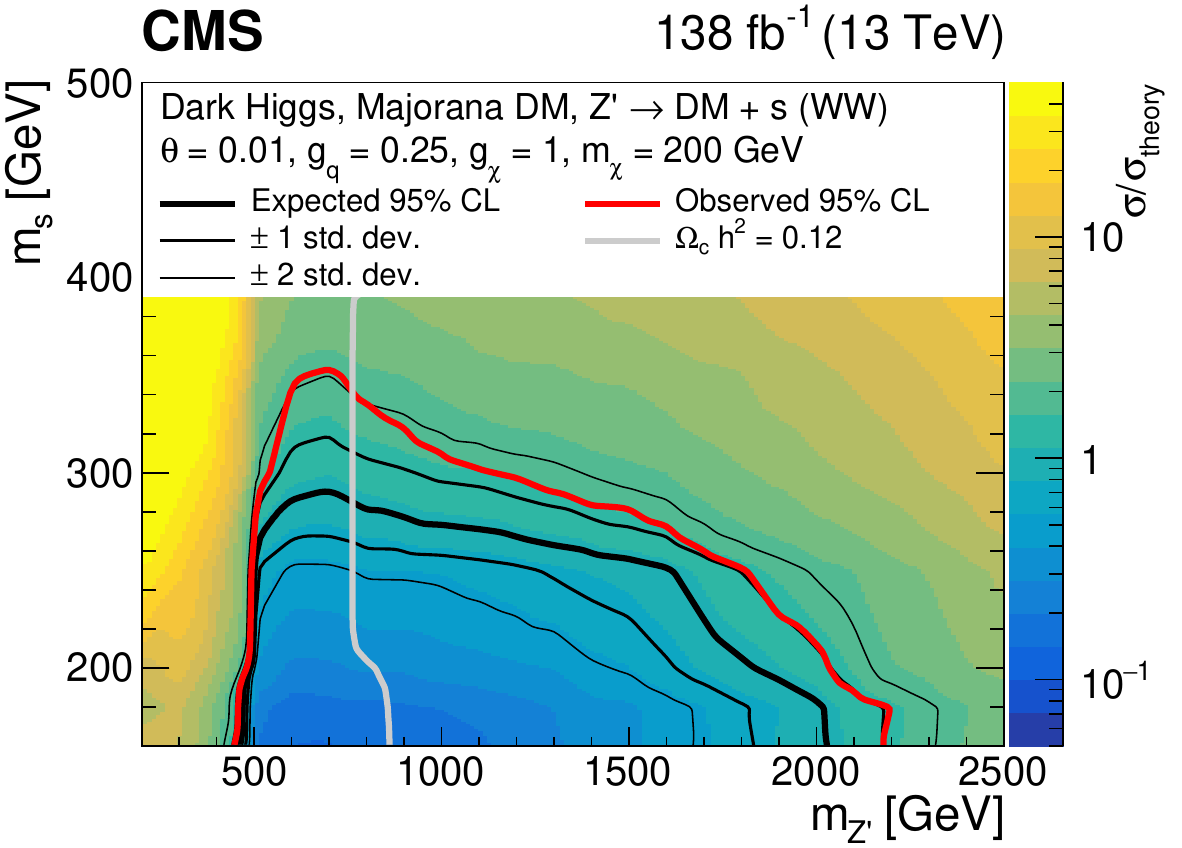}}
    \caption{Observed (red lines) and expected (black lines) 95\% \CL exclusion limits for the dark-Higgs boson model in terms of \mHdark (written as s in the figure) and \mZprime for $\mDM=150\GeV$ (upper) and 200\GeV (lower) (where \mDM is written as $m_{\chi}$ and \gDM is written as $g_{\chi}$ in the figure). The gray line indicates where the model parameters produce exactly the observed thermal-relic density. In the upper plot, the area above the upper gray line and below the lower gray line is excluded. In the lower plot, the area to the right of the gray line is excluded. Figure taken from Ref.~\cite{CMS:2023dof}.}
    \label{fig:darkhiggsww}
\end{figure}

\cmsParagraph{Dark-Higgs boson portal\label{par:darkHiggsPortalLimits}} Exclusion limits at 95\% \CL are set on DM production in the context of the $\PZpr\Hdark$ model, where the dark Higgs boson radiates from a \PZpr mediator and \mHdark is above the $\PW\PW$ mass threshold. They are presented in Fig.~\ref{fig:darkhiggsww} for different values of $\mDM$. The most stringent limits are obtained for $\mDM = 200\GeV$, excluding dark-Higgs boson masses up to $\approx$350\GeV at \mZprime masses of 700\GeV, and up to $\mZprime \approx 2200\GeV$ for $\mHdark = 160\GeV$.

As shown in the \cmsLeft plot of Fig.~\ref{fig:darkhiggsww}, which shows the exclusion boundaries for $\mDM=150\GeV$, the sensitivity sharply drops for the case that $\mHdark>2\mDM$, because then \Hdark predominantly decays to two DM particles and not to a pair of \PW bosons.

\begin{figure}[thb!]
    \centering
    \includegraphics[width=0.7\textwidth]{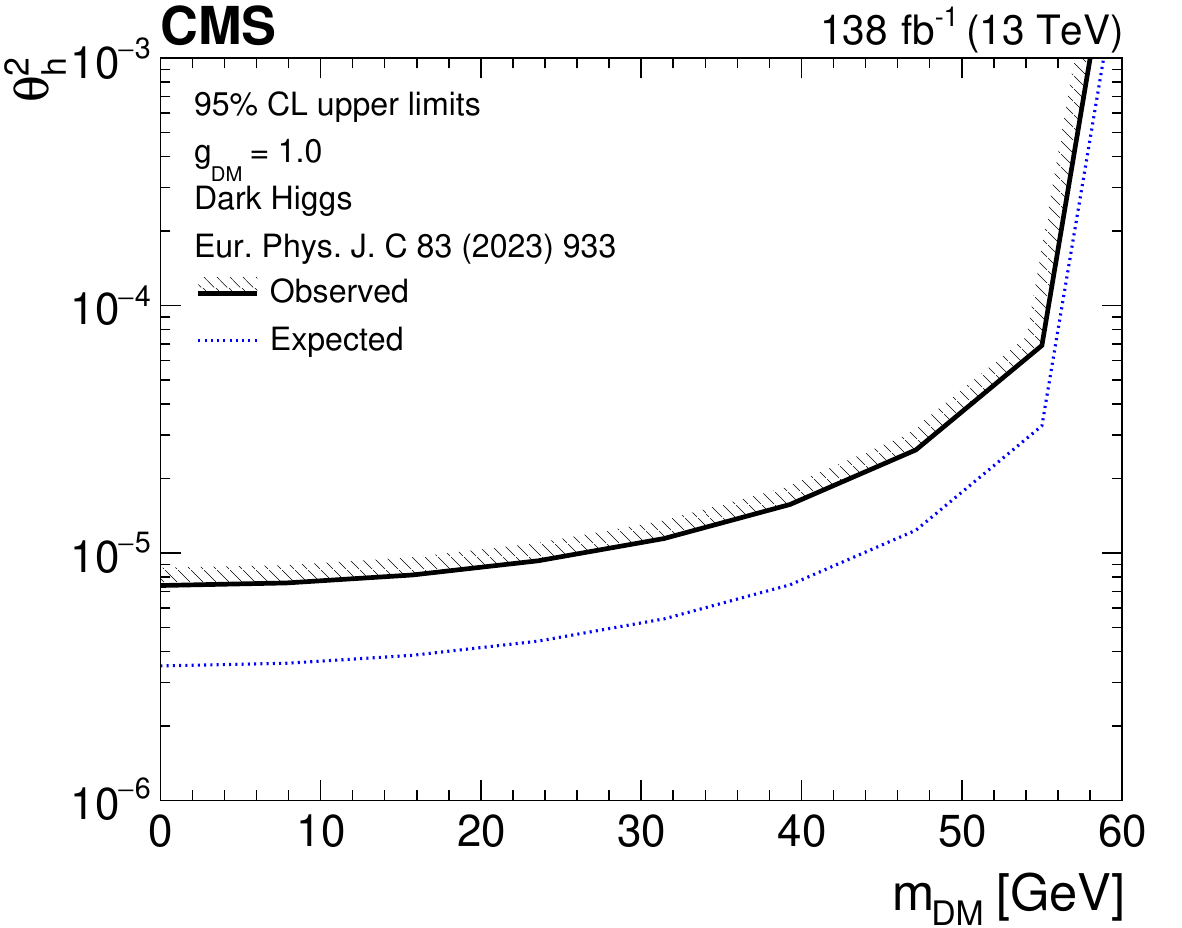}
    \caption{95\% CL upper limits on the mixing parameter $\thetaH^{2}$ from the \hinv analysis~\cite{CMS:2023sdw} (Section~\ref{sec:higgs_invisible}) interpreted with a dark-Higgs boson model.}
    \label{fig:darkhiggsthetaH}
\end{figure}

The limits on \brhinv can be reinterpreted to place limits on the mixing parameter \thetaH, as shown in Fig.~\ref{fig:darkhiggsthetaH}. The exclusion worsens as \mDM approaches $m_{\PH}/2$. These results are largely independent of the dark-Higgs boson mass. In these limits, the predicted thermal-relic density for the prescribed couplings and fermionic DM yields an overabundance of DM by several orders of magnitude~\cite{Krnjaic:2015mbs}.

\begin{figure}[htbp!]
    \centering
    \includegraphics[width=0.47\textwidth]{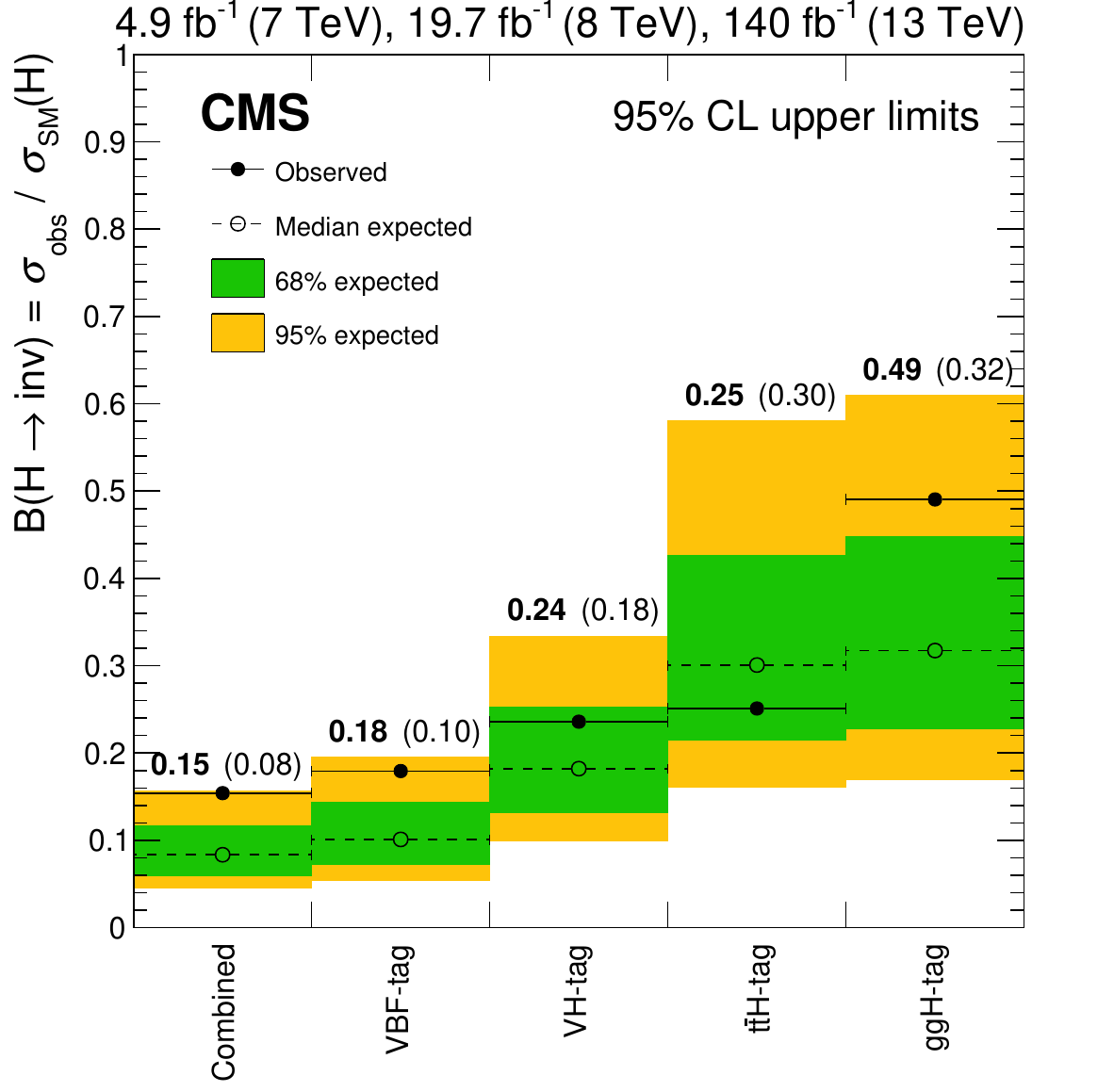}
    \raisebox{0.25em}{\includegraphics[width=0.49\textwidth]{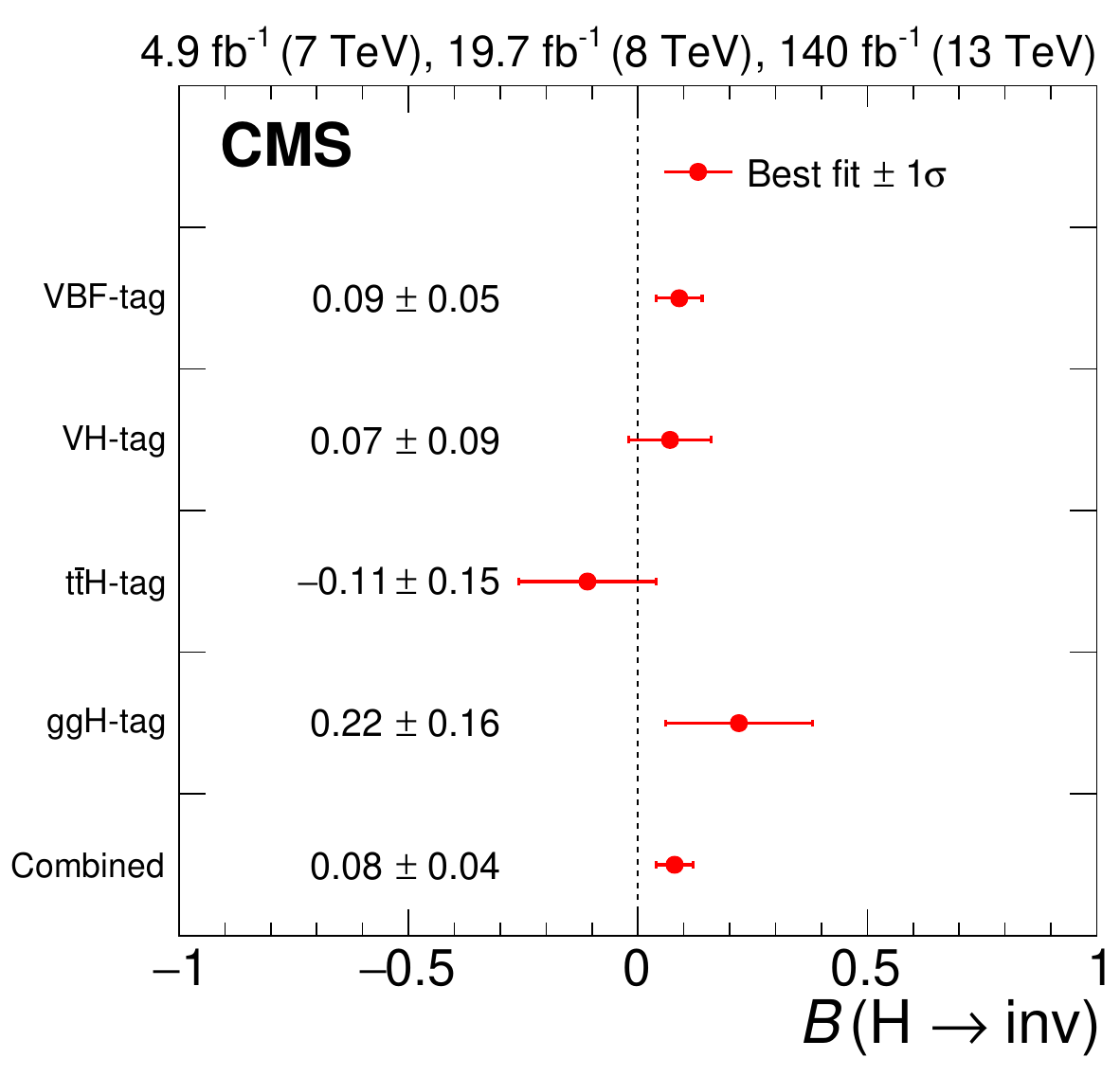}}
    \caption{Results on \brhinv, shown separately for each Higgs boson production mode as tagged by the input analyses, as well as combined across modes. Left: observed and expected upper limits on \brhinv at 95\% \CL. Right: best-fit estimates of \brhinv. Figure adapted from Ref.~\cite{CMS:2023sdw}.}
    \label{fig:higgs_portal_comb}
\end{figure}

\cmsParagraph{Higgs boson portal\label{sec:higgsPortalResults}} The individual 95\% \CL limits on \brhinv are reported in Table~\ref{tab:llscan_vals_comb} and are presented in Fig.~\ref{fig:higgs_portal_comb}. The VBF category drives the upper limit on \brhinv because of the sizable VBF production cross section and a large signal selection efficiency. The combined 95\% \CL upper limit on \brhinv of 0.15 (0.08 expected) is obtained using Run~1 (2011--2012) and Run~2 (2015--2018) data. 

\begin{table}[htb]
\centering
\topcaption{The observed best-fit estimates of \brhinv, for each analysis channel in the combination, and the 95\% \CL observed and expected (exp) upper limits on \brhinv. Table adapted from Ref.~\cite{CMS:2023sdw}.}
\begin{tabular}{lcc}
Channel & Best-fit \brhinv & Upper limits on \brhinv at 95\% \CL \\\hline
Combined & $\hphantom{-}0.08\pm0.04$ & 0.15 (0.08 exp) \\
\vbf-tag & $\hphantom{-}0.09\pm0.05$ & 0.18 (0.10 exp) \\
\vh-tag & $\hphantom{-}0.07\pm0.09$ & 0.24 (0.18 exp) \\
\ttbarh-tag & $-0.11\pm0.15$ & 0.25 (0.30 exp) \\
\ggH-tag & $\hphantom{-}0.22\pm0.16$ & 0.49 (0.32 exp) \\
\end{tabular}
\label{tab:llscan_vals_comb}
\end{table}

\begin{figure}[htbp!]
    \centering
    \includegraphics[width=0.6\textwidth]{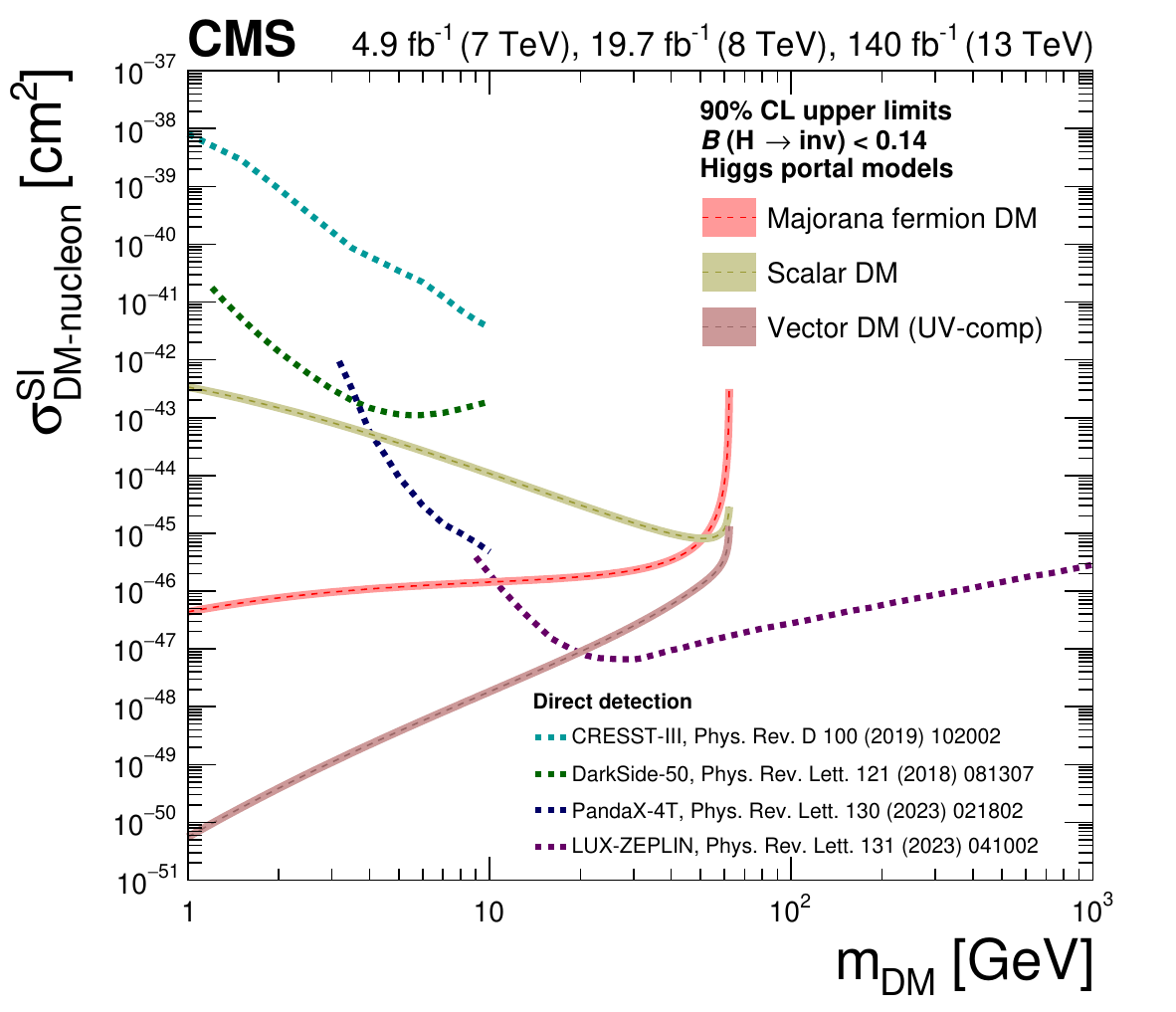}
    \caption{Translation of the exclusion limits on \brhinv into  90\% \CL upper limits on the spin-independent DM-nucleon scattering cross section~\cite{CMS:2023sdw}, and comparison with results from the CRESST-III~\cite{CRESST:2019jnq}, DarkSide-50~\cite{DarkSide-50:2022qzh}, PandaX-4T~\cite{PandaX:2022aac}, and LUX-ZEPLIN~\cite{LZ:2022lsv}
    experiments. Figure adapted from Ref.~\cite{CMS:2023sdw}.}
    \label{fig:spin_independent_DM_cross_section}
\end{figure}

Searches for DM at DD experiments can be interpreted in terms of Higgs portal models, assuming the DM particle interacts with an atomic nucleus via the exchange of a Higgs boson. We compare the sensitivity of the CMS search for invisible Higgs boson decays with the sensitivity of DM searches at DD experiments in Fig.~\ref{fig:spin_independent_DM_cross_section}. An ultraviolet-complete model~\cite{Baek:2012se} is considered for vector DM in addition to the EFT-based fermionic (Majorana) and scalar DM scenarios. Based on these assumptions, and the assumptions within DD about the local DM density,  our collider-based search sensitivity complements that of DD experiments for DM masses of a few {\GeVns}.

\begin{figure}[htbp!]
    \centering
    \includegraphics[width=0.7\textwidth]{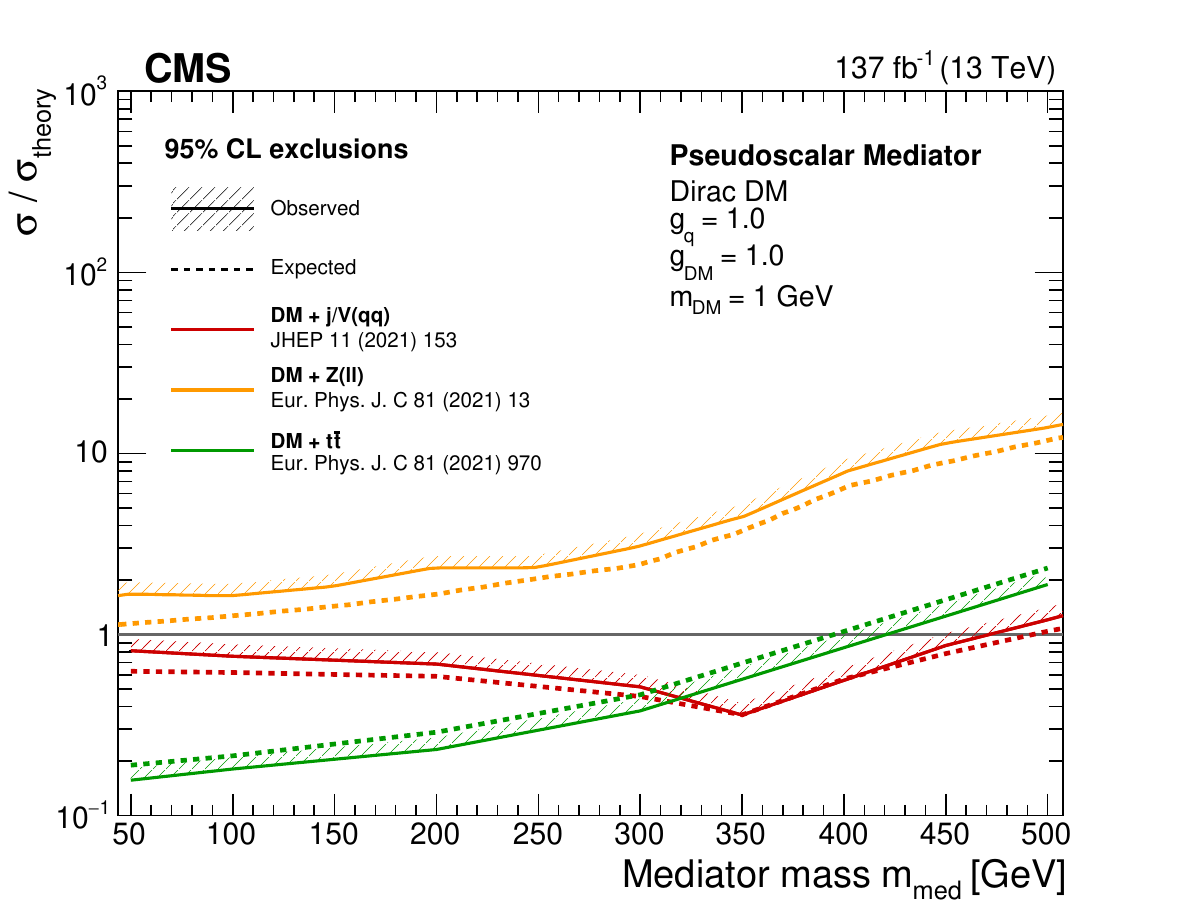}
    \caption{Observed (solid lines) and expected (dashed lines) 95\% \CL exclusion limits for the pseudoscalar model in terms of \mMed for different \ptmiss-based DM searches from CMS~\cite{CMS:2021far,CMS:2020ulv,CMS:2021eha}. The hashed areas indicate the direction of the excluded area from the observed limits. Following the recommendation of the LHC DM Working Group~\cite{Boveia:2016mrp,Albert:2017onk}, the exclusions are computed for a universal quark coupling of $\gq = 1.0$ and for a DM coupling of $\gDM = 1.0$. The exclusion away from $\sigma / \sigma_{\text{theory}} =1$ only applies to coupling combinations that yield the same kinematic distributions as the benchmark model considered here.}
    \label{fig:summary_pseudo}
\end{figure}

\cmsParagraph{Pseudoscalar portal} A summary table and a plot for 95\% \CL observed exclusion limits on \mMed for different \ptmiss-based DM searches from CMS in the pseudoscalar model are presented in Table~\ref{tab:summary_pscalar} and Fig.~\ref{fig:summary_pseudo}. Due to enhanced cross sections in the pseudoscalar case, the monojet search is more sensitive for scalar mediators and excludes mediator masses smaller than 0.47\TeV. The same reasoning in terms of production cross section as \mMed approaches $2m_\PQt$ applies as for the scalar case, which is why the most stringent limit is found for $\mMed=\mA\approx350\GeV$. The search for {\ttbar}+\ptmiss excludes pseudoscalar mediators lighter than 0.42\TeV. These exclusion limits are obtained under the assumption that $\gDM=\gq=1.0$.

\begin{table}[htbp]
\centering
\topcaption{Summary of 95\% \CL observed exclusion limits on $\mMed = \mA$ for \mbox{\ptmiss-based} DM searches from CMS in the pseudoscalar model. Following the recommendation of the LHC DM Working Group~\cite{Boveia:2016mrp,Albert:2017onk}, the exclusions are computed for a universal quark coupling of $\gq = 1.0$ and for a DM coupling of $\gDM = 1.0$. Each search listed here used data corresponding to $\Lint=137\fbinv$.}
\renewcommand{\arraystretch}{1.2}
\begin{tabular}{lccl}
\multirow{2}{*}{Reference} & \multirow{2}{*}{Channel} & 95\% \CL lower limit & \multirow{2}{*}{Notes} \\
 & & on \mA [{\TeVns}] & \\
\hline
\cite{CMS:2021far} & Monojet        &   $0.47$    & \\
\cite{CMS:2020ulv} & Mono-\PZ       &   \NA              &
Excludes $\sigma/\sigma_{\text{theory}} = 1.65$ for $\mA = 100\GeV$
\\
\cite{CMS:2021eha} & {\ttbar}+\ptmiss    &   $0.42$    & \\
\label{tab:summary_pscalar}
\end{tabular}
\end{table}

\begin{figure}[htbp!]
    \centering
    \includegraphics[width=0.49\textwidth]{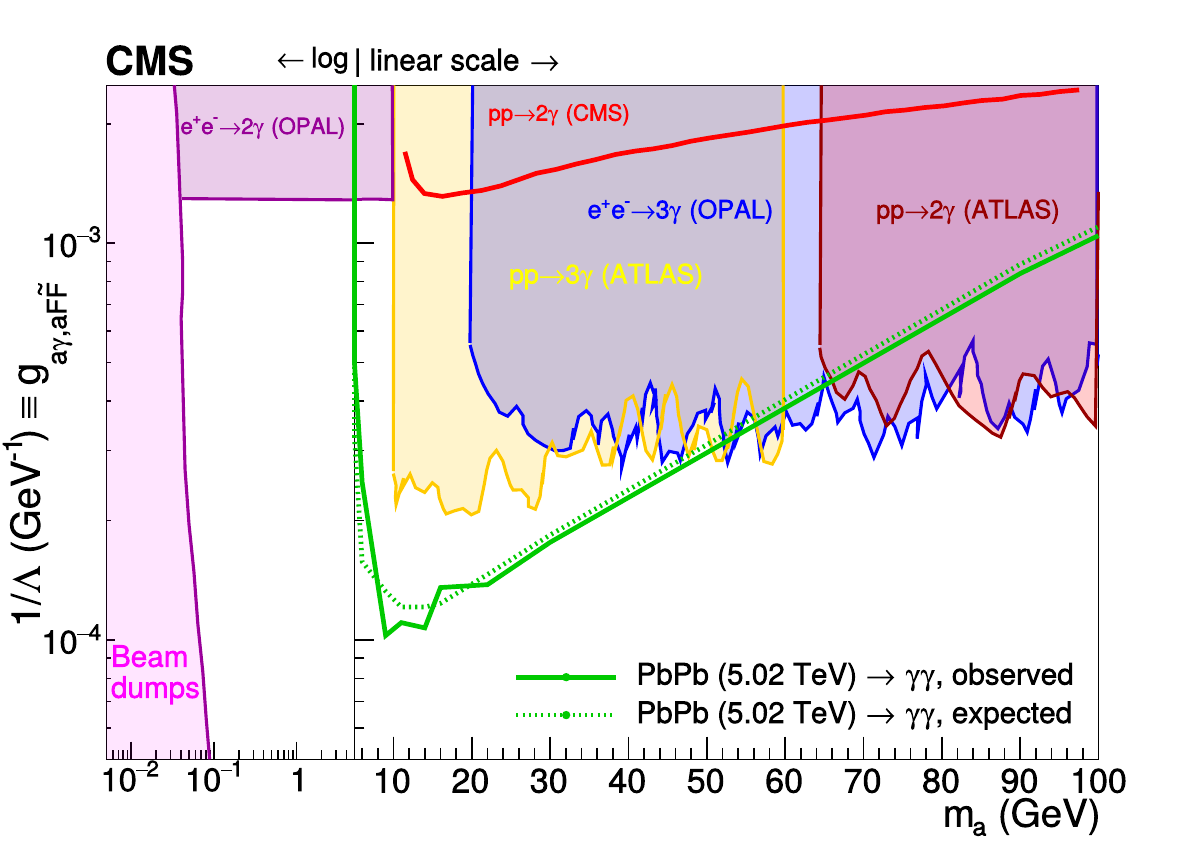}
    \includegraphics[width=0.49\textwidth]{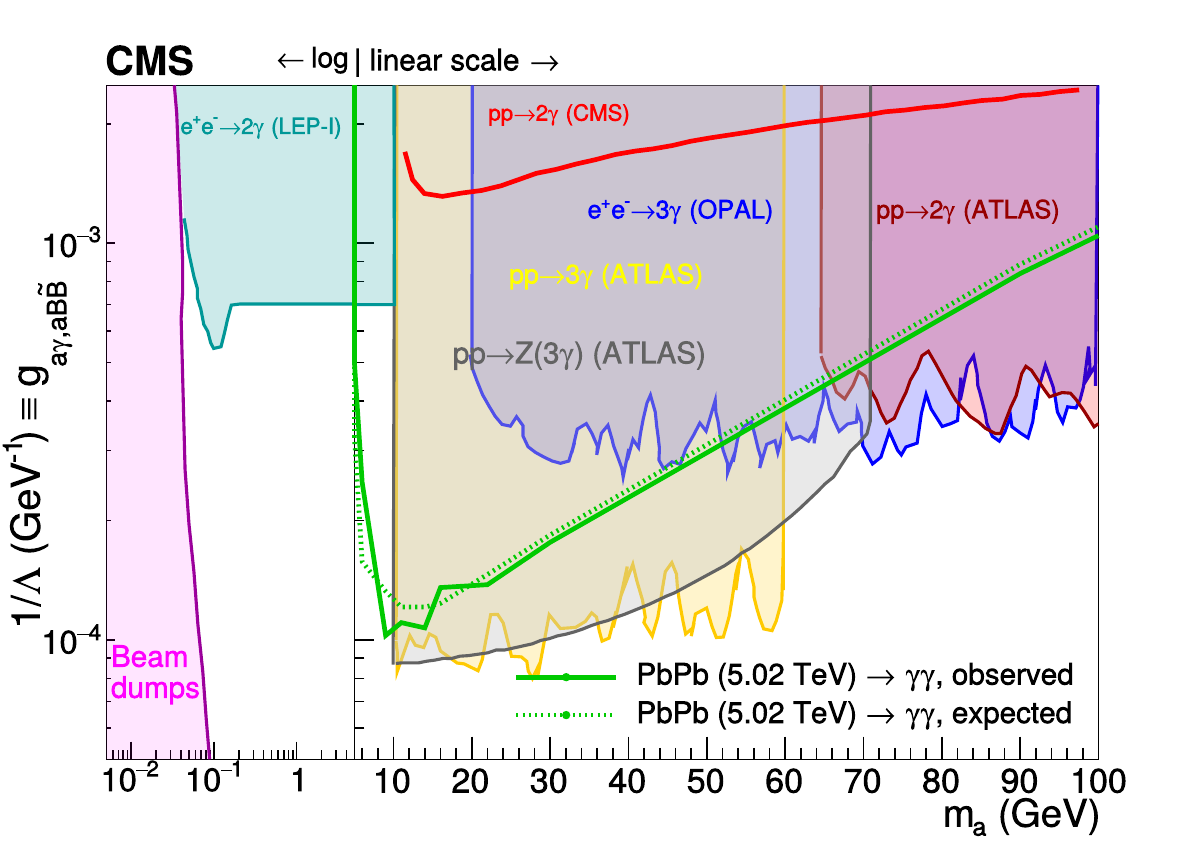}
    \caption{Observed (solid green lines) and expected (dashed green lines) exclusions at 95\% \CL in the ALP-photon coupling versus ALP mass plane, assuming ALP coupling to photons only (\cmsLeft) and including also the hypercharge coupling and thus processes involving the \PZ boson (\cmsRight). The limits shown are derived in Refs.~\cite{Knapen:2016moh,limits_lep} from measurements at beam dumps~\cite{Dobrich:2015jyk}, in electron-positron collisions at LEP-I~\cite{limits_lep} and LEP-II~\cite{limits_opal}, and in $\Pp\Pp$ collisions at the LHC~\cite{Chatrchyan:2012tv,limits_atlas_2gamma,limits_atlas_3gamma}, and they are compared to the PbPb limits from Ref.~\cite{FSQ-16-012}. Figures taken from Ref.~\cite{FSQ-16-012}.}
    \label{fig:ALPresults}
\end{figure}

\cmsParagraph{Axion-like particle portal\label{sec:ALPportal}} The CMS Collaboration has searched for ALPs (Section~\ref{subsec:ALPs}) that couple to photons in PbPb UPCs~\cite{FSQ-16-012}, as described in Section~\ref{sec:FSQ-16-012}. The results are shown in Fig.~\ref{fig:ALPresults}. Two scenarios are considered where the ALP couples to photons alone or also to hypercharge. Constraints on the ALP coupling to the photons only, assuming $\BR(\Pa\to\PGg\PGg)=100\%$, are the most stringent CMS exclusion limits so far when $m_{\Pa}$ is in the 5--50\GeV range.

\subsubsection{Fermion portal}

\begin{figure}[htbp!]
    \centering
    \includegraphics[width=0.85\textwidth]{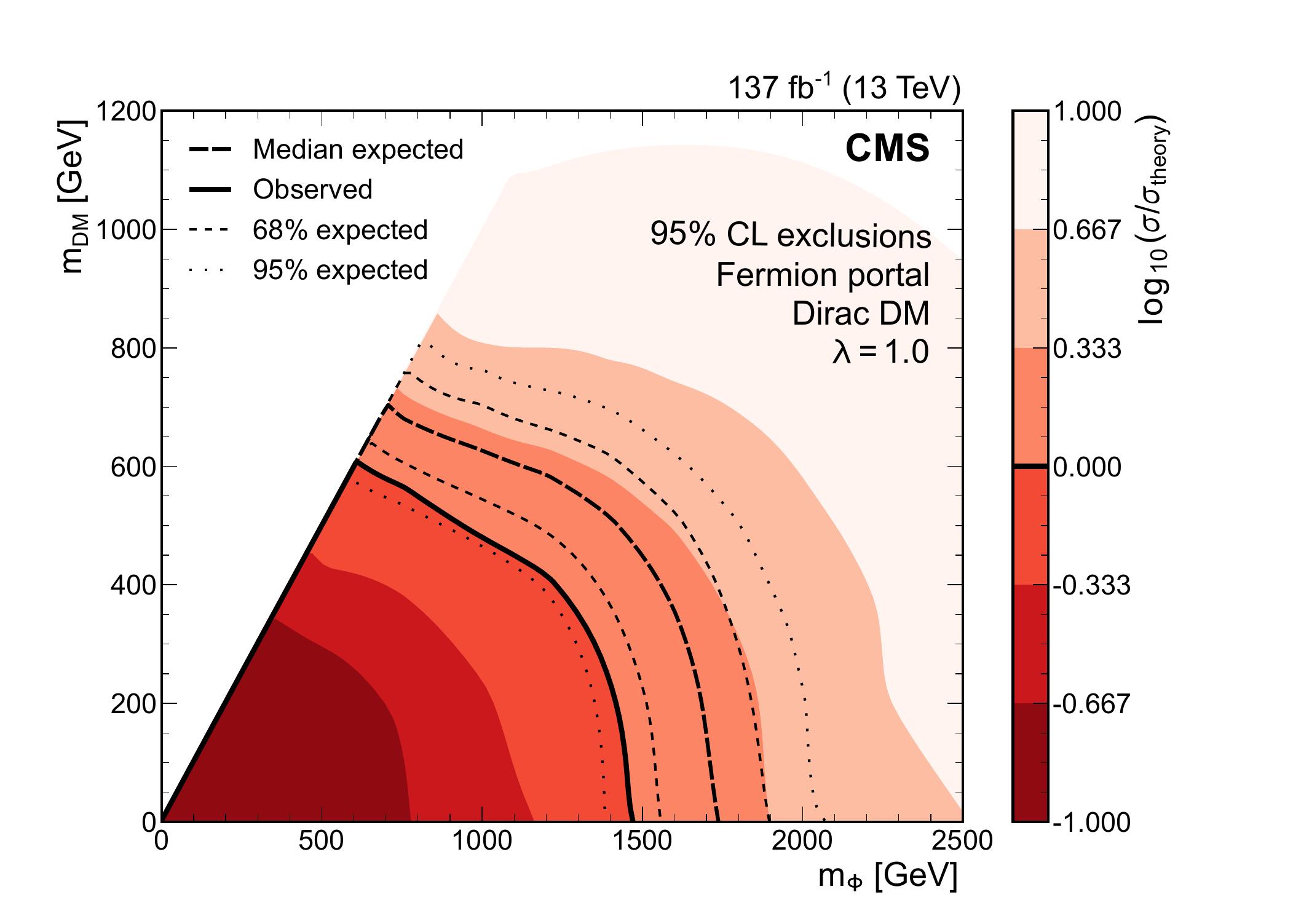}
    \caption{Observed (solid line) and expected (dashed lines) exclusions at 95\% \CL in the $m_{\Pbifun}$-\mDM plane for the fermion portal model scenario obtained from the monojet search performed using data collected in 2016--2018. Figure adapted from Ref.~\cite{CMS:2021far}.}
    \label{fig:monojet_fermionportal}
\end{figure}

Figure~\ref{fig:monojet_fermionportal} presents 95\% \CL limits for the fermion portal model, obtained from the monojet search~\cite{CMS:2021far}. In the specific model probed, the mediator \Pbifun couples to DM particles and right-handed \PQu quarks with coupling strength $\lambda=1$. Exclusions are presented in terms of the DM mass and the mass of the mediator. We note that resonant squark searches can hypothetically outperform monojet searches here, but such a search with CMS data has not been performed recently.

\subsection{Extended dark sectors}

\subsubsection{The 2HDM+a scenario}
\label{sec:2hdma}
\begin{figure}[htbp!]
    \centering
    \includegraphics[width=0.85\textwidth]{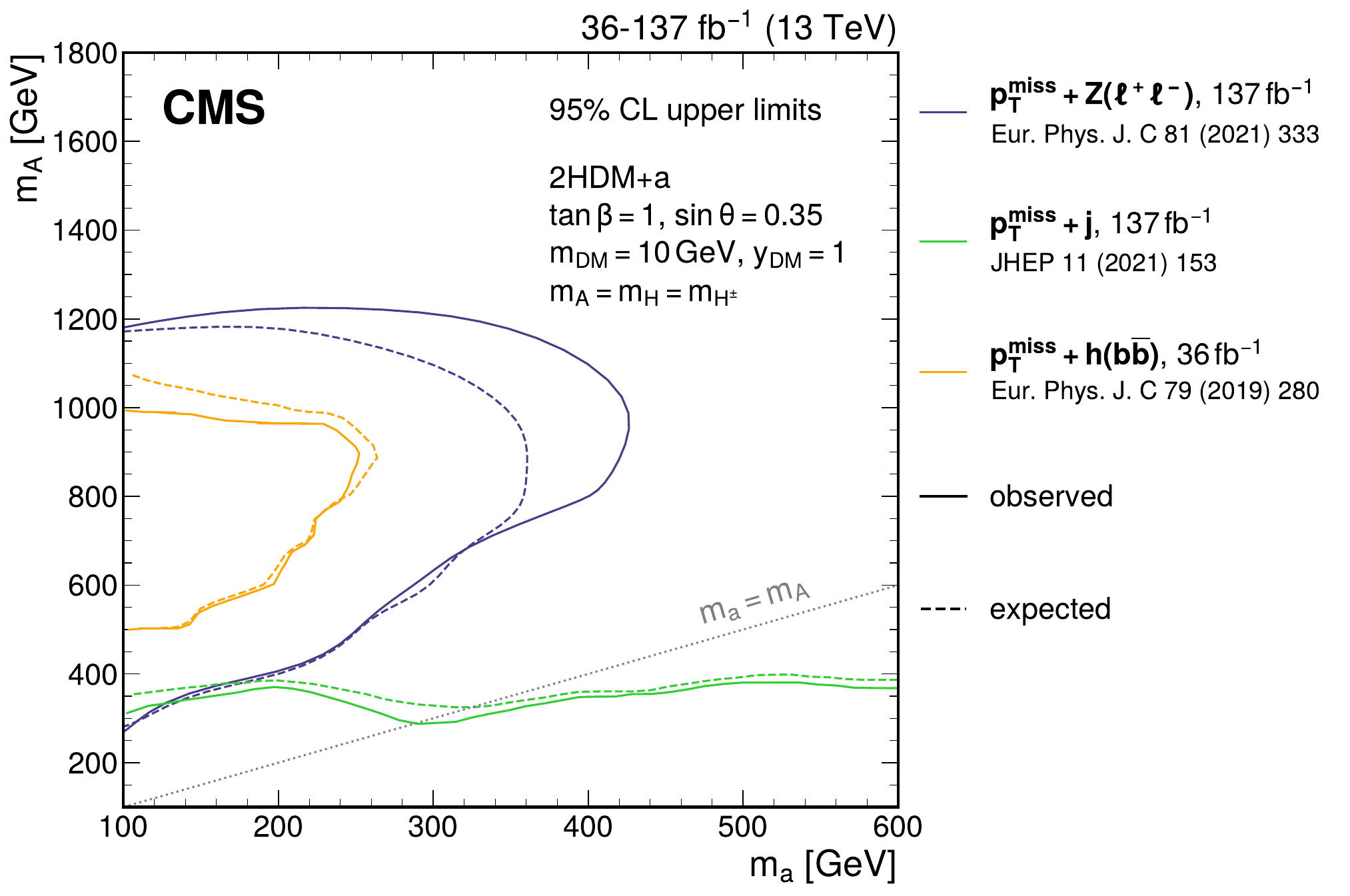}
    \caption{Observed (solid lines) and expected (dashed lines) exclusion regions at 95\% \CL in the $m_{\Pa}$-$m_{\PA}$ plane for the 2HDM+a scenario arising from various ``mono-X'' searches performed using data collected in 2016--2018~\cite{CMS:2020ulv,CMS:2021far,CMS:2018zjv}. Following the recommendation of the LHC DM Working Group~\cite{Boveia:2016mrp,Albert:2017onk}, the projection is performed for values of the other parameters as follows: $m_{\PH}=m_{\PA}=m_{\PH^{\pm}}$, $\sin\theta=0.35$, $\tan\beta=1$, $\mDM=10\GeV$, and $\yDM=1$.}
    \label{fig:summary_2hdma_mamA}
\end{figure}

This section presents results interpreted in the \mbox{2HDM+a}, as described in Section~\ref{sec:2HDM}. A summary table and a plot for the 95\% \CL observed exclusion limits in the $m_{\Pa}$-$m_{\PA}$ plane for different \ptmiss-based DM searches from CMS are presented in Table~\ref{tab:summary_2hdm} and Fig.~\ref{fig:summary_2hdma_mamA}, respectively. From the figure it can be seen that the mono-\PZ analysis sets exclusion limits that depend on the ratio of the pseudoscalar masses $m_{\PA}/m_{\Pa}$; this is because the process is dominated by resonant production of the heavy scalar $\PH$ and subsequent decay $\PH \to \PZ \Pa$; an analogous situation occurs in the mono-\PH analysis, with the $\PA \to \PH \Pa$ channel being dominant instead. On the other hand, the exclusion limit set by the monojet analysis is almost independent of $m_{\PA}$; this is because in this case the process reduces to the simplified model case with \ptmiss+ISR, and the heavier pseudoscalar plays essentially no role.

\begin{table}[htbp]
\centering
\topcaption{Summary of 95\% \CL observed exclusion limits in the heavy pseudoscalar mass $\mA$ for \mbox{\ptmiss-based} DM searches from CMS in the 2HDM+a scenario. Following the recommendation of the LHC DM Working Group~\cite{Boveia:2016mrp,Albert:2017onk}, the projection is performed for values of the other parameters as follows: $m_{\PH}=m_{\PA}=m_{\PH^{\pm}}$, $\sin\theta=0.35$, $\tan\beta=1$, $\mDM=10\GeV$, and $\yDM=1$. Each search listed here used data corresponding to \Lint=137\fbinv.}
\renewcommand{\arraystretch}{1.2}
\begin{tabular}{lcc}
Reference & Channel & 95\% \CL lower limit on \mA [{\TeVns}]  \Bstrut \\
\hline
\cite{CMS:2020ulv} & Mono-\PZ   &   $1.2\hphantom{5} $ \\
\cite{CMS:2021far} & Monojet    &   $0.39$ \\
\cite{CMS:2018zjv} & Mono-\PH  &   $1.0\hphantom{5}$  \\
\label{tab:summary_2hdm}
\end{tabular}
\end{table}

\begin{figure}[htbp!]
    \centering
    \includegraphics[width=0.85\textwidth]{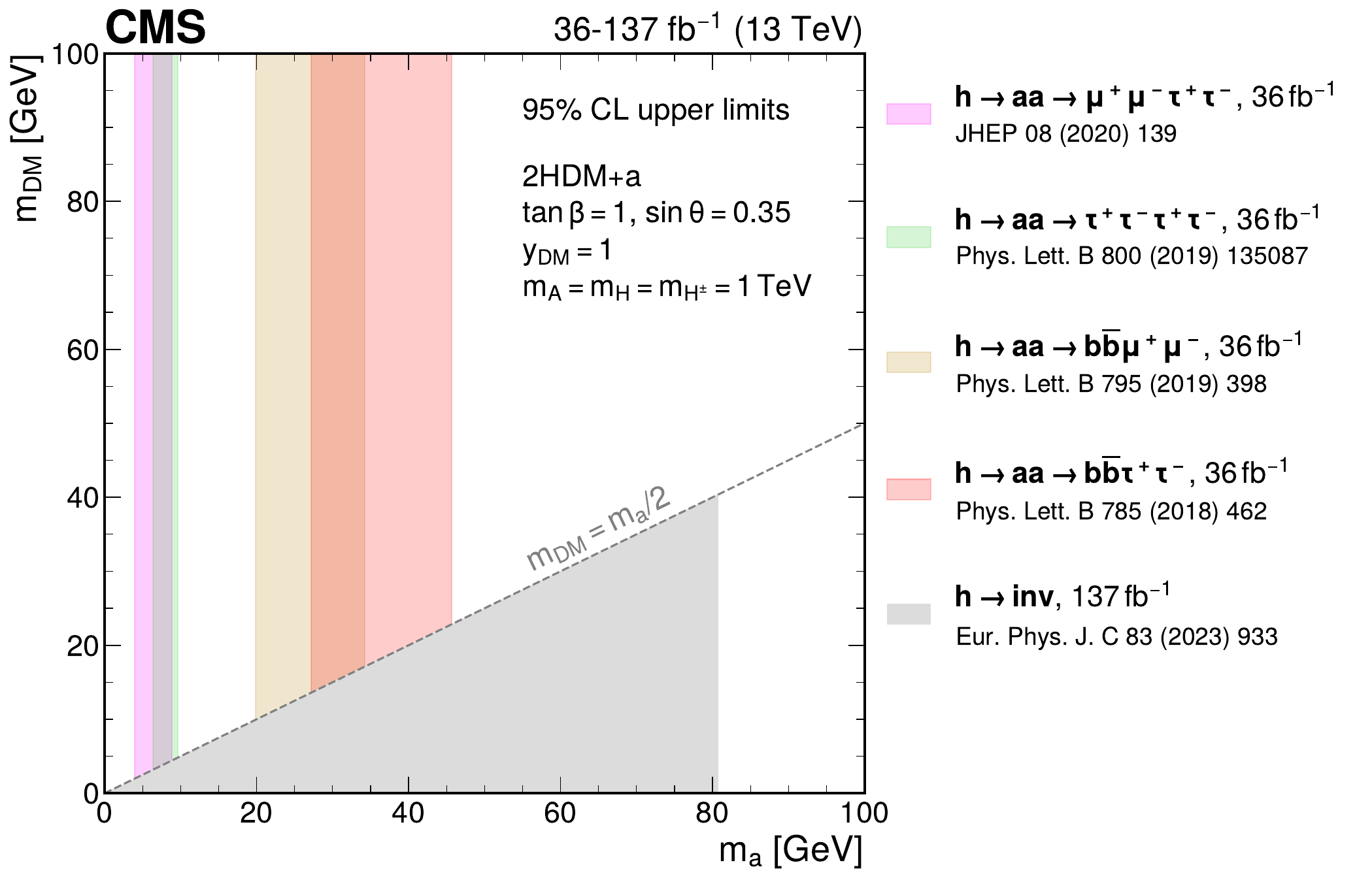}
    \caption{Exclusion regions at 95\% \CL in the $m_{\Pa}$-\mDM plane for the 2HDM+a scenario arising from searches for exotic and invisible decays of the 125\GeV Higgs boson performed using data collected in 2016--2018~\cite{CMS:2020ffa,CMS:2019spf,CMS:2018nsh,CMS:2018zvv,CMS:2023sdw}. Following the recommendation of the LHC DM Working Group~\cite{Boveia:2016mrp,Albert:2017onk}, the projection is performed for values of the other parameters as follows: $m_{\PH}=m_{\PA}=m_{\PH^{\pm}}=1\TeV$, $\sin\theta=0.35$, $\tan\beta=1$, and $\yDM=1$. The branching fractions of the pseudoscalar boson to SM and DM particles are computed using the \textsc{MadWidth}~\cite{Alwall:2014bza} functionality within \MGvATNLO.
    }
    \label{fig:summary_h-aa-2hdma}
\end{figure}

Figure~\ref{fig:summary_h-aa-2hdma} summarizes searches for the \mbox{2HDM+a} scenario that approach the problem from the viewpoint of exotic decays of the 125\GeV Higgs boson instead. If the $\Pa \to \PDM\PDM$ decay is not kinematically allowed, searches for the visible products of the $\PH \to \Pa\Pa$ process are the most stringent. Otherwise, the interpretation of the Higgs boson invisible decay limits in terms of the \mbox{2HDM+a} scenario gives the strongest limits.

\subsubsection{Hidden Abelian Higgs model and supersymmetry}

\begin{figure}[htbp!]
    \centering
    \includegraphics[width=0.7\linewidth]{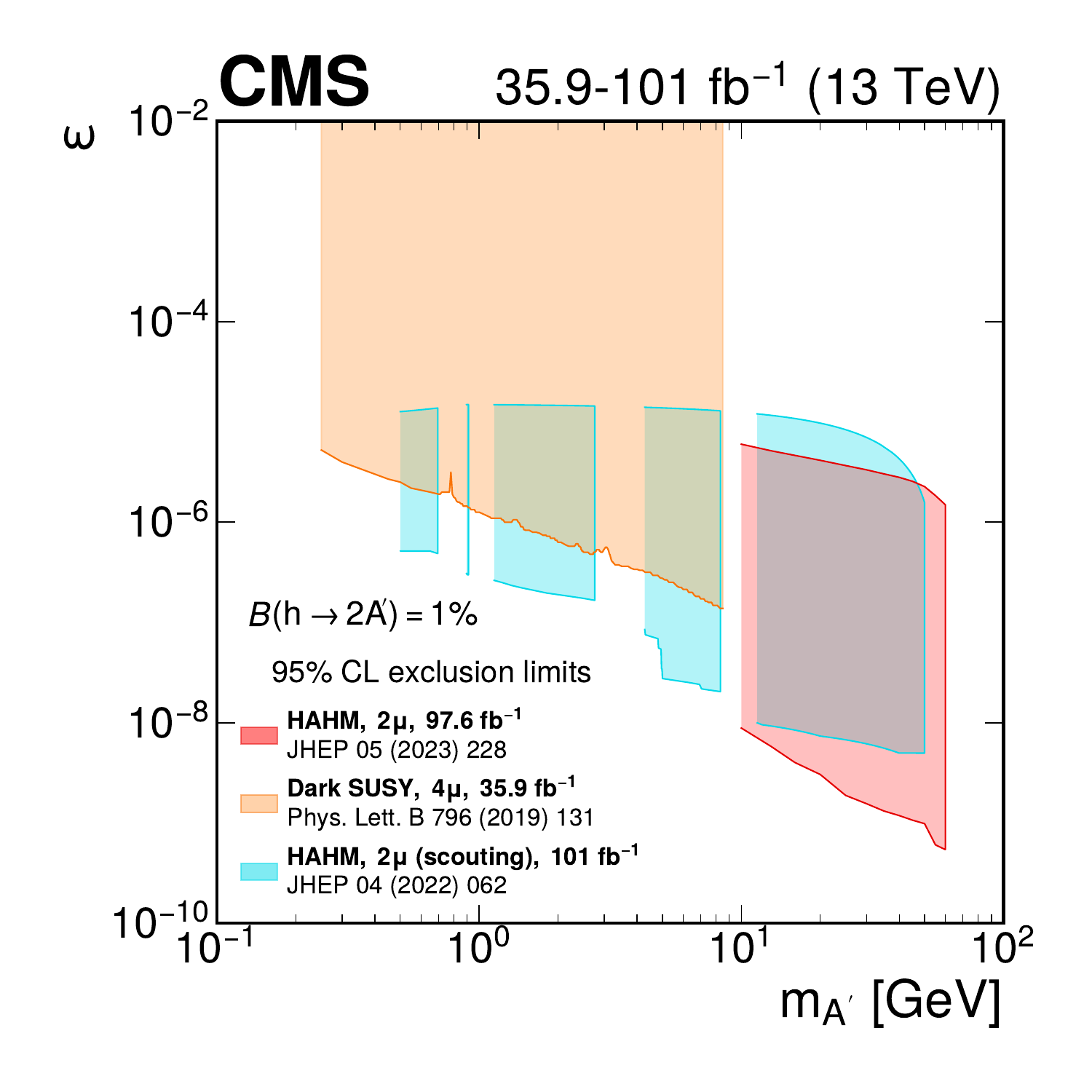}
    \caption{Observed 95\% \CL exclusion contours in the plane defined by the kinetic mixing parameter ($\epsilon$) and the mass of the new dark boson. A summary of Run~2 CMS searches that target displaced signatures is presented. Two of those searches, namely Refs.~\cite{EXO-21-006} (red) and \cite{EXO-20-014} (blue), consider the HAHM signal and use a final state with at least two muons $(2\Pgm+X)$, and the latter one uses data scouting. The third search (orange)~\cite{CMS-HIG-18-003} uses a final state with at least four muons $(4\Pgm+X)$ and a dark SUSY signal scenario.}
    \label{fig:CMS_Longlived_DarSector}
\end{figure}

\cmsParagraph{Hidden Abelian Higgs model and dark supersymmetry \label{sec:darkSUSYHAHM}} Results interpreted in the HAHM and a dark SUSY scenario, as described in Sections~\ref{sec:HAHM} and \ref{sec:DarkSUSY}, respectively, are presented in this section. Figure~\ref{fig:CMS_Longlived_DarSector} shows a summary of dark boson results for analyses that target LLPs, in contrast to the dark photon summary with prompt analyses shown in Section~\ref{sec:darkPhotonPortalResults}. Three analyses are covered in this figure. The first is a search for displaced dimuons~\cite{EXO-21-006} with a HAHM signal benchmark (Section~\ref{sec:HAHM}). The second analysis, which uses the same benchmark model, is a search for displaced dimuon resonances with data scouting~\cite{EXO-20-014}. The third search evaluates the CMS sensitivity to prompt and displaced dimuons in final states with $4\Pgm+X$ in the context of a dark SUSY signal scenario (Section~\ref{sec:DarkSUSY})~\cite{CMS-HIG-18-003}. For all three searches, $\BR(\Ph\to2\PAprime)=1\%$ is assumed. The $\Lint$ used for each analysis varies depending on the available triggers and data sets at that time.

\begin{figure*}[htbp!]
\centering
\includegraphics[width=0.8\linewidth]{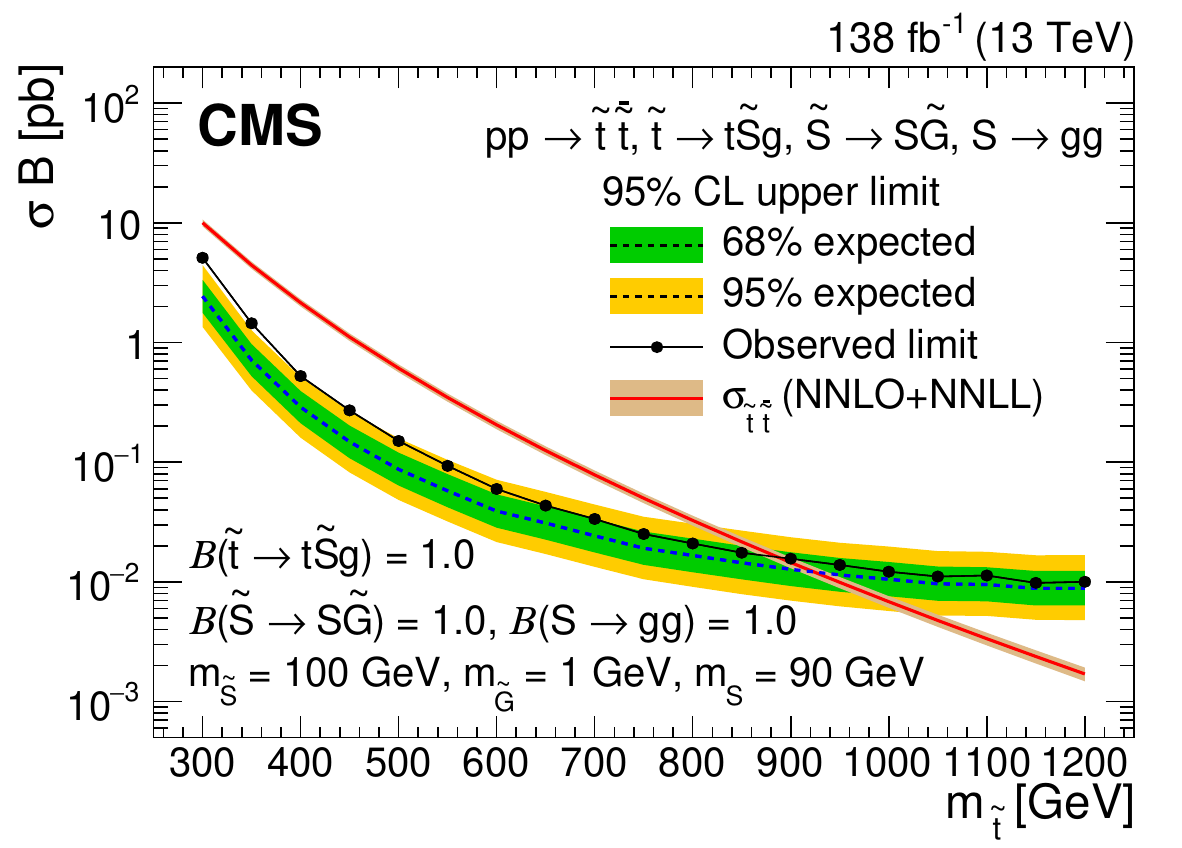}
\includegraphics[width=0.8\linewidth]{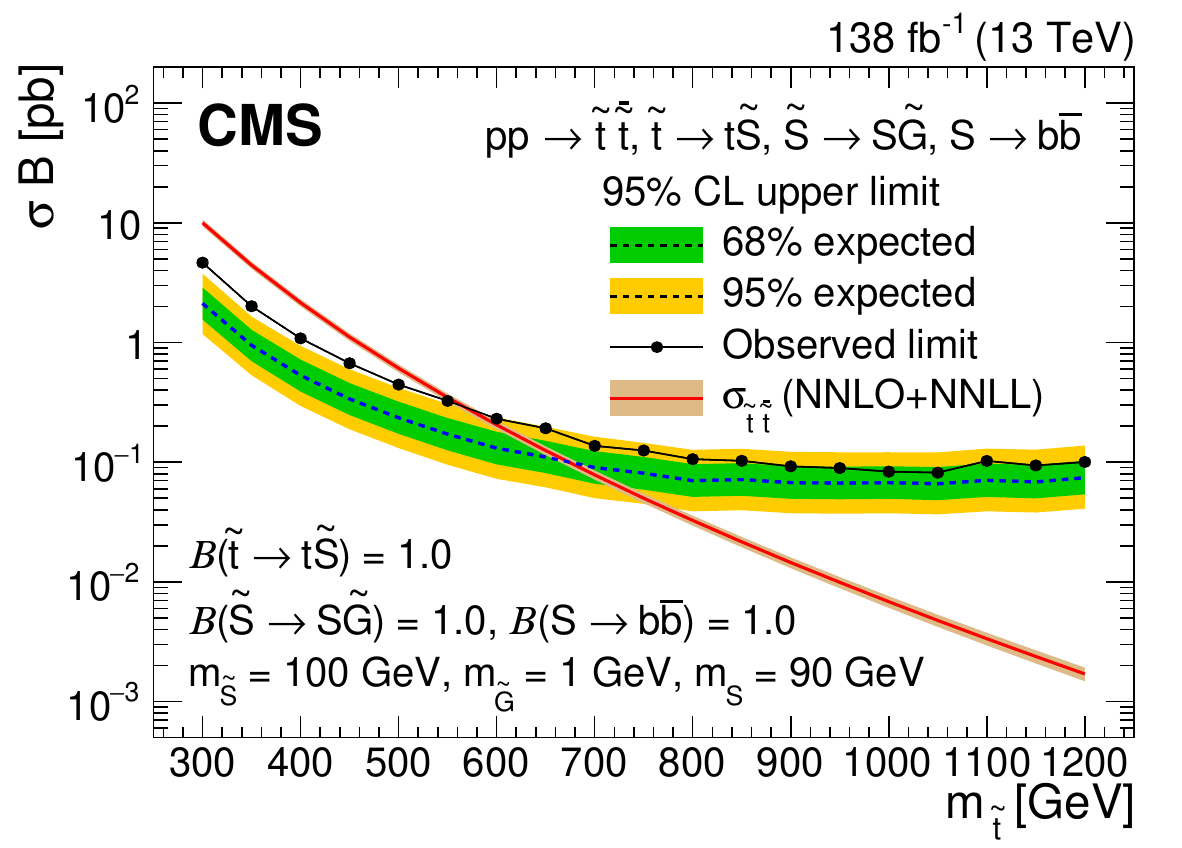}
\caption{Expected and observed 95\% \CL upper limit on the product of the top squark pair production cross section and branching fraction in terms of the top squark mass for the stealth SYY SUSY model (upper) and stealth SHH SUSY model (lower). Particle masses and branching fractions assumed for the model are included. The expected cross section is computed at NNLO accuracy, improved by using the summation of soft gluons at next-to-next-to-leading logarithmic (NNLL) order, and is shown in the red curve. Upper figure adapted from Ref.~\cite{SUS-19-004}.}
\label{fig:stealth}
\end{figure*}

\cmsParagraph{Stealth supersymmetry\label{sec:sensStealth}} Stealth SUSY models are detailed in Section~\ref{sec:theory_stealth}. The stealth SUSY search described in Section~\ref{sec:signatures_stealth} targets top squark pair production with decays via the stealth vector portal, and the limits on this model are shown in the upper plot in Fig.~\ref{fig:stealth}. Other portals such as a Higgs portal are also possible. The main difference between these two scenarios is that the six gluons in the stealth vector portal are replaced by four \PQb quarks in the Higgs portal, resulting in a reduction of the number of jets in the event. However, the Higgs portal still features many more jets, as shown in Fig.~\ref{fig:stealth_diagrams}, than the dominant \ttbar background, and thus, sensitivity to this model is still expected for this search.

The lower plot in Fig.~\ref{fig:stealth} shows the expected and observed 95\% \CL upper limit on the product of the top squark pair production cross section and branching fraction via the Higgs portal in terms of the top squark mass. The branching fractions are assumed to be 100\% for the chosen decay chain: $\sTop \to \PQt \PSS$, $\PSS\to \PS \sGra$, and $\PS\to \bbbar$. The observed (expected) mass exclusion is found to be 570 (670)\GeV, compared to 870 (920)\GeV for the vector portal. The sensitivity can be improved by explicitly taking advantage of the additional \PQb quarks expected from decays via the Higgs portal.

\begin{figure*}[htbp!]
\centering
\includegraphics[width=0.49\linewidth]{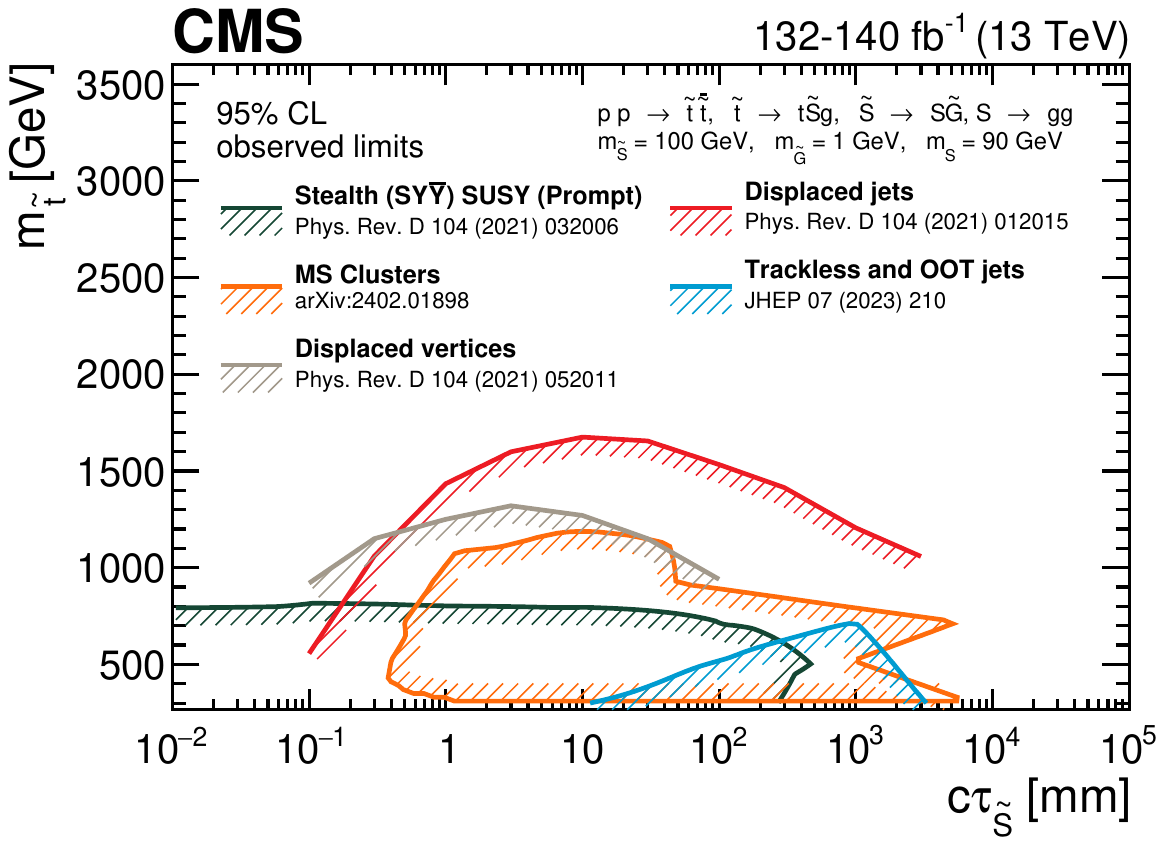}
\includegraphics[width=0.49\linewidth]{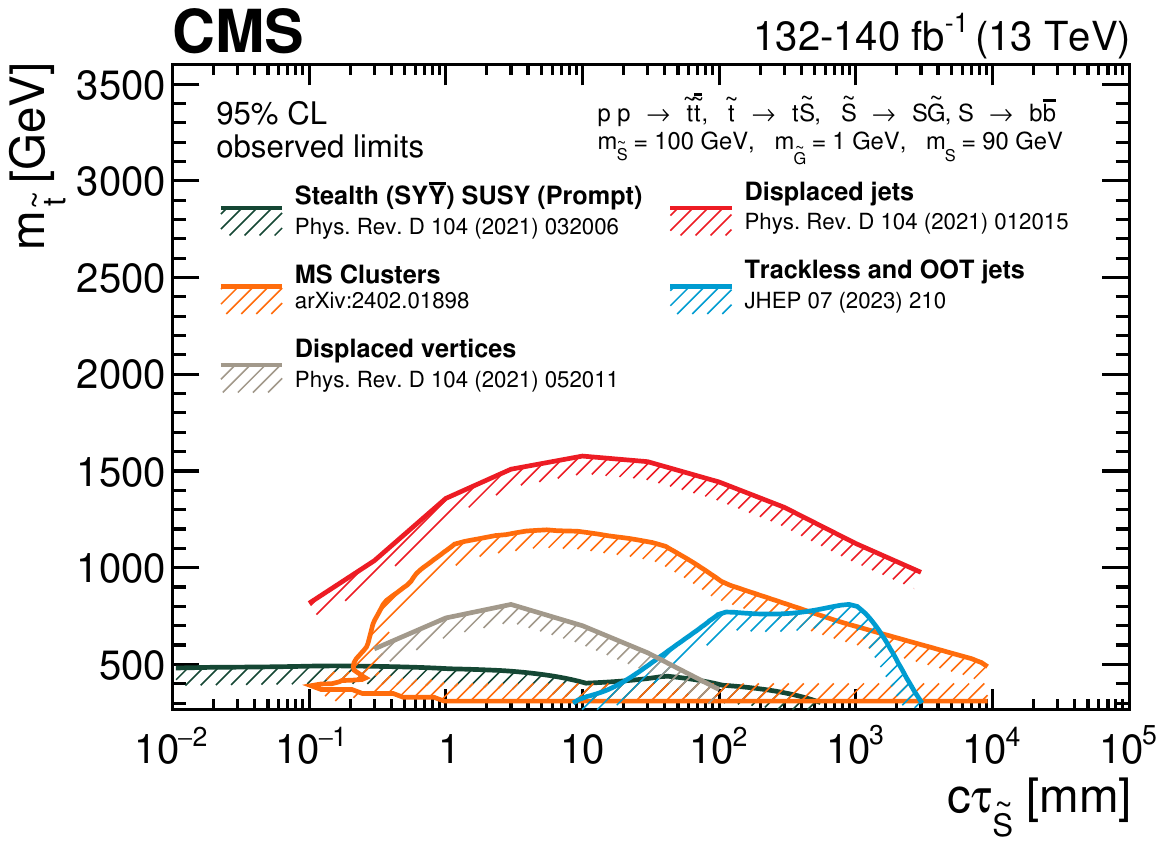}
\includegraphics[width=0.49\linewidth]{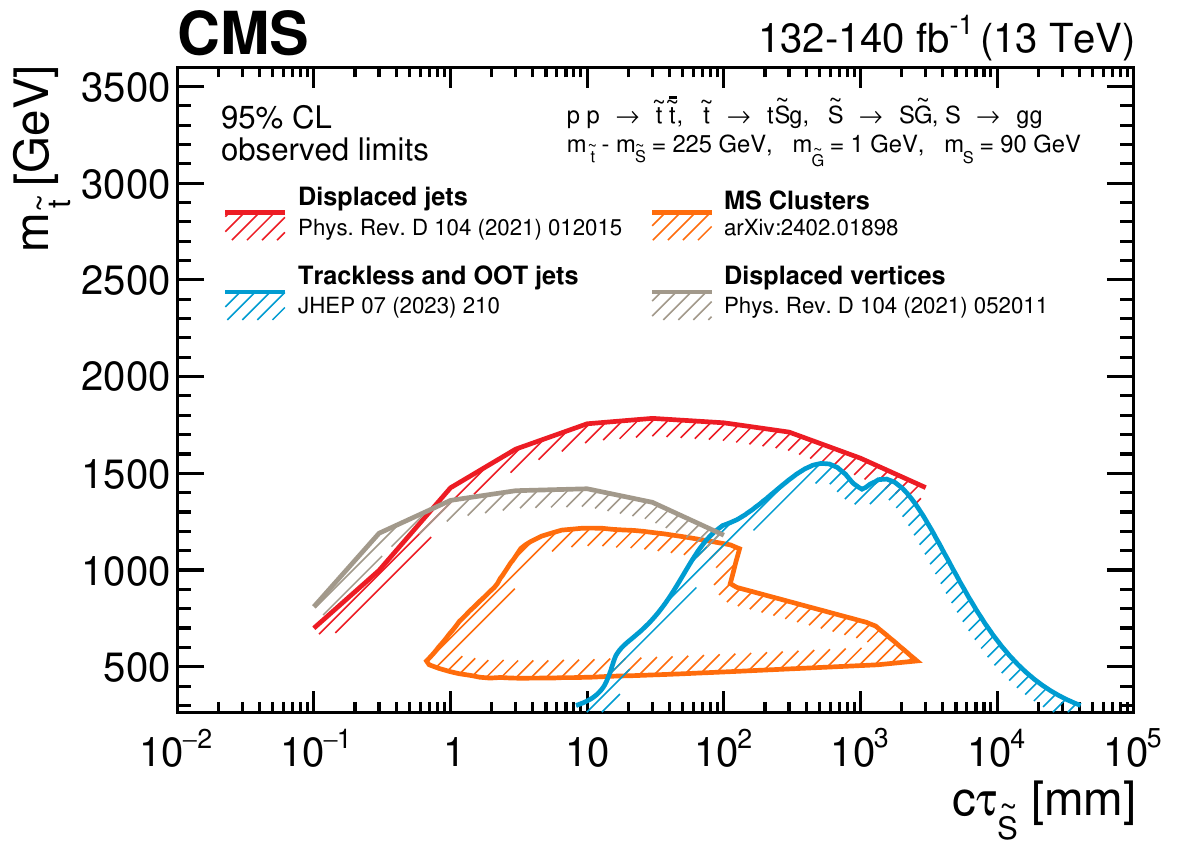}
\includegraphics[width=0.49\linewidth]{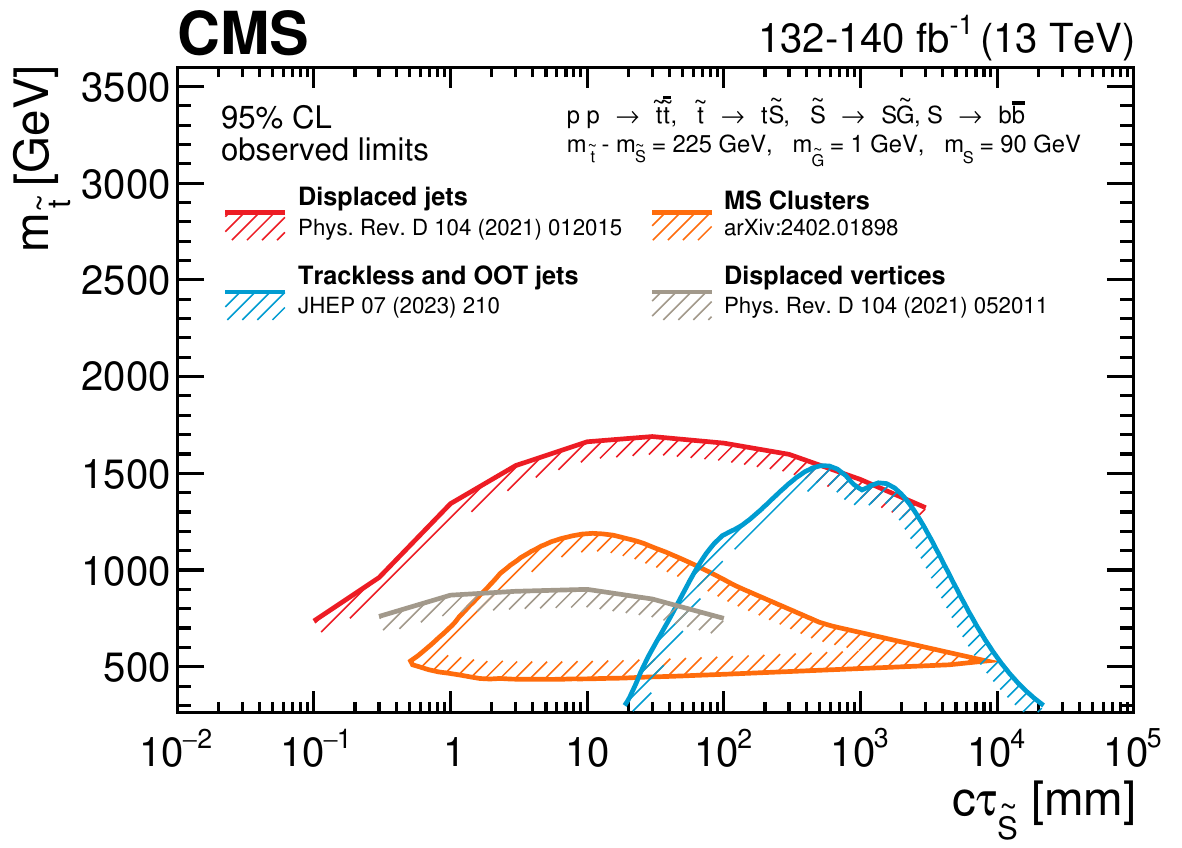}
\caption{Observed 95\% \CL exclusions of the product of the top squark pair production cross section and branching fraction as functions of the top squark mass and proper decay length of the singlino for the stealth \SYY (\cmsLeft) and stealth SHH (\cmsRight) SUSY model where the mass of the singlino is $100\GeV$ (upper) and $m_{\sTop}-225\GeV$ (lower). Exclusions are for the stealth SUSY search~\cite{SUS-19-004} (dark green), the displaced vertices search~\cite{EXO-19-013} (gray), the displaced-jets search~\cite{CMS:2020iwv} (red), the trackless- and OOT jets search~\cite{CMS-PAS-EXO-21-014} (blue), and muon system showers search (MS clusters)~\cite{CMS:2023arc} (orange). The hatching direction on each contour denotes the region of excluded 2D phase space that is bounded by the respective contour. Note that the displaced-jets search has no sensitivity less than $c\tau_{\PSS}=0.1\mm$, the DVs search has no sensitivity less than $c\tau_{\PSS}=0.1$ (0.3)\mm for the \SYY (SHH) model, and the stealth SUSY search has no sensitivity to either stealth SUSY model when $m_{\sTop} - m_{\PSS} = 225\GeV$ because the jets are likely to be outside of the detector acceptance. Additionally, for the specific result here, the muon system showers search only uses the CSCs component of the muon system.}
\label{fig:stealth_syy_shh_2D}
\end{figure*}

Considering the \SYY and Higgs portal stealth SUSY models discussed above, if the singlino is long lived, then dedicated LLP searches could be sensitive to these SUSY models. In addition to the stealth SUSY search~\cite{SUS-19-004}, four LLP-style searches, including the displaced-jets search~\cite{CMS:2020iwv}, the DVs search~\cite{EXO-19-013}, the trackless- and OOT jets search~\cite{CMS-PAS-EXO-21-014}, and the muon system showers search (MS clusters)~\cite{CMS:2023arc} reinterpret their analyses for these stealth SUSY models, where the proper decay length of the singlino ($c\tau_{\PSS}$) ranges from 0.01\mm to 1000\mm. Figure~\ref{fig:stealth_syy_shh_2D} shows observed exclusions on the product of the top squark pair production cross section and branching fraction in terms of the top squark mass and proper decay length of the singlino for the \SYY and Higgs portal versions of the stealth SUSY model. Two singlino mass scenarios are considered: where $m_{\PSS}=100\GeV$ and where $m_{\PSS}=m_{\sTop}-225\GeV$. The branching fractions are assumed to be 100\% for the decay chain for either the \SYY ($\sTop \to \PQt \PSS \Pg$, $\PSS\to \PS \sGra$, and $\PS\to \Pg\Pg$) or Higgs portal ($\sTop \to \PQt \PSS$, $\PSS\to \PS \sGra$, and $\PS\to \PQb\PAQb$). Each exclusion contour bounds (bounding direction denoted by hatching) the 2D parameter space that is excluded by the respective search.

\subsubsection{Inelastic dark matter}
\begin{figure}[htbp!]
\centering
\includegraphics[width=0.7\linewidth]{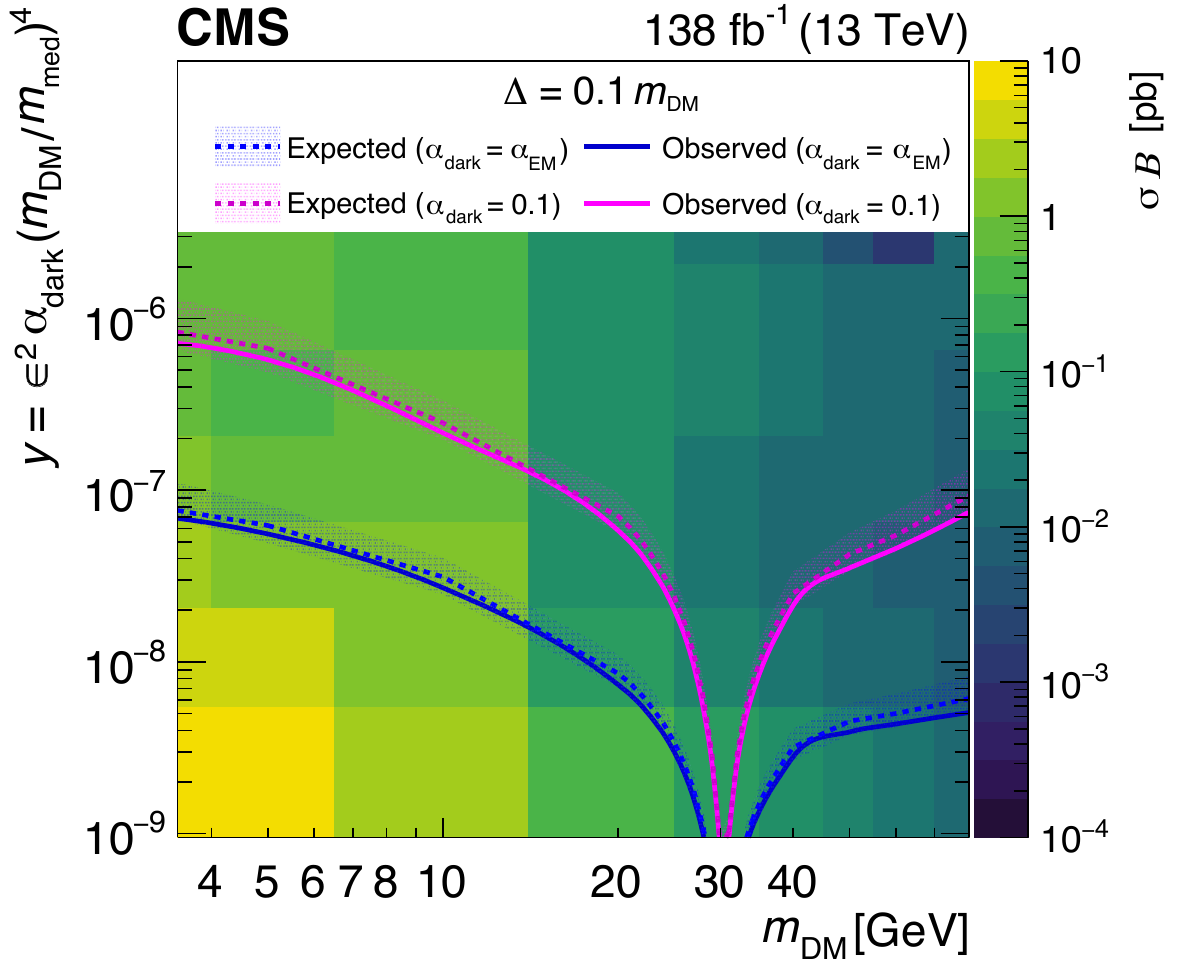}
\caption{Two-dimensional exclusion surface in the search for IDM, assuming $\Delta = 0.1 \, \mDM$, in terms of the DM mass \mDM and the signal strength $y$, with $\mMed = 3 \, \mDM$. Filled histograms denote observed limits on $\sigma(\Pp\Pp \to \PAprime \to \PDM_2 \PDM_1) \, \BR(\PDM_2 \to \PDM_1 \PGmp \PGmm)$. Solid (dashed) curves denote the observed (expected) exclusion limits at 95\%~CL, with 68\%~CL uncertainty bands around the expectation. Regions above the curves are excluded, depending on the \aDark hypothesis: $\aDark = \alpha_{\text{EM}}$ (dark blue) or 0.1 (light magenta). The sensitivity is higher in the region near $\mDM \approx 30\GeV$ or $\mMed \approx 90\GeV$ because of the \PAprime mixing with the \PZ boson in that mass range. Figure adapted from Ref.~\cite{CMS-PAS-EXO-20-010}.}
\label{fig:iDM_limits}
\end{figure}

The first dedicated collider search for IDM has been conducted by the CMS Collaboration~\cite{CMS-PAS-EXO-20-010} and is described in Section~\ref{parag:inelDMEXO20010}. No evidence for the signal is observed. Limits at 95\%~\CL are set on the product of the DM production cross section and decay branching fraction of the excited state $\sigma(\Pp\Pp \to \PAprime \to \PDM_2 \PDM_1) \, \BR(\PDM_2 \to \PDM_1 \mu^+ \mu^-)$. These limits can be translated into limits on the interaction strength $y$ and the DM particle mass \mDM, in terms of the mass split $\Delta$ between the DM states and the coupling strength \aDark of the DS gauge interaction. That translation has a strong dependency both on \aDark itself and on the dark photon (mediator) mass \mMed, therefore the results, shown in Fig.~\ref{fig:iDM_limits} for the 10\% mass-split scenario, are presented for the recommended $\mDM = \mMed/3$ choice and for two \aDark hypotheses. For $\Delta = 0.1\,\mDM$, at $\mDM = 3$ and 80\GeV respectively, values of $y$ greater than ${\approx}10^{-7}$--$10^{-6}$ are excluded for the $\aDark = 0.1$ hypothesis. Conversely, for the $\aDark = \alpha_{\text{EM}}$ hypothesis, values of $y$ greater than ${\approx}10^{-8}$--$10^{-7}$ are excluded for the same \mDM values. The $\PAprime$-\PZ resonance effect greatly improves the limits when $\mDM\simeq 30\GeV$.

\subsubsection{Hidden valleys}
\label{sec:res-hidden-valleys}
This section presents results interpreted in terms of dark QCD models, as described in Section~\ref{sec:darkqcd}.

\cmsParagraph{Semivisible jets\label{par:section7_svj}} As explained in Section~\ref{sec:EXO-19-020}, we reinterpret the dijet resonance search and monojet DM search for the SVJ model.
For the dijet resonance search, following Ref.~\cite{CMS:2019gwf}, the background estimation from control regions (CRs) in data is used for signals with $\mZprime \geq 3\TeV$, while the analytic fit-based background estimation is used for lower \mZprime. For the interpretation of the monojet search, we use the \MADANALYSIS implementation~\cite{DVN/IRF7ZL_2021}.

\begin{figure*}[t]
\centering
\includegraphics[width=1\linewidth]{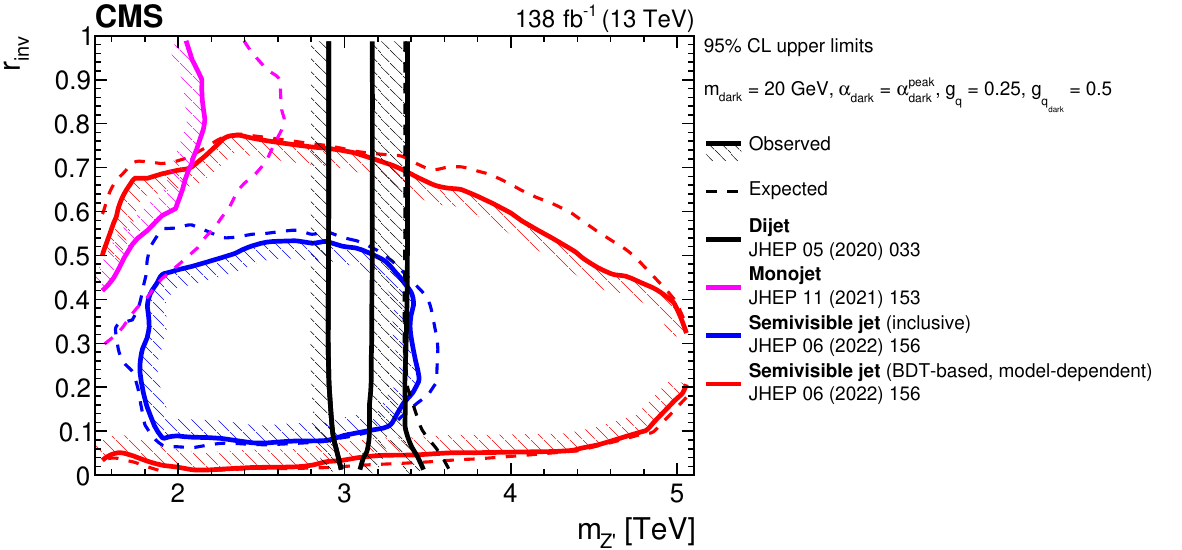}
\caption{Observed and expected 95\% \CL excluded regions of the \mZprime-\rinv plane from the dedicated SVJ search~\cite{CMS:2021dzg}, the dijet search~\cite{CMS:2019gwf} (Section~\ref{sec:EXO-19-012}), and the monojet search~\cite{CMS:2021far} (Section~\ref{sec:EXO-20-004}). The hashed areas indicate the direction of the excluded area from the observed limits. The calculation of \gqdark for this signal model is described in Section~\ref{sec:svjtheory}.
}
\label{fig:svj_reinterp}
\end{figure*}

The results from both interpretations are compared to the results from the dedicated SVJ search, with and without the BDT tagger, in Fig.~\ref{fig:svj_reinterp}. The complementary sensitivities of each strategy are clearly visible. The monojet search is more sensitive for large \rinv values, and the standard DM interpretation of the dijet search, effectively considering only \ZprimeToQQ events, also provides good sensitivity in this region. Accounting for the combination of effects of SVJ model parameters on observables used in the monojet search, the most stringent exclusion is found for $\rinv = 0.8$, as this maximizes the overall selection efficiency for SVJ signals. For very small \rinv values, the reinterpreted dijet search provides the best sensitivity. At intermediate \rinv values, the dedicated SVJ search is the most sensitive, especially when the BDT is used to identify SVJs, though the latter strategy introduces more model dependence. The advantage of the dedicated strategy would increase with the branching fraction for \ZprimeToDark, which grows for larger \gqdark or smaller \gq values.

\begin{figure*}[htbp]
\centering
\includegraphics[width=1\linewidth]{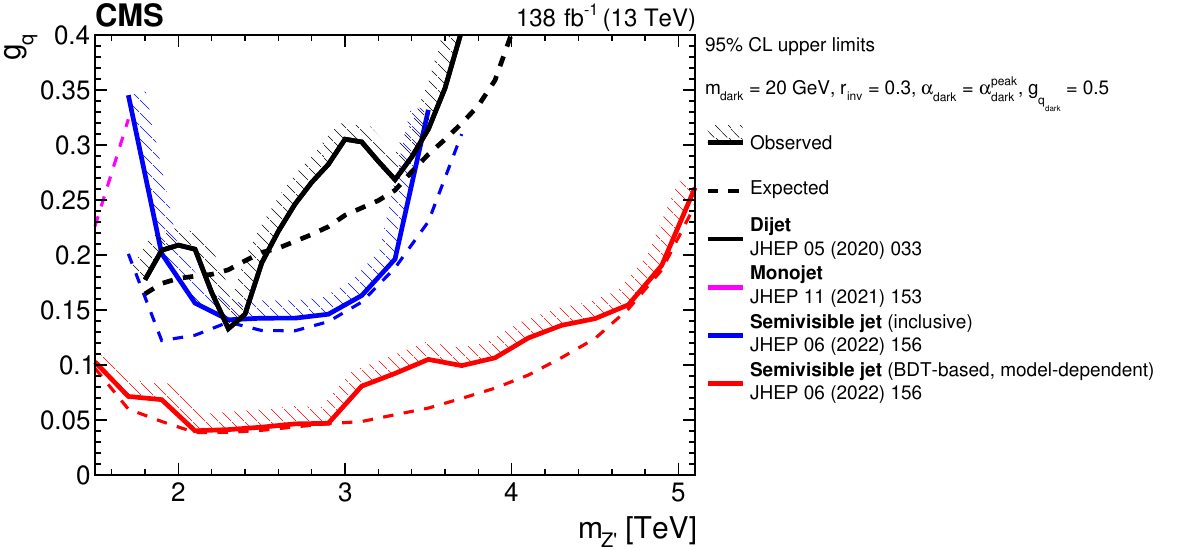}
\includegraphics[width=1\linewidth]{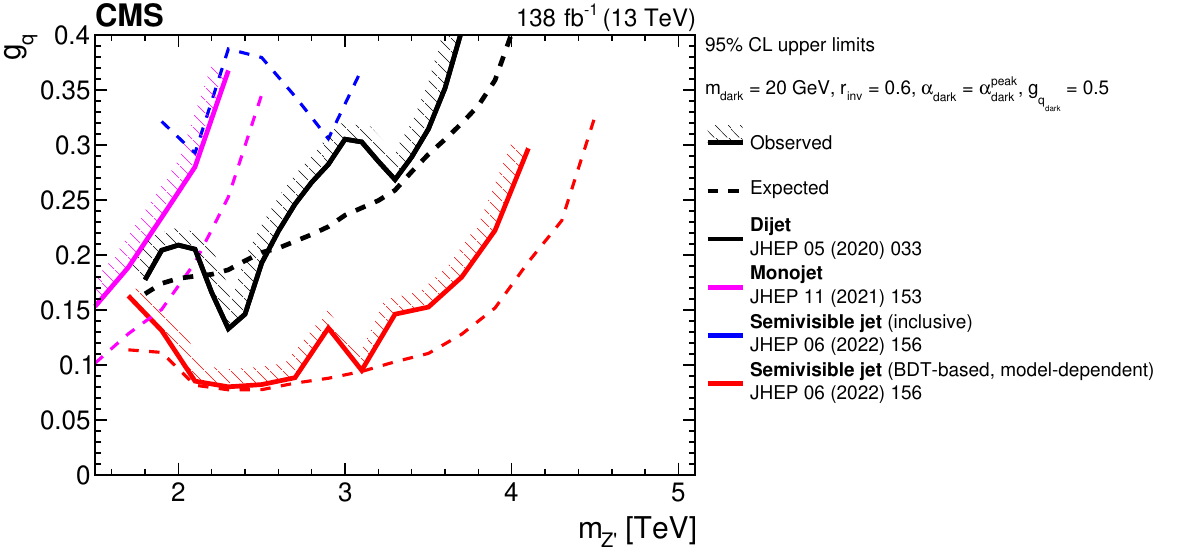}
\caption{Observed and expected 95\% \CL exclusion limits on \gq for SVJ signals from the dedicated SVJ search~\cite{CMS:2021dzg}, the dijet search~\cite{CMS:2019gwf}, and the monojet search~\cite{CMS:2021far}, for $\rinv=0.3$ (upper) and $\rinv=0.6$ (lower). The hashed areas indicate the direction of the excluded area from the observed limits. The observed limits from the monojet search in the upper plot and the inclusive SVJ search in the lower plot are outside the range of validity of the narrow-width approximation, so they are not shown. The calculation of \gqdark for this signal model is described in Section~\ref{sec:svjtheory}.
}
\label{fig:svj_gq_limit}
\end{figure*}

These cross section limits can be interpreted as limits on \gq for fixed parameter values $\gqdark = 0.5$ and $\mDark = 20\GeV$, following the procedure described in Section~\ref{par:DMsimp_zprime_result}. For both the SVJ search and the interpretation of the monojet search, the initial and final states for the procedure are $\qqbar$ and $\Pqdark\Paqdark$, respectively. Those searches do not depend strongly on the \PZpr boson width within the narrow-width regime, because the resolutions of the search variables are intrinsically limited by the information lost in \ptmiss. In contrast, the resolution of the dijet mass used in the dijet search is small enough that even minor increases in the mediator width are visible~\cite{CMS:2018mgb}. Therefore, the existing \gq exclusion from the dijet search is used directly; though this underestimates the exclusion at small \rinv, SVJ events do not contribute to the dijet search limit for $\rinv \gtrsim 0.1$, so this is a reasonable approximation in the majority of the signal model parameter space. Figure~\ref{fig:svj_gq_limit} shows the excluded values of \gq for SVJ signals from all searches for two representative values $\rinv = 0.3$ and 0.6. Only values that satisfy the narrow-width approximation $\Gamma_{\PZpr}/\mZprime < 10\%$ are shown. For $\rinv = 0.3$, the acceptance of the SVJ search is maximized, and even without the BDT tagger, it provides the strongest exclusions for a wide range of \PZpr boson masses. For $\rinv = 0.6$, the BDT-based SVJ search still provides a strong exclusion even as the search acceptance decreases, while the monojet search has the best exclusion at small \mZprime. The $\rinv = 1$ case is equivalent to the vector DM simplified model, so the coupling exclusion from the dijet and monojet searches can be seen in Fig.~\ref{fig:summary_gq}.

\begin{figure}[htbp!]
    \centering
    \includegraphics[width=0.95\textwidth]{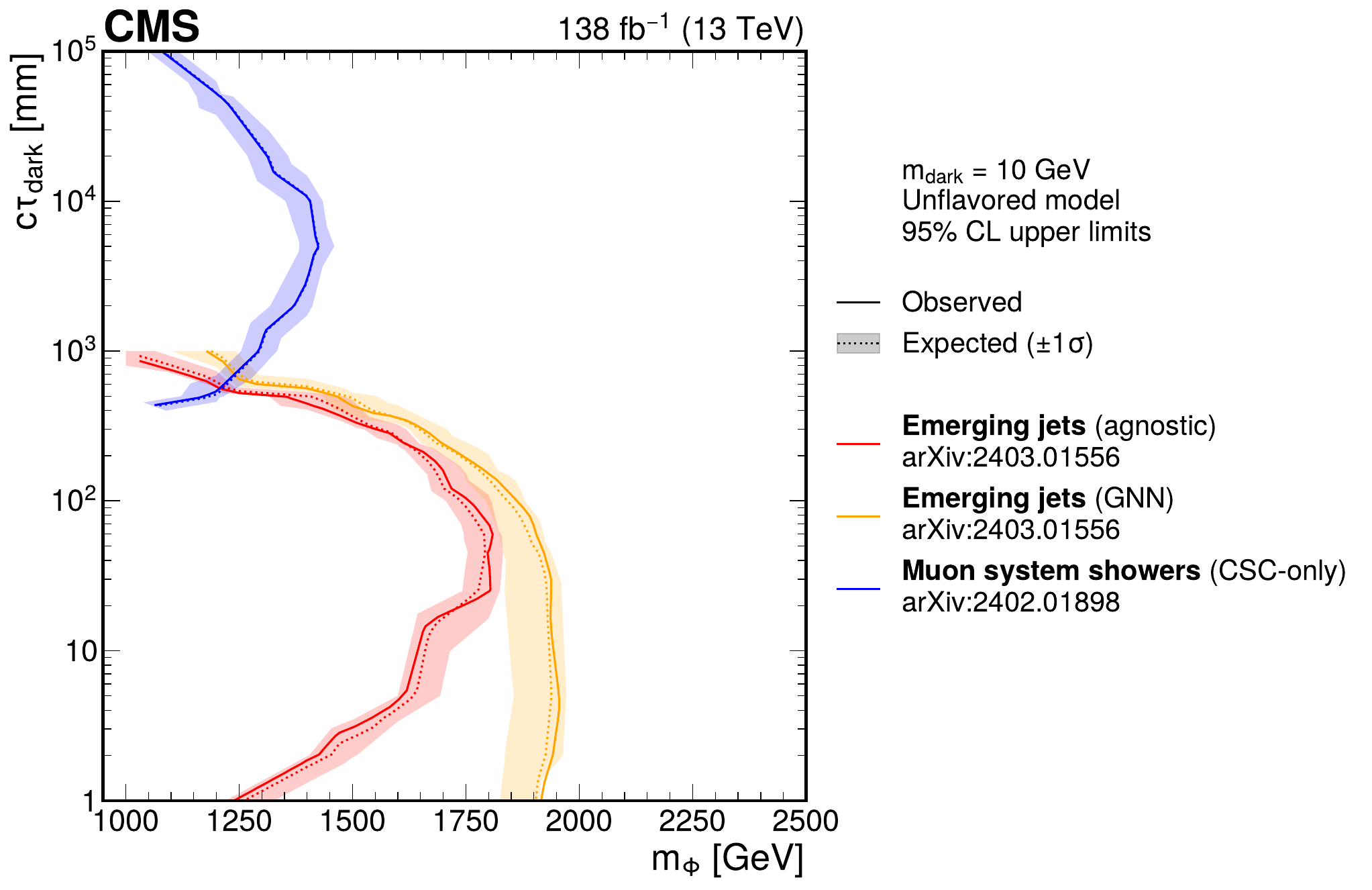}
    \includegraphics[width=0.95\textwidth]{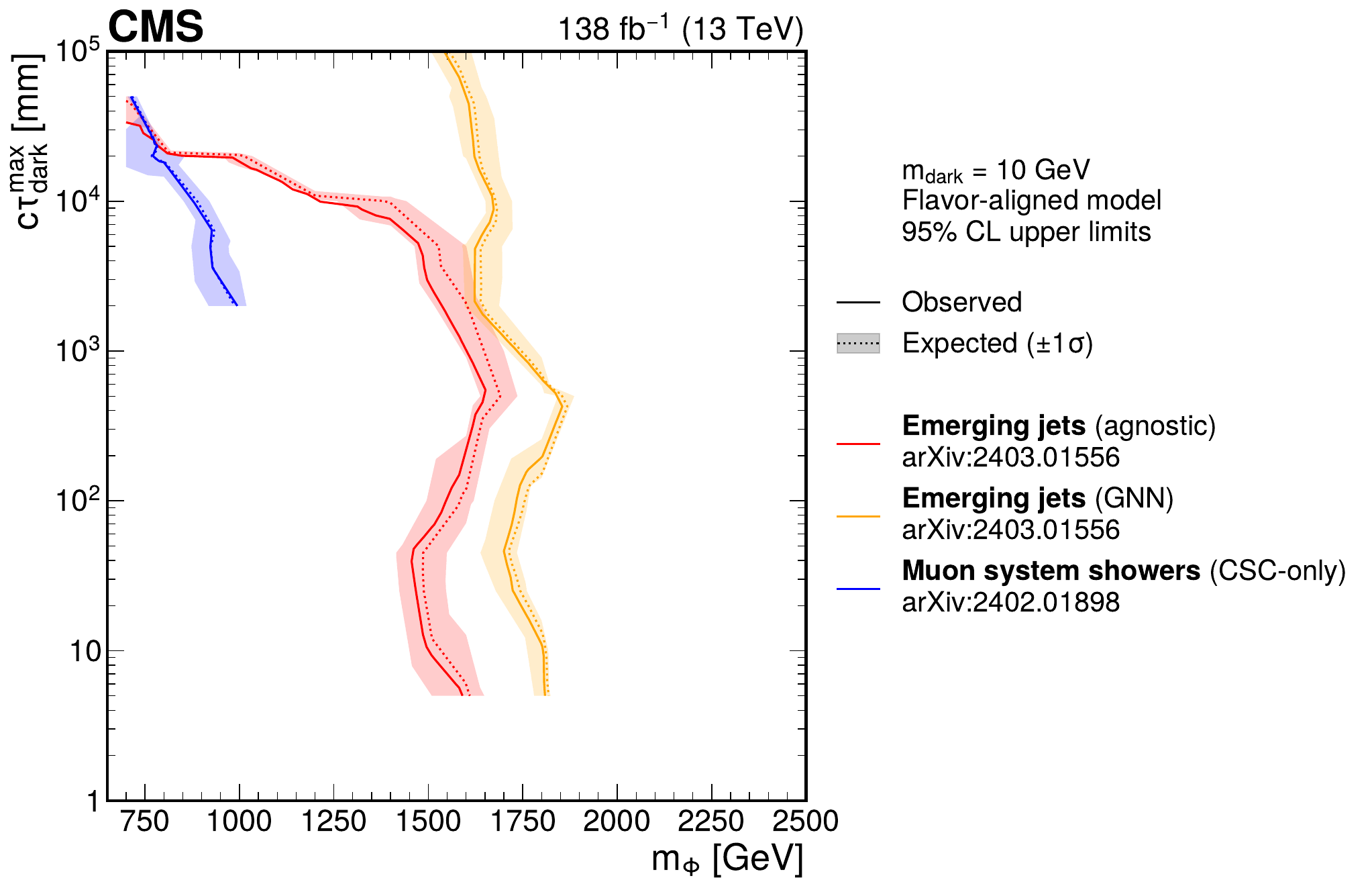}
    \caption{Observed and expected 95\% \CL exclusion limits from the track-based~\cite{CMS:2024emj} and muon detector shower-based~\cite{CMS:2023arc} searches for pair production of a bifundamental mediator that decays into a jet and an emerging jet, for $\mDark=10\GeV$ and various choices of \Pbifun masses and \Ppidark proper decay lengths, in the unflavored model (upper) and the flavor-aligned model (lower).}
    \label{fig:emj_reinterp}
\end{figure}

\cmsParagraph{Emerging jets\label{par:emjsens}} The track-based EJ search and the muon detector shower search (Sections~\ref{subsec:emj_run1} and~\ref{sec:msclusters}) have complementary sensitivity to EJ signatures, targeting smaller and larger lifetimes, respectively. The exclusion limits for unflavored and flavor-aligned EJ models from both searches are shown in Fig.~\ref{fig:emj_reinterp} for signals with the dark-meson mass $\mDark=10\GeV$. For the dedicated EJ search, the results from both the model-agnostic EJ tagger and the model-dependent GNN tagger are shown. For the muon detector shower search, results are obtained by clustering CSC hits. The sensitivity of the muon detector shower search to the flavor-aligned model is reduced because this model has a broader spread of lifetimes and therefore fewer particles reach the muon detectors.

Other LLP searches are not sensitive to EJ models, for various reasons. The searches for delayed jets (Section~\ref{sec:EXO-19-001}) and trackless jets (Section~\ref{sec:EXO-21-014}) use timing measurements that rely on the exotic particles being sufficiently delayed, and EJs do not satisfy this requirement. The displaced-jet search (Section~\ref{parag:disp_jets}) uses triggers that require at most two prompt tracks to be associated with the jets, which rejects most EJs because they contain tracks with a broader mix of displacements. The DV search (Section~\ref{sec:EXO-19-013}) relies on reconstructing DVs, which is inefficient for EJs, as each vertex tends to have only a few tracks associated with it.

\begin{figure}[htbp!]
    \centering
    \includegraphics[width=0.45\textwidth]{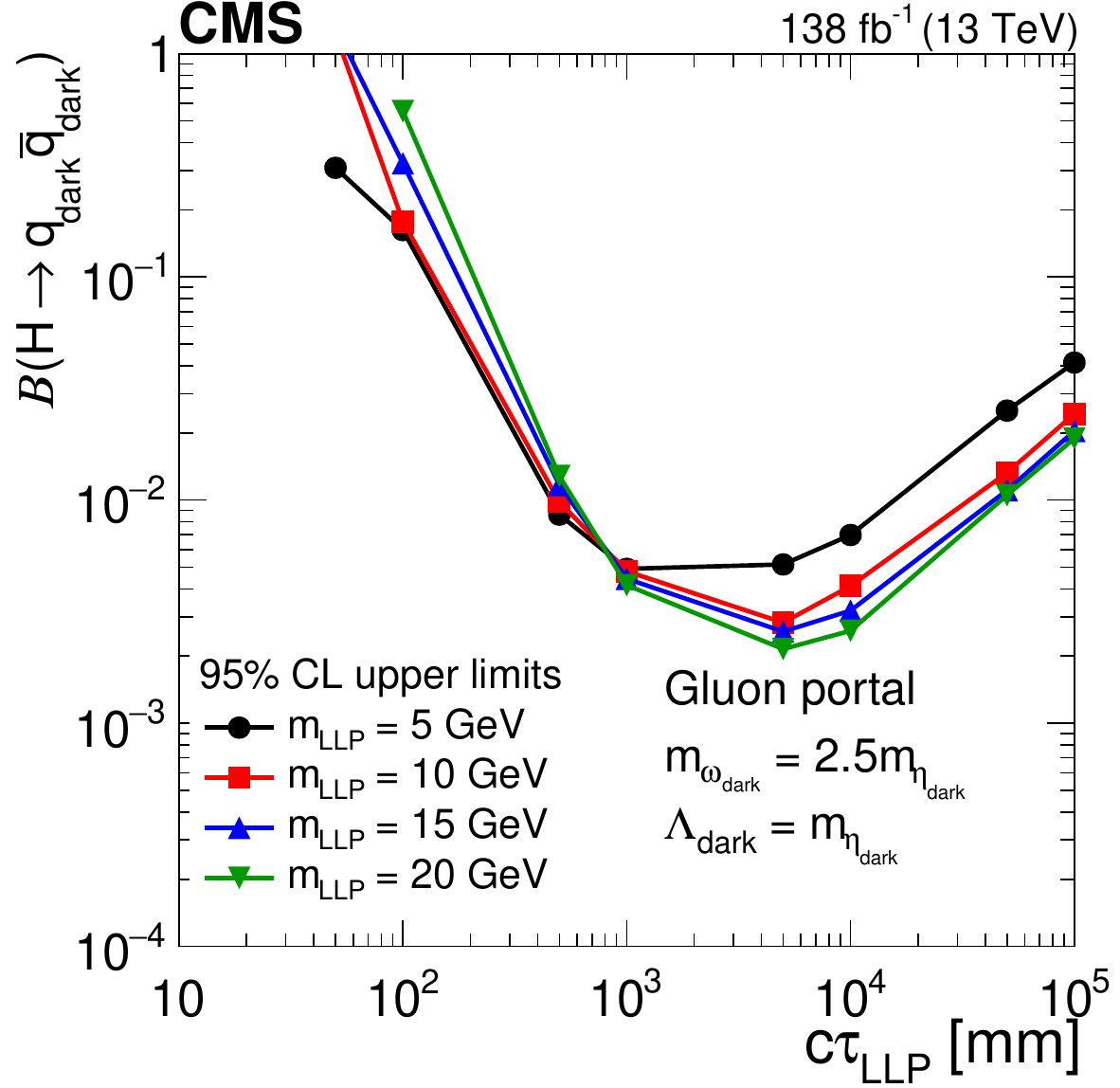}
    \includegraphics[width=0.45\textwidth]{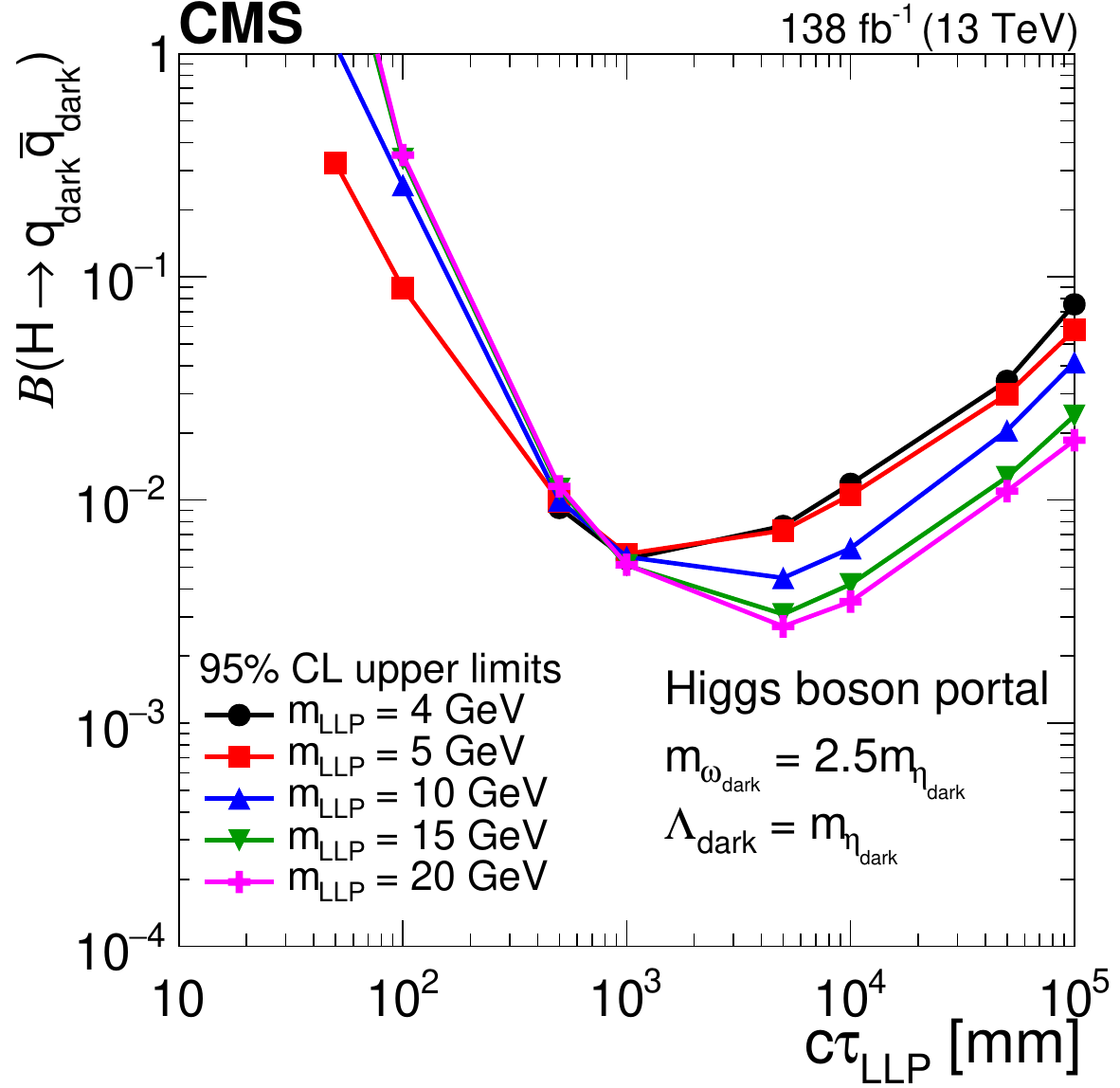}
    \caption{Observed 95\% \CL exclusion limits on the branching fraction of the Higgs boson decay into DS hadrons, $\Psi$, for the search for neutral decays in the muon system (Section~\ref{sec:msclusters}).  Sensitivity for the gluon (\cmsLeft) and Higgs boson (\cmsRight) DS decay portals are shown. The model parameters considered here are $\momegadark=2.5\metadark$, $\lamDark =\metadark$. Figure adapted from Ref.~\cite{CMS:2023arc}.}
    \label{fig:clustersdarkshowers}
\end{figure}

The search for neutral decays in the muon system (Section~\ref{sec:msclusters}) is also interpreted using the decay portal EJ models. The representative exclusion limits for two decay portals, the gluon and the Higgs boson, are shown in Fig.~\ref{fig:clustersdarkshowers}.

\begin{figure}[htbp!]
    \centering
    \includegraphics[width=0.8\linewidth]{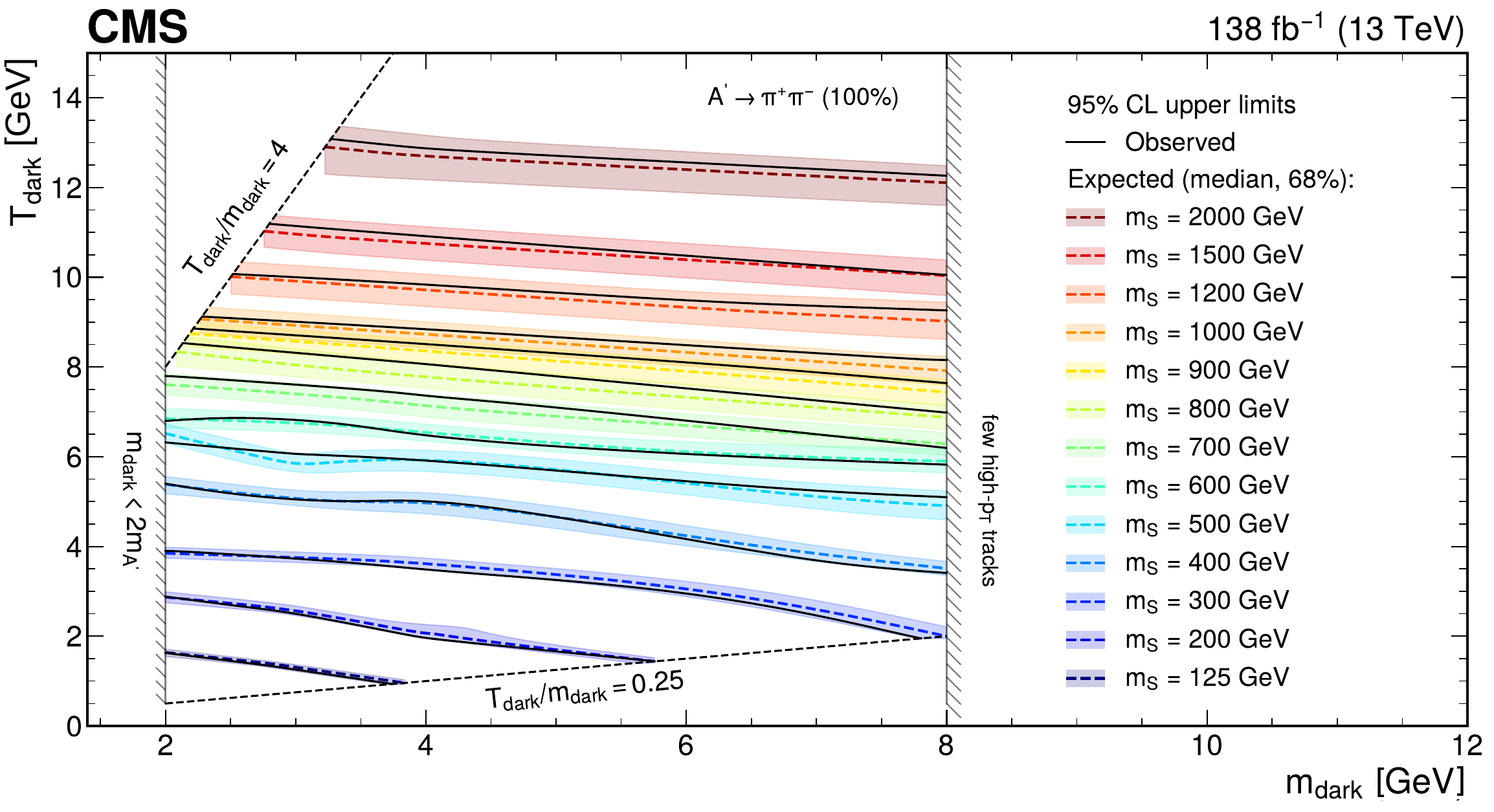}
    \caption{Observed and expected 95\% \CL excluded regions in the SUEP search (Section~\ref{sec:SUEPsearch}) in \mDark--\TD for each \mS value, considering the case with $\mAprime=1.0\GeV$ ($\PAprime \to \pipi$ with $\BR=100\%$). The regions below the lines are excluded. Figure taken from Ref.~\cite{CMS-PAS-EXO-23-002}.}
    \label{fig:3D_limits}
\end{figure}

\cmsParagraph{Soft unclustered energy patterns\label{par:suepSens}} The SUEP search (Section~\ref{sec:SUEPsearch}) is interpreted in terms of limits on the production cross section for different values of the signal model parameters: the mediator mass \mS, the dark-meson mass \mDark, and the temperature \TD. The excluded ranges in the \mS--\mDark--\mAprime--\TD parameter space, where \mAprime is the mass of the dark photon, are obtained by comparing the expected and observed cross section limits to the theoretical signal cross section, assuming the mediator is produced through gluon-gluon fusion with couplings similar to the SM Higgs boson~\cite{deFlorian:2016spz}. Figure~\ref{fig:3D_limits} shows the exclusions for all \mS values in the plane of \mDark and \TD with $\mAprime=1.0\GeV$. Similar exclusions are obtained for other \mAprime values and their corresponding decay patterns. In the signal models with the highest track multiplicity, corresponding to the most SUEP-like signatures and arising when $\mS/\TD \approx \mS/\mDark \approx 100$, the most stringent limits are set.

\begin{figure*}[htbp!]
\centering
\includegraphics[width=0.8\linewidth]{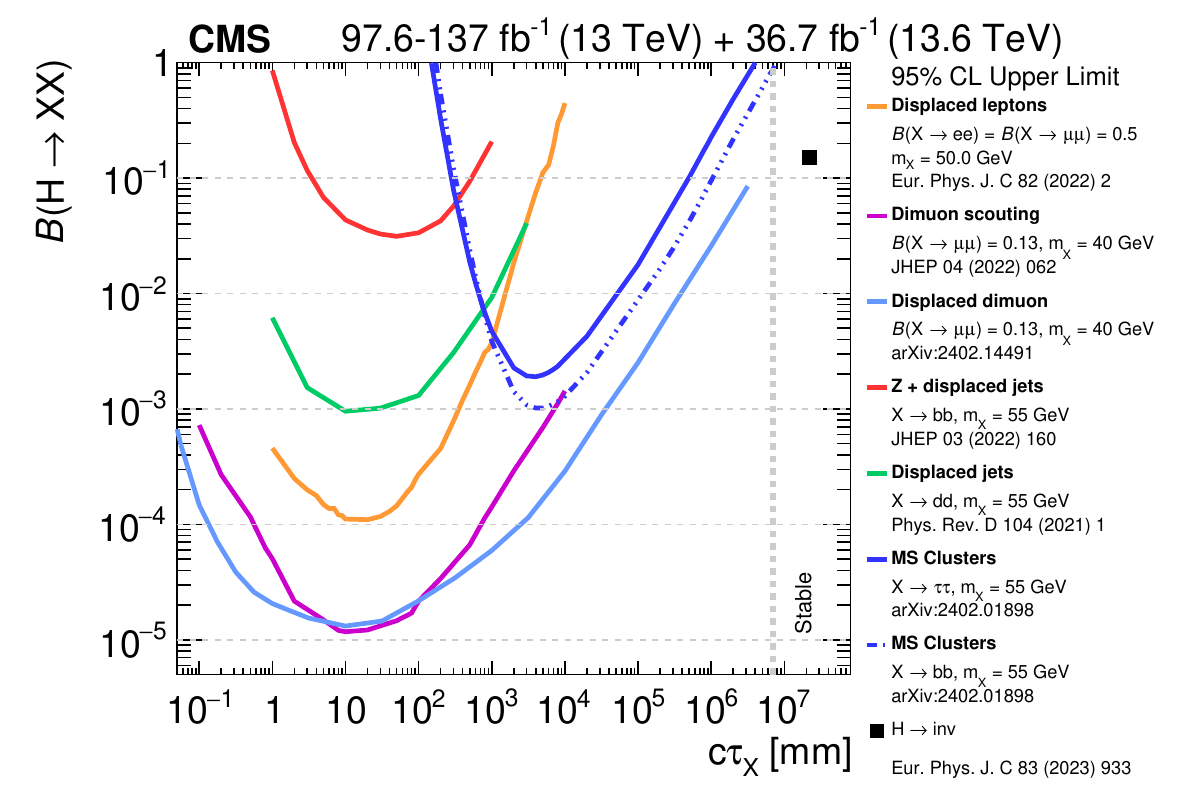}\\
\caption{Observed 95\% \CL upper limits on the branching fraction of Higgs bosons decaying into LLPs with masses between 40 and 55\GeV~\cite{EXO-18-003,EXO-20-014,CMS-PAS-EXO-23-014,CMS:2021yhb,CMS:2020iwv,CMS:2023arc,CMS:2023sdw}. The LLP mass and decay assumptions are given in the legend.}
\label{fig:higgsllp_high}
\end{figure*}

\begin{figure*}[htbp!]
\centering
\includegraphics[width=0.8\linewidth]{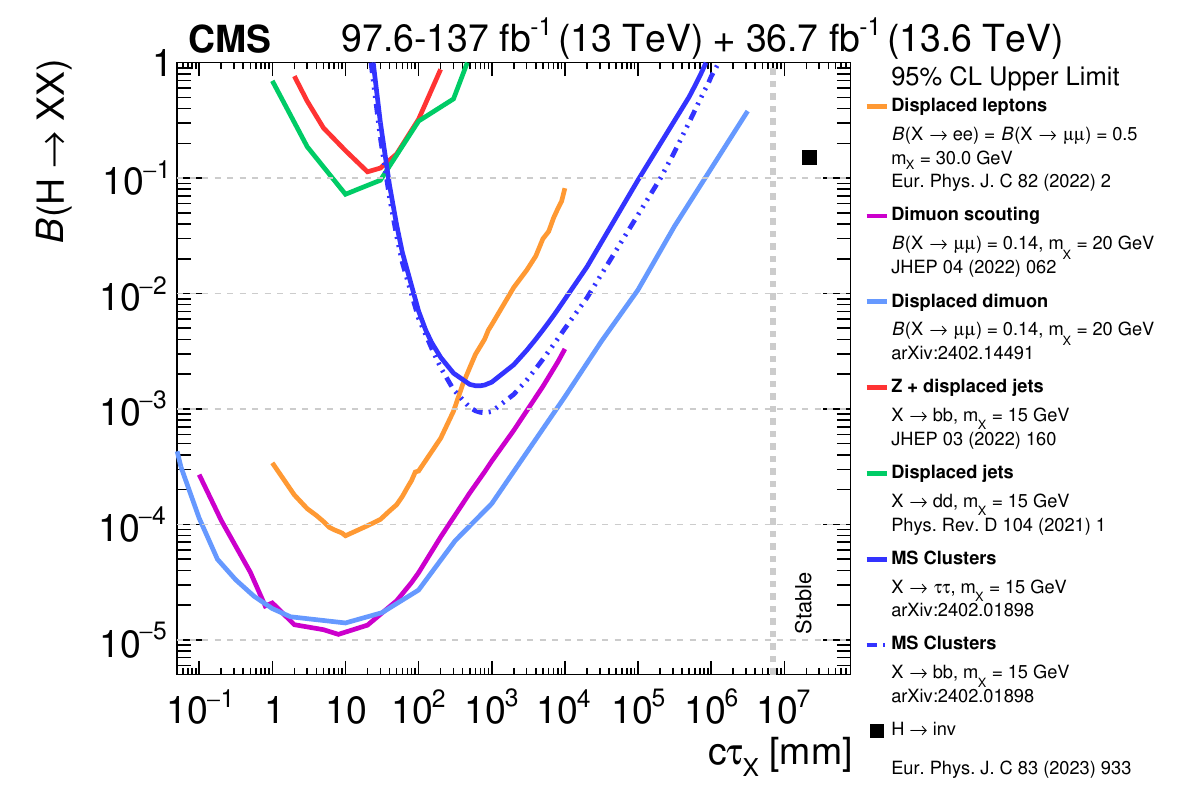}\\
\caption{Observed 95\% \CL upper limits on the branching fraction of Higgs bosons decaying into LLPs with masses between 15 and 30\GeV~\cite{EXO-18-003,EXO-20-014,CMS-PAS-EXO-23-014,CMS:2021yhb,CMS:2020iwv,CMS:2023arc,CMS:2023sdw}. The LLP mass and decay assumptions are given in the legend.}
\label{fig:higgsllp_low}
\end{figure*}

\begin{figure*}[htbp!]
\centering
\includegraphics[width=0.8\linewidth]{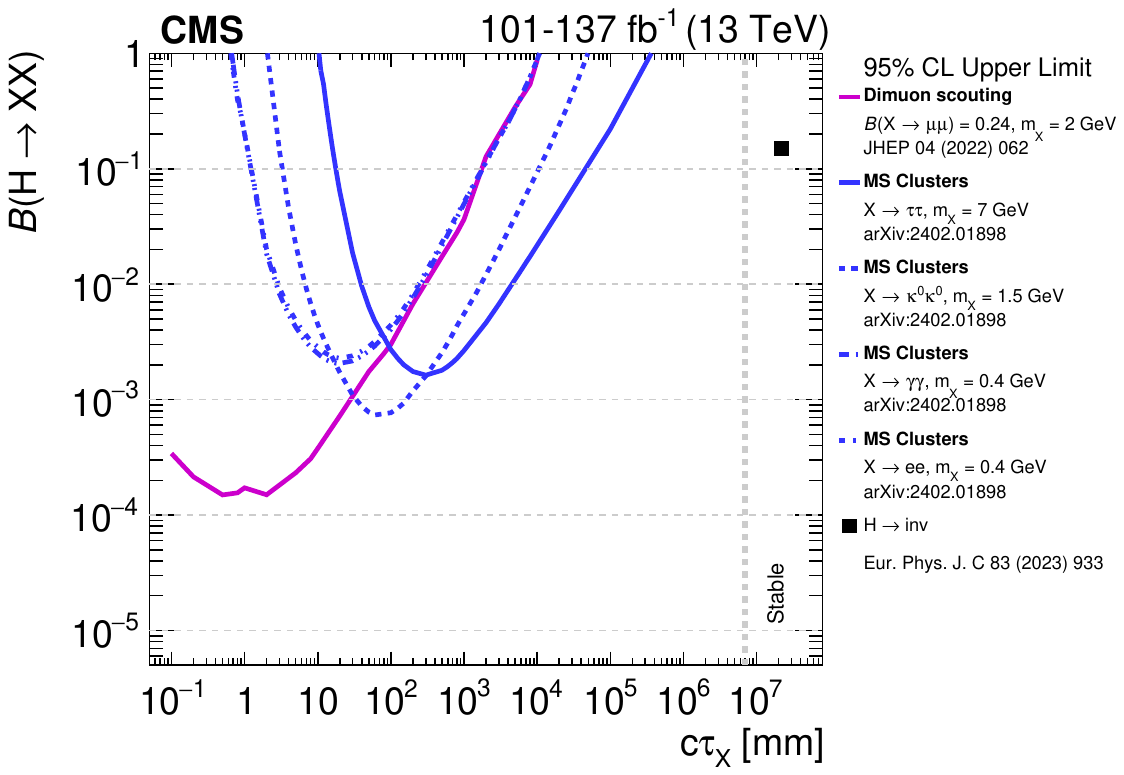}
\caption{Observed 95\% \CL upper limits on the branching fraction of Higgs bosons decaying into LLPs with masses between 0.4 and 7\GeV~\cite{EXO-20-014,CMS:2023arc}. The LLP decay assumptions are given in the legend.}
\label{fig:higgsllp_vlow}
\end{figure*}

\cmsParagraph{Higgs boson decays into long-lived particles\label{sec:higgsLLPSens}} Exotic decays of the Higgs boson into LLPs are well motivated in a variety of models, such as those motivated by neutral naturalness, as described in Section~\ref{sec:neutnat}. Several CMS searches have been reinterpreted in a scenario in which an exotic Higgs boson is produced in $\Pp\Pp$ collisions and then decays into two LLPs, here denoted \PX (as shown in the \cmsRight diagram in Fig.~\ref{fig:2hdma_diagrams}). These interpretations are shown in Figs.~\ref{fig:higgsllp_high}, \ref{fig:higgsllp_low}, and \ref{fig:higgsllp_vlow}. Figure~\ref{fig:higgsllp_high} shows the upper limits on the branching fraction of Higgs bosons decaying into LLPs with masses between 40 and 55\GeV, as functions of the LLP proper decay length. Figure~\ref{fig:higgsllp_low} shows the same but for masses between 15 and 30\GeV, and Fig.~\ref{fig:higgsllp_vlow} shows the same but for masses between 0.4 and 7\GeV. The LLP mass and decay assumptions are given in the legends of the figures. The dedicated LLP searches presented in these figures probe a variety of final states, and LLP masses and lifetimes. These are shown to provide complementary sensitivity for lifetimes between 0.1 and $10^6$\unit{mm} that far exceeds the inclusive limit achieved by the \hinv analysis. The sensitivity to models with smaller lifetimes is limited by the background from \PB meson decays and the resolution of the tracker. The reach in this regime could be extended further by considering the sensitivity for prompt searches to \Pb quark final states.

\cmsParagraph{Heavy long-lived particles\label{sec:heavyLLPs}} Dark sectors may have complex constituents including \TeV scale scalar and vector bosons that decay into LLPs in the DS as well as to DM candidate particles~\cite{Albouy:2022cin}. This can include scenarios motivated by neutral naturalness, as described in Section~\ref{sec:neutnat}. The LLPs may be boosted if their mass is significantly less than the parent particle. These particles can typically decay both to displaced leptonic and hadronic final states. The displaced signatures that can be reconstructed range from a few microns to several meters. In addition, the final state may include significant \ptmiss from either decays of LLPs outside of acceptance or from invisible particles produced in the decays. 

Given the wide range of potential signatures, multiple search strategies have been employed to provide sensitivity, as detailed in Section~\ref{sec:signatures_llp}. Many of these searches were originally designed to achieve sensitivity to supersymmetric models or lower energy signatures, such as decays of the SM Higgs boson. However, excellent sensitivity is achieved by these searches for DS models, such as decays of heavy \PZpr and heavy \Hdark bosons to LLPs. Sensitivities for leptonic final states are shown in Refs.~\cite{EXO-21-006,EXO-18-003}. Hadronic final states are considered below. The \PZpr model is used to probe the sensitivity to DSs with \TeV-scale production of LLPs while the \Hdark model is used to probe sensitivity to DSs with masses of order $100\GeV$.

\begin{figure*}[htbp!]
\centering
\includegraphics[width=0.9\linewidth]{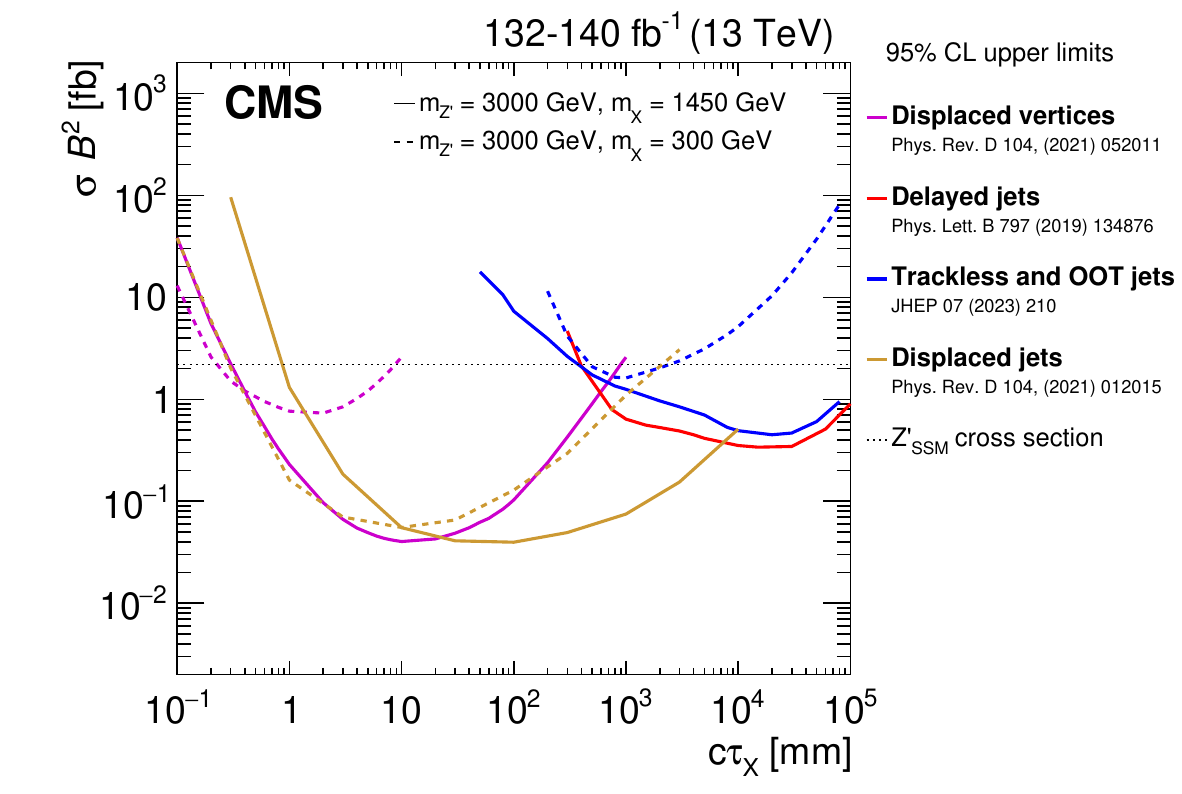}
\includegraphics[width=0.9\linewidth]{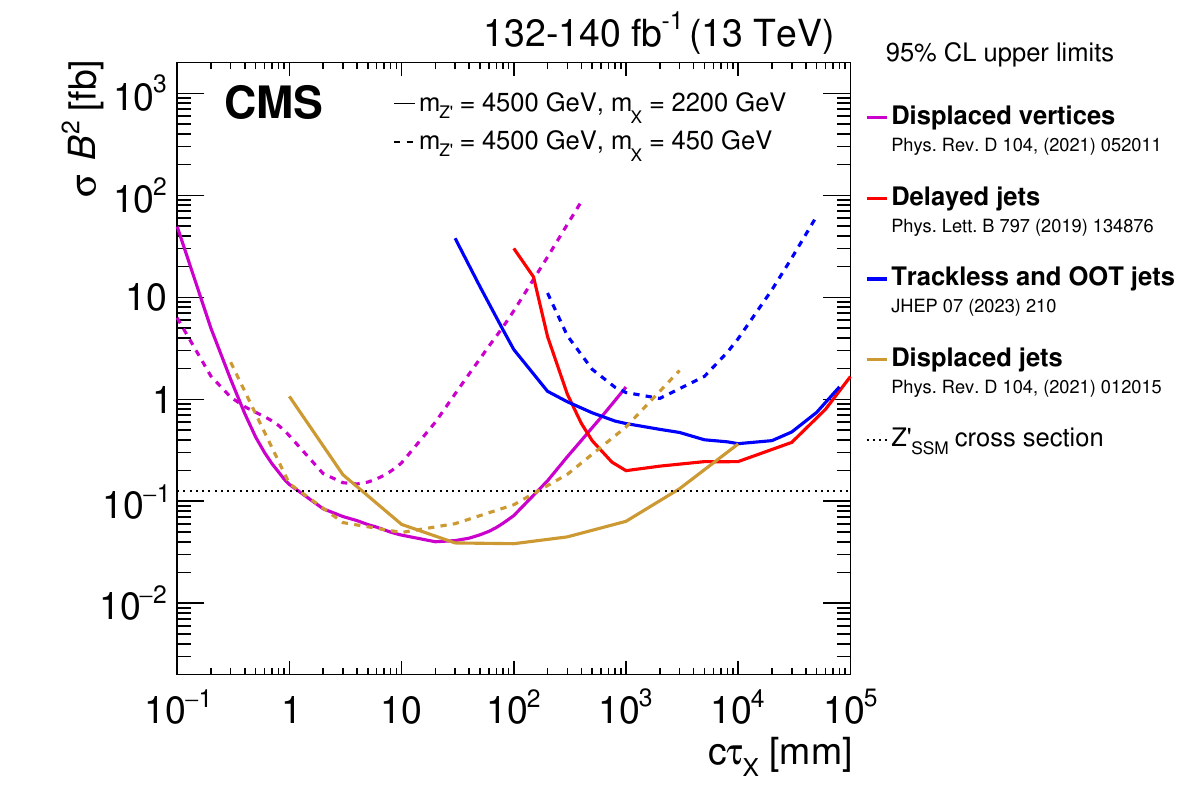}
\caption{Observed 95\% \CL exclusion limits for \PZpr bosons decaying into LLPs with fully hadronic final states, for a \PZpr boson mass of 3000\GeV (upper) and 4500\GeV (lower). Analyses employing different strategies are shown to have complementary lifetime sensitivity~\cite{EXO-19-013,EXO-19-001,CMS-PAS-EXO-21-014,CMS:2020iwv}. The theoretical cross section assumes the \PZpr boson has SM-like couplings to SM quarks~\cite{Altarelli:1989ff}.
}  
\label{fig:zprimeLL_4b}
\end{figure*}

\begin{figure*}[htbp!]
\centering
\includegraphics[width=0.9\linewidth]{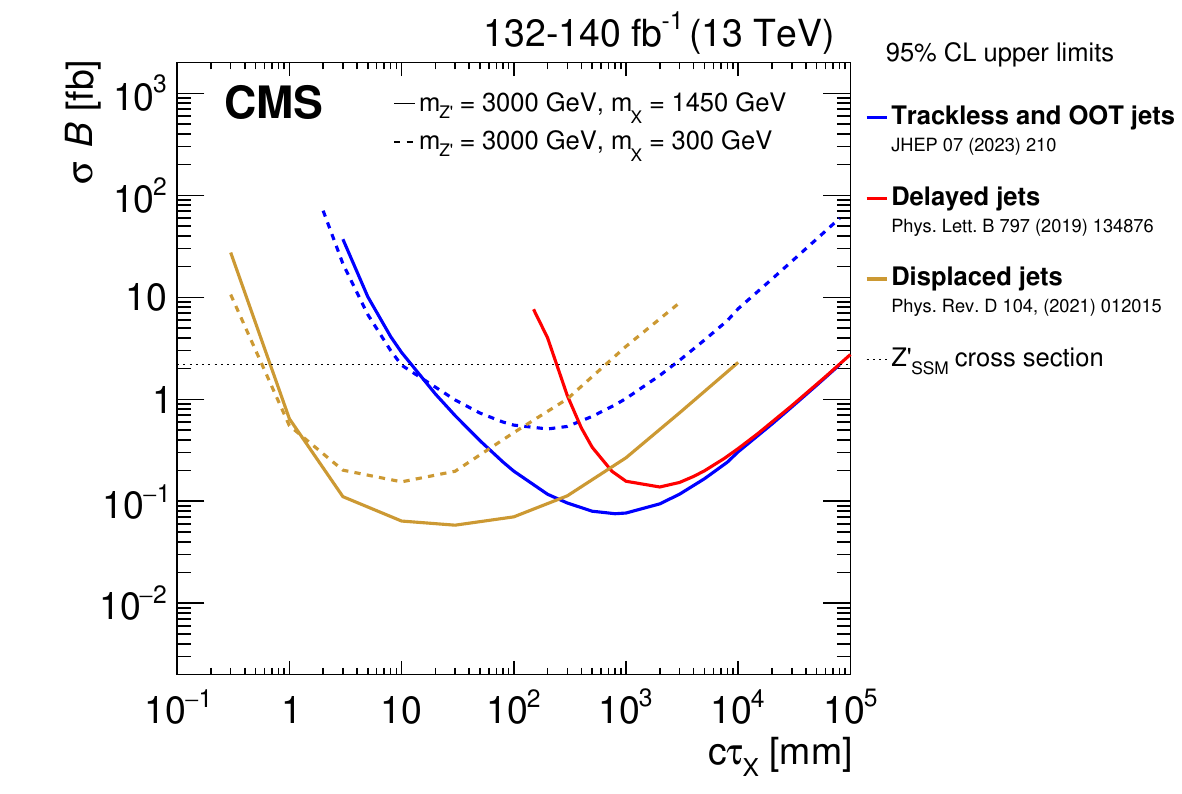}
\includegraphics[width=0.9\linewidth]{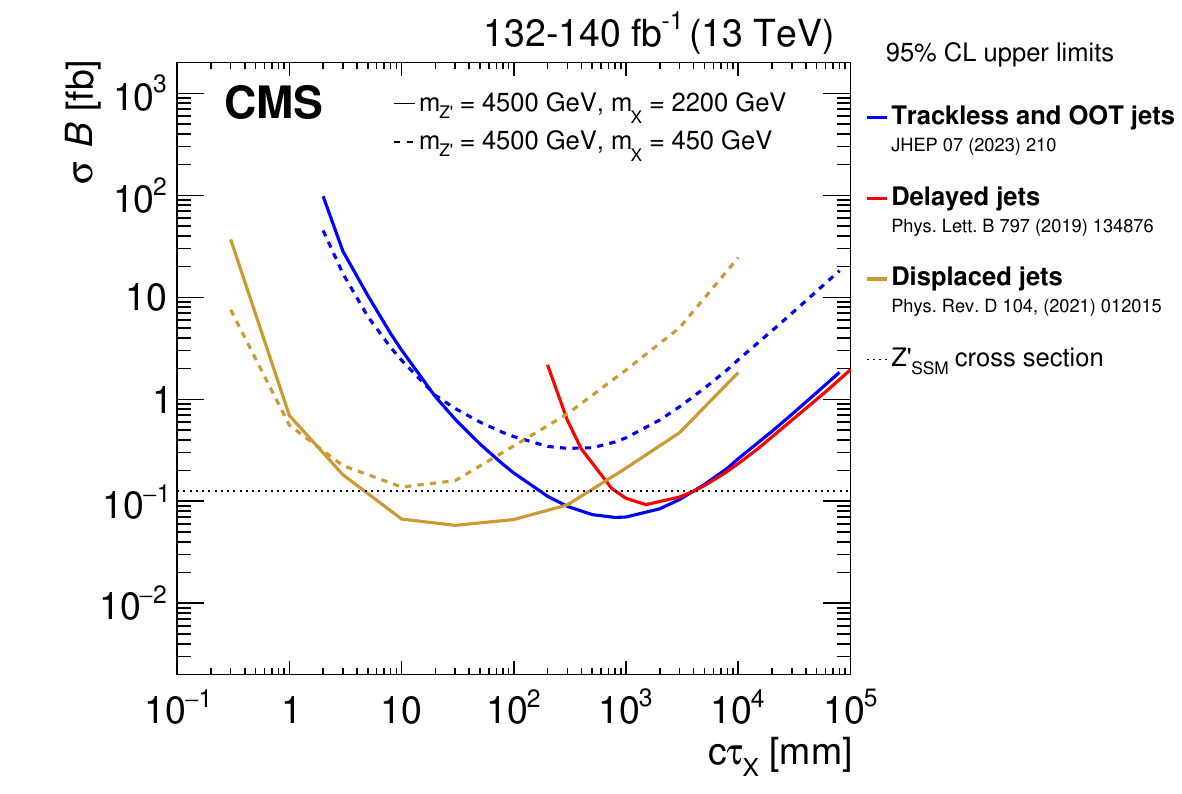}
\caption{Observed 95\% \CL exclusion limits for \PZpr bosons decaying into LLPs with hadronic plus \ptmiss final states, for a \PZpr boson mass of 3000\GeV (upper) and 4500\GeV (lower). Analyses employing different strategies are shown to have complementary lifetime sensitivity~\cite{EXO-19-001,CMS-PAS-EXO-21-014,CMS:2020iwv}. The theoretical cross section assumes the \PZpr boson has SM-like couplings to SM quarks~\cite{Altarelli:1989ff}.
}
\label{fig:zprimeLL_2b2nu}
\end{figure*}

The exclusion limits for several CMS LLP searches to \PZpr bosons decaying into a pair of LLPs are shown in Fig.~\ref{fig:zprimeLL_4b} for \PZpr boson masses of 3000 and 4500\GeV. The use of multiple search techniques provides extensive lifetime coverage. The DV search has the best sensitivity for lower lifetimes as it uses the tracker while the calorimetry and muon system based searches have optimal sensitivity for longer lifetimes. To probe spectra with DM candidates, models in which the \PZpr boson decays into an LLP and a DM candidate are considered in Fig.~\ref{fig:zprimeLL_2b2nu}. As \ptmiss is significantly increased, searches using \ptmiss show substantially improved reach compared to signal models in which the \PZpr boson decays into two LLPs.

\begin{figure*}[htbp!]
\centering
\includegraphics[width=0.9\linewidth]{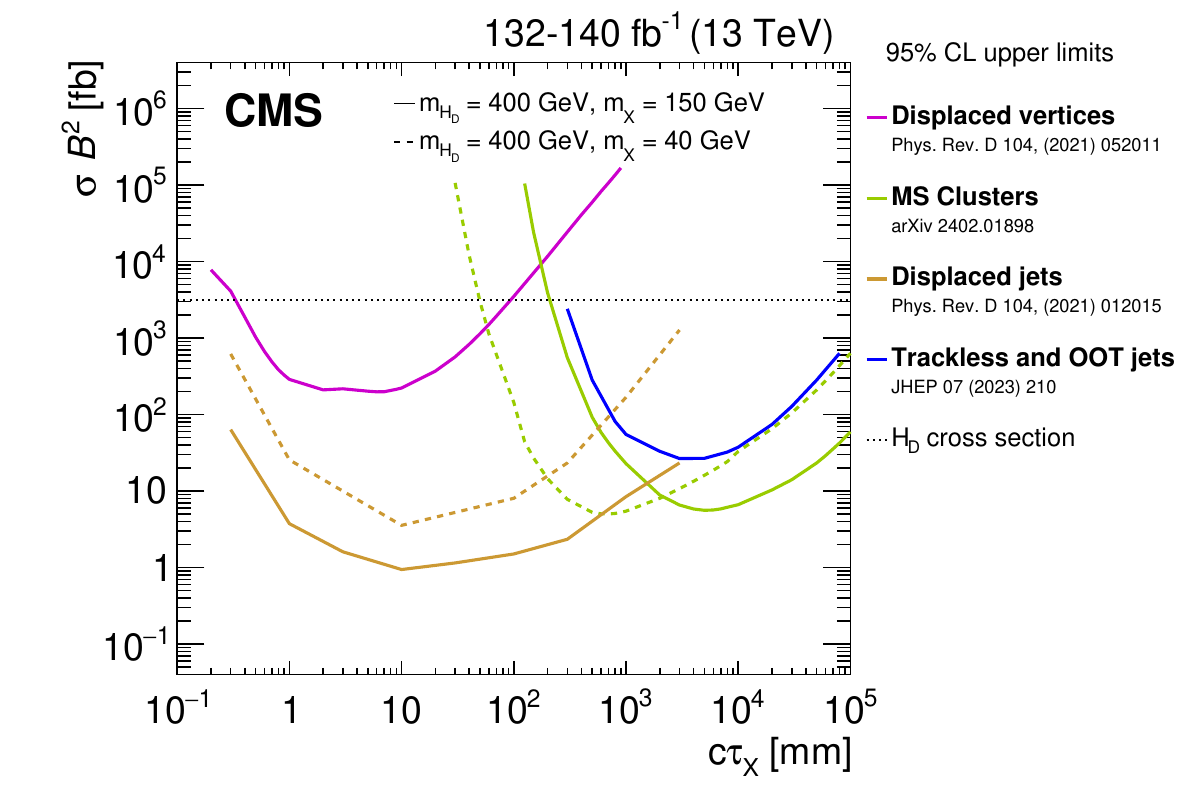}
\includegraphics[width=0.9\linewidth]{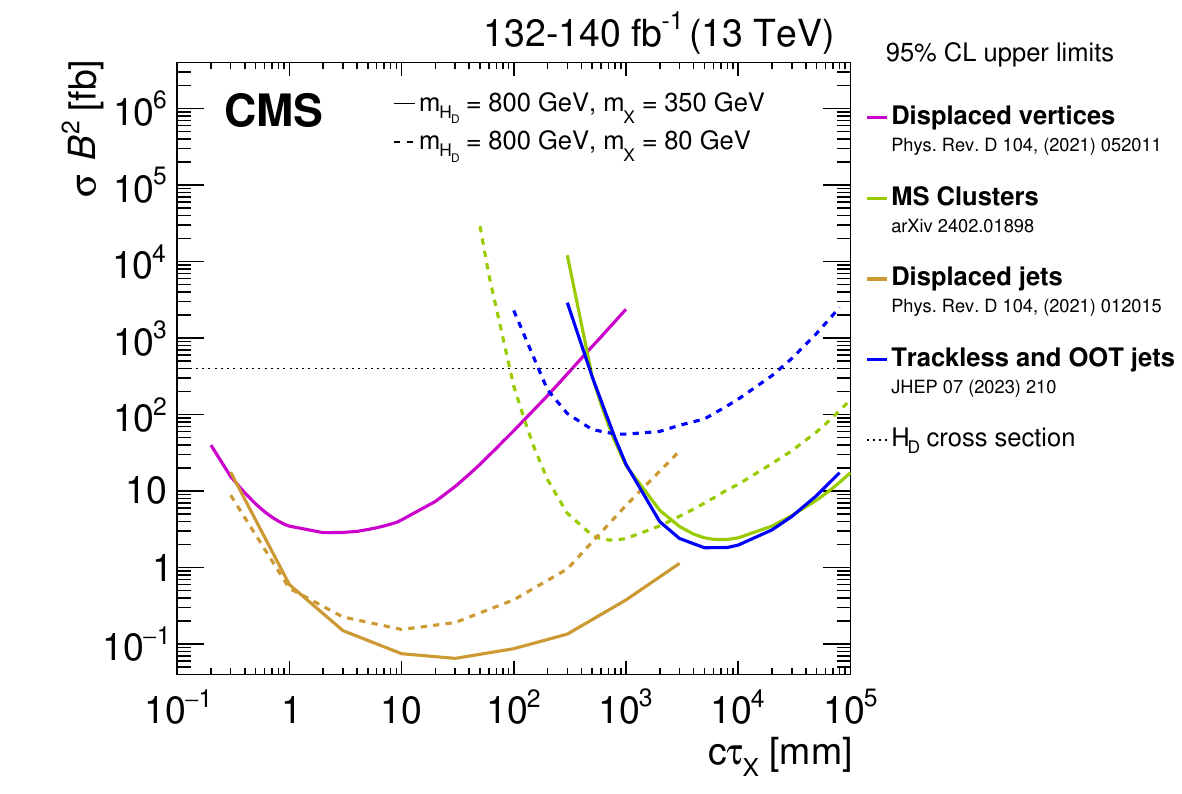}
\caption{Observed 95\% \CL exclusion limits for \Hdark decaying into LLPs with a fully hadronic final state, for a \Hdark mass of 400\GeV (upper) and 800\GeV (lower). The \Hdark production cross section is calculated using point-like effective theory and assumes the mediator is produced through gluon-gluon fusion with couplings similar to the SM Higgs boson~\cite{deFlorian:2016spz}. Analyses employing different strategies are shown to have complementary lifetime sensitivity~\cite{EXO-19-013,CMS:2023arc,CMS:2020iwv,CMS-PAS-EXO-21-014}.
}  
\label{fig:higgsLLP_4b}
\end{figure*}

\begin{figure*}[htbp!]
\centering
\includegraphics[width=0.9\linewidth]{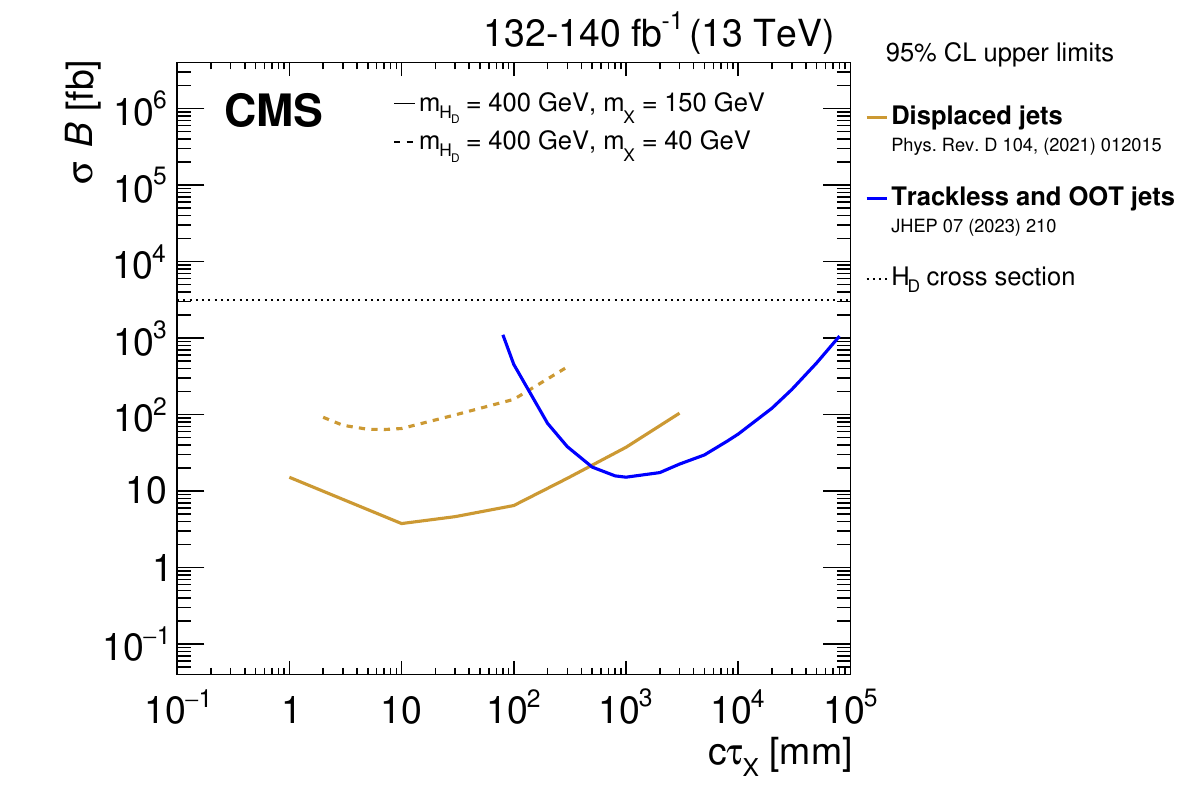}
\includegraphics[width=0.9\linewidth]{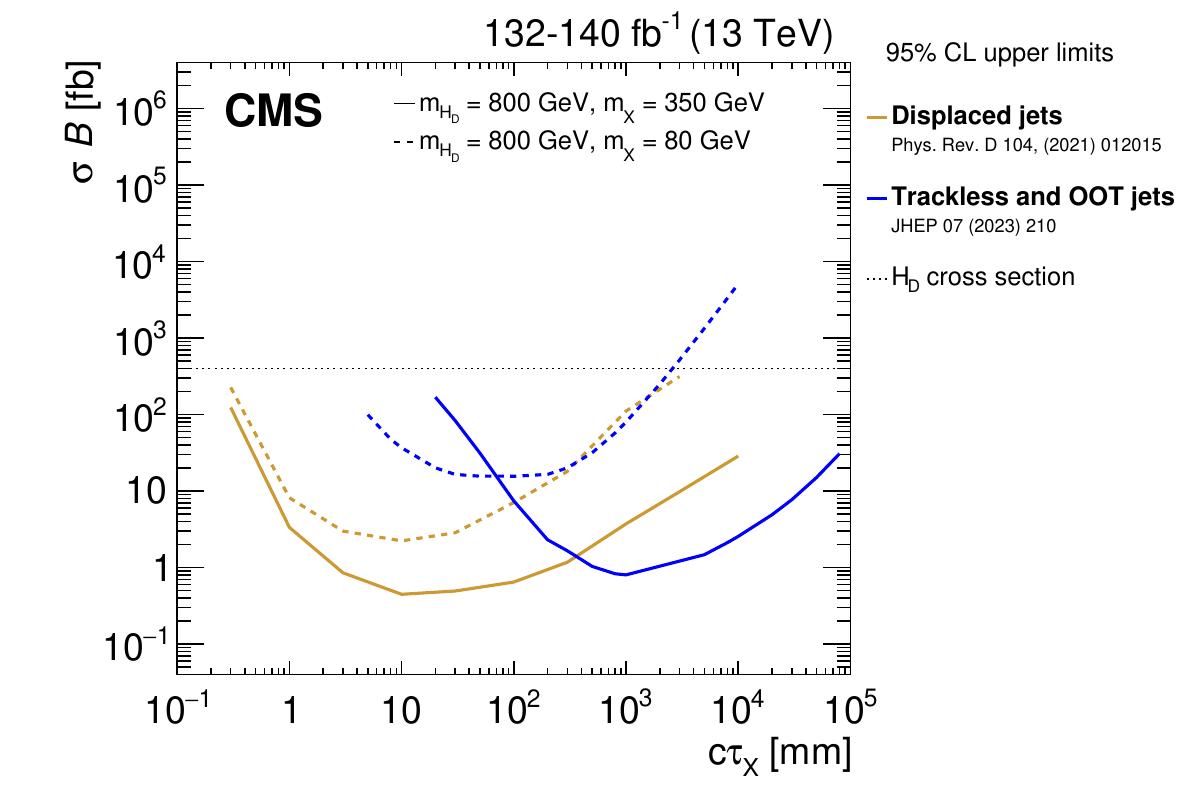}
\caption{Observed 95\% \CL exclusion limits for \Hdark decaying into LLPs with a hadronic plus \ptmiss final state, for a \Hdark mass of 400\GeV (upper) and 800\GeV (lower). The \Hdark production cross section is calculated using point-like effective theory and assumes the mediator is produced through gluon-gluon fusion with couplings similar to the SM Higgs boson~\cite{deFlorian:2016spz}. Analyses employing different strategies are shown to have complementary lifetime sensitivity~\cite{CMS:2020iwv,CMS-PAS-EXO-21-014}.
}  
\label{fig:higgsLLP_2b2nu}
\end{figure*}

The exclusion limits for several CMS LLP searches for \Hdark decaying into a pair of LLPs are shown in Fig.~\ref{fig:higgsLLP_4b}  for \Hdark masses of 400 and 800\GeV, respectively. The use of multiple search techniques is again shown to provide extensive lifetime coverage. It can also be seen that the large energy thresholds used for the DV search cause a stronger dependence of the sensitivity on the mass of \Hdark compared to the muon system search. To probe spectra with DM candidates, models in which the \Hdark decays into an LLP and a DM candidate are considered in Fig.~\ref{fig:higgsLLP_2b2nu}. As \ptmiss is significantly increased for such signatures, searches using \ptmiss show substantially improved reach.


\section{Experimental methods}
\label{sec:commonExperimentalChallenges}

Searches for DS physics use experimental methods that are applicable to many types of signatures. Various methods are employed, shared, and continually improved across different analyses. 

In this Section, we describe standard CMS event reconstruction techniques like tracking and vertexing (Section~\ref{sec:trackingAndVertexing}) and the particle-flow (PF) algorithm (Section~\ref{sec:PF}). We also describe the reconstruction of objects that are particularly relevant for DS searches, including jet substructure (Section~\ref{sec:jetsAndSubstructure}), \ptmiss (Section~\ref{sec:ptmiss}), LLPs (Section~\ref{sec:LLPreconstruction}), and the precision proton spectrometer (PPS, Section~\ref{subsec:PPS}).

Given the prominence of the \ptmiss variable in DS searches, and the sizeable effect of pileup on that observable, pileup mitigation methods (Section~\ref{sec:met}) are developed to make its estimation as robust as possible. Similarly, the composite nature of the \ptmiss means that every component must correspond to a real, well-reconstructed particle arising from the collision. To mitigate the effects of detector and reconstruction software failures, which can lead to artificial \ptmiss, filters for spurious events (Section~\ref{sec:metfilters}) are developed and used in the analyses. 

The design, deployment, monitoring, and characterization of trigger algorithms are fundamental components of all CMS analyses. Certain DS signatures introduce unique features that necessitate extensions to the standard trigger and data acquisition paradigm. This new data-taking paradigm is discussed in Section~\ref{sec:trigger}.

Analysis techniques for heavy ion collisions are discussed in Section~\ref{subsec:HI}.

Finally, searches for DS physics often probe the tails of distributions, e.g., of \ptmiss for mono-X searches using simplified models, or of displacements in searches for LLPs; equally challenging, many of the searches for extended DSs look for subtle signatures in multijet events. All of these signatures are notoriously hard to model accurately using simulation. To obtain a robust and reliable background estimation, many DS searches therefore employ methods based on CRs in data or other techniques based on collision data to estimate the major background contributions. A few such methods shared among many of the search efforts are discussed in Section~\ref{sec:background}. These methods based on CRs in data are employed only when simulation-based methods are not sufficient, as significant challenges must be overcome, such as finding appropriate and representative CRs with adequate amounts of data and accounting for signal contamination.

\subsection{Tracking and vertexing}\label{sec:trackingAndVertexing}
The track reconstruction starts from the hit reconstruction, where the signal above specific thresholds in pixel and strip channels are clustered into hits. The initial estimation of the hit position is determined by the charge and the position of the cluster and is corrected for the Lorentz drift in the magnetic field. This initial estimation of the hit position is utilized in the following steps of seed generation and track finding. 

In the seed generation, the initial possible track candidates are formed, which serve as the starting points for the propagation using the Kalman filter~\cite{Fruhwirth:1987fm}. The CMS detector utilizes an iterative tracking process~\cite{CMS:2014pgm,Run2TrackingDPNote}, with each iteration starting from a specific group of seeds. The seeds are formed using two, three, or four hits in the different layers of the pixel detector and the strip detector. The earlier iterations utilize the hits in the pixel detector to target prompt tracks, while the later iterations focus more on the tracks with larger displacements. After each iteration, hits associated with reconstructed tracks are removed. In this way, the tracking at CMS becomes efficient for reconstructing tracks with different displacements, with high efficiency up to about 60\cm. 

After the seeds belonging to a given iteration are formed, a combinatorial track finder based on the Kalman filter is applied, where the track candidates produced by the seeds are extrapolated to the next compatible layers using the Kalman filter. After the extrapolation reaches the final layer, track fitting is achieved by updating the track parameters through the smoothing step of the Kalman filter. The track candidates with too many missing hits or with $\pt$ below some threshold specific to a given iteration are dropped. Since all the seeds are extrapolated at the same time, there could be some tracks with significant overlaps. When two tracks share more than 19\% of the hits, the one with a smaller number of hits is removed; if both tracks have the same number of hits, the one with a larger $\chi^{2}$ is discarded. 

The primary vertex (PV) is taken to be the vertex corresponding to the hardest scattering in the event, evaluated using tracking information alone, as described in Section~9.4.1 of Ref.~\cite{Contardo:2015bmq}. The silicon tracker used in 2016 measured charged particles within the range $\abs{\eta} < 2.5$. For nonisolated particles of $1 < \pt < 10\GeV$ and $\abs{\eta} < 1.4$, the track resolutions were typically 1.5\% in \pt and 25--90 (45--150)\mum in the transverse (longitudinal) impact parameter~\cite{CMS:2014pgm}. At the start of 2017, a new pixel detector was installed~\cite{Phase1Pixel}; the upgraded tracker measured particles up to $\abs{\eta} < 3.0$ with typical resolutions of 1.5\% in \pt and 20--75\mum in the transverse impact parameter~\cite{DP-2020-049} for nonisolated particles of $1 < \pt < 10\GeV$. 

\subsection{Particle flow}\label{sec:PF}
The PF algorithm~\cite{CMS:2017yfk} aims to reconstruct and identify each individual particle in an event, with an optimized combination of information from the various elements of the CMS detector. The energy of photons is obtained from the ECAL measurement. The energy of electrons is determined from a combination of the electron momentum at the primary interaction vertex as determined by the tracker, the energy of the corresponding ECAL cluster, and the energy sum of all bremsstrahlung photons spatially compatible with originating from the electron track. The energy of muons is obtained from the curvature of the corresponding track. The energy of charged hadrons is determined from a combination of their momentum measured in the tracker and the matching ECAL and HCAL energy deposits, corrected for the response function of the calorimeters to hadronic showers. Finally, the energy of neutral hadrons is obtained from the corresponding corrected ECAL and HCAL energies.

\subsection{Jets and substructure}\label{sec:jetsAndSubstructure}
For each event, hadronic jets are clustered from these reconstructed particles using the infrared- and collinear-safe anti-\kt algorithm~\cite{Cacciari:2008gp, Cacciari:2011ma} with a distance parameter of 0.4 (AK4 jets) or 0.8 (AK8 jets). Some analyses also use the Cambridge--Aachen algorithm~\cite{Dokshitzer:1997in} with a distance parameter of 1.5 (CA15 jets). Any jet with a distance parameter ${>}0.4$ is referred to as a ``large-radius jet''. Jet momentum is determined as the vectorial sum of all particle momenta in the jet, and is found from simulation to be, on average, within 5--10\% of the true momentum over the entire \pt spectrum and detector acceptance. Additional $\Pp\Pp$ interactions within the same or nearby bunch crossings, known as pileup (PU), can contribute additional tracks and calorimetric energy depositions, increasing the apparent jet momentum. To mitigate this effect, tracks identified to be originating from PU vertices are discarded and an offset correction is applied to correct for remaining contributions~\cite{CMS:2020ebo}. Jet energy corrections are derived from simulation studies so that the average measured energy of jets becomes identical to that of particle-level jets. In situ measurements of the momentum balance in dijet, $\phojet$, $\PZ$+jet, and multijet events are used to determine any residual differences between the jet energy scale in data and in simulation, and appropriate corrections are made~\cite{CMS:2016lmd}. Additional selection criteria are applied to each jet to remove jets potentially dominated by instrumental effects or reconstruction failures~\cite{CMS:2020ebo}.

If a resonance is much heavier than its decay products, the decay products are highly Lorentz boosted. This results in very collimated sprays of particles from those decay products, where hadronic decays cannot be reconstructed into individual small-radius jets, but are merged into one large-radius jet. In order to remove soft and wide-angle radiation in these jets, jet substructure~\cite{Marzani:2019hun} or jet grooming techniques such as trimming~\cite{Krohn:2009th} and soft drop~\cite{softdrop} are applied. Jet trimming is a method that removes sources of contamination by exploiting the difference in scale between the hard emissions of final state radiation and the relatively soft emissions from initial-state radiation (ISR). This algorithm begins with seed jets that are reclustered using the anti-\kt algorithm and then trimmed according to the subjet \pt. The soft-drop algorithm removes soft and wide-angle radiation from the jet by reclustering the large-radius jet with the Cambridge--Aachen algorithm and testing if $\min(\ptsub{i},\ptsub{j}) > z_{\text{cut}} \ptsub{i+j} (\DR_{ij} / R)^{\beta}$ in each declustering step. The standard parameters used in the CMS experiment are $z_{\text{cut}} = 0.1$ and $\beta = 0$. The hardest branch is followed until the soft-drop requirement is fulfilled, where the procedure stops. As a consequence, at most two soft-drop subjets are defined by this procedure. The mass is calculated as the invariant mass of the two subjets and is called the soft-drop mass \msd.

\subsection{\texorpdfstring{\ptmiss}{Missing transverse momentum}}\label{sec:ptmiss}
The missing transverse momentum vector \ptvecmiss is computed as the negative vector sum of the transverse momenta of all the PF candidates in an event, and its magnitude is denoted as \ptmiss~\cite{CMS:2019ctu}. The \ptvecmiss is modified to account for corrections to the energy scale of the reconstructed jets in the event. The reconstruction of \ptmiss, a key parameter in many DS searches, poses a significant challenge in the high-PU environment of the LHC. CMS has made concerted efforts to characterize the detector response and resolution to optimize the measurement of \ptmiss, as detailed in Sections~\ref{sec:met} and \ref{sec:metfilters}. Additional variables that represent aspects of the event global activity are also defined and used throughout the analyses. The total hadronic transverse momentum \HT is defined as the scalar \pt sum of all jets that meet certain selection criteria. While the details of the selection may vary among different analyses, a common definition is to use all jets with $\PT > 30\GeV$ and $\abs{\eta} < 3.0$. The missing hadronic transverse momentum (missing \HT, \HTmiss) is similarly defined as the magnitude of the vector \ptvec sum of all jets. In the same vein, the hadronic recoil $\vec{u}$ is defined as the vector \ptvec sum of all PF candidates except for those identified with the decay products of an EW boson. It is often used as an ancillary variable to approximate the behavior of the \ptmiss.

\subsection{Long-lived particle reconstruction} \label{sec:LLPreconstruction}

\begin{figure}
    \centering
    \includegraphics[width=0.99\textwidth]{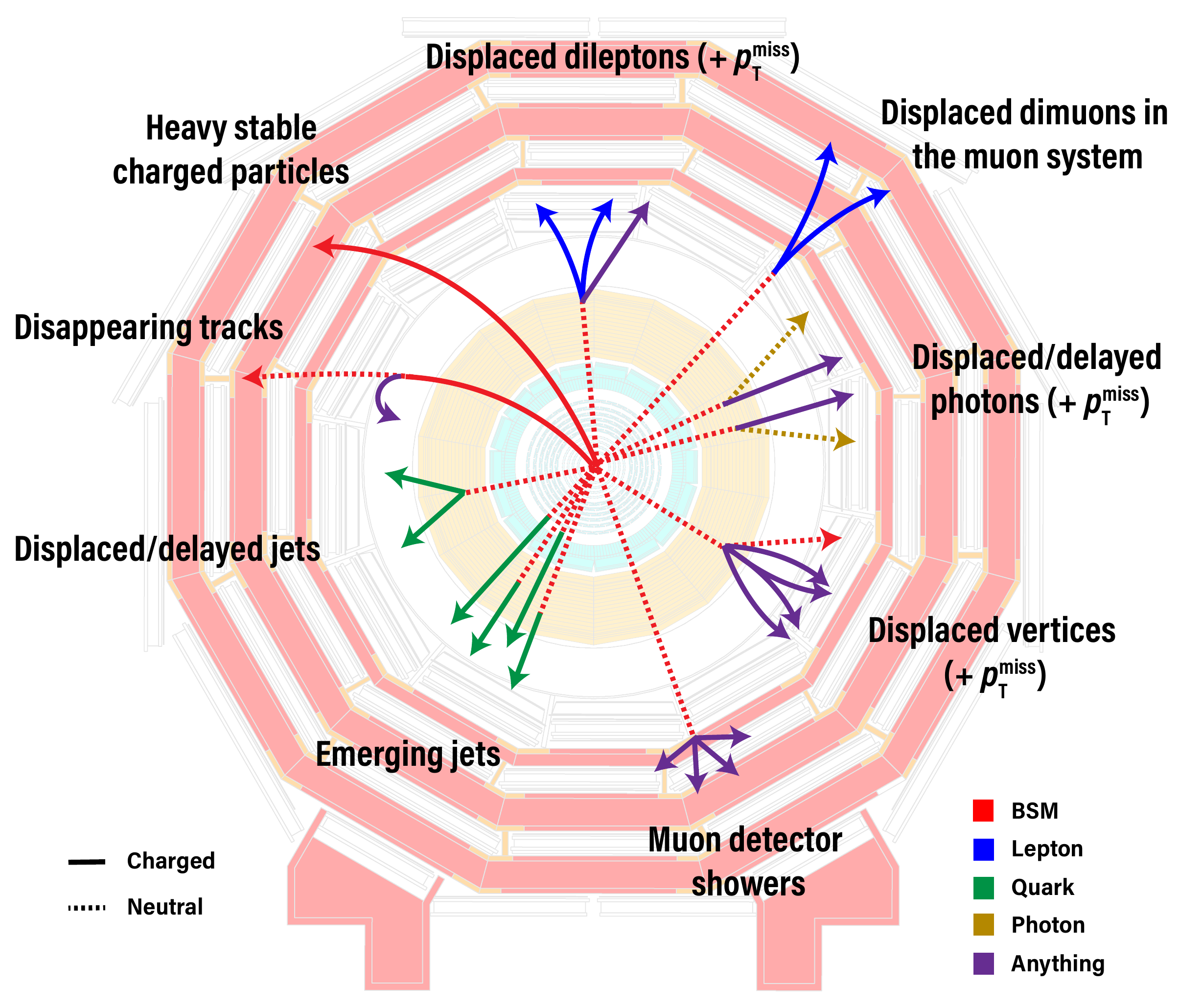}
    \caption{Sketch of several LLP signatures probed by the CMS experiment. Figure adapted from Ref.~\cite{Alimena:2019zri}.}
    \label{fig:LLPsignatures}
\end{figure}

When produced at the LHC, LLPs have a distinct experimental signature: they can decay far from the primary $\Pp\Pp$ interaction vertex but within the detector, or even completely pass through the detector before decaying. Figure~\ref{fig:LLPsignatures} shows a sketch of several LLP signatures probed by the CMS experiment. Some specific examples of LLP signatures include displaced and delayed leptons, photons, and jets; disappearing tracks; and nonstandard tracks produced by monopoles or heavy stable charged particles. In the context of DS searches, the reconstruction and identification of LLPs depend on their intrinsic properties, such as mass, charge, and lifetime~\cite{Alimena:2019zri,Lee:2018pag}. Standard triggers, object reconstruction, and background estimation are often inadequate for LLP searches because they are designed for promptly decaying particles that originate near the interaction point, and custom techniques are often needed to analyze the data if one considers particles with significant displacement. However, some CMS LLP searches use standard PF objects, for example, if they target small displacements in the pixel detector or if they reconstruct prompt objects that are produced in association with LLPs.

Here we describe specific offline object reconstruction techniques that are used to identify long-lived and displaced particles in CMS. Section~\ref{sec:displacedTracking} describes the dedicated tracking and vertexing algorithms used for displaced particles, and the options available to mitigate challenges coming from tracks that do not belong to the DV. Section~\ref{sec:jetTagging} illustrates how a dedicated displaced-jets tagger improves the identification of such displaced jets amid jets from SM background processes. Then, out-of-time (OOT) jets and delayed calorimetry are described in Section~\ref{sec:delayedCalo}, and the dedicated reconstruction algorithms for displaced muons are described in Section~\ref{sec:displacedMuons}. In addition, the reconstruction of a new signature coming from particles that shower in the muon system is described in Section~\ref{sec:mdShowers}. Finally, the measurement of \dEdx for LLPs that produce anomalous ionization signals in the tracker is described in Section~\ref{dEdx}.

The dedicated reconstruction techniques detailed in this section employ almost every CMS subdetector from the tracker to the muon system to provide sensitivity to displaced particles with displacements from a few tens of microns to several meters. For smaller displacements, the performance of LLP searches is limited by the resolution of the tracker as well as large amounts of SM backgrounds from \PB meson decays.

\subsubsection{Displaced tracking and vertexing}
\label{sec:displacedTracking}

Displaced tracking and displaced vertexing are important handles to identify LLPs decaying inside the inner tracking system of CMS. As explained in Section~\ref{sec:trackingAndVertexing}, displaced tracking is included in the last iterations of the iterative tracking process. This iterative tracking approach is also available in the HLT system of CMS. Although HLT tracking has a degraded performance compared to offline tracking and is usually limited to some specific regions of interest, such as regions around jets, it enables us to develop and implement dedicated LLP triggers for displaced jets and displaced leptons, greatly enlarging the coverage of the LLP searches at CMS~\cite{LLPRun3TriggerDPNote}, as shown throughout Section~\ref{sec:reinterpretationAndResults}. 

\begin{figure}
    \centering
    \includegraphics[width=0.5\textwidth]{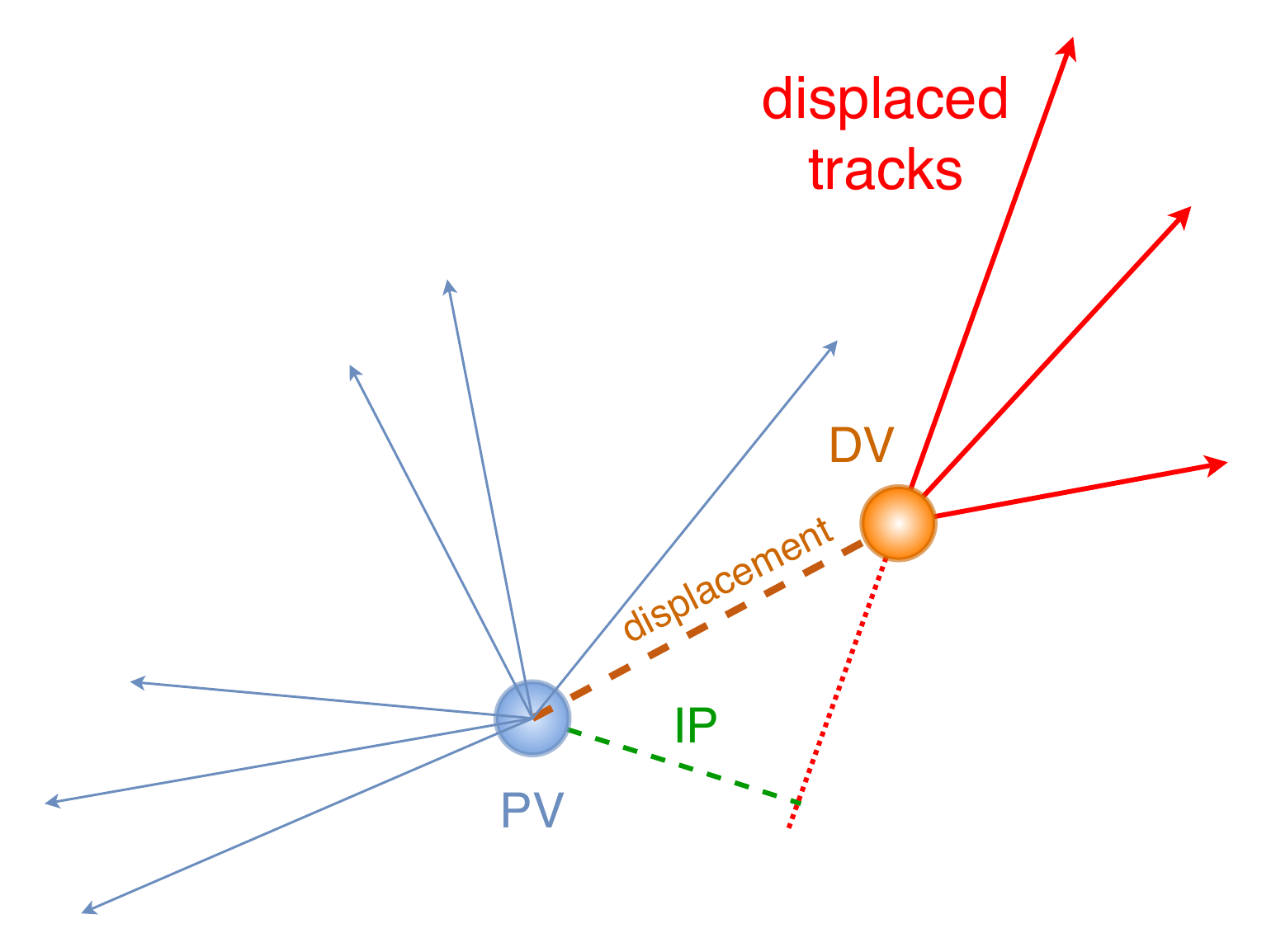}
    \caption{Illustration of the appearance of a displaced vertex (DV) from the decay of a long-lived particle resulting in charged-particle tracks that are displaced with respect to the primary interaction vertex (PV), and hence can have large impact parameter (IP) values. 
In BSM searches, LLPs have very long lifetimes compared to SM particles, leading to large displacements of the secondary vertices. Figure adapted from Ref.~\cite{CMS:2017wtu}.}
    \label{fig:secondaryVertex}
\end{figure}

Beyond the track reconstruction, displaced vertexing using the reconstructed tracks is also a powerful tool to further discriminate exotic LLP signatures from SM background processes. The ``inclusive vertex finder'', which is the standard DV reconstruction algorithm at CMS~\cite{CMS:2017wtu}, is tuned for reconstructing decays of heavy-flavor hadrons arising from SM processes through their secondary vertex and is not efficient in reconstructing exotic LLP decays. Therefore dedicated DV reconstruction algorithms are used in exotic LLP searches, which significantly improve the signal-to-background discrimination. The vector pointing from the PV to the point of closest approach of a DV track is referred to as the impact parameter (IP) vector. Figure~\ref{fig:secondaryVertex} illustrates these concepts.

In general, for vertex reconstruction tasks, it can be proven mathematically that the Kalman filter provides the optimal performance assuming Gaussian noise and no outlier tracks, which are the tracks that do not belong to the vertex but are used in the fitting. In reality, however, the presence of outlier tracks is inevitable, owing to the dense tracking environment associated with the $\Pp\Pp$ collisions at CMS, especially when searching for DVs accompanied by hadronic decays. Several approaches have been adopted in CMS LLP searches to address this challenge of searching for DVs amidst outlier tracks. 

One approach to filtering outlier tracks in the DV reconstruction is to start with all possible pairs of preselected tracks, which serve as the initial vertex candidates. The vertex candidates are then iteratively merged when they share tracks and have a small distance significance between the two vertices. After each merging, the new vertex candidate is refitted using the Kalman filter, and the vertex candidates with large $\chi^{2}$ per degree of freedom are dropped. In this way, the input track candidates are automatically partitioned into different vertices during the vertex reconstruction process, while minimizing potential contamination from outlier tracks. This method is employed by several searches for DVs within the beam pipe~\cite{EXO-19-013,CMS:2024trg}.

Another powerful technique to tackle the outlier-track contamination issue is the adaptive vertex fitter (AVF)~\cite{Fruhwirth:2007hz}, which is used in the inclusive displaced-jets search~\cite{CMS:2020iwv}. The AVF is a combination of the Kalman filter and the deterministic annealing algorithm. The deterministic annealing algorithm is inspired by an analogy to statistical physics, where the clustering of objects (\eg, tracks) can be viewed as a phase transition when the system experiences a gradual reduction of ``temperature", known as the annealing process~\cite{Rose:1998dzq}. Through the annealing process, one can better escape from the local minima caused by outliers and reach the true solution of a given optimization problem. In AVF, during the fitting, each track is assigned a weight according to its distance significance with respect to the vertex candidates and a given temperature $T$, which controls the shape of the weight function:

\begin{equation}\label{eq:avf}
w_{\text{track}_{i}}\equiv\frac{\exp(-\chi_{i}^{2}/2T)}{\exp(-\chi_{i}^{2}/2T)+\exp(-\chi_{c}^{2}/2T)}, \quad \chi_{i}^{2} = d^{2}_{i}(\mathbf{x}_{i}, \mathbf{v})/\sigma^{2}_{i} \, 
\end{equation}
where $\chi_{c}^{2}$ defines a threshold such that a track with larger $\chi_{i}^{2}$ is more likely to be an outlier than to have its position $\mathbf{x}$ associated with the vertex with position $\mathbf{v}$. The Kalman filter is then applied iteratively using the weighted track candidates. At each iteration, a specific value of $T$ is chosen, starting at 256, and decreasing iteration by iteration until it reaches 1. The values of $T$ are chosen such that the vertex reconstruction has good efficiency and resolution. In this way, the outlier tracks with large $\chi_{i}^{2}$ are downweighted after each iteration, which leads to a vertex fitting that is robust against the contamination of outlier tracks.

Pileup mitigation is another important consideration that analysts must consider when vertexing displaced objects. In the case of displaced-jet searches, it has become conventional to select the vertex with the assistance of the $\alpha_{\text{max}}$ parameter~\cite{Sirunyan:2017jdo}, shown here for a particular vertex $v_i$ and jet $j$:
\begin{equation}
    \alpha_{\text{max}}(\mathrm{j}) \equiv \max_{v_i} \alpha({v_i}, \mathrm{j})= \max_{v_i}{\left[\frac{\sum_{\text{tracks} \in v_i \cap \mathrm{j}} \pt}{\sum_{\text{tracks} \in j}\pt} \right]}.
\end{equation}
The parameter $\alpha_{\text{max}}$ takes the maximum of the ratio of the summed track \pt for all tracks associated with the jet and a particular vertex $v_i$ to the total summed \pt for all tracks associated with the jet in consideration, where $v_i$ are the candidate primary vertices. Tracks are associated with the jet geometrically, e.g., by defining a $\Delta R$ requirement that is consistent with the type of jet used in the analysis. The tracks are associated with a vertex based on their weight, calculated for a given vertex as in Eq.~\eqref{eq:avf}. The individual values of $\alpha$ for a given vertex $v_i$ range from 0 to 1, where $\alpha \approx 0$ is most consistent with displaced jets and $\alpha \approx 1$ is most consistent with prompt jets from the PV. The value of $\alpha$ for PU jets is within the range of 0 to 1 for a given vertex. To avoid selecting these jets, one takes the maximum of the alpha values for all vertices in the event.

\subsubsection{Displaced-jet tagger}
\label{sec:jetTagging}

Jets displaced from the $\Pp\Pp$ collision region, and arising from the decay of LLPs, are a key experimental signature for many theoretical extensions to the SM~\cite{ArkaniHamed:2004fb, Giudice:2004tc, Hewett:2004nw, Barbier:2004ez, Strassler:2006ri}.

In the displaced-jets search~\cite{CMS:2020iwv}, a dedicated algorithm was deployed to reconstruct the DV arising from LLP decays, using the displaced tracks associated with a dijet system. The properties of the associated tracks and DV can provide the discrimination power to distinguish LLP signals from SM backgrounds. A displaced-jets tagger is built using these properties based on a gradient-boosted decision tree (GBDT).

\begin{figure}
    \centering
    \includegraphics{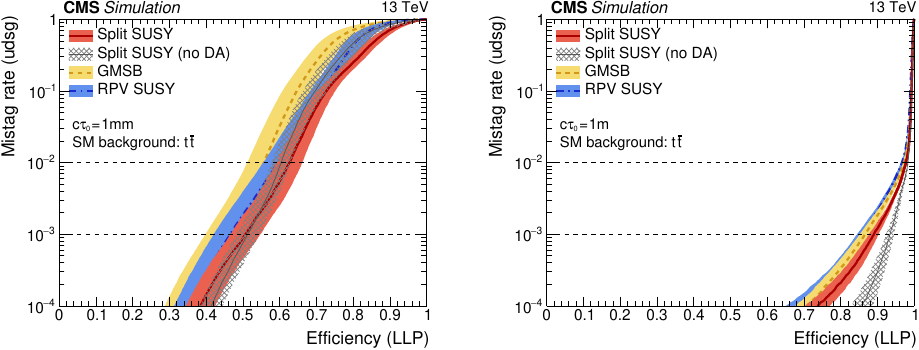}
    \caption{The ROC curves illustrating the displaced jet tagger performance for a split SUSY benchmark model with a long-lived gluino with mass 2000\GeV that decays into a quark-antiquark pair and a neutralino with mass 0\GeV (solid line), a GMSB SUSY benchmark model with a long-lived gluino with a mass of 2500\GeV that decays into a gluon and a light gravitino of mass 1\unit{keV} (dashed line), and an $R$-parity violating (RPV) SUSY~\cite{Barbier:2004ez} benchmark model with a long-lived top squark with a mass of 1200\GeV that decays into a bottom quark and a charged lepton (dot-dashed line), assuming $\ctau$ values of 1\mm (\cmsLeft) and 1\unit{m} (\cmsRight). The thin line with hatched shading indicates the performance obtained with a DNN training using split SUSY samples but without domain adaptation (DA). Figure taken from Ref.~\cite{CMS:2019dqq}.}
    \label{fig:displacedJetDNN}
\end{figure}

A deep neural network (DNN) has also been designed to identify displaced jets~\cite{CMS:2019dqq}. The DNN architecture is inspired by the CMS \textsc{DeepJet} algorithm~\cite{Stoye_2018, CMS-DP-2018-058} that identifies jets originating from the hadronization of \PQb quarks. The DNN provides a multiclass classification scheme similar to the \textsc{DeepJet} algorithm but it also accommodates the ``LLP jet'' class. The network is trained using simulated events, which are typically drawn from the relevant parameter space of simplified models. Given that the experimental signature for a displaced jet depends strongly on the lifetime of the LLP, a parameterized approach~\cite{Baldi:2016fzo} is adopted by using the lifetime parameter as an input to the DNN. This approach permits hypothesis testing over several orders of magnitude of lifetimes using a single network. Another key design feature of the DNN is the use of domain adaptation~\cite{ganin2014unsupervised}, along with the use of training examples taken from LHC data, to ensure a similar classification performance in simulation and $\Pp\Pp$ collision data. These two improvements, namely using domain adaptation and the lifetime as an input to the DNN, were made on top of the original \textsc{DeepJet} architecture. The performance of the tagger is model- and lifetime-dependent, but it can typically provide a rejection factor in excess of 10\,000 for jets from SM processes while maintaining a large signal efficiency (e.g., ${\gtrsim}10\%$) for LLPs with proper decay lengths in the millimeter range.
Figure~\ref{fig:displacedJetDNN} shows the receiver operating characteristic (ROC) curves for the DNN, for a number of SUSY models that contain an LLP and for two choices of lifetimes, $\ctau = 1\mm$ and 1\unit{m}. The ROC curves are shown for background rejection rates ${>}10^{-4}$ because the working points only use this region.

\subsubsection{Delayed calorimetry}
\label{sec:delayedCalo}

\begin{figure}
    \centering
    \includegraphics[width=0.8\textwidth]{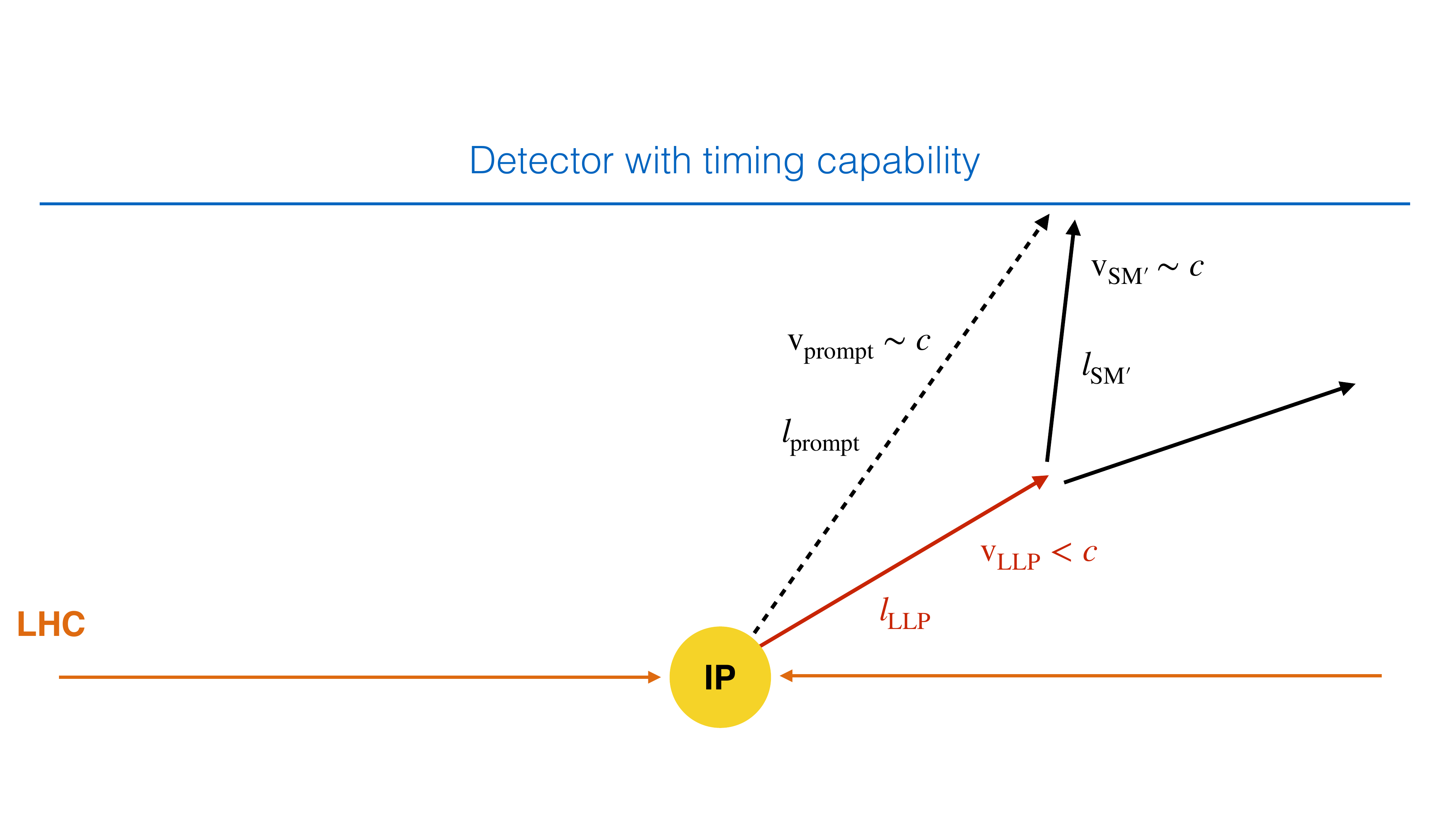}
    \caption{Illustration of contributions to the delay of particles that originate from LLP decays. For prompt decays, the path length to reach a particular location on the detector with timing capability ($l_{\text{prompt}}$), is smaller than the path length for a deposit originating from an LLP decay ($l_{\mathrm{LLP}}$+$l_{\mathrm{SM'}}$). In addition, the velocity of the light SM particles ($\rm{v}_{\text{prompt}}$) will be close to that of light while the velocity of the LLP ($\rm{v}_{\rm{LLP}}$) can be significantly lower. These factors lead to substantial delays for the decay products of LLPs, which can be exploited to improve sensitivity.}
    \label{fig:timingDiagram}
\end{figure}

The time resolution of the CMS calorimeter cells is around 400~ps for the ECAL~\cite{Chatrchyan:2009aj}, and a few \unit{ns} for the HCAL~\cite{Mans:1481837}. 
This performance makes timing an excellent discriminant to identify energy deposits from slow-moving particles that arrive out of time. As shown with a generic timing detector sketch in Fig.~\ref{fig:timingDiagram}, these deposits can be delayed for two reasons: the extended path length to reach the calorimetry as compared with deposits from particles originating from the interaction point, and heavy LLPs can travel with a velocity significantly smaller than that of light. The heavier the mass and longer the lifetime of the LLP, the longer it will take to reach the detector and deposit calorimeter energy.

The CMS Collaboration has carried out two analyses that exploit the fact that LLPs decaying into hadrons near the calorimeter surface can be identified as OOT jets~\cite{EXO-19-001, CMS-PAS-EXO-21-014}. The ECAL crystals associated with the jet can be used to define a new variable, the jet time, as the energy-weighted sum of the arrival times of measured pulses. The effective jet time resolution, taking into account clock jitter, size of the collision beam spot and calibration effects, ranges from 400--600\unit{ps} for jets with \pt ranging from 30--150\GeV. Any difference in the simulation of the time resolution~\cite{delRe:2015hla} is corrected by selecting dedicated CRs in the data.

\subsubsection{Displaced muons}
\label{sec:displacedMuons}

A detailed description of the CMS muon reconstruction algorithms and their performance has been given in Refs.~\cite{CMS:2013vyz,CMS:2012nsv,CMS:2018rym}. Here, we will briefly summarize how muons from $\Pp\Pp$ collisions are reconstructed in CMS in general and then describe the specifics of displaced-muon reconstruction.

In general, muons from $\Pp\Pp$ collisions in CMS are reconstructed using a combination of information from the tracker and the muon system. The muon system chambers are assembled into four ``stations'' at increasing distances from the interaction point; each station provides reconstructed hits in several detection planes, which are combined into track segments, forming the basis of muon track reconstruction in the muon system. ``Standalone muon tracks'' are built along the muon's trajectory using a Kalman filter technique~\cite{Fruhwirth:1987fm} that exploits track segments from the muon subdetectors (DTs, CSCs, and RPCs). Independently, ``tracker muon tracks'' are built by propagating tracker tracks to the muon system with loose matching to DT or CSC segments. If at least one muon segment matches the extrapolated track, the tracker track qualifies as a tracker muon track. Finally, ``global muon tracks'' are built by matching standalone muon tracks with tracker tracks. In contrast to tracker muons, global muon trajectories are determined from a combined Kalman filter fit using both tracker and muon system information.

\begin{figure}[htbp]
\centering
\includegraphics[width=.5\textwidth]{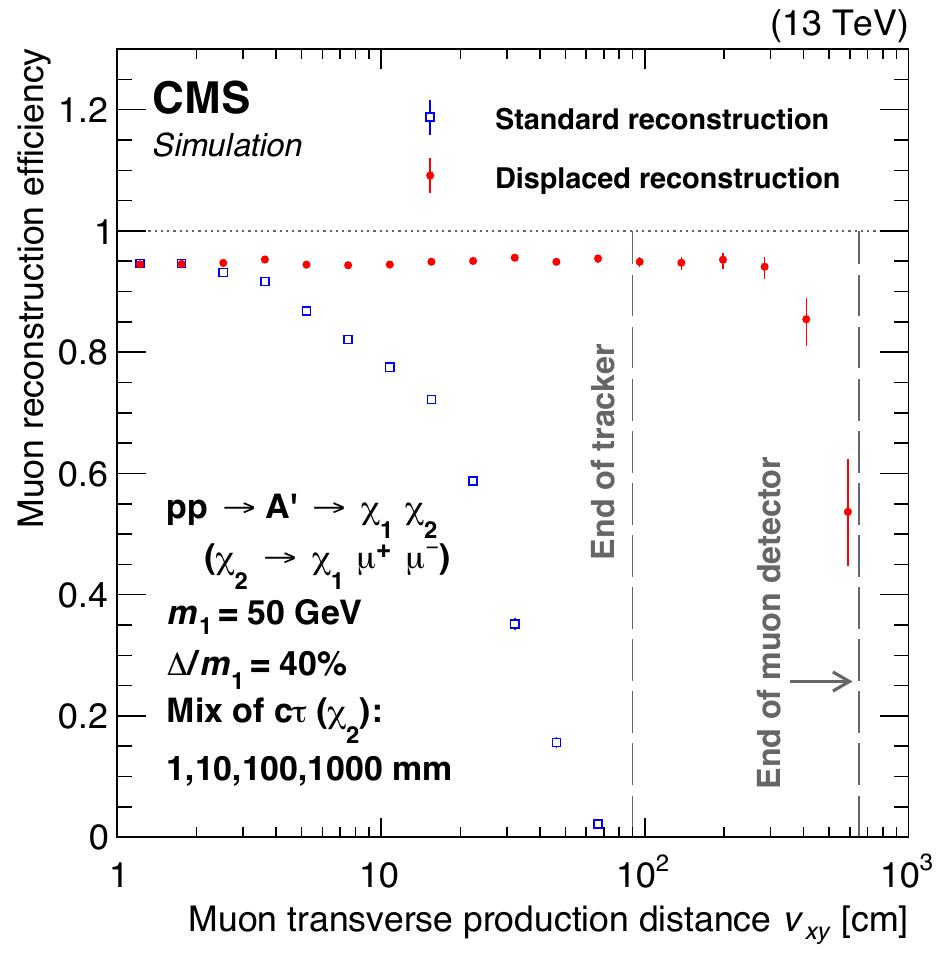}
\caption{Simulated muon reconstruction efficiency of standard global muon (blue squares) and DSA (red circles) track reconstruction algorithms as a function of transverse vertex displacement $v_{xy}$, for the IDM model discussed in Section~\ref{sec:iDM}. The two dashed vertical gray lines denote the ends of the fiducial tracker and muon detector regions, respectively. Figure taken from Ref.~\cite{CMS-PAS-EXO-20-010}.}
\label{fig:DSAeff}
\end{figure}

For displaced muons coming from decays of LLPs, the muon reconstruction algorithm that provides the best performance depends on how displaced the muon is from the interaction point. Muons produced relatively near the interaction point can be accurately reconstructed using the tracker muon or global muon reconstruction algorithms developed for prompt muons. The efficiency of these algorithms, however, rapidly decreases as the distance between the interaction point and the muon origin increases; the efficiency drops to zero for muons produced in the outer tracker layers and beyond. On the other hand, such muons are still efficiently reconstructed by the standalone muon reconstruction algorithms. These standalone muon algorithms reconstruct muons with displacements up to a few meters, but they have poorer spatial and momentum resolution than muons reconstructed using more precise information from the silicon tracker. The standalone muon reconstruction, as well as all prompt muon reconstruction algorithms, use beamspot constraints that bias the kinematics of displaced muons that are reconstructed. A ``displaced standalone'' (DSA) muon track reconstruction algorithm was developed for displaced muons~\cite{CMS-DP-2015-015,CMS:2018rym,exo-16-004}. The DSA muon track algorithm uses only hits in the muon chambers and has the beamspot constraints removed from all stages of the muon reconstruction procedure. Thus, DSA tracks provide the largest efficiency and best resolution for displaced muons, out of all the available standalone muon track algorithms. It maintains a muon reconstruction efficiency of 0.95 up to a muon transverse production distance of 300\unit{cm}, as compared with standard algorithms, where the efficiency steeply declines after 10\unit{cm}, as shown in Fig.~\ref{fig:DSAeff}~\cite{CMS-PAS-EXO-20-010}.

Several analyses~\cite{EXO-21-006,CMS-PAS-EXO-20-010} use displaced muons spanning a wide range of displacements, and take advantage of multiple muon reconstruction algorithms. For example, an attempt to match DSA tracks with global or tracker muons is made, and if such a match is found, the global or tracker muon is used for further analysis, while if not, the original DSA track is used. As a result of this matching procedure, much of the $\Pp\Pp$ collision background is eliminated and the sensitivity to LLP decays in the tracker is greatly increased because tracker and global muons have much better spatial and momentum resolution than standalone muons.

\subsubsection{Muon detector showers}
\label{sec:mdShowers}

LLPs that decay in the muon detectors could either be reconstructed as displaced muons, which are described in the previous section, or as muon detector showers. Owing to the design of the CMS muon detectors, which are composed of detector planes interleaved with the steel layers of the magnet flux-return yoke, LLPs that decay into any nonmuon particles within or just prior to the muon detectors can induce hadronic and electromagnetic showers, giving rise to a high hit multiplicity in localized detector regions. This signature uses the muon detector as a sampling calorimeter to identify displaced showers produced by LLPs that decay into hadrons, electrons, photons, or \PGt leptons. Additionally, due to the large amount of shielding from the calorimeters, solenoid, and steel flux-return yoke, requiring the presence of such a signature in an event reduces the otherwise large contributions from background processes.

To reconstruct the decays of LLPs in the muon detector, the muon detector hits are clustered in $\eta$ and the azimuthal angle $\phi$ using the \textsc{dbscan} algorithm~\cite{dbscan}, which groups hits by high-density regions. 
    
The cluster reconstruction efficiency strongly depends on the LLP decay position. The efficiency is largest when the LLP decays near the edge of the shielding material, where there is enough material to induce the shower, but not so much that it stops the shower secondaries. The cluster reconstruction efficiency also depends on whether the LLP decays hadronically or leptonically. In general, hadronic showers have larger efficiency, because they are more likely to penetrate through the steel in between stations, while showers induced from electromagnetic decays generally occupy just one station and are stopped by the steel between stations. When the LLP decays near or in the CSCs, the inclusive CSC cluster reconstruction efficiency is approximately 80\% for fully hadronic decays, 55\% for $\PGtp\PGtm$ decays, and 35\% for fully leptonic decays. When the LLP decays close to or in the DTs, the inclusive DT cluster reconstruction efficiency is approximately 80\% for fully hadronic decays, 60\% for $\PGtp\PGtm$ decays, and 45\% for fully leptonic decays.

\subsubsection{\texorpdfstring{\dEdx}{dE/dx}} \label{dEdx}

Studying anomalous ionization in the tracker provides a powerful tool to search for various LLP signals. For example, heavy charged particles are characterized by low speeds, inferred from time-of-flight measurements in muon chambers in case of sufficiently long lifetimes and large ionization signals in the tracker. In addition, anomalously low amounts of ionization in the tracker could indicate the presence of fractionally charged particles.

Each layer of the silicon pixel and strip trackers of CMS provides a measurement of the charge deposit, which is transformed into a \dEdx measurement after the application of a conversion factor from charge to energy and division by the path length. Dedicated estimators and discriminators have been designed to combine the set of \dEdx measurements in the most appropriate way. 

The \Ih estimator, first used in a CMS search reported in Ref.~\cite{Khachatryan:2011ts}, is defined as
\begin{equation}
    \centering
        \Ih= \biggl( \cfrac{1}{N} \sum_j^{N} c_{j}^{-2} \biggr)^{-1/2}.
    \label{eq:HarmonicEstimator}
\end{equation}
This harmonic estimator is intended to provide the most probable value for the different \dEdx ($c_{j}$) measurements that follow a Vavilov~\cite{Vavilov:1957zz} or Landau~\cite{Landau:1944if} distribution. The sum in Eq.~\eqref{eq:HarmonicEstimator} includes all of the measurements along a track that have passed a cleaning procedure to discard measurements from atypical cluster deposit distributions and deposits too close to module edges. The \Ih estimator is preferred to a simple measurement average as it is very robust against upward fluctuations in $c_{j}$. It is, however, sensitive to downward fluctuations, which are unlikely to randomly occur. This \Ih estimator has been used for example to search for heavy charged particles considered as stable at the scale of the CMS detector~\cite{Khachatryan:2016sfv}, and also for charged particles with much shorter lifetimes leading to disappearing track signatures~\cite{disappearing_track,CMS:2020atg}.

\begin{figure}[htbp]
\centering
\includegraphics[width=.6\textwidth]{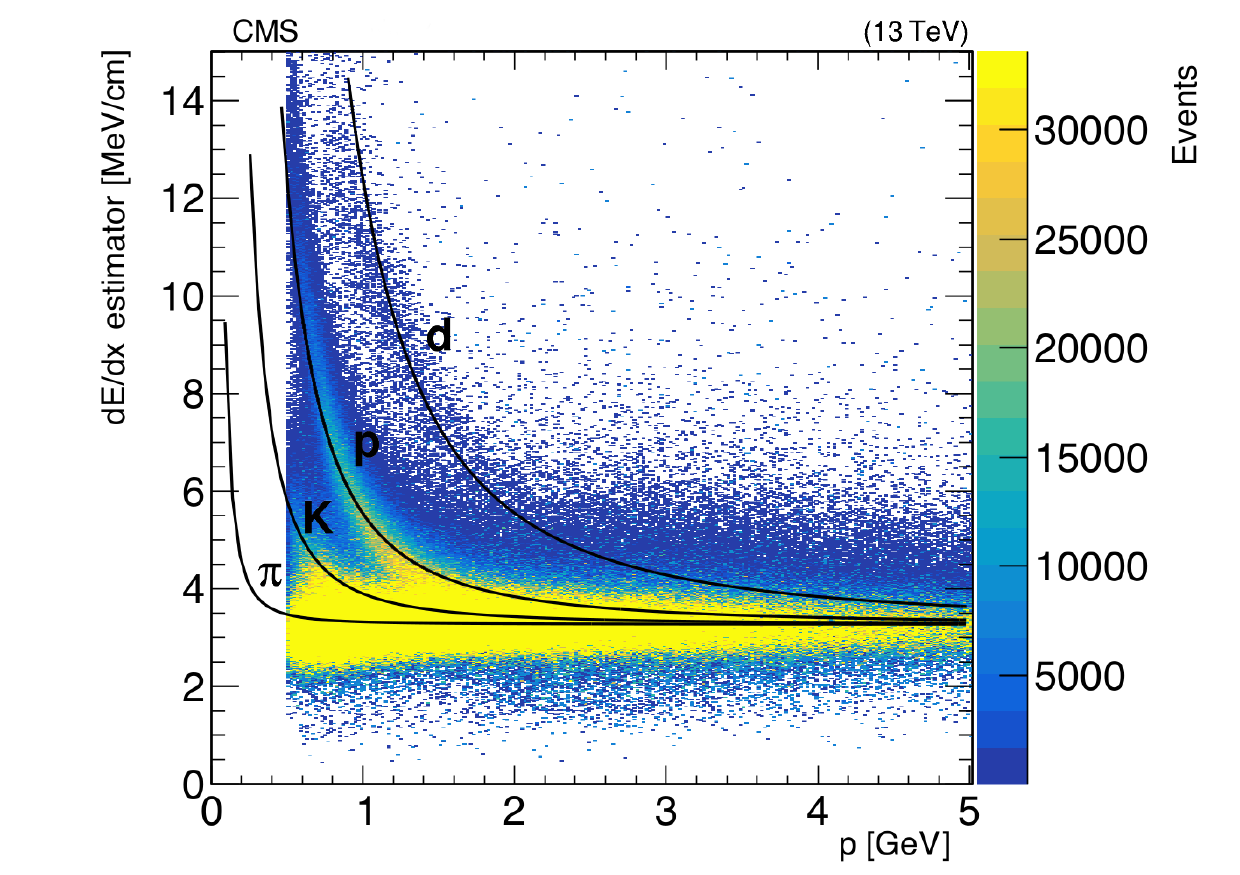}
\caption{Distribution of the \Ih estimator, computed using \dEdx measurements in the silicon strip tracker, versus the track momentum, using the data recorded in 2017 during the LHC Run~2. Expected \dEdx losses for pion, kaon, proton, and deuteron particles are shown as black lines. Tracks with $\pt < 0.5\GeV$ are not included in this plot.}
\label{fig:dedx}
\end{figure}

In addition, the \Ih estimator provides an estimate of the mass of the LLP candidate under the $Q=1\mathrm{e}$ hypothesis. It uses an approximate Bethe--Bloch parameterization in the non-ultra-relativistic regime that relates the measured ionization to the particle mass $m$ and the track momentum $p$:
\begin{equation}
    \Ih = K\cfrac{m^2}{p^2}+C,
\end{equation}
where the empirical parameters $K$ and $C$ extracted from low-momentum tracks in the range $0.5 < p < 5\GeV$. Figure~\ref{fig:dedx} shows this parametrization for the pions, kaons, protons, and deuterons at small momenta. 

In addition to the \Ih estimator, two independent discriminators are defined in Eqs.~(\ref{eq:fpix}-\ref{eq:ias}): \FiPixels, which uses only the \dEdx pixel detector information, and \GiStrip based on \dEdx measurements in the strip tracker, where the $i$ subscript refers to ionization. Both discriminators are designed to distinguish LLP signal events (with output values close to 1) from background events (with values close to 0). 

The \FiPixels discriminator is defined as
\begin{equation}
\FiPixels = 1 -  \prod_{j=1}^n P_j \sum_{m=0}^{n-1} \frac{[-\ln( \prod_{j=1}^n P_j)]^m}{m!},
        \label{eq:fpix}
\end{equation}
where $n$ is the number of measurements in the silicon pixel detector, excluding the first barrel layer, and $P_j$ is the probability that the minimum ionizing particle would produce a charge larger than or equal to the $j$--th measurement as predicted by a detailed simulation (called \textsc{PixelAV}~\cite{Chiochia:2004qh}) calibrated to data.

The \GiStrip discriminator is defined as
\begin{equation}
    \GiStrip = \frac{3}{N} \left(
    \frac{1}{12N} + \sum_{j=1}^N
    \left[
    P_j \left( P_j - \frac{2j-1}{2N} \right)^2 \right] \right),
    \label{eq:ias}
\end{equation}
where $N$ is the number of charge measurements in the silicon strip tracker, $P_j$ is the probability for a minimum ionizing particle to produce a charge smaller or equal to the $j$--th charge measurement for the observed path length in the detector, and the sum is over the track measurements ordered in terms of increasing $P_j$. These $P_j$ probabilities are determined using \dEdx templates in bins of path length values. The templates vary with detector module geometry and event PU. The probabilities are determined using data when used for data and determined using simulation when used for simulation.

These kinds of estimators can also address searches for particles with an electric charge different from unity~\cite{HSCP_run1}. For signals with a charge lower than unity, characterized in that case by a small \dEdx deposit, a large number of \dEdx measurements below a given threshold can be used to separate signal and background~\cite{CMS-PAS-EXO-19-006}.

\subsection{Precision proton spectrometer reconstruction}
\label{subsec:PPS}

The CMS-TOTEM PPS~\cite{CMS:2014sdw} is a system of near-beam tracking and timing detectors, located in ``Roman pots'' at about 200\unit{m} on both sides of the CMS interaction point. The Roman pots are movable near-beam devices that allow the detectors to be moved near (within a few \unit{mm}) to the beam, directly into the beam vacuum pipe. The PPS is designed to search for the process $\Pp\Pp \to \Pp\Pp+\PX$ where the system \PX can involve SM or DS final states. It allows the measurement of the 4-momenta of scattered protons and their time-of-flight from the interaction point during standard running conditions in regular high-luminosity fills.

The proton momenta are measured by two tracking stations on each arm of the spectrometer. With the PPS setup, protons that lose approximately 3--15\% of their momentum can be measured. This translates into an acceptance for the system \PX with a mass starting at $m_{\PX}\simeq 300\GeV$ and up to about 2\TeV. The fractional momentum loss $\xi$ of the protons can be measured from the proton track positions and angles (details can be found in Ref.~\cite{TOTEM:2022vox}). The timing information that can be used to measure the longitudinal coordinate of the vertex via time-of-flight and suppress the background from PU is not used in the analyses discussed below.

A search for new physics in central exclusive production (CEP) using the PPS and the missing-mass technique was presented in Ref.~\cite{CMS:2023roj} and will be described in Section~\ref{sec:EXO-19-009}.

\subsection{Pileup mitigation}
\label{sec:met}

Searches for invisible particles rely on an accurate description of \ptmiss. Similarly, jet substructure, as it is used, \eg, in searches for dark QCD signatures, is highly susceptible to particles from pileup collisions. The CMS Collaboration has developed several widely used techniques for mitigating the impact of PU. One of these techniques, known as charged-hadron subtraction (CHS)~\cite{CMS:2019ctu}, has served as the standard method for PU mitigation in jet reconstruction since the beginning of Run~2. The CHS algorithm operates by excluding charged particles associated with reconstructed vertices from PU collisions during the jet clustering process. To address the impact of neutral PU particles in jets, an event-by-event jet-area-based correction is applied to the jet four-momenta. Additionally, a technique for identifying PU-related jets (PU jet ID) is used to reject jets primarily composed of particles originating from PU interactions.

\begin{figure}[htbp]
    \centering
    \includegraphics{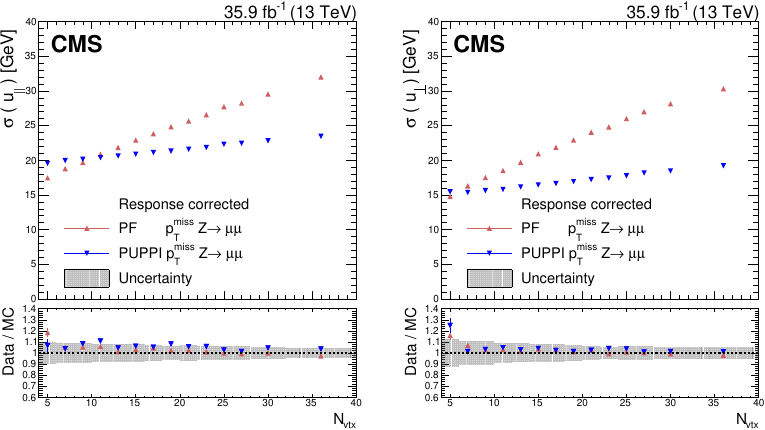}
    \caption{Upper panels: PUPPI and PF \ptmiss resolution of $u_\parallel$ (\cmsLeft) and $u_{\!\perp}$ (\cmsRight) components of the hadronic recoil as a function of $N_{\textrm{vtx}}$, in $\PZ \to \PGm\PGm$ data. 
    Lower panels: data-to-simulation ratio. 
    Systematic uncertainties are represented by the shaded band.
    Figure taken from Ref.~\cite{CMS:2019ctu}.}
    \label{fig:METvsPU}
\end{figure}

However, all these techniques have limitations when it comes to effectively removing PU contributions from neutral particles. For instance, the jet-area-based correction acts on the entire jet and is incapable of entirely eliminating PU contributions from jet shape or jet substructure observables. To address this limitation, a new PU mitigation technique, known as PU-per-particle identification (PUPPI)~\cite{CMS:2020ebo,Bertolini:2014bba}, has been introduced. This algorithm works at the particle level and builds upon the preexisting CHS algorithm. The PUPPI algorithm computes the probability that a neutral particle originates from PU, based on the distribution of charged PU particles in its vicinity, and adjusts the energy of the neutral particle based on its respective probability. As a result, objects formed from hadrons, such as jets, \ptmiss, and lepton isolation, demonstrate reduced dependency on PU when PUPPI is employed~\cite{CMS:2019ctu}. The improved performance of the resolution of the PUPPI hadronic recoil in $\PZ \to \PGm\PGm$ processes with respect to PU effects, represented by the number of reconstructed vertices $\mathrm{N}_{\mathrm{vtx}}$ is shown in Fig.~\ref{fig:METvsPU}; the hadronic recoil vector is divided into components parallel ($u_\parallel$) and perpendicular ($u_{\!\perp}$) to the boson axis.

Searches for LLPs must often employ dedicated strategies for PU mitigation to avoid a significant impact on the selection efficiency. These are discussed in Section~\ref{sec:LLPreconstruction}. 

\subsection{Filters for spurious events}
\label{sec:metfilters}

Spurious events can occur because of a variety of reconstruction failures, detector malfunctions, or noncollision backgrounds and have anomalous high-\ptmiss measurements, thus posing an obstacle for DS signatures that feature high-\ptmiss. Such events are rejected by dedicated event filters that tag the events with \ptmiss from genuine physics processes and remove more than 85--90\% of these spurious high-\ptmiss events with a mistagging rate of less than 0.1\%~\cite{CMS:2019ctu}. These filters allow the removal of events with ``artificial \ptmiss'' arising from: interactions of machine-induced background particles moving along the beam direction, known as ``beam halo'', with the hadronic calorimeter; significant noise in the HCAL barrel and endcaps, detected by distinctive geometrical patterns of the readout electronics and by the usage of pulse shape and timing information; spurious signals in ECAL arising from sources such as anomalous large pulses in the endcap $5\times 5$ crystal groups (supercrystals) and inoperative readout electronics; and high-\PT particle tracking failures leading to poorly measured PF muons and charged hadrons.

In the case that artificial \ptmiss is the dominant source of background, custom filters optimized for a particular kinematic phase space may be needed~\cite{CMS:2021dzg}. For instance, when requiring that the jet momentum aligns with \ptmiss, over 40\% of the QCD multijet background originates from events with artificial \ptvecmiss caused by nonfunctional calorimeter cells. These events were not consistently detected by the dedicated filters mentioned earlier and additional analysis requirements were developed and employed.

\subsection{Triggers and data scouting}
\label{sec:trigger}

The standard paradigm for CMS operation is defined by: triggered operation of the detector, the recording of the full raw data of the readout electronics, and prompt reconstruction of the events followed by multiple data reduction passes in order to serve the needs of the physics analysis groups. On the other hand, the acquisition of collision events at CMS is constrained, directly and indirectly, by a number of factors: the electronics of the detector subsystems, the bandwidth and local data buffer available to the DAQ system, and the processing and storage capabilities available to the offline and computing system. The flexibilization of different components of the aforementioned paradigm allows for tradeoffs to be made. The recording of events using preprocessed, lighter data formats instead of full raw data is called \emph{data scouting}, and allows an increase of the data acquisition rate whilst keeping the bandwidth low. \emph{Data parking}, on the other hand, refers to data that is not reconstructed promptly after acquisition but instead is ``parked'' to be reconstructed later when the Tier-0 computing resources are idle. Data scouting and parking are the subject of their own Report~\cite{EXO-23-007}.

Models featuring DS physics predict a wide variety of final states in $\Pp\Pp$ collisions. Many triggers (as discussed in Section~\ref{subsec:standardTriggers}) are correspondingly developed to target these experimental signatures, which include \ptmiss arising from stable particles that do not interact with the detector, leptons produced at the $\Pp\Pp$ interaction point (prompt) or away from it (displaced), and standard or unconventional jet signatures created via enriched DS dynamics. While CMS successfully targets a range of these models, challenges arise in obtaining sensitivity to theories with exotic topologies, particularly those featuring new low-mass states in the DS. These states are generally difficult to probe because of trigger limitations. Decays of such low-mass DS states into SM particles lead to final-state particles that have either very low momentum (soft particles) or are very collinear, depending on the Lorentz boost in the laboratory frame. Both situations present triggering challenges. If these DS states are instead stable within the detector volume, they induce a soft \ptmiss spectrum that is also difficult to use for triggering unless combined with energetic ISR jets, leading to loss of signal acceptance.

Several techniques are employed in CMS to address these challenges and improve sensitivity to DS models with exotic signatures. We discuss the use of data scouting (Section~\ref{subsec:scouting}) to expand the range of low-mass DS particles that can be probed in CMS, after describing the relevant standard triggers available in CMS during the Run~2 data-taking period.

\subsubsection{Standard triggers} \label{subsec:standardTriggers}

Event selection in CMS starts with a two-tiered trigger system, as discussed in Section~\ref{sec:detector}. Standard triggers save the acquired data in a raw format that represents the complete information of the detector readout electronics. The advantage of saving the data in this format is that they can be reconstructed multiple times, profiting from more accurate calibrations that usually only become available later in the running period. The trade-off is the large size of the data volume, of order 1\unit{MB/event}. Thus the trigger system must balance the selection efficiency for signal events with the background rejection rate, which is correlated with the trigger output bandwidth. Since the HLT runs an optimized version of the full reconstruction software, a number of dedicated reconstruction techniques described later in this section are also implemented in the HLT.

Prompt electrons and muons are relatively easy to trigger on, especially when they are well-isolated from other activity in the event. Analyses targeting these signatures usually employ general-purpose lepton triggers. For example, in 2018, the isolated single-electron trigger required $\pt> 32\GeV$, and the dielectron trigger required $\pt> 25\GeV$ for both electrons. Likewise, the general-purpose isolated single-muon trigger required $\pt> 24\GeV$, and the isolated dimuon trigger required $\pt> 17\,(8)\GeV$ for the largest (second-largest) \pt muon. Trigger algorithms that forsake the isolation requirements are also available, but their minimum \pt thresholds have to be set higher in order to keep trigger rates manageable. As explained in Section~\ref{sec:LLPreconstruction}, both kinds of algorithms are less effective for displaced leptons, for which dedicated triggers were developed. For signatures with tau leptons and \PQb-tagged jets, the most common strategy is to use the standard reconstruction and identification techniques for the tau lepton or \PQb jet itself and then design a dedicated trigger algorithm focusing on the final state as a whole.

The more challenging signatures are those with only photons or hadronic jets in the final state. Stringent kinematic thresholds are applied to the trigger algorithms to keep the rates within the allocated bandwidth. Dedicated triggers featuring special reconstruction algorithms for displaced or delayed objects are again deployed.

\begin{figure}[htbp]
   \centering
   \includegraphics[width=0.6\textwidth]{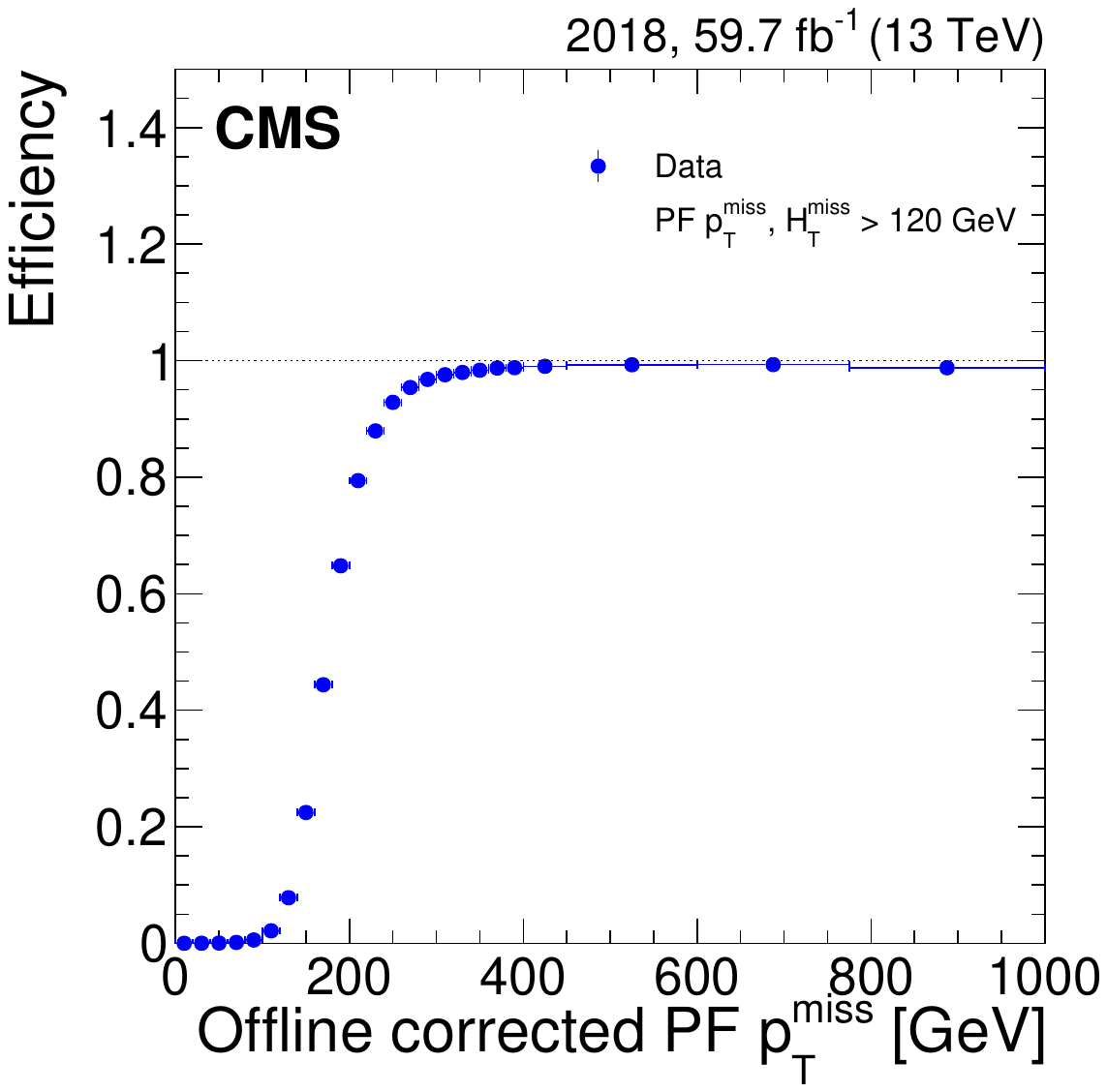}
   \caption{The event selection efficiency for requiring HLT thresholds of 120\GeV in both \ptmiss and \HTmiss as a function of the offline corrected \ptmiss, which takes into account jet energy scale corrections.}
   \label{fig:METturnon}
\end{figure}

Finally, an all-purpose \ptmiss trigger is available to select events where a particle such as a DM candidate produced in the collision escapes the CMS detector and leaves no signal. As described in Section~\ref{sec:ptmiss}, this signature is extremely sensitive to experimental conditions such as detector calibrations and PU. The trigger requirement relies on an online calculation of \ptmiss that is based on all PF candidates reconstructed at the HLT except for muons. It is usually combined with an \HTmiss requirement, where jets are subjected to stringent identification requirements. The kinematic thresholds for these algorithms are \ptmiss and $\HTmiss > 110$ (120)\GeV in 2016--2017 (2018) data. Unavoidable discrepancies exist between the online (trigger level) and offline reconstruction of \ptmiss, because the latter benefits from additional subdetector information and improved calibrations. The effect of those discrepancies is shown in the efficiency curve in Fig.~\ref{fig:METturnon}. After requiring online that both the \ptmiss and \HTmiss are ${>}120\GeV$, the offline corrected \ptmiss only reaches $\sim$95\% efficiency when it is ${>}250\GeV$. Table~\ref{tab:summaryTriggers} displays a subset of the trigger algorithms deployed in CMS during 2018 that select events based on the presence of one or two physics objects.
The complete CMS HLT event selection comprises $\sim\mathcal{O}(700)$ trigger algorithms, including those for alignment/calibration, monitoring, and backup.

\begin{table}[htbp]
   \centering
   \topcaption{Summary of \PT (or \ET) requirements (in {\GeVns}) of a subset of the HLT algorithms deployed in CMS during 2018, for trigger paths based on one or two physics objects. One \PT threshold value is given for the single-object triggers, and two \PT threshold values are given for the di-object triggers. Triggers with isolated leptons are labeled ``iso.'', and have generally lower kinematical thresholds than the corresponding algorithms that do not impose isolation requirements on leptons. The ``1-prong'' note for the tau lepton trigger refers to a selection targeting the \Pgt decay into a single charged particle + neutrals. The ``barrel'' note for the photon trigger refers to a photon reconstructed solely within the barrel section of the ECAL. The ``AK4'' and ``AK8'' notes refer to jets reconstructed with the anti-\kt algorithm and a distance parameter of 0.4 and 0.8, respectively~\cite{Cacciari:2008gp}; the mass threshold is applied to \mtrim, the trimmed jet mass~\cite{Krohn:2009th}. The ``\PQb tags'' note refers to the number of jets that are \PQb-tagged with the \textsc{DeepCSV} algorithm~\cite{CMS:2017wtu}.
 }
   \cmsTable
   {\renewcommand{\arraystretch}{1.2}
\begin{tabular}{@{} lccccccc @{}}
	& \multicolumn{7}{c}{\textbf{Single-object triggers}}\\
	\cline{2-8}
	& \Pe		& \Pgm		& \Pgt (iso.)	& \Pgg					& Jet													&	\ptmiss	& \HT\\
	& 32 (iso.)	& 24 (iso.)	& 180			& 110 (iso., barrel)	& 500 (AK4) 					& 120	& 1050 \\
	& 115 		& 50		& 				& 200 					& 400, $m_\text{trim}>30$ (AK8) & 		& 330 + 4 jets, 3 \PQb tags\\[12pt]
	& \multicolumn{7}{c}{\textbf{Di-object triggers}}\\
	\cline{2-8}
	& \Pe		& \Pgm		& \Pgt (iso.)	& \Pgg					& Jet													&	\ptmiss	& \HT\\
 \raisebox{1.5ex}{\Pe} &	 
		  \shortstack{\Tstrut23, 12 (iso.)\\25, 25} &  
		\shortstack{23, 12 (iso.)\\27, 37} & 
		\raisebox{1.5ex}{24 (iso.), 30} &	
		&
		\shortstack{30 (iso.), 35\\50, 165} 	& &
            \shortstack{28 (iso.), 150\\15, 600}\\
\raisebox{1.5ex}{\Pgm}	& 
		\shortstack{23, 12 (iso.)\\27, 37} &
		\shortstack{17, 8 (iso.), $m_{\Pgm{}\Pgm} > 3.8$\\37, 27} &	
		\raisebox{1.5ex}{20 (iso.), 27} & \raisebox{1.5ex}{17, 30}\\
\Pgt (iso.)	& &	& 35, 35	&	&	&	50 (1-prong), 100	\\
\raisebox{1.5ex}{\Pgg} &	&   &   &
            \shortstack{30, 18 (iso.)\\ 70, 70} \\
\ptmiss	&		&	&	&	&	&	& 100, 500\Bstrut\\
	\end{tabular}
   }
   \label{tab:summaryTriggers}
\end{table}

\subsubsection{Data scouting}

\label{subsec:scouting}

The fundamental rate limitation in CMS is the total amount of data that can be transferred to storage at once, not the number of events that can be stored. A powerful technique to increase the event rate involves decreasing the information stored per event, thereby releasing some of the data bandwidth to store more events. This technique is termed ``data scouting'' in CMS and has been deployed since Run~1. Here, we give a brief overview of data scouting, as it is relevant for some DS searches.

In Run~2, two scouting strategies are defined: one focusing on final states involving muons, and the other on hadronic final states. The ``muon scouting'' data set saves only muon information per event, apart from limited event-level information. This drastically reduces the event size from roughly 1\unit{MB} to about 9\unit{kB} in 2018, enabling muon triggers with much lower momentum thresholds at the same instantaneous luminosity, which is the ratio of the rate of events produced per unit of time to the cross section for a given process. The muon pair (dimuon) scouting trigger requires each muon to have $\pt > 3\GeV$ at the HLT, compared to the standard CMS dimuon trigger requirements of $\pt > 17\GeV$ for the first muon and $\pt > 8\GeV$ for the second; in both cases, muons are required to be isolated. The trigger rate goes up to about 6\unit{kHz}, allowing us to record 200 times more dimuon events than we otherwise would.

Several analyses have exploited the muon scouting data set to enhance sensitivity to low-mass physics. Searches for prompt~\cite{CMS-PAS-EXO-21-005} and displaced~\cite{CMS:2019buh} resonances decaying to muon pairs obtain some of the most stringent exclusion limits on dark photon production for few-\GeV dark photon masses. Model-independent searches such as the one in Ref.~\cite{CMS-PAS-EXO-21-005} also employ muon scouting data to enable the investigation of additional DS models, such as the 2HDM+a framework.

A second scouting strategy in Run~2 collects only jet-related information per event. This data set, termed ``PF scouting'', enables a considerable reduction in the jet trigger \HT thresholds, expanding the range of low-mass jet-related searches feasible in CMS. The PF scouting trigger sets a requirement of $\HT > 410\GeV$ at the software level, computed by considering jets with $\pt > 40\GeV$, compared with the standard trigger requirement of $\HT > 1050\GeV$. By storing only jet-related information in the event, the event size is reduced from 1\unit{MB} to about 15\unit{kB}, and the trigger rate is increased to about 2\unit{kHz}. For comparison, the rate of the data set that comprises all standard jet triggers is close to 400\unit{Hz}.

The Run~2 jet scouting technique has been used to enhance the low-mass sensitivity to several dijet, trijet, and multijet analyses~\cite{EXO-19-004,PhysRevD.99.012010}. For example, a search for dijet resonances~\cite{EXO-19-004} attained a dijet mass sensitivity as low as 350\GeV, compared to about 500\GeV when using the standard triggers. A more detailed description of DS analyses that feature scouting data sets can be found in Sections~\ref{paragraph:EXO-19-018}, \ref{paragraph:EXO-21-005}, \ref{par:EXO-19-004}, and \ref{paragraph:EXO-20-014}.

\subsection{Techniques for heavy ion data}
\label{subsec:HI}

One of the main goals of the LHC as an energy-frontier $\Pp\Pp$ collider is to discover new massive particles and/or FIPs. In addition to $\Pp\Pp$ collisions, the LHC also provides high-energy HI collisions, and in particular lead-lead (PbPb) collisions, which are key tools to study the properties of the quark-gluon plasma.

Typically, one does not consider HI collisions as a place to look for BSM physics. They are characterized by a very large number of outgoing particles (a charged-particle multiplicity more than two orders of magnitude larger than in $\Pp\Pp$ collisions~\cite{ALICE:2015juo}), which makes tracking much more challenging. Moreover, the integrated luminosity (\Lint), which is the instantaneous luminosity integrated over some time period, or equivalently, the number of events of interest divided by the cross section of the process, for PbPb collisions was 390\,$\mu$b$^{-1}$ and 1650\,$\mu$b$^{-1}$, respectively in 2015 and 2018, which is many orders of magnitude smaller than for $\Pp\Pp$ collisions. 

However, a fraction of HI interactions takes place with no overlap between the two nuclei. In such ultra-peripheral collisions (UPCs), the two ions only interact through the electromagnetic force, \ie, an exchange of photons, producing very low multiplicity events. Additionally, HI runs are tuned to yield no PU, which further simplifies tracking. Overall, UPCs result in signatures with extremely well-isolated objects, suitable for BSM physics searches, \eg, searches for ALPs.

The CMS experiment is well equipped to record and investigate both $\Pp\Pp$ and HI collisions. The main challenges in UPCs of heavy ions from the experimental perspective are related to triggering and detector noise. For instance, in light-by-light scattering (discussed in Section~\ref{sec:FSQ-16-012}) the final state consists exclusively of two low-energy photons. Since there is no other activity in the detector, one cannot rely on associated tracks, muons, jets, or \ptmiss to trigger the measurement. Instead, the photons themselves have to be used for triggering. For such rare processes, it is crucial to lower the photon energy requirement for both triggering and offline reconstruction as much as possible, which enters the regime where calorimeter noise becomes significant. As an example, a recent light-by-light scattering analysis, described in Section~\ref{sec:FSQ-16-012}, triggers on diphotons with transverse energy ${>}2\GeV$. The noise in the barrel region of the ECAL is at the level of $\approx$0.7\GeV. However, in the endcap region it can get as large as $\approx$6\GeV.

These challenges associated with UPCs are typically addressed by carefully studying triggering and reconstruction efficiencies with tag-and-probe techniques, as well as masking regions of the detector where noise levels are too large to perform the analysis. This strategy yields satisfactory results, allowing CMS to observe evidence for light-by-light scattering and derive the most stringent limits, at the time of publication, on the production of ALPs with masses between 5 and 50\GeV~\cite{FSQ-16-012}.

\subsection{Background estimation strategies and statistical methods}
\label{sec:background}

The accurate estimation of the background is paramount to the sensitivity of BSM searches, and DS searches are no exception. As a prototype analysis, we consider the search for a monojet plus \ptmiss signal, which has as one of its dominant backgrounds the \Zvvjets processes. The differential yield of this background in the signal region (SR) can be connected to the yield of \zlljets events in a suitably defined CR, in events with two well-reconstructed leptons with an invariant mass close to that of the \PZ boson. Measurements of the lepton trigger, reconstruction, and identification efficiencies can be used to estimate the true \zll yield, and from that the \Zvv yield. The procedure can be repeated for each relevant background in the SR, with different efficiencies being accounted for in each case. This is the essence of the transfer factor technique, which is described in more detail in Section~\ref{sec:transferFactor}.

In case there is only one relevant background in the SR of the search, other methods are available. One example would be the search for dijet resonances, which is completely dominated by the background from SM events composed uniquely of jets produced through the strong interaction, referred to as QCD multijet events. In that search, the standard approach is to model the dijet invariant mass distribution by an empirical, smooth function, fit it to the SR data, and search for localized, high-significance excesses (``bumps’’) of the fit that would indicate the signal’s presence. This approach is called the bump hunt and is described in Section~\ref{sec:bumpHunt}. On the other hand, if the posited signal does not give rise to localized deviations in the search variable, the bump hunt may fail to reveal its presence. An alternative approach is to define an auxiliary variable and divide the data into statistically independent sets based on disjoint ranges of the two variables; the definition of the sets must be such that the majority of the signal populates only one of them, again defined as the SR. In the limit that the variables are uncorrelated, the ratios between the yields of the CRs can be used to predict the background yield in the SR. This is the ABCD method, which is described in more detail in Section~\ref{sec:abcd}.

For most of the searches presented in this Report, the \CLs method~\cite{Read1,Junk:1999kv} is used to obtain a limit at 95\% confidence level (\CL) using the profile likelihood test statistic~\cite{Cowan:2010js}, often in the asymptotic approximation. The CMS statistical analysis tool \textsc{Combine}~\cite{CMS:2024onh} is used to compute these limits. The robustness and precision of the estimation of contributions from SM background processes determine the sensitivity of searches for new physics. Historically, simulated background events obtained with the Monte Carlo (MC) method have been used most of the time to seed templates for background contributions in the SR and obtain uncertainties. These systematic uncertainties in the MC distributions are represented as nuisance parameters that are adjusted in a maximum likelihood fit, based on the observed data distribution, to obtain the final background model. In many cases, however, methods based on CRs in data or more sophisticated background estimation strategies are employed to quantify background contributions in the SRs. In the following, a few of the common methods of background estimation either fully based on CRs in data, or partially based on data and assisted through the simulation, are briefly introduced.

\subsubsection{Transfer factor technique} \label{sec:transferFactor}

The underlying idea of using transfer factors (TFs) to predict background contributions is to measure ratios of yields for processes across regions, rather than calibrate the absolute background shape. As a consequence, if the two samples used to build the ratios are impacted by a specific systematic uncertainty in the same or a similar way, its effect largely cancels out and does not affect the ratios. For instance, it is conceivable to assume that an event sample with a jet recoiling against a dilepton system and an event sample featuring a jet recoiling against a single photon will have the same uncertainties affecting the measurement of the jet, \ie, jet energy scale and resolution. Thus, the ratio of the (differential) yields in these two samples is largely unaffected by jet uncertainties, while being affected by lepton/photon identification and scale uncertainties.

This strategy is particularly powerful when applied in mono-X-type analyses, where the \ptmiss spectrum is a powerful shape discriminator between the BSM signal and the SM background and is typically used for signal extraction. Because of the symmetry of the various SM V+jets processes, the main background contribution in the SR coming from the \Zvvjets process can be calibrated utilizing CRs enriched in \zlljets, \Wlvjets, and \phojets events. By excluding the leptons and photons from the computation of \ptmiss in the CR, the so-called hadronic recoil becomes a proxy for the \ptmiss spectrum in the SR. 

A binned likelihood fit to the data is performed simultaneously in different CRs and in the SRs to estimate the dominant \Zvvjets and \Wlvjets backgrounds in each \ptmiss bin. 

The part of the likelihood function constraining the \Zvvjets background in the monojet analysis in Ref.~\cite{CMS:2021far}, which is representative of other mono-X-type searches, is given as:
\begin{equation}
\begin{aligned}
\mathcal{L}_{\textrm{c}}(\boldsymbol{\mu}^{\Zvv}, \boldsymbol{\mu}, \boldsymbol{\theta}) &=
\prod_{i} P\left(d^{\gamma}_{i} |B^{\gamma}_{i}(\boldsymbol{\theta}) +\frac{ \mu^{\Zvv}_{i} }{R^{\gamma}_{i}(\boldsymbol{\theta})}   \right) \\
&~\times \prod_{i} P\left(d^{\PZ}_{i}|B^{\PZ}_{i}(\boldsymbol{\theta}) +\frac{\mu^{\Zvv}_{i} }{R^{\PZ}_{i}     (\boldsymbol{\theta})}   \right ) \\
&~\times \prod_{i} P\left(d^{\PW}_{i}|B^{\PW}_{i}(\boldsymbol{\theta}) +\frac{f_{i}(\boldsymbol{\theta})\mu^{\Zvv}_{i}}{R^{\PW}_{i}(\boldsymbol{\theta})} \right)\\
&~\times \prod_{i} P\left(d_{i}     |B_{i}(\boldsymbol{\theta}) + (1+f_{i}(\boldsymbol{\theta})) \mu^{\Zvv}_{i}  + \mu S_{i}(\boldsymbol{\theta})\right ). 
\end{aligned}
\end{equation}

In the above likelihood function, $P(n|x)$ is the Poisson probability of observing $n$ events when $x$ are expected, $d^{\gamma/\PZ/\PW}_{i}$ is the observed number of events in each bin of the photon, dimuon/dielectron, and single-muon/single-electron CRs, and $B^{\gamma/\PZ/\PW}_{i}$ is the background in the respective CRs. The systematic uncertainties are modeled with nuisance parameters ($\boldsymbol{\theta}$), which enter the likelihood as additive perturbations to the TFs $R^{\gamma/\PZ/\PW}_{i}$. Each $\boldsymbol{\theta}$ parameter has an associated Gaussian constraint term in the full likelihood. The parameter $\boldsymbol{\mu}^{\Zvv}$ represents the yield of the \Zvv background in the SR and is left freely floating in the maximum likelihood fit. The function $f_{i}(\boldsymbol{\theta})$ is the TF between the \Zvvjets and \Wlvjets backgrounds in the SR and acts as a constraint between these backgrounds. The likelihood also includes the SR, with $B_{i}$ representing all the background estimates from simulation, $S$ representing the nominal signal prediction, and $\mu$ being the signal strength parameter also left floating in the case of an $S+B$ fit ($\mu=0$ otherwise).

In this likelihood, the expected numbers of \Zvvjets events in each bin of \ptmiss are the free parameters of the fit. Transfer factors, derived from simulation, are used to link the yields of the \zlljets, \Wlvjets, and \phojets processes in the CRs with the \Zvvjets and \Wlvjets background estimates in the SR. These TFs are defined as the ratio of expected (from simulation) yields of the target process in the SR and the process being measured in the CR, e.g.:
\begin{equation}
    R_i^\PZ=\frac{N_{i,\text{MC}}^{\PZ(\Pgm\Pgm)}}{N_{i,\text{MC}}^{\Zvv}}.
\end{equation}
To estimate the \Wlvjets background in the SR, the TFs between the  \Wlvjets background estimates in the SR and the $\PW(\PGm\PGn_\PGm)$+jets and $\PW(\Pe\PGn_\Pe)$+jets event yields in the single-lepton CRs are constructed. These TFs take into account the impact of lepton acceptances and efficiencies, lepton veto efficiencies, and the difference in the trigger efficiencies in the case of the single-electron CR.

The \Zvv~background prediction in the SR is connected to the yields of $\PZ(\Pgm\Pgm)$ and $\PZ(\Pe\Pe)$ events in the dilepton CRs. The associated TFs account for the differences in the branching fraction of $\PZ$ bosons to charged leptons relative to neutrinos and the impact of lepton acceptance and selection efficiencies. In the case of dielectron events, the TF also takes into account the difference in the trigger efficiencies. The resulting constraint on the \Zvvjets process from the dilepton CRs is limited by the statistical uncertainty in the dilepton CRs because of the large difference in branching fractions between \PZ boson decays into neutrinos and \PZ boson decays into electrons and muons.

The \phojets CR is also used to predict the \Zvvjets process in the SR through a TF, which accounts for the difference in the cross sections of the \phojets and \Zvvjets processes, the effect of acceptance and efficiency of identifying photons along with the difference in the efficiencies of the photon and \ptmiss triggers. The addition of the \phojets CR mitigates the impact of the limited statistical power of the dilepton constraint, because of the larger production cross section of \phojets process compared to that of \Zvvjets process.

Finally, a TF is also defined to connect the \Zvvjets and \Wlvjets background yields in the SR, to further benefit from the larger statistical power that the \Wlvjets background provides, making it possible to experimentally constrain \Zvvjets production at large \ptmiss.

These TFs rely on an accurate prediction of the ratio of $\PZ$+jets, $\PW$+jets, and \phojets cross sections. Therefore, leading order (LO) simulations for these processes are corrected using boson \pt-dependent next-to-LO (NLO) QCD K-factors derived using \MGvATNLO. They are also corrected using \pt-dependent higher-order EW corrections extracted from theoretical calculations~~\cite{Denner:2009gj,Denner:2011vu,Denner:2012ts,Kuhn:2005gv,Kallweit:2014xda,Kallweit:2015dum}. The higher-order corrections are found to improve the data-to-simulation agreement for both the absolute prediction of the individual $\PZ$+jets, $\PW$+jets, and \phojets processes, and their respective ratios.

\subsubsection{Bump-hunt technique}\label{sec:bumpHunt}

Searches for localized excesses in invariant mass distributions are an honored tradition in high-energy physics, with the discovery of the \PJGy~\cite{E598:1974sol,SLAC-SP-017:1974ind} and \PGU~\cite{E288:1977xhf} mesons, the \PZ boson~\cite{UA1:1983mne,UA2:1983mlz} and, more recently, the Higgs boson~\cite{PRLB2012-716-1-1,CMS-PAPERS-HIG-12-028,Chatrchyan:1529865,CMS:2022dwd,arXiv:2207.00092}. Any new mediator particle \PX predicted in BSM scenarios has several experimental observables, including its rest mass $m_{\PX}$, its decay width $\Gamma_{\PX}$, and its production cross section $\sigma_{\PX}$. If the mediator decays into SM particles or a mixture of SM and DM particles, its rest mass can be measured by determining the energy and angle of emission of all its decay products. The mass spectrum of its decay products is expected to show an increase in the number of event counts at the ``resonance'' $m_{\PX}$ value because of the enhancement in the production cross section from the propagator of a massive mediator. The width of the resonance, or ``bump'' in the reconstructed mass spectrum, will depend on the decay interactions and the detector resolution that measures the decay products. For strong (or strong-like) interactions, with short lifetimes, the resonance shape may be wide (larger than the experimental resolution). Its shape can be approximated by a Breit--Wigner function for the intrinsic line shape, convoluted with a Gaussian function for the resolution. Parton luminosities are greater for masses below the resonance peak, such that the Breit--Wigner shape can present a significant ``shoulder`` on the lower tail. This effect may be significant near the kinematic threshold of $m_{\PX}$ production.

In some cases, a full reconstruction of $m_{\PX}$ is impossible since the decays include invisible particles from DM candidates. In those cases, it is important to include the \ptvecmiss in the definition of the reconstructed $m_{\PX}$, such as \mTii~\cite{Barr:2009wu} or the razor variable $R$~\cite{Rogan:2010kb,CMS:2012yib}. For example, in the case of SVJs $\PZpr \to \Pqdark\Paqdark$, cf. Section~\ref{sec:svjtheory}, the invariant mass of the reconstructed (visible) jets \mjj is a worse proxy for \mZprime than the transverse mass \mT defined to include the \ptvecmiss~\cite{CMS:2021dzg}:

\begin{equation}
\mT^{2} = \left[\etsub{jj} + \ETmiss\right]^{2} - \left[\ptvecsub{jj} + \ptvecmiss\right]^{2} = \mjj^{2}+2\ptmiss\left(\sqrt{\mjj^{2}+\ptsub{jj}^2} - \ptsub{jj}\text{cos}(\phijjmiss)\right).\label{eq:mt}
\end{equation}

Here, \mjj is the invariant mass of the system composed of the two largest-\pt large-radius jets, and $\ptvecsub{jj}$ is the vector sum of their \ptvec. The quantity ${\etsub{jj}^{2} = \mjj^{2} + \abs{\ptvecsub{jj}}^{2}}$, while it is assumed that the system carrying the \ptmiss is massless, \ie, ${\ETmiss = \ptmiss}$. This enables the simplification in the second line of Eq.~\eqref{eq:mt}, with \phijjmiss as the azimuthal angle between the dijet system and the \ptvecmiss. In this case, \mT is much closer to \mZprime than \mjj: it has better resolution and its peak reproduces \mZprime more accurately. 

The estimation of the background is critical when looking for a bump in the reconstructed mass spectrum. In contrast to the signal, the background (typically QCD multijet) spectrum is smoothly falling. Despite the progress of QCD multijet MC generators with NLO and next-to-NLO (NNLO) accuracy, the mass spectra obtained from MC generators tend not to agree very well with the data in both shape and normalization. This is caused by the large theoretical uncertainties (such as nonperturbative effects, parton distribution functions [PDFs], and the renormalization and factorization scales) and experimental uncertainties (such as the jet energy scale and resolution smearing), which can be even more pronounced in final states with large \ptvecmiss where misreconstructed SM jets are the dominant background. Therefore, many searches estimate the QCD multijet background parametrically, directly from data. The fit can include templates from signal (at different mass values) or a parameterized signal function, and other components for background. If no significant deviation from the background-only hypothesis is found, limits on the cross section as a function of $m_{\mathrm{X}}$ can be set. Using the data to describe the background solves the problem of poor modeling of detector effects in novel signatures, although limited event counts at large invariant mass may become a problem. This strategy works if there are no features introduced by the calibration or other experimental effects, and it is important to have a well-controlled, smoothly falling background spectrum.

At the LHC, several families of fit functions have been used to model the QCD multijets background, which are called the ``dijet function'' ($f_\text{\!dijet}$ and its enhanced version $f_\text{\!dijet2}$) and the ``UA2 function'' ($f_\text{\!UA2}$):
\begin{equation}
\begin{aligned}
f_\text{\!dijet}(x)  &= \frac{p_0(1-x)^{p_1}}{x^{p_2+p_3\ln x + p_4\ln^2x}} ,\\
f_\text{\!dijet2}(x) &= \frac{p_0(1-x)^{p_1+p_2\ln x+p_3\ln^2x}}{x^{p_4+p_5\ln x + p_6\ln^2x}} ,\\
f_\text{\!UA2}(x)    &= \frac{p_0 \re^{-p_1x-p_2x^2}}{x^{p_3[1+p_4\ln x + p_5\ln^2x]}}.  
\end{aligned}
\end{equation}
Here $x$ is the reconstructed mass divided by $\sqrt{s}$. These families of functions have been found in the past to fit the observed QCD spectrum in hadron colliders~\cite{UA2:1990gao,CDF:2008ieg,ATLAS:2010ivc,CMS:2010qze,CMS:2021dzg}. The number of parameters $p_N$ used in each function must be optimized in each case. The Fisher test~\cite{fishertest,Lomax:2012xyz} can determine if adding a new parameter to a function improves the fit to a given distribution. Two functions (one with fewer parameters than the other) are fit to the same distribution and the value  
$$F_{\text{test}}=\frac{(q_1-q_2)/(n_2-n_1)}{q_2/(n_{\text{bins}}-n_2)}$$
is calculated, where $q_i$, $n_i$ refer to the goodness-of-fit measurement and number of parameters in each function ($n_1<n_2$), and $n_{\text{bins}}$ is the number of bins in the distribution. The goodness-of-fit parameter is usually the $\chi^2$ value, which has been observed to give more stable results than the residual sum of squares. The value of $F_{\text{test}}$ is then compared to $F_{\text{crit}}$, which is defined by $\int_{F_{\text{crit}}}^{\infty}F_{\text{dist}}dx = \alpha_{\text{crit}}$, where $F_{\text{dist}}$ is an $F$-distribution with $n_2-n_1$ and $n_{\text{bins}}-n_2$ degrees of freedom and $\alpha_{\text{crit}}=0.05$. If $F_{\text{test}} > F_{\text{crit}}$, the function with more parameters ($n_2$) provides a better fit than the function with fewer parameters ($n_1$). The value of $\alpha_{\text{crit}}$ may be adjusted depending on the result of the bias tests, described next, and the stability of the results. 

This way of estimating the background from a fit to the data will typically be one of the largest experimental uncertainties in the statistical analysis to extract the signal. We typically assign the statistical uncertainty in the fit parameters as a background shape systematic uncertainty, and this tends to be large for large values of the reconstructed mass. It is also very important to test alternate functions to describe the QCD multijet background and check if using them introduces a bias in the results because the data in reality follows a different distribution from what was chosen for the fit. Some analyses use discrete profiling to estimate the uncertainty from different background functions and possible bias~\cite{Dauncey:2014xga}. 
Some possible alternate functions are listed here~\cite{EXO-19-004,CMS:2022usq,Harris:2011bh}: 
\begin{equation}
\begin{aligned}
f_\text{\!polynomial}(x)  &= \frac{p_0}{(1+p_1x+p_2x^2+p_3x^3)^{p_4}} ,\\
f_\text{\!extended polynomial}(x) &= \frac{p_0(1-x)^{p_1}(1+p_2x+p_3x^2)}{x^{p_4+p_5\ln x}} ,\\
f_\text{\!power-law times exponential}(x) &= \frac{p_0 e^{-p_{1}x}}{x^{p_2}} ,\\
f_\text{\!other}(x)    &= \frac{p_0 (1-x^{1/3})^{p_1}}{x^{p_2}}. 
\end{aligned}
\end{equation}

A self-closure test can be performed by generating pseudo-experiments with the main background function and fitting them with the same function to extract a signal measurement. Of course, the result here should be zero signal, but the spread in the results measures how robust the main function is to data fluctuations. This can be compared (and the corresponding uncertainty estimated) with a bias-closure test in which the main function is used to fit pseudo-experiments now generated with the alternate function. The results again should yield zero signal, and will tell us if our choice of background function has any potential to bias our results: if the self and bias-closure tests agree within their uncertainties, then no additional systematic uncertainties need to be included for this background estimation method. In addition, one can perform similar tests by injecting signal in both tests at the time of generating the pseudo-experiments and observing if the sensitivity to the signal also behaves similarly in both cases. 

An alternative strategy to model the background without empirical functions is to measure the observed distribution in a CR and derive correction factors from simulation to account for differences between the CR and the SR. This method can have smaller uncertainties than methods using empirical functions, but it can only be employed when the CR is not biased by trigger requirements.

\subsubsection{The \texorpdfstring{``ABCD''}{"ABCD"} method}
\label{sec:abcd}

\begin{figure}[hbtp]
\centering
\includegraphics[width=0.5\textwidth]{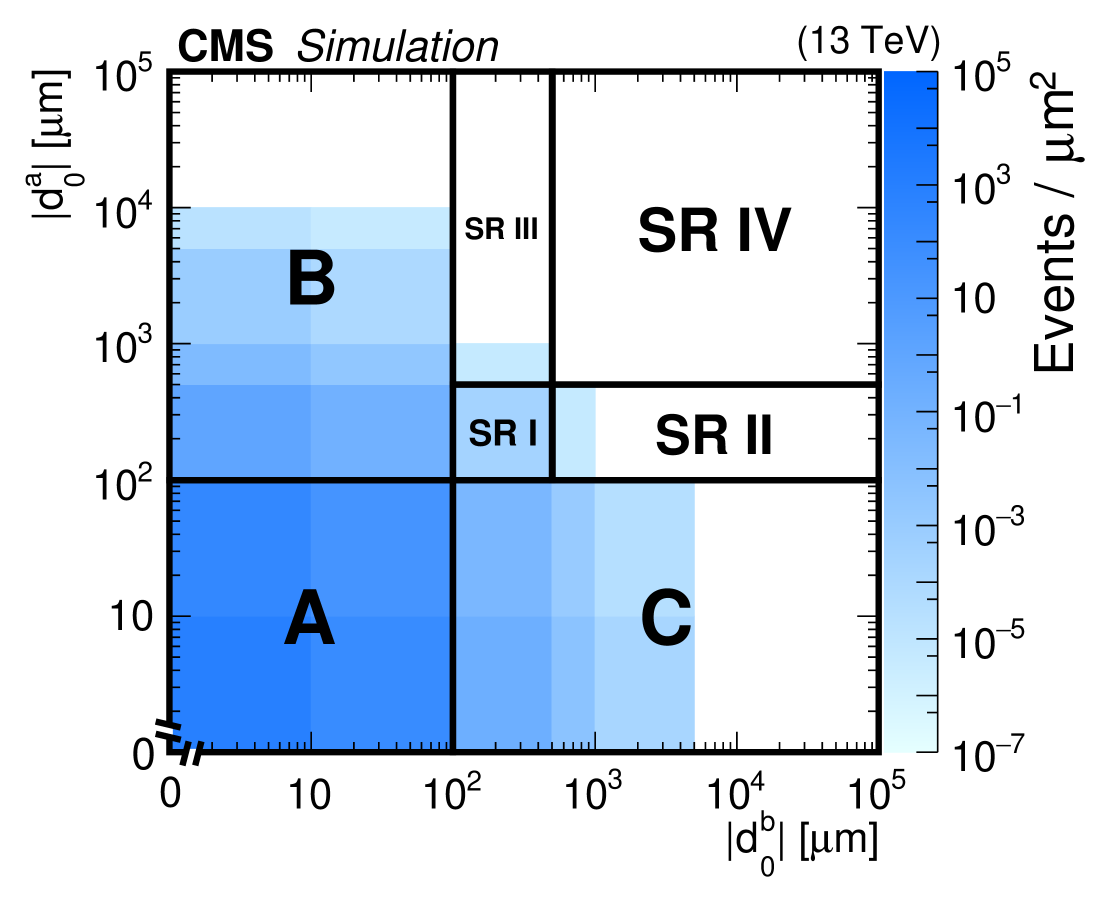}
\caption{
A diagram of the ABCD method, shown for illustration on simulated background events in a search for LLPs that decay to displaced leptons. The CRs are regions A, B, and C. There are four SRs, labeled I--IV, in this search. Figure taken from Ref.~\cite{EXO-18-003}.
}
\label{fig:EXO_18_003_ABCD}
\end{figure}

Background estimations based on CRs in data are often used for more reliable descriptions of backgrounds. One of the most widely used such methods is the matrix (``ABCD'') method, which was first introduced in Ref.~\cite{PhysRevD.44.29}. An example of how this method is used in a CMS analysis is shown in Fig.~\ref{fig:EXO_18_003_ABCD}. The ABCD method uses two independent variables to define four statistically independent regions, including the SR D and CRs A, B, and C. The two variables that are used to define the ABCD plane need to be statistically independent for the background process, allowing the prediction of the background yield in the SR to be constrained by the background yield in CRs A, B, and C: $N_{\mathrm{D}} = N_{\mathrm{B}} N_{\mathrm{C}}/N_{\mathrm{A}}$, where $N_{\mathrm{X}}$ is the number of background events in region X. Ideally, the CRs should be enriched with background events and devoid of signal events.

In cases where there is potential signal contamination in the CRs, a binned maximum likelihood fit is performed simultaneously in the four bins, with the signal strength included as a floating parameter. The background yields in the four regions are constrained to obey the standard ABCD relationship. This is possible because the background yields in the four regions require only three parameters to be fully described, given the independence of the two variables defining the ABCD plane. Thus one degree of freedom remains, which is used to fit the signal strength across all regions. Systematic uncertainties that impact the signal and background yields are treated as nuisance parameters with log-normal probability density functions.

Potential small correlations between the two variables defining the ABCD plane can be understood and controlled with additional validation regions adjacent to the SR~\cite{Choi:2019mip}. This method is called the ``extended ABCD method.'' These regions are located in between the corresponding CR and the SR in the ABCD plane and provide a path to estimate the correlation between the two observables. 

Additionally, CMS explores the usage of machine-learning-assisted ABCD techniques to derive discriminators that are decorrelated from a variable of interest or from another discriminator following the distance correlation technique proposed in Ref.~\cite{Kasieczka:2020yyl}.

\section{Data set and signal simulation}
\label{sec:datasetAndSignalSimulation}

Most of the analyses presented in this Report use the Run~2 $\Pp\Pp$ collision data sample, corresponding to \Lint~up to 140\fbinv at $\sqrt{s} = 13\TeV$, collected by the CMS detector in 2016--2018. The \Lint~for the 2016, 2017, and 2018 data-taking years have 1.2--2.5\% individual uncertainties~\cite{CMS-LUM-17-003,CMS-PAS-LUM-17-004,CMS-PAS-LUM-18-002}, while the overall uncertainty in \Lint~for the 2016--2018 period is 1.6\%. Some analyses use Run~1 $\Pp\Pp$ collision data, taken in 2010--2012 with $\sqrt{s} = 7$ and 8\TeV, or Run~3 $\Pp\Pp$ collision data, taken in 2022 with $\sqrt{s}=13.6\TeV$. Finally, some analyses use Run~2 HI collision data, namely, PbPb collisions taken in 2015 with $\sqrtsNN = 5.02\TeV$.

Data sets of simulated events, for both the SM background and BSM signals, are used by the searches to optimize the analysis criteria for sensitivity as well to check the agreement with data for basic kinematic variables. The simulation of collision events is implemented through a fixed-order perturbative calculation of up to four noncollinear high-\pt partons for the QCD terms, supplemented with a description of the underlying event, parton showering, multiparton interactions and hadronization. The perturbative calculation step is usually performed by a matrix-element calculator and event generator; versions 2.2.2 and 2.6.5 of the \MGvATNLO~\cite{Alwall:2014hca} package are used for almost all the analyses presented in this Report, and \POWHEG v2~\cite{Nason:2004rx,Frixione:2007vw,Alioli:2010xd} is also used for certain processes, primarily single top, \ttbar, and Higgs boson production. The next step is, in turn, usually implemented by the \PYTHIA8~\cite{Sjostrand:2014zea} generator. The combination of the two steps requires a matching procedure to avoid double-counting of processes in the combination, with the exact recipe depending on the order of the perturbative calculation. The MLM matching~\cite{Alwall:2007fs} is used for LO calculations, while the FxFx~\cite{Frederix:2012ps} and \POWHEG~\cite{Frixione:2007vw} methods are used for NLO. The PDFs are used to map the simulated colliding protons to the initial-state partons that are present in the matrix-element calculation; conversely, the \PYTHIA parameters are adjusted to a set of values that better describe the observed dynamics of high-energy proton collisions, which is referred to as a tune. By the end of Run~2, most analyses discussed in this Report converged in the usage of the NNPDF3.1 NNLO PDFs~\cite{NNPDF:2017mvq} and the CP5 tune~\cite{CMS:2019csb}. The simulation of specific new physics models may differ in particular aspects of these steps. If so, the simulation details can be found in the original publications of the corresponding searches.

The detector response to simulated particles is modeled using the \GEANTfour software~\cite{Agostinelli:2002hh}. Custom simulations of the detector electronics are used to produce readouts similar to those observed in data, in a process known as digitization. Pileup interactions are also included in the simulation. The simulated samples are corrected to make the PU distribution match the distribution in data as closely as possible. Event generation of new physics processes may need modifications to any of the steps of the simulation. The most notable case is the treatment of LLPs; the mass, charge, interactions, and lifetime of those particles are relayed to \GEANTfour, in a manner consistent with its treatment by the previous steps.

When simulating dark QCD models, the dedicated HV module in \PYTHIA8 is used for showering and hadronization in the DS. \PYTHIA version 8.230 or higher is used to access important features, such as the running of the dark coupling. In earlier versions, these features were added by patching the source code~\cite{EMJ2:Schwaller:2015gea}. Additional modifications to \PYTHIA are required to simulate the flavored emerging jet model~\cite{Renner:2018fhh}. The SUEPs are simulated using a custom \PYTHIA module that produces dark hadrons according to a Boltzmann distribution~\cite{Knapen:2016hky}. Dark-hadron properties, including branching fractions and lifetimes, are computed separately and specified in the \PYTHIA configuration as needed for each signal model. In particular, \rinv for SVJ models is implemented by reducing the branching fractions to SM quarks for all dark hadron species; dark hadrons that do not decay into SM quarks are marked as stable. Because stable dark hadrons must be produced in pairs (in order to conserve quantum numbers), events with an odd number of stable dark hadrons are rejected. For the dark QCD signal models studied in this Report, \PYTHIA is used for the LO matrix-element calculations as well as hadronization and showering. For other models, such as those requiring processes not implemented in \PYTHIA or more accurate simulation of ISR, DS particles produced by \MGvATNLO can be interfaced with \PYTHIA for hadronization and showering in both the DS and the SM~\cite{Cohen:2017pzm}.

In the following results, for some models, we present a minimum allowed coupling that will satisfy thermal-relic density constraints, assuming thermal freeze-out of DM. Typically, there is a minimum allowed coupling between the standard model and the DS. For couplings smaller than the minimum, which would have earlier freeze-out times, the DM production in the early universe would exceed the observed DM, as measured by the Planck experiment~\cite{Planck:2018vyg}. The minimum coupling can be determined by computing the thermal-relic density for various coupling values and scanning over the range of values to yield the smallest that satisfies the observed constraint. To perform the thermal-relic density calculation, we use the \MADDM~3.0 software framework~\cite {Ambrogi:2018jqj} with the appropriate \MGvATNLO signal models for the quoted searches. For fixed DM and mediator masses and a fixed DM coupling (typically $\gDM=1.0$), the thermal-relic density follows a parabolic form, allowing the minimum allowed coupling to be determined through a coupling scan.

\section{Signatures}
\label{sec:signatures}

The CMS Collaboration has a broad program of searches for models of BSM physics that provide DM candidates; an overview of the theoretical framework for these models is provided in Section~\ref{sec:theoreticalFramework}. In this section, we briefly discuss the details of each search and the signatures of the models targeted. The sensitivity of a broad range of signatures to DSs is probed, and no significant excess of events is observed over the background predictions.

These searches are categorized by their final states. In their simplest form, DM particles would interact weakly in the detector and thus be largely invisible, with the events featuring large \ptmiss. However, the DM particle could be accompanied by a promptly-produced, visible particle, which can be used to trigger on the events. These associated objects could either come from ISR, which has the advantage of being universally possible but suppressed by additional powers of the coupling, or arise through the same production mechanism as DM, as in t-channel mediators, where the extra jets are produced in mediator decay. We call such final states ``invisible, prompt final states'', and discuss them in Section~\ref{sec:signatures_invis}.

If the mediator particle can decay into invisible DM final states, then it may also be able to decay into SM particles that are visible in the detector. Such final states are dubbed ``visible, prompt final states'' and are discussed in Section~\ref{sec:signatures_vis}.

Furthermore, LLPs can be DM candidates or be produced in association with the DM. We discuss final states involving displaced and long-lived signatures in Section~\ref{sec:signatures_llp}.

It is notable that many general categories of theoretical models can potentially present any of these final states. For example, strongly coupled DSs can produce SVJs with invisible final states, SUEPs with visible final states, EJs with displaced final states, or potentially mixtures of these novel objects. Further, there may be deep connections between different final states: any mediator produced via an SM process can decay into the same SM particles, leading to a visible final state. Therefore, investigation of the visible final state can help exclude other final states without depending on the detailed phenomenology. These sorts of connections between final states is made more explicit in Section~\ref{sec:reinterpretationAndResults}, where the results are shown. These considerations motivate the breadth and continued expansion of the CMS search program, as the nature of DM remains unknown.

\subsection{Invisible final states} \label{sec:signatures_invis}

As described in Section~\ref{sec:met}, if DM particles are produced but do not interact with the CMS detector, these ``invisible'' particles can be deduced through the use of \ptvecmiss. The CMS DM searches for invisible final states include mono-X searches (Section~\ref{sec:monoX}), searches for the Higgs boson decaying into BSM invisible states (Section~\ref{sec:higgs_invisible}), and searches for SVJs (Section~\ref{sec:svj}). 

\subsubsection{Mono-X searches}
\label{sec:monoX}

Many theoretical models predict the production of DM particles that are not directly detectable in LHC collisions. If these final state particles recoil with large transverse momentum against other detectable SM particles, the result is a transverse momentum imbalance in a collision event, \ptvecmiss. This type of event topology is rarely produced in SM processes and therefore enables a highly sensitive search for DM candidates. The resulting signature yields a final state denoted as X+\ptmiss, \ie, the ``mono-X'' signature, where `X' is the recoiling SM object, such as a jet, vector boson, photon, top quark, or Higgs boson. Depending on the exact nature of the SM-DM interaction, the recoiling SM object can either come from ISR off the incoming partons, or be a part of the new interaction. The different mono-X channels thus either provide complementary probes of DM production (\eg, mono-\PH vs. monojet), or serve as cross-checks of one another (\eg, monojet vs. monophoton). The sensitivities of these searches to a range of simplified DM models are shown in Section~\ref{sec:simpDSResults}. The sensitivity to the 2HDM+a extended DS scenario is shown in Section~\ref{sec:2hdma}. The sensitivity to the SVJ signature is shown in Section~\ref{par:section7_svj}. 

\cmsParagraph{Search for monojet and hadronically-decaying mono-V dark matter\label{sec:EXO-20-004}} In mono-X searches, one of the most sensitive approaches is to use energetic hadronic jets accompanying the invisible particles to select signal candidates. The experimental signature therefore comprises one or more energetic jets and large \ptmiss. While the \ptmiss is the intrinsic result of BSM or SM particles escaping a detector without leaving any trace, the hadronic jets may be produced in the hard-scattering process as ISR (reconstructed as an AK4 jet), as the hadronic decay products of a Lorentz-boosted \PW or \PZ boson (reconstructed as a single large-radius jet with a characteristic substructure), or, as in the case of the fermion portal model, in the decay of a new mediator. These final states are commonly referred to as monojet and mono-V, respectively.

\begin{figure}[htbp!]
\centering
\includegraphics[width=0.57\linewidth]{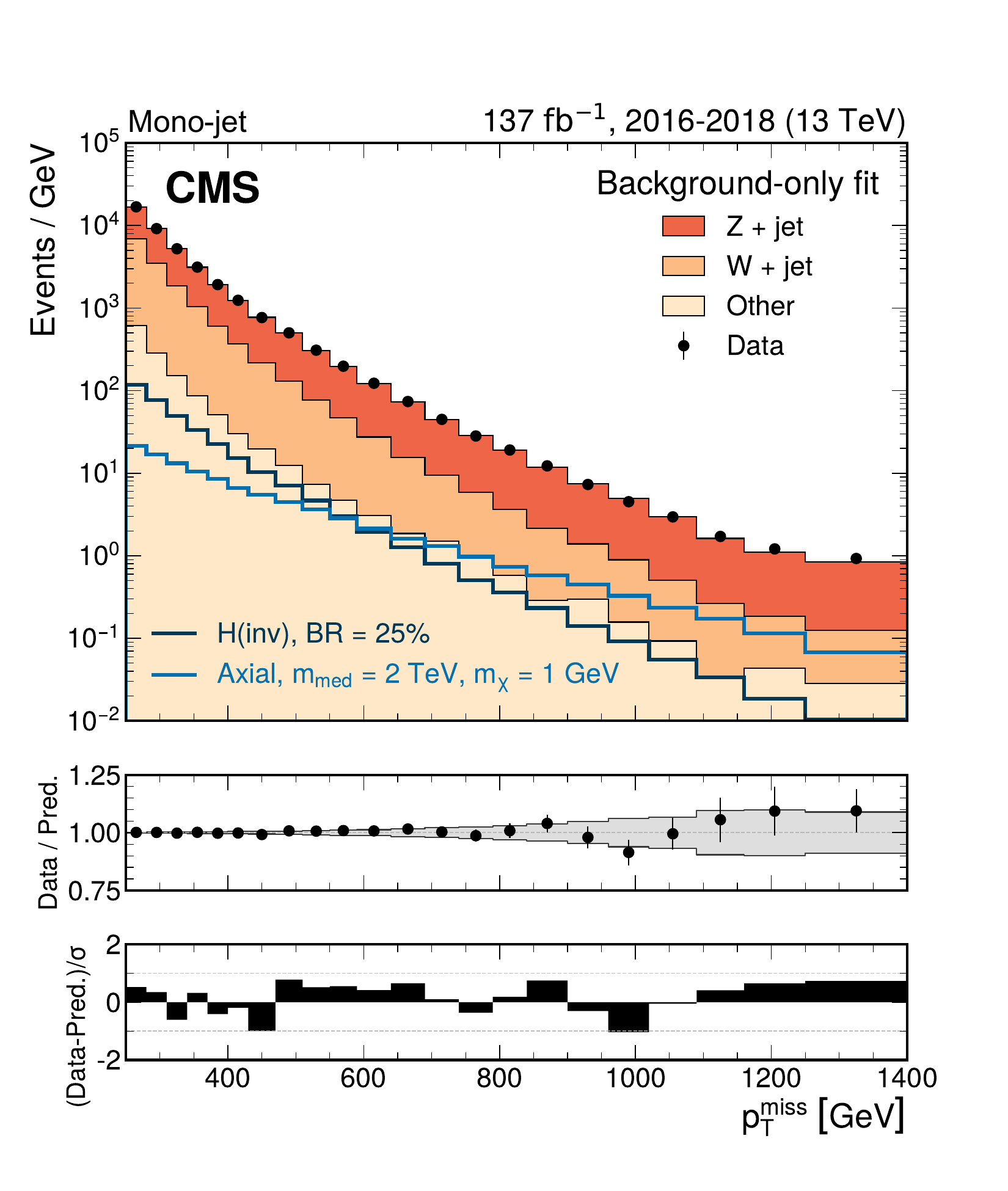}
\caption{Comparison of \ptmiss between data and the background prediction in the monojet SR after the simultaneous fit for the full Run~2 data set. The upper panel shows the \ptmiss distribution, the middle panel shows the ratio of the data to the prediction, and the lower panel shows the ratio of the data minus the prediction, all divided by the uncertainty. The axial vector signal and ${\PH(\text{inv})}$ signal are shown, the second of which is described in Section~\ref{sec:higgs_invisible}. Figure taken from Ref.~\cite{CMS:2021far}.}
\label{fig:monojet_dist}
\end{figure}

A search for monojet and mono-V signatures is presented in Ref.~\cite{CMS:2021far} and uses a data sample corresponding to $\Lint=101\fbinv$, collected in 2017--2018. The analysis defines signal categories for events with and without an identified V candidate. In both signal categories, the signal is expected to show up as an excess of events over the background at large values of \ptmiss. The leading SM background contributions in the SRs originate from $\Zvvjets$ and $\Wlvjets$ production, which is estimated using the TF technique described in Section~\ref{sec:transferFactor}. A statistical combination is performed with the results of an earlier analysis~\cite{CMS:2018Monojet}, which used data collected in 2016, corresponding to $\Lint=36\fbinv$. For illustrative purposes, the distribution of \ptmiss in the monojet SR including contributions from the full Run~2 data set is presented in Fig.~\ref{fig:monojet_dist}.

\cmsParagraph{Search for new physics in leptonically decaying \texorpdfstring{\PZ}{Z} boson events\label{sec:EXO-19-003}}The mono-\PZ final state can yield a dilepton signature if a \PZ boson is produced in $\Pp\Pp$ collisions, recoils against DM or other BSM invisible particles, and subsequently decays into two oppositely charged leptons ($\Pell^{+}{\Pell^{-}}$, where $\Pell=\Pe$ or \PGm). A search for DM~\cite{CMS:2020ulv} was performed using events with a leptonically decaying \PZ boson and large \ptmiss, in a data sample collected in 2016--2018 corresponding to $\Lint=137\fbinv$.

Several SM processes can contribute to the mono-\PZ signature. The most important backgrounds come from diboson processes: $\PW\PZ\to\Pell\PGn\Pell\Pell$ where one charged lepton escapes detection, $\PZ\PZ\to\Pell\Pell\PGn\PGn$, and $\PW\PW\to\Pell\Pell\PGn\PGn$. There can also be contributions produced by decays of top quarks in \ttbar or $\PQt\PW$ events. Smaller contributions may come from triple vector boson processes. The DY production of lepton pairs, $\PZ/\PGg^*\to\Pell\Pell$, has no intrinsic source of \ptmiss but can still mimic a mono-\PZ signature when the momentum of the recoiling system is poorly measured. A simultaneous maximum likelihood fit to the \ptmiss or \mT distributions in the SR and CRs constrains the background normalizations and their uncertainties. 

\begin{figure}[htbp!]
\centering
\includegraphics[width=0.49\linewidth]{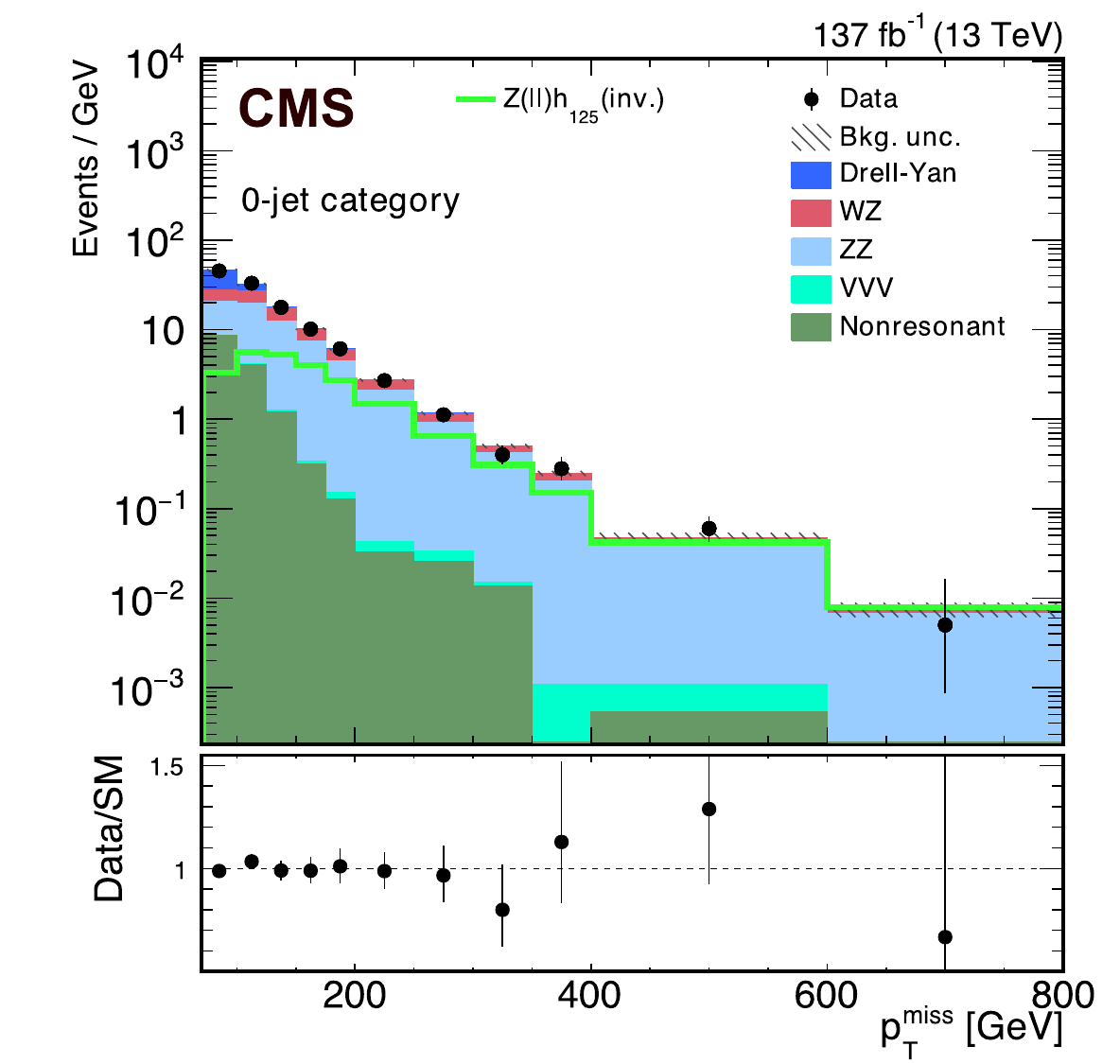}
\includegraphics[width=0.49\linewidth]{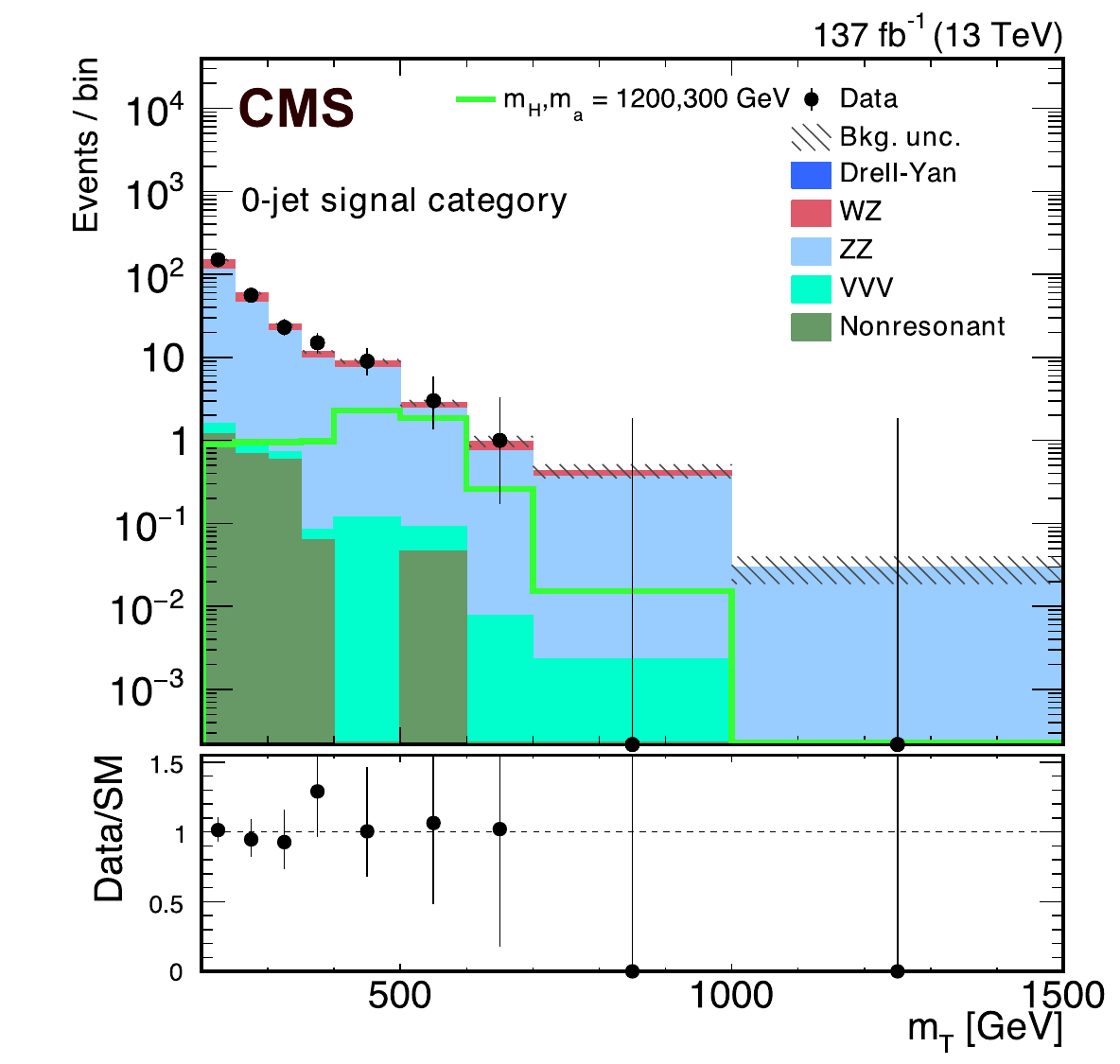}
\caption{The \ptmiss (\cmsLeft) and \mT (\cmsRight) distributions for events in the SR in the 0-jet final state, in the search for new physics in leptonically decaying \PZ boson events. The uncertainty band includes both statistical and systematic components. Figures adapted from Ref.~\cite{CMS:2020ulv}.}
\label{fig:ptmiss_mt_sr_dist_monoz}
\end{figure}

The results of the search are shown in Sections~\ref{sec:simpDSResults} and~\ref{sec:2hdma}. The distributions in \ptmiss and \mT for events in the SR are presented for the 0-jet final state in Fig.~\ref{fig:ptmiss_mt_sr_dist_monoz}.

\cmsParagraph{Search for \texorpdfstring{mono-\PQt}{mono-t} events} The associated production of neutrinos and a single top quark is heavily suppressed in the SM~\cite{PhysRevD.2.1285,Andrea:2011ws}. This renders the signature of large \ptmiss and a single top quark an excellent search channel for DM production (``mono-\PQt''). Here, DM candidates might be created from the decay of a new vector or axial-vector mediator V that could be produced via flavor-changing neutral currents (FCNCs), creating a common vertex with an incoming light quark and an outgoing, single top quark. The final-state objects are significantly boosted, giving large \ptmiss and recoiling, collimated top quark decay products. 

The analysis presented in Ref.~\cite{CMS:2018gbj} uses data corresponding to $\Lint=36\fbinv$, collected in 2016, to search for the mono-\PQt signature. The analysis focuses on hadronic decays of the top quark that can be reconstructed as a single CA15 jet, whose features subsequently are used for signal isolation and rejection of reducible background processes. Jets with a reconstructed $\pt > 250\GeV$ are considered in the analysis. This allows the analysis to be sensitive also to top quarks in the intermediate $\pt$ regime where the decay products are not contained in a ``standard'' large-radius jet with a distance parameter $0.8$. Generalized energy correlation functions (ECFs)~\cite{Larkoski:2013eya,Moult:2016cvt} are calculated for the leading CA15 jet in the event and are used for selecting jets with a substructure compatible with the one expected from fully merged top quark decays. This is done with the help of a boosted decision tree (BDT) algorithm~\cite{Friedman:2001wbq}, which exploits 11 ECF ratios to maximize the discrimination between simulated top quark jets and QCD (\ie, quark- or gluon-initiated) jets~\cite{CMS:2020poo}. The BDT algorithm is calibrated using two data CRs enriched in events from $\ttbar \to \Pell$+jets and DY+jets production, respectively.

\begin{figure}
\centering
\includegraphics[width=0.55\textwidth]{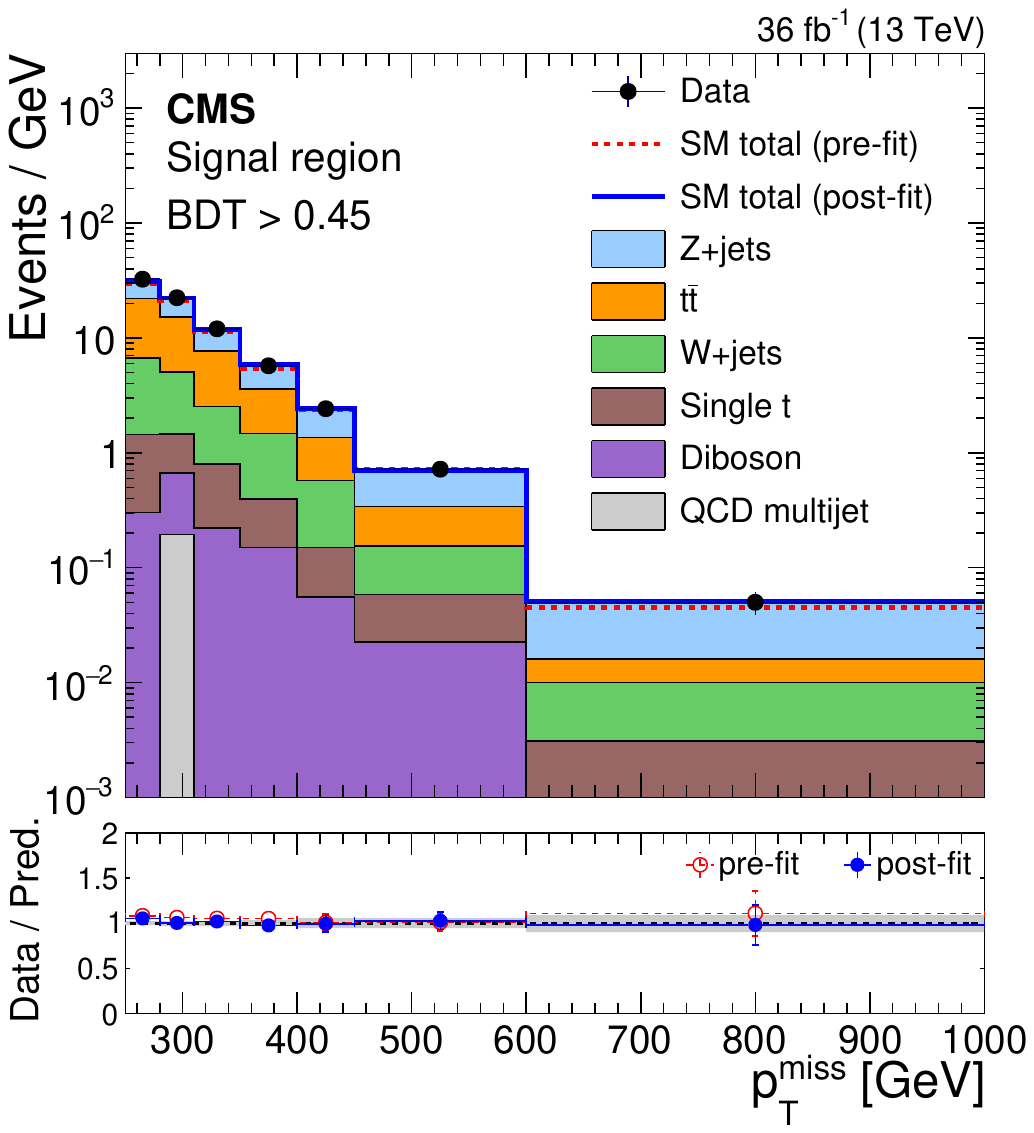}
\caption{Distribution of \ptmiss from SM backgrounds and data in the SR after simultaneously fitting the SR and all CRs, in the search for mono-\PQt events. Each bin shows the event yields divided by the width of the bin. The figure corresponds to the tight category of the SR. The stacked histograms show the individual fitted SM background contributions. The blue solid (red dashed) line represents the sum of the SM background contributions normalized to their fitted yields (to the prediction). The lower panel shows the ratio of data to fitted prediction. The gray band on the ratio indicates the one standard deviation uncertainty on the prediction after propagating all the systematic uncertainties and their correlations in the fit. Figure taken from Ref.~\cite{CMS:2018gbj}.}
\label{fig:EXO-16-051}
\end{figure}

The SM background \ptmiss spectrum is estimated using the TF technique described in Section~\ref{sec:transferFactor}, relying on multiple CRs to constrain the \Zvvjets and \Wlvjets backgrounds in the SR. The \ptmiss distribution after the fit is shown for a representative category in Fig.~\ref{fig:EXO-16-051}.

\cmsParagraph{Search for monophoton events\label{sec:EXO-16-053}} The photon and large  \ptmiss (monophoton) final state has the advantage of high purity and selection efficiency. A search for DM~\cite{CMS:2018ffd} was performed in the monophoton final state using data corresponding to $\Lint=36\fbinv$, collected in 2016.  
The most significant SM background processes in this channel are the $\PZ\PGg$ (where the \PZ boson decays into a pair of neutrinos) and $\PW\PGg$ (where the \PW boson decays into a charged lepton and a neutrino) diboson processes. Together these processes account for 70\% of the SM background. Other SM background processes include $\PW\to\Pell\PGn$ (where \Pell is misidentified as a photon), \phojets, QCD multijet events (with a jet misidentified as a photon), $\PGg\PGg$, \ttbar{}\PGg, t\PGg, VV\PGg (where V refers to a \PW or a \PZ boson) and $\zll$+\PGg. This channel also has an additional small background contribution from noncollision sources such as beam-halo interactions and spikes, which are measured as isolated, high-energy deposits arising from instrumental effects in the ECAL. 

\begin{figure}
\centering
\includegraphics[width=0.55\textwidth]{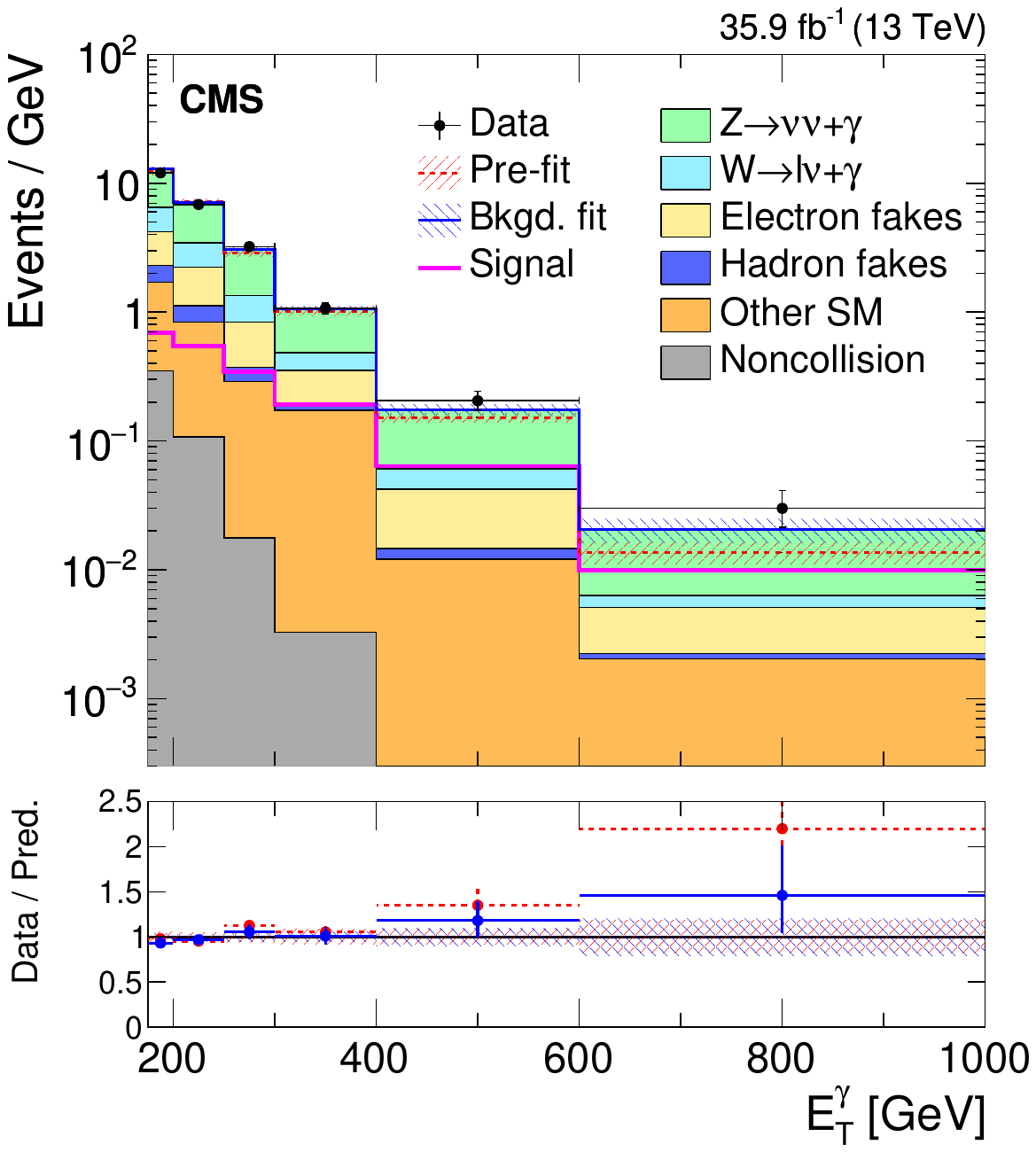}
\caption{Observed $\ET^\PGg$ distribution in a SR compared with the post-fit background expectations for various SM processes, in the search for monophoton events. The last bin of the distribution includes all events with $\ET^\PGg> 1000\GeV$. The expected background distributions are evaluated after performing a combined fit to the data in all the control samples and the SR. The ratios of data with the pre-fit background prediction (red dashed line) and post-fit background prediction (blue solid line) are shown in the lower panel. The bands in the lower panel show the post-fit uncertainty after combining all the systematic uncertainties. The expected signal distribution from a 1\TeV vector mediator decaying into 1\GeV DM particles is overlaid. Figure adapted from Ref.~\cite{CMS:2018ffd}.}
\label{fig:EXO-16-053}
\end{figure}

The $\PZ\PGg$ and $\PW\PGg$ backgrounds are estimated using observed data in the four mutually exclusive CRs using TF techniques, as described in Section~\ref{sec:transferFactor}. The potential signal contribution is extracted from the data via the simultaneous fit to the $\ET^\PGg$ distribution in the signal and CRs. The $\ET^\PGg$ distribution after the fit is shown for a representative category in Fig.~\ref{fig:EXO-16-053}.

\cmsParagraph{Searches for dark matter in Higgs boson associated production\label{sec:EXO-18-011}} The Higgs boson discovery at the LHC opened a new window into mono-X searches for new BSM physics processes through the $\PH$+$\ptmiss$ signature. Owing to the small Yukawa couplings to light quarks and gluons, the ISR of the Higgs boson is suppressed, but it can be produced in the case of a new interaction with DM particles. Thus, the mono-\PH production can be either a result of final-state radiation of DM particles or of a BSM interaction of DM particles with the Higgs boson, typically via a mediator particle.

\begin{figure}
\centering
\includegraphics[width=0.55\textwidth]{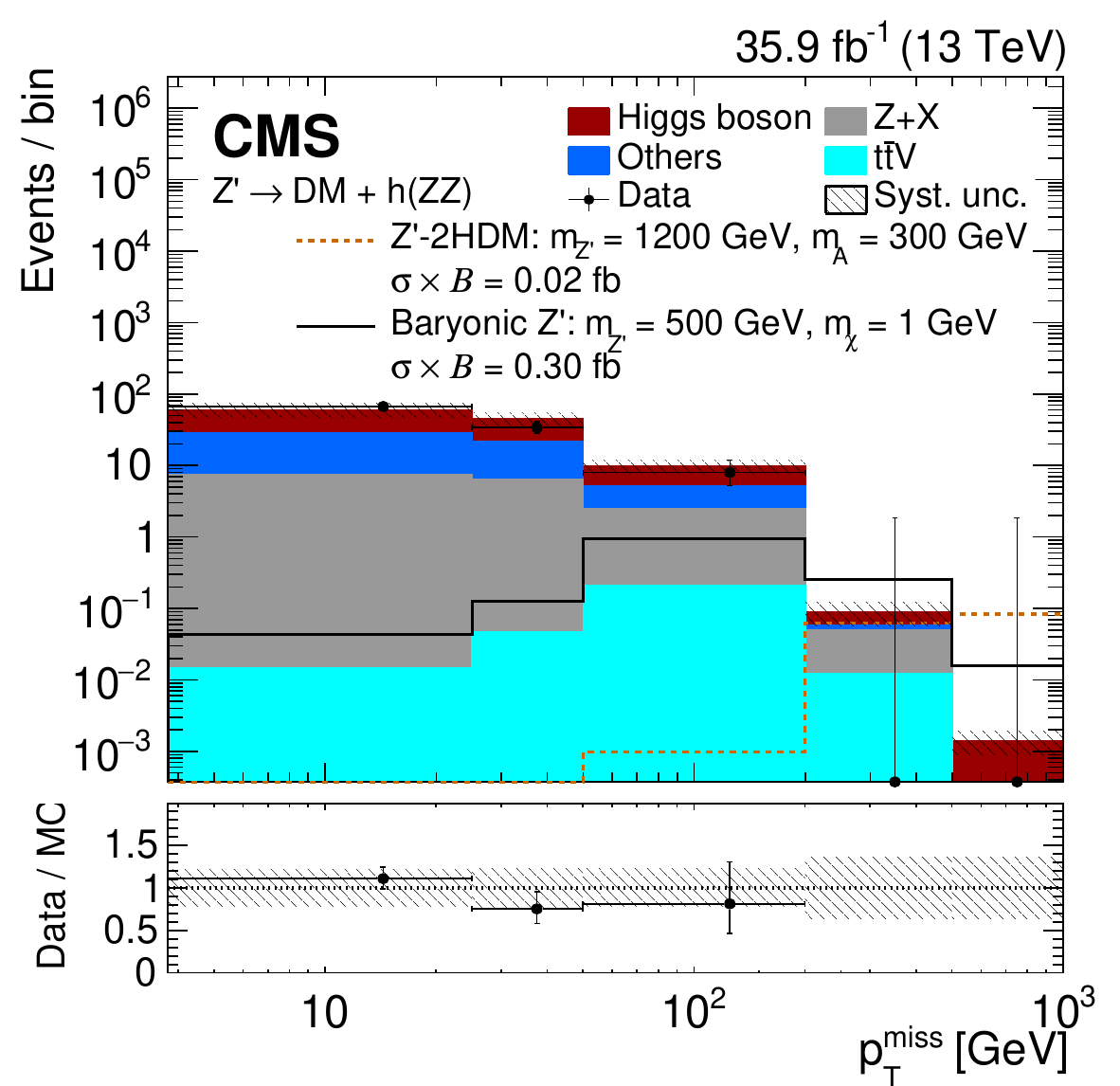}
\caption{The \ptmiss distribution for the expected background and observed events in data in the $\PH\to\PZ\PZ$ analysis. Two signal benchmarks, corresponding to the \PZpr-2HDM (dotted orange line) and baryonic \PZpr (solid black line) model are superimposed. The signal is normalized to the product of cross section and branching fraction, where \BR represents the $\PH\to\PZ\PZ$ branching fraction. The systematic uncertainties are shown by the hatched band. The ratios of the data and the sum of all the SM backgrounds are shown in the bottom panels. Figure taken from Ref.~\cite{CMS:2019ykj}.}
\label{fig:EXO-18-011}
\end{figure}

A search for DM particles~\cite{CMS:2019ykj} was performed using events with a Higgs boson candidate and large \ptmiss. The search is performed in five Higgs boson decay channels: $\PH \to \PQb\PAQb$, $\PGg\PGg$, $\PGtp\PGtm$, $\PWp\PWm$, and $\PZ\PZ$. The analyses are based on a data sample corresponding to $\Lint=36\fbinv$, collected in 2016. The $\PH \to \PQb\PAQb$ channel has also been probed in a dedicated search~\cite{CMS:2018zjv}. The statistical combination of the five decay modes is performed in order to improve the overall sensitivity. The $\PH \to \PQb\PAQb$ channel provides the highest sensitivity thanks to the large branching fraction and manageable background in the large-\ptmiss region. The $\PH \to \PGg\PGg$ and $\PH\to \PZ\PZ$ channels provide better resolution in the reconstructed Higgs boson invariant mass, while the $\PH\to \PGtp\PGtm$, $\PH\to \PWp\PWm$, and $\PH\to \PZ\PZ$ channels benefit from lower SM backgrounds, which results in a higher sensitivity to signals with smaller \ptmiss values. The \ptmiss distribution after the fit is shown for the $\PH\to\PZ\PZ$ analysis in Fig.~\ref{fig:EXO-18-011}.

All analyses exploit the mass reconstruction values of the Higgs boson decay products and are parameterized in different categories distinguished by the \ptmiss values to separate the signal and backgrounds. 

\cmsParagraph{Search for dark matter in a dark \texorpdfstring{Higgs+\ptmiss}{Higgs + missing transverse momentum} channel\label{sec:EXO-21-012}} The mono-\PH signature can also be used to probe DS models that include a dark Higgs boson \Hdark. A search~\cite{CMS:2023dof} using data corresponding to $\Lint=137\fbinv$ was performed for a dark-Higgs model where the \PZpr and \Hdark bosons are mediators between the DS and the SM. Through radiation of a dark Higgs boson \Hdark from a \PZpr or a \PGc particle, the $\Hdark$+$\PGc\PGc$ signature can provide collider probes to the DS. When the mass of \Hdark exceeds 160\GeV, the $\PW\PW$ channel is the dominant decay channel of the dark-Higgs model. The experimental signature is therefore $\PW\PW$ production with the presence of additional \ptmiss from the DM particles. This phase space has been explored in the dilepton and lepton+jets channels of the $\PW\PW$ decay.

\begin{figure}
\centering
\includegraphics[width=0.55\textwidth]{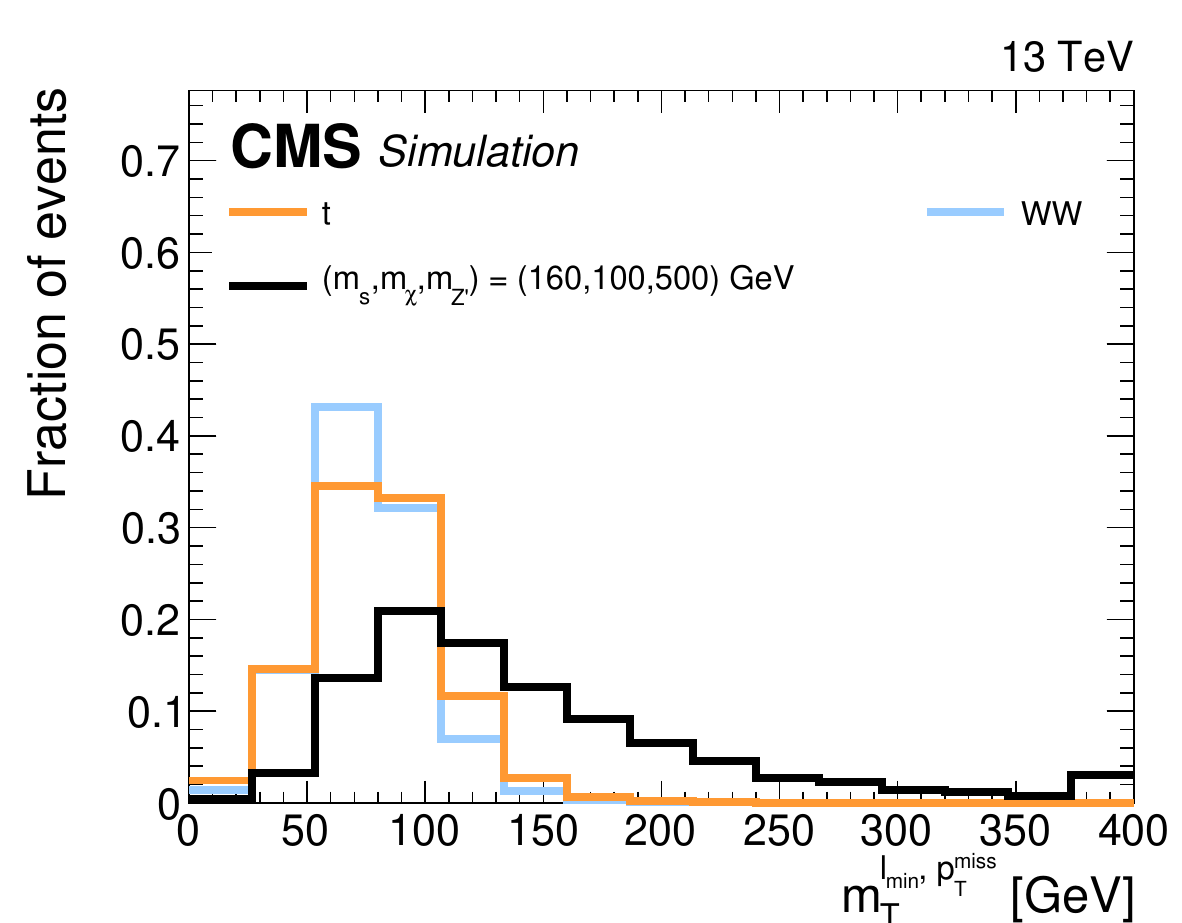}
\caption{Normalized distribution of the transverse mass of the trailing lepton plus missing transverse momentum system in the dilepton channel of the dark Higgs+\ptmiss search, for a signal with $\mHdark = 160\GeV$ (denoted as \mS in the figure), $\mDM = 100\GeV$ (denoted as $m_\PDM$ in the figure), and $m_{\PZpr} = 500\GeV$ (black), after the event selection criteria are applied. Predictions for the two main backgrounds of the analysis, nonresonant $\PW\PW$ and top quark production, are shown as blue and orange solid lines, respectively. The last bin includes the overflow. Figure taken from Ref.~\cite{CMS:2023dof}.}
\label{fig:mt_darkhiggs}
\end{figure}

Kinematic relations between the $\PW\PW$ remnants and the \ptmiss proved to be the crucial factors in distinguishing SM backgrounds from the $\Hdark$+$\PGc\PGc$ signature. More specifically, the transverse mass \mT of the lepton (the trailing lepton in the case of the dilepton channel) combined with \ptmiss is essential for discriminating signal events. For the background, \mT cannot exceed $m_{\PW}$ except for resolution effects, whereas the enlarged amount of \ptmiss in the signal tends to produce higher \mT values as shown in Fig.~\ref{fig:mt_darkhiggs}, taken from Ref.~\cite{CMS:2023dof}. Other vital aspects are the angles between the visible particles and \ptvecmiss. Compared to SM processes, the visible decay products of the $\PW\PW$ pair in dark-Higgs production tend to be more collimated with one another, while they tend to be back-to-back with \ptvecmiss. 

\subsubsection{Searches for the Higgs boson decaying into invisible final states} 
\label{sec:higgs_invisible}

In the SM, the Higgs boson can only decay invisibly (\hinv) via ${\PH\to\PZ\PZ^{*}\to4\PGn}$, with an expected branching fraction of about 0.1\%~\cite{Dittmaier:2011ti}. The CMS detector lacks the systematic precision and sufficient data to probe such a small branching fraction. Invisible background signatures can occur from particles falling outside of the detector's acceptance or from particles that are misreconstructed or mismeasured, which are rare phenomena that occur in background events with high production cross sections relative to \hinv events. In many DM models, however, the \hinv branching fraction, \brhinv, is $\mathcal{O}$(10\%). Therefore, measuring a sufficiently stringent upper limit on \brhinv can constrain the contribution from Higgs boson decays into DM candidates.

The \vbf production channel provides the highest sensitivity to \hinv events, as described in Section~\ref{para:HIG-20-003} and Ref.~\cite{CMS:2022qva}. The 2016--2018 results are combined with the results of an earlier CMS publication using Run~1 and 2015 data~\cite{CMS:2016dhk}, as described in Section~\ref{para:HIG-21-007} and Ref.~\cite{CMS:2023sdw}.

Several DS models predict a massless dark photon \PGgdark that couples to the Higgs boson, leading to $\PH \to \PGg \PGgdark$ decays and a partially visible final state. Two CMS searches for this decay mode are described in Section~\ref{para:HiggsToZd} and Refs.~\cite{CMS:2019ajt,CMS:2020krr}.

The sensitivities of the analyses described in this section are shown for simplified DS models containing a Higgs portal and a dark-Higgs portal in Sections ~\ref{par:darkHiggsPortalLimits} and ~\ref{sec:higgsPortalResults}, respectively. The sensitivity of these analyses to the 2HDM+a scenario is shown in Section~\ref{sec:2hdma}.

\cmsParagraph{Search for \texorpdfstring{\hinv}{Higgs boson decay into invisible final states} produced via vector boson fusion}
\label{para:HIG-20-003}

\begin{figure}[htbp!]
\centering
\includegraphics[width=0.48\linewidth]{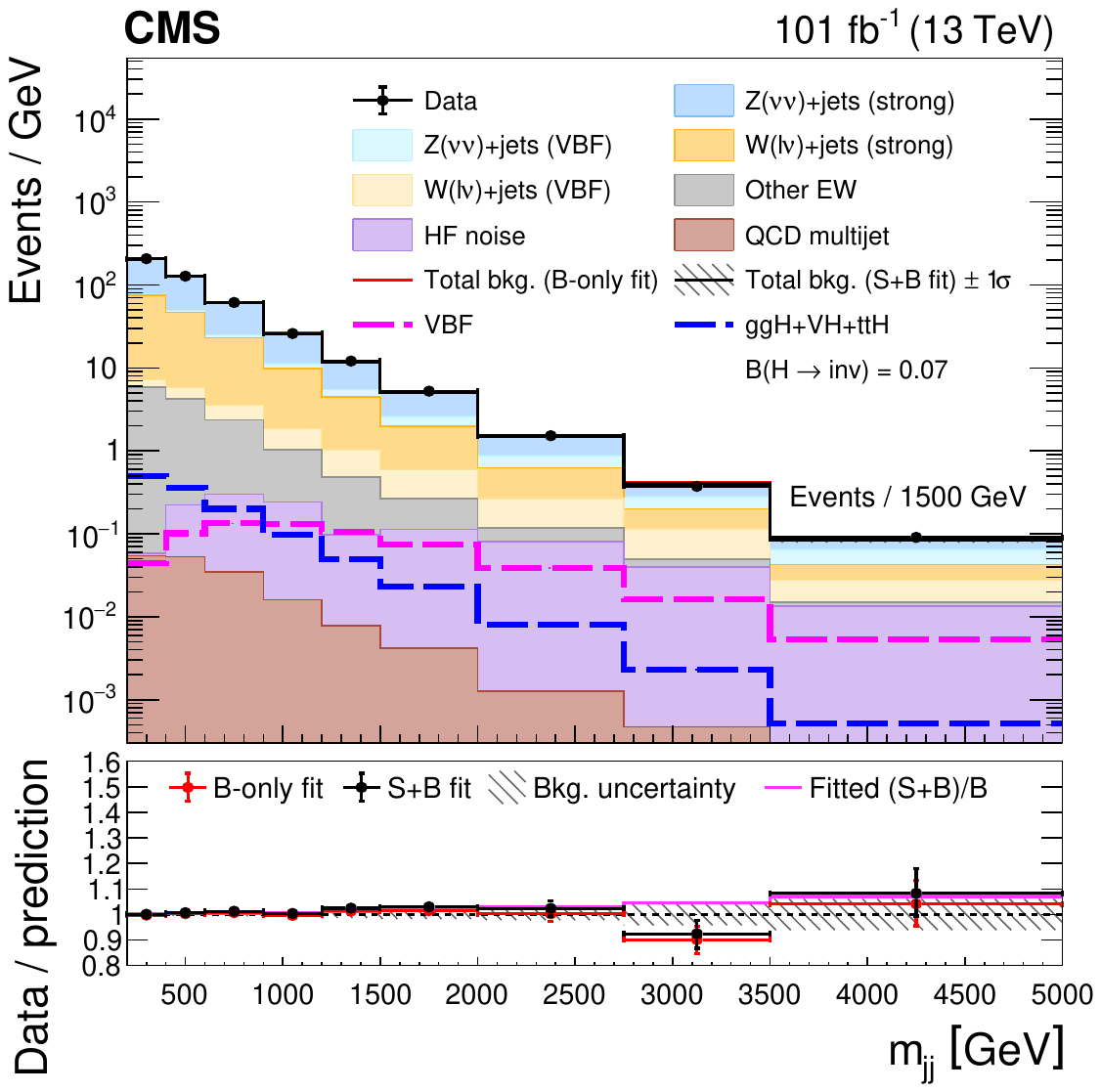}
\includegraphics[width=0.48\linewidth]{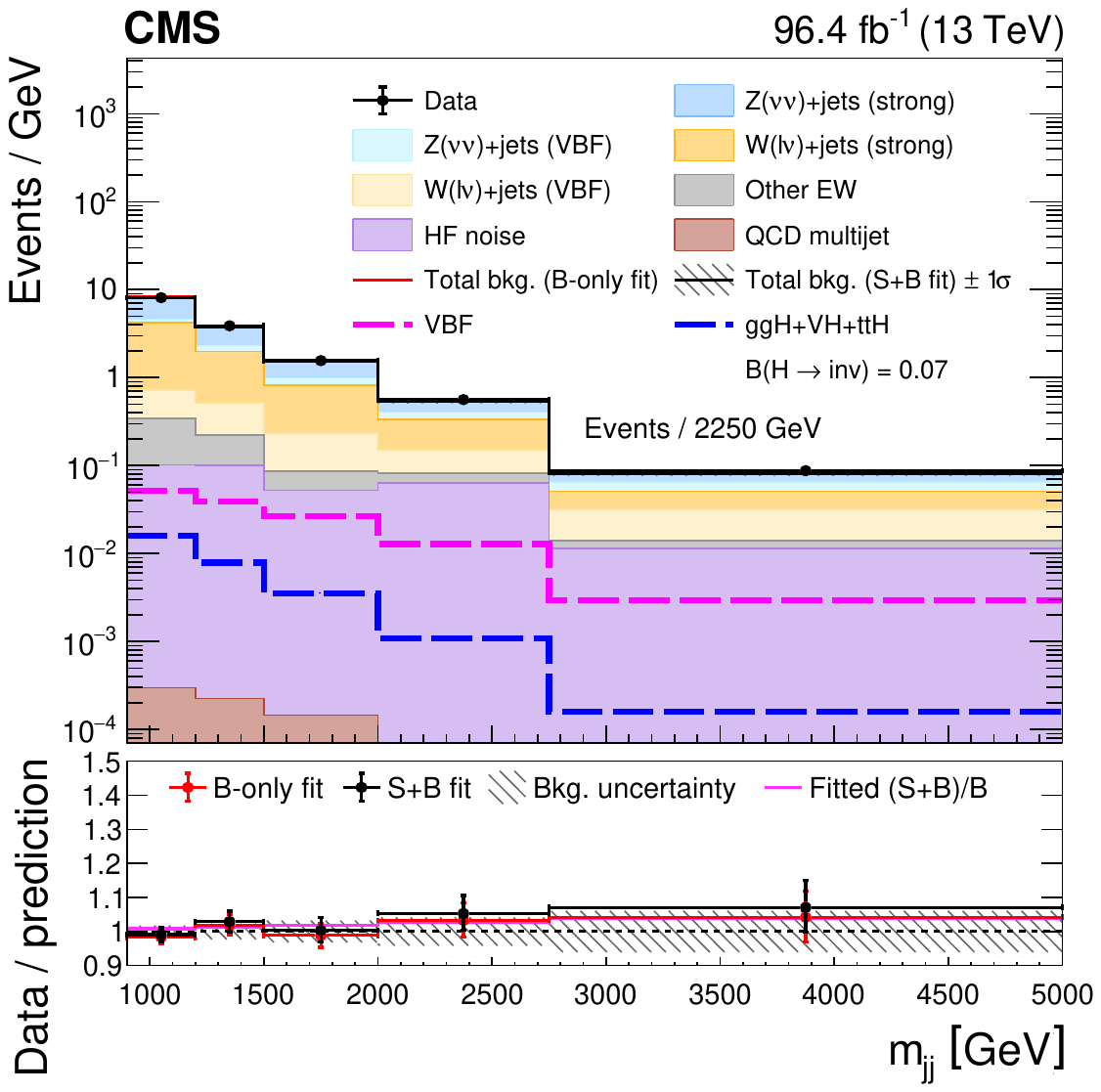}
\caption{Distributions of the dijet pair invariant mass in the SRs of the search for \hinv produced via vector boson fusion, for the high missing transverse momentum category (\cmsLeft) and for the dijet-based category (\cmsRight). The signal processes are scaled by the fitted value of \brhinv, shown in the legend. The background contributions are estimated from the fit to the data (S+B fit). The total background estimated from a fit assuming \brhinv = 0 (B-only fit) is also shown. The yields from the 2017 and 2018 samples are summed and the correlations between their uncertainties are neglected. The last bin of each distribution integrates events above the bin threshold divided by the bin width. Figures adapted from Ref.~\cite{CMS:2022qva}.}
\label{fig:vbf_hinv_mjj}
\end{figure}

Because of its large production cross section~\cite{deFlorian:2016spz} and distinctive event topology, the \vbf production mechanism drives the overall sensitivity in the direct search for invisible decays of the Higgs boson. A search for an invisibly decaying Higgs boson, produced in the \vbf mode, is performed with $\Pp\Pp$ collision data corresponding to $\Lint=101\fbinv$ at $\sqrt{s}=13\TeV$, collected in 2017--2018~\cite{CMS:2022qva}. The invariant mass of the jet pair produced by \vbf, \mjj, is used as a discriminating variable to separate the signal and the dominant backgrounds arising from \vh production in association with two jets (\Vjets). The dijet invariant mass \mjj is shown in Fig.~\ref{fig:vbf_hinv_mjj} in the SRs for the signal and the dominant backgrounds. 

\cmsParagraph{Search for \texorpdfstring{\hinv}{Higgs boson decay into invisible final states} in events with a \texorpdfstring{\ttbar}{tt} pair or a vector boson and combination of all \texorpdfstring{\hinv}{Higgs boson decay into invisible final states} searches\label{para:HIG-21-007}} A search for an invisibly decaying \PH produced in association with a \ttbar quark pair or a \PV boson (\ttbarh/resolved~\vh), where the associated particles decay into a fully hadronic final state, is reported in Ref.~\cite{CMS:2023sdw}. The search uses LHC $\Pp\Pp$ collision data collected during the years 2016--2018, corresponding to $\Lint=138\fbinv$ at $\sqrt{s}=13\TeV$. The \vh production analysis looks only at topologies in which the presence of the \PV boson is inferred from well-separated decay products, complementing the previous \vh search with merged decay products arising from boosted \PV bosons described in Section~\ref{sec:EXO-20-004} and Ref.~\cite{CMS:2021far} (monojet/mono-V). The \ttbarh production mechanism is investigated using final states containing \PQb jets and conditionally boosted \PQt quarks or \PW bosons. (The categories always require a \PQb jet and either contain a boosted top quark or a \PW boson, or they do not contain either boosted object.) The signal is extracted from a fit to the hadronic recoil distribution of selected events, where the hadronic recoil is defined as the vector sum of \ptvecmiss and the \ptvec of any selected charged lepton(s) or photon in an event. Two main sources of background dominate in the SR. The first is events with invisible \PZ boson decays and visible jets (\zinv). The second, referred to as the lost-lepton background, \llost, where \Pell stands for either an \Pe or \PGm, includes events from \ttbarjets and \wjets production processes where one or more leptons are misreconstructed, excluded by the phase space selection, or fall outside the detector acceptance.

A combined likelihood fit is performed across all Run~1 and Run~2 \hinv analyses reported by CMS, presented in Table~\ref{tab:final_state_breakdown}. The results are presented in Section~\ref{sec:higgsPortalResults}.

\begin{table*}[htb]
\centering
\topcaption{Data sets, respective integrated luminosities, and relevant publications for each \hinv production mode across Run~1 and Run~2. For some data-taking periods, no \hinv searches have been performed for the given production mode. Table adapted from Ref.~\cite{CMS:2023sdw}.}
\begin{tabular}{lllll}
Analysis tag & Production mode & \multicolumn{3}{c}{$\Lint$ [{\fbinv}]} \\
 & & 7\TeV & 8\TeV & 13\TeV (Run~2) \\\hline
\vbf-tagged & \vbf & \NA & 19.2~\cite{CMS:2014gab} & 140~\cite{CMS:2016dhk,CMS:2022qva} \\ [\cmsTabSkip]

\multirow{4}{*}{\vh-tagged} & $\PZ(\Pell\Pell)\PH$ & 4.9~\cite{CMS:2014gab} & 19.7~\cite{CMS:2014gab} & 140~\cite{CMS:2016dhk,CMS:2020ulv} \\
 & ${\PZ(\bbbar)\PH}$ & \NA & 18.9~\cite{CMS:2014gab} & \NA \\
 & ${\PV(\text{jj})\PH}$ & \NA & 19.7~\cite{CMS:2016xus} & 140~\cite{CMS:2016dhk,CMS:2023sdw} \\
 & Boosted \vh & \NA & \NA & 138~\cite{CMS:2021far} \\ [\cmsTabSkip]

\multirow{2}{*}{\ttbarh-tagged} & \ttbarh(hadronic) & \NA & \NA & 138~\cite{CMS:2023sdw} \\
& \ttbarh(leptonic) & \NA & \NA & 138~\cite{CMS:2019ysk,CMS:2020pyk,CMS:2021eha} \\ [\cmsTabSkip]

\ggH-tagged & \ggH & \NA & 19.7~\cite{CMS:2016xus} & 140~\cite{CMS:2016dhk,CMS:2021far} \\
\end{tabular}
\label{tab:final_state_breakdown}
\end{table*}

\cmsParagraph{Search for dark photons in Higgs boson decays into an undetected particle and a photon\label{para:HiggsToZd}} Two analyses searching for a scalar Higgs boson $\PH$ decaying into an undetected particle and a photon are reported in Refs.~\cite{CMS:2019ajt,CMS:2020krr}. These analyses use $\Pp\Pp$ collision data collected at $\sqrt{s}=13\TeV$ in 2016--2018, corresponding to $\Lint=138\fbinv$. Several BSM scenarios predict Higgs boson decays into undetected particles and photons~\cite{Curtin:2013fra,Djouadi:1997gw,Petersson:2012dp}. In these searches, the target final states are $\cPZ(\to \Pell\Pell) \PH(\to \PGg\PGgdark)$ and $\Pq\Pq\PH(\to \PGg \PGgdark)$, where $\Pell = \Pe, \Pgm$, the final-state quarks (\Pq) arise from the \vbf process, and \PGgdark is a massless dark photon that couples to the Higgs boson through a DS~\cite{Gabrielli:2013jka,Gabrielli:2014oya,Biswas:2016jsh,Biswas:2017anm}. The dark photon \PGgdark escapes undetected in the CMS detector. 

In the first analysis, which uses associated $\PZ\PH$ production~\cite{CMS:2019ajt}, the leptonic decays of the $\cPZ$ boson, consisting of two oppositely charged same-flavor high-$\pt$ isolated leptons, are used to tag the Higgs boson candidate events. Signal events are furthermore required to have large \ptmiss from the undetectable particle, an isolated high-$\pt$ photon (\ptg), and no more than two jets.

\begin{figure}
\centering
\includegraphics[width=0.55\textwidth]{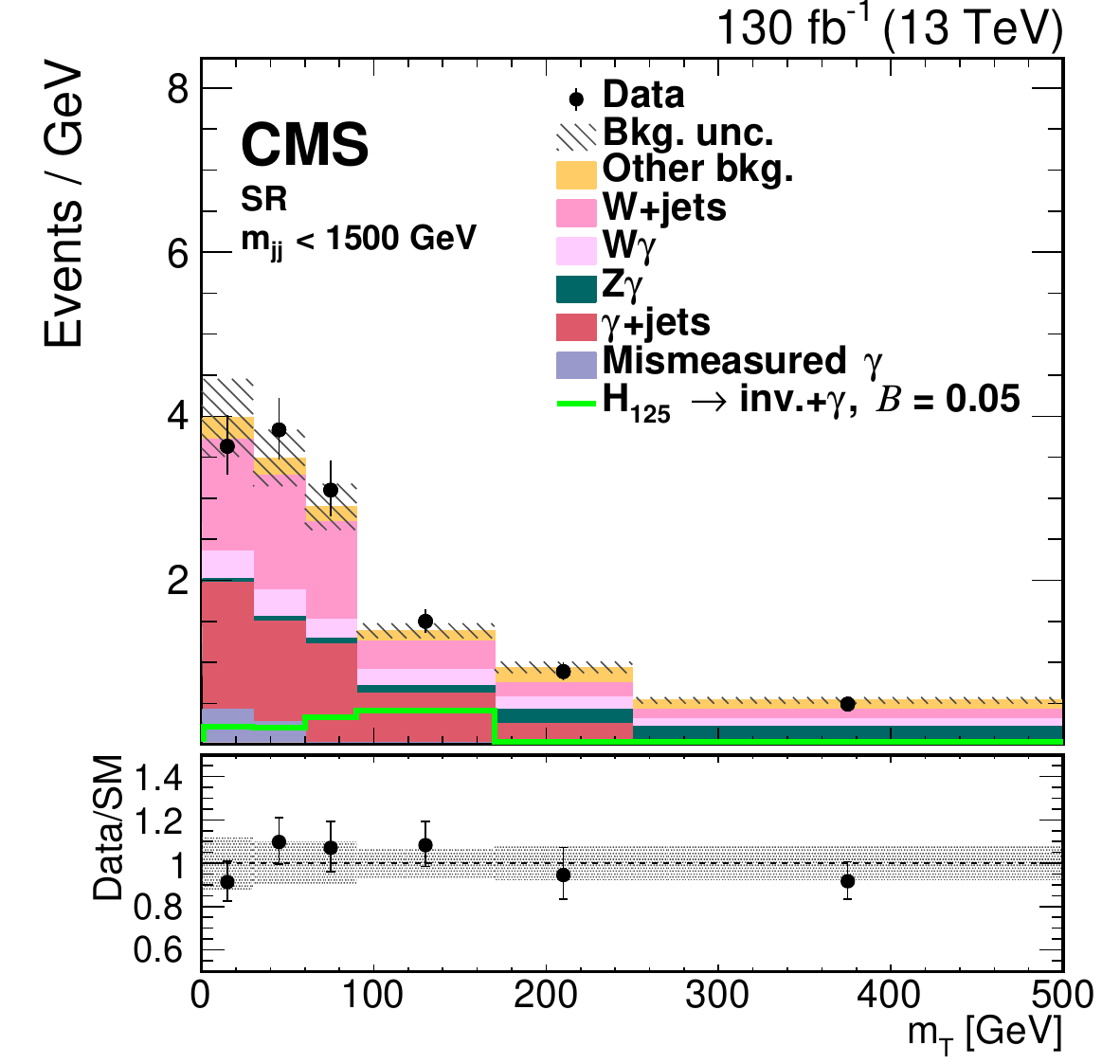}
\caption{The \mT distribution from the simultaneous fit for events with $\mjj < 1500\GeV$ in the SRs of the search for dark photons in Higgs boson decays, in the VBF production mode. The category other background includes contributions from {\PZ}+jets, nonprompt, top quark, VV, and VVV processes. Overflow events are included in the last bin. The shaded bands represent the combination of the statistical and systematic uncertainties in the predicted yields. The light green line, illustrating the possible contribution expected from inclusive SM Higgs boson production, assumes a branching fraction of 5\% for $\hinv$+$\gamma$ decays. The lower panel shows a per-bin ratio of the data yield and the background expectation. The shaded band corresponds to the combined systematic and statistical uncertainty in the background expectation. Figure taken from Ref.~\cite{CMS:2020krr}.}
\label{fig:EXO-20-005}
\end{figure}

In the \vbf production mode~\cite{CMS:2020krr}, a Higgs boson is accompanied by two jets that exhibit a large separation in pseudorapidity ($\detajj$) and a large dijet mass ($\mjj$). The invisible particle together with the photon produced in the Higgs boson decay can recoil with $\pt$ against the \vbf dijet system, resulting in an event with a large \ptmiss, which is used to select signal-enriched samples. After applying the selection, a binned maximum likelihood fit to \mT is performed to discriminate between the signal and the remaining background processes. The \mT distribution after the fit is shown for a representative category in Fig.~\ref{fig:EXO-20-005}.

\subsubsection{Signatures from hidden valley models}
\label{sec:svj}

As described in Section~\ref{sec:darkqcd}, some HV models predict unique signatures from a QCD-like force in the DS with corresponding dark quarks ($\Pqdark\Paqdark$). When produced at the LHC, dark quarks shower and hadronize in the DS giving rise to dark jets made of stable and unstable dark hadrons. The stable dark bound states do not interact with the detector. If the unstable ones decay promptly to SM quarks, this leads to an SVJ made of collimated visible and invisible particles.

\begin{figure}[htbp!]
\centering
\includegraphics[width=0.5\linewidth]{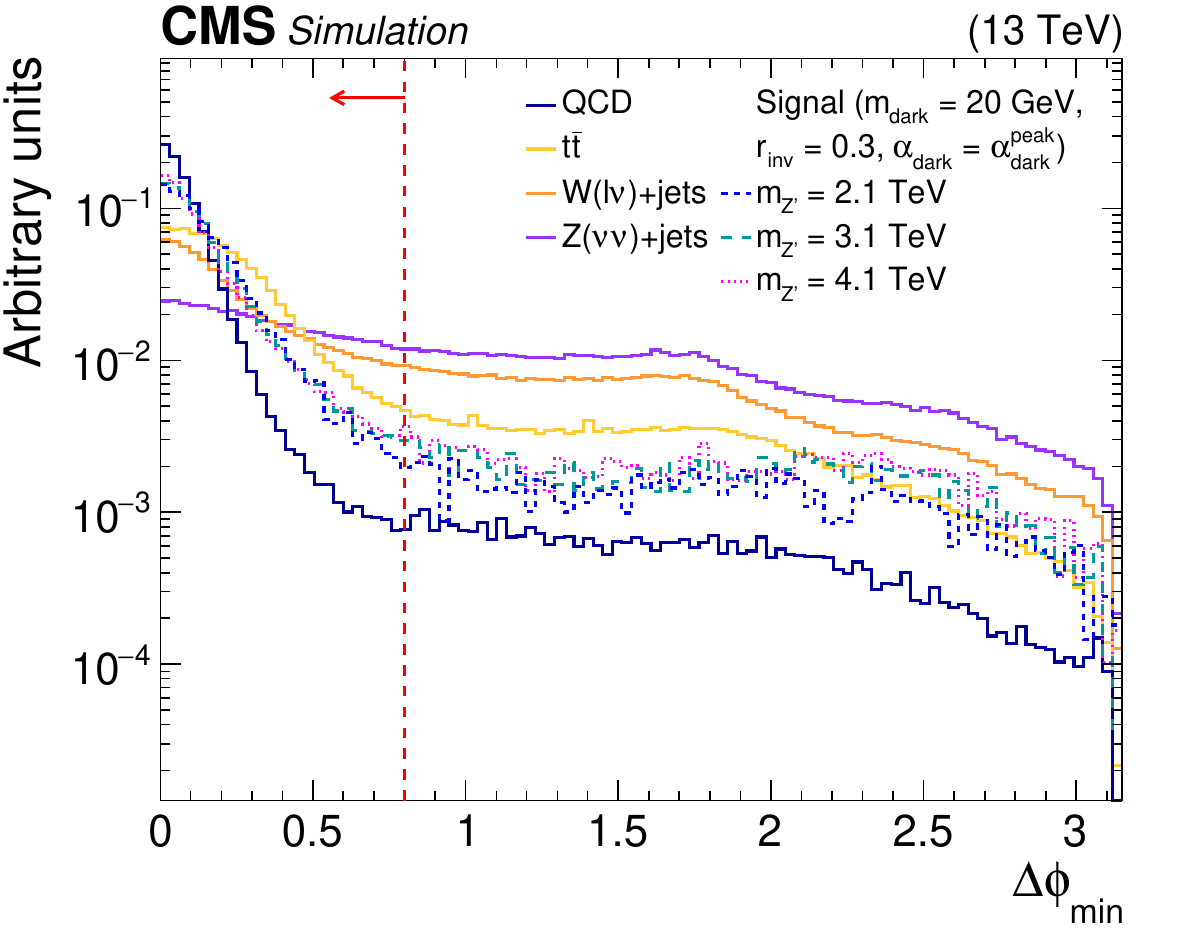}
\caption{The normalized distribution of the minimum azimuthal angle between the \ptvecmiss and each of the two leading jets (\mindphi) for simulated SM backgrounds and several SVJ signal models. The red vertical
dotted line indicates the selection requirement on this variable. Figure taken from Ref.~\cite{CMS:2021dzg}.}
\label{fig:EXO-19-020_dphi}
\end{figure}

\cmsParagraph{Search for semivisible jets\label{sec:EXO-19-020}} A search is performed~\cite{CMS:2021dzg} for SVJs using data collected during Run~2 and corresponding to $\Lint=138\fbinv$. Resonant production of a leptophobic \PZpr mediator decaying into dark quarks, $\qqbar \to \PZpr \to \Pqdark\Paqdark$, leads to a final state with two SVJs. The \ptvecmiss is aligned with one of the jets, as shown in Fig.~\ref{fig:EXO-19-020_dphi}, and has a moderate magnitude. Both jets carry a fraction of invisible momentum, leading to a partial cancellation when the jets are back-to-back. The SVJs are expected to be larger than typical SM jets, because they arise from a double parton shower and hadronization process: first in the DS and later in the SM sector. Depending on the parameter of the model, the signature can vary significantly. We assess models with $1.5 \leq \mZprime \leq 5.1\TeV$, $1 \leq \mDark \leq 100\GeV$, and $0 \leq \rinv \leq 1$. Because of the invisible momentum carried by stable dark hadrons, the mass of the mediator cannot be fully reconstructed. Instead, a bump hunt is performed using the transverse mass \mT of the dijet system and the \ptmiss. The SM backgrounds, dominated by QCD multijets with artificial \ptmiss but also including significant fractions of \ttbar, \wjets, and \zjets processes with genuine \ptmiss from neutrinos, are expected to have steeply falling \mT distributions. Two versions of the search are performed: an inclusive search using only selections on event-level kinematic variables, and a model-dependent search using a BDT trained on specific signal models to identify SVJs.

\begin{figure*}[t]
\centering
\includegraphics[width=0.49\linewidth]{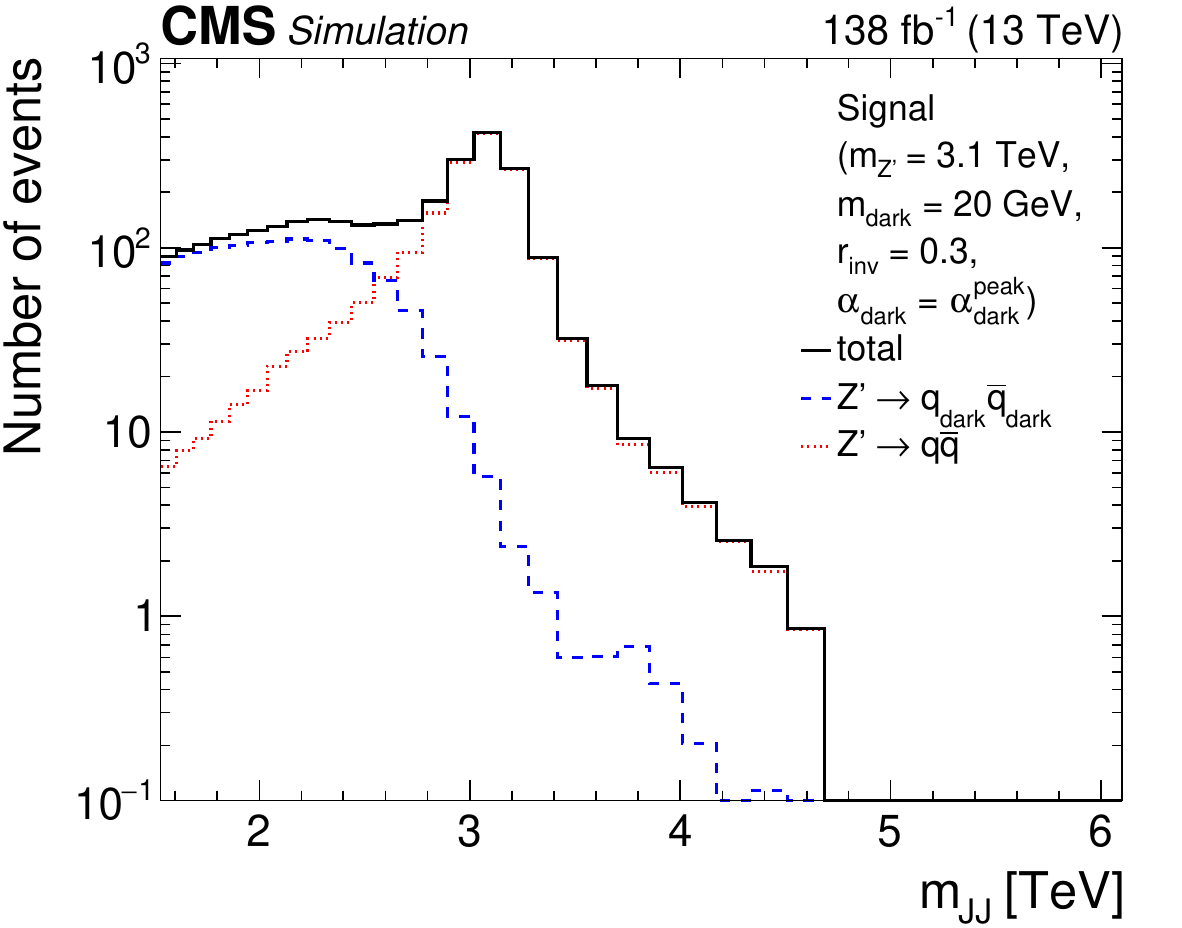}
\includegraphics[width=0.49\linewidth]{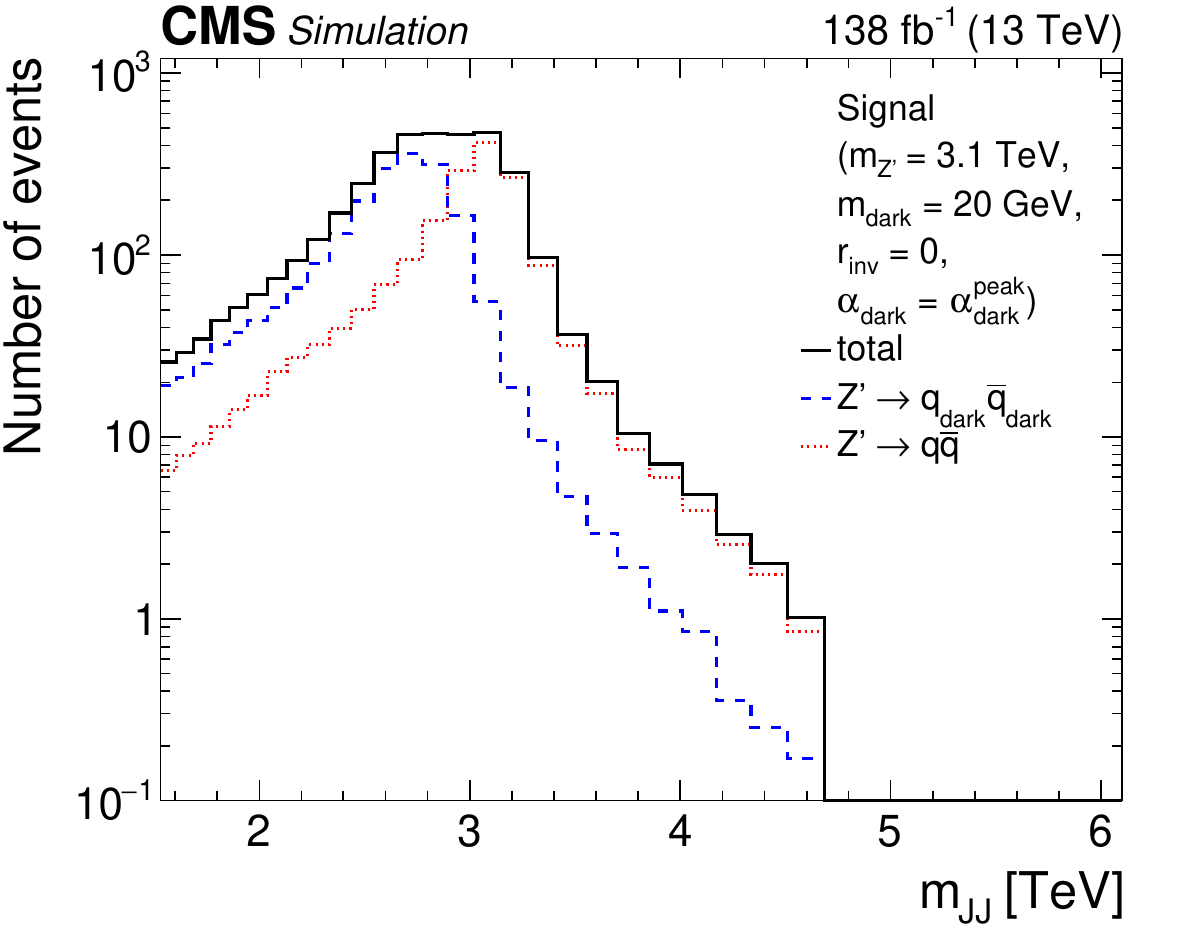}
\caption{The dijet mass distributions for the combination of \ZprimeToDark and \ZprimeToQQ events, for $\rinv=0.3$ (\cmsLeft) and $\rinv=0.0$ (\cmsRight), in SVJ signal models.
}
\label{fig:svj_dijet_hists}
\end{figure*}

The SVJ models with extreme values of \rinv, close to 0 or 1, overlap with the phase space of dijet resonance searches (Section~\ref{sec:EXO-19-012}) and monojet DM searches (Section~\ref{sec:EXO-20-004}). Hence, we can reinterpret these two searches for the SVJ signal model. Accordingly, the DM coupling in the dark QCD model is set to $\gqdark = 0.5$ in order for the \PZpr boson to have width and branching fractions consistent with the LHC DM Working Group benchmark model for simplified DM, as noted in Section~\ref{sec:darkqcd}. Because both possible final states have visible components, Fig.~\ref{fig:svj_dijet_hists} shows the dijet mass distributions from \ZprimeToDark and \ZprimeToQQ, both individually and summed, in the correct proportions for the specified coupling values. For $\rinv = 0.3$, the \ZprimeToDark events have relatively lower dijet mass values, so they do not contribute substantially to the sensitivity of a dijet resonance search, which remains dominated by \ZprimeToQQ events. However, for $\rinv = 0.0$, the two contributions to the dijet mass distribution are similar enough that the summed distribution is enhanced around the resonant peak, providing correspondingly greater sensitivity. The remaining minor degradation in the \ZprimeToDark dijet mass distribution primarily occurs because of the presence of neutrinos from decays of heavy-flavor hadrons, which are more likely to be produced in SVJs than in SM jets.

\begin{figure*}[t]
\centering
\includegraphics[width=0.5\linewidth]{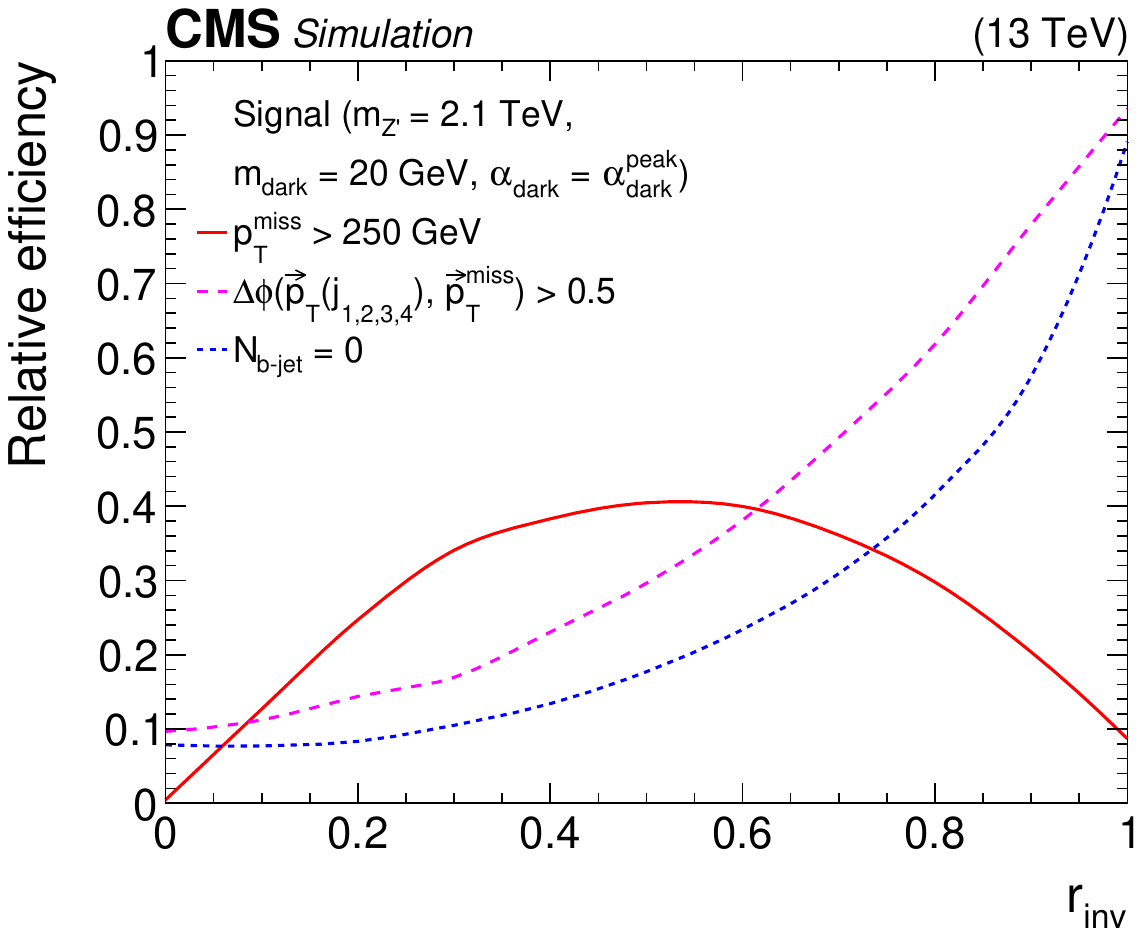}
\caption{The relative efficiencies of several selection criteria from the monojet search for SVJ signals. The efficiencies of the $\dphifull$ and $\nbjets$ requirements are evaluated after the $\ptmiss > 250\GeV$ requirement. The uncertainty in the simulation is negligible.
}
\label{fig:svj_monojet_eff}
\end{figure*}

For a interpretation of the monojet DM search for the SVJ model, it is important to note that the efficiency of triggering on \ptmiss, which imposes an offline requirement of $\ptmiss>250\GeV$, is maximized for $\rinv = 0.5$, as shown in Fig.~\ref{fig:svj_monojet_eff}. At higher \rinv values, the majority of dark hadrons are stable and invisible, leading to increased cancellation of the invisible momenta of the two jets from the \PZpr boson decay, which correspondingly reduces the transverse component. However, the efficiencies of several other requirements are maximized for $\rinv = 1.0$: $\dphifull>0.5$ for the leading four jets with $\pt>30\GeV$, and $\nbjets=0$ considering all jets with $\pt>20\GeV$. As \rinv increases, fewer dark hadrons decay into visible particles, decreasing the number of possible reconstructed jets in each event; since visible and invisible dark hadrons are produced together in collimated sprays, any reconstructed jets may be aligned with \ptmiss. At $\rinv = 1.0$, the only visible particles in the signal events come from ISR. SVJs tend to be enriched in \PQb hadrons because of the higher mass scale of dark hadrons compared to SM quarks, which enables them to decay into \bbbar pairs. In the models considered here, $\mDark = 20\GeV$, leading to $\BR(\Prhodark \to \bbbar) = 0.2$ and $\BR(\Ppidark \to \bbbar) = 0.94$. The signal model specifies that \Prhodark are produced 75\% of the time, leading to an overall branching fraction of 0.385 for any unstable dark hadron to decay into \PQb quarks. The relative efficiencies for these requirements are also presented in Fig.~\ref{fig:svj_monojet_eff}.

We present results for the two interpretations in Section~\ref{par:section7_svj}.

\subsection{Fully visible and prompt signatures} \label{sec:signatures_vis}

In addition to searching for decays into invisible final states as described in Section~\ref{sec:signatures_invis}, the DS can also be probed by searching for decays of the mediator to SM particles and fully visible final states. For example, we can search for mediators that decay into pairs of leptons or jets. These searches provide results that are complementary to the invisible decays described above. We organize this section into searches for low-mass resonances (Section~\ref{sec:lowMass}), \ie, resonances below several hundreds of {\GeVns}; searches for high-mass resonances (Section~\ref{sec:highMass}), \ie, resonances above several hundred {\GeVns}; and searches with other prompt and visible signatures that do not easily fit into these two categories (Section~\ref{sec:otherSig}). 

\subsubsection{Low-mass resonance searches} \label{sec:lowMass}

Searches for low-mass dijet resonances~\cite{CMS:2019gwf} are strongly limited by the trigger bandwidth because of overwhelming background rates, as discussed in Section~\ref{sec:trigger}. The triggers, listed in Table~\ref{table:dijet-triggers}, result in a lower threshold of 1.8\TeV on the resonance masses probed by conventional dijet resonance searches. 

\begin{table}[htbp]
    \centering
    \topcaption{Trigger thresholds for various jet-based triggers in Run~2. All values are in {\GeVns}.}
    \begin{tabular}{lccc}
       Trigger  &  2016 & 2017 & 2018 \\
        \hline
       $\HT$  & 800, 900 & 1050 & 1050 \\
       AK4 PF jet \pt & 450 & 500 & 500 \\
       AK8 PF jet \pt & 450 & 500 & 500 \\
       AK8 PF jet \pt (\mtrim) & 360 (30) & 400 (30) & 400 (30) \\
       Single AK4 calo jet \pt & 500 & 500 & 500 \\
    \end{tabular}
    \label{table:dijet-triggers}
\end{table}

The CMS Collaboration has utilized a number of techniques to circumvent this limitation:
\begin{itemize}
    \item  Resonances with masses as small as 600\GeV can be probed with the data scouting technique~\cite{CMS:2018mgb}, wherein the trigger thresholds are lower by saving to disk only high-level physics objects, \ie, jets clustered from calorimeter towers or particle flow candidates, rather than the full detector readout. 
    \item Online $\PQb$~tagging has been used to allow jet energy thresholds to be reduced at the trigger level. At the HLT, \PQb-tagged jets with lower \pt thresholds between 70 and 160\GeV are employed, depending on \PQb-jet multiplicity and pseudorapidity. This allows sensitivity to resonance masses as small as 325\GeV~\cite{EXO-16-057}. 
    \item Resonances with masses as small as 10\GeV can be probed by requiring significant ISR, either in the form of jets~\cite{CMS-PAPERS-EXO-17-001, CMS-PAPERS-EXO-18-012, CMS-PAPERS-EXO-17-024} or photons~\cite{CMS-PAPERS-EXO-17-027}. In this topology, acceptable trigger rates are achieved by placing selection criteria on variables that are not strongly correlated with the resonance mass, e.g., the ISR object momenta. The dijet system itself is significantly boosted and hence is reconstructed as a single large-radius jet (AK8 or CA15) with a two-pronged substructure. These searches are described below.
\end{itemize}

Additionally, several DS models predict the existence of a resonance that can decay into pairs of SM leptons. Searches for low-mass resonances decaying into a pair of muons are described in Ref.~\cite{CMS:2019buh} and Sections~\ref{paragraph:EXO-19-018} and ~\ref{paragraph:EXO-21-005}. Related to these, a search where the dark photon is long lived is also presented in Section~\ref{paragraph:HIG-18-003}.

The sensitivities of the searches described in this section to a range of simplified DM models are shown in Section~\ref{sec:simpDSResults}.

\cmsParagraph{Search for low-mass vector resonances decaying into quark-antiquark pairs\label{par:EXO-18-012}} The most recent search for low-mass, boosted dijet resonances, which uses data from 2016 and 2017 corresponding to $\Lint=77\fbinv$, is described in Ref.~\cite{CMS-PAPERS-EXO-18-012}. The analysis searches for new, spin-1 \PZpr bosons decaying into quark-antiquark pairs recoiling against ISR, targeting a mass range of $50<\mZprime<450\GeV$. The \PZpr bosons are assumed to couple equally to all flavors of quarks, with a universal coupling constant $\gq$. The trigger selects AK8 jets with $\pt>380\,(400)\GeV$ in 2016 (2017) and a trimmed mass (described in Section~\ref{sec:jetsAndSubstructure}) greater than 30\GeV; the trigger has good efficiency for \PZpr boson masses greater than 50\GeV, which sets the lower bound on the search range. The analysis uses offline AK8 and CA15 jets, depending on the signal mass considered. The \pt requirements for offline AK8 jets are $\pt>500\,(525)\GeV$ in 2016 (2017) data and $\pt>575\GeV$ for CA15 jets. Jet substructure techniques are used to distinguish the signal from the backgrounds, which include QCD multijets, $\PQt\PAQt$, and $\PW/\PZ$+jets. The signal resonance is identified using the soft-drop mass variable \msd~\cite{softdrop}, which removes soft and wide-angle radiation from the jet. The soft drop algorithm reduces the mass of jets from QCD, where the mass arises in part from soft gluon radiation, while preserving the mass of two-pronged signal jets. The variable $N_2^1$, defined using ratios of ECFs~\cite{Moult:2016cvt}, is used to reject QCD events; two-pronged jets tend to have a lower value of $N_2^1$ than QCD jets.

\begin{figure}
\centering
\includegraphics[width=0.55\textwidth]{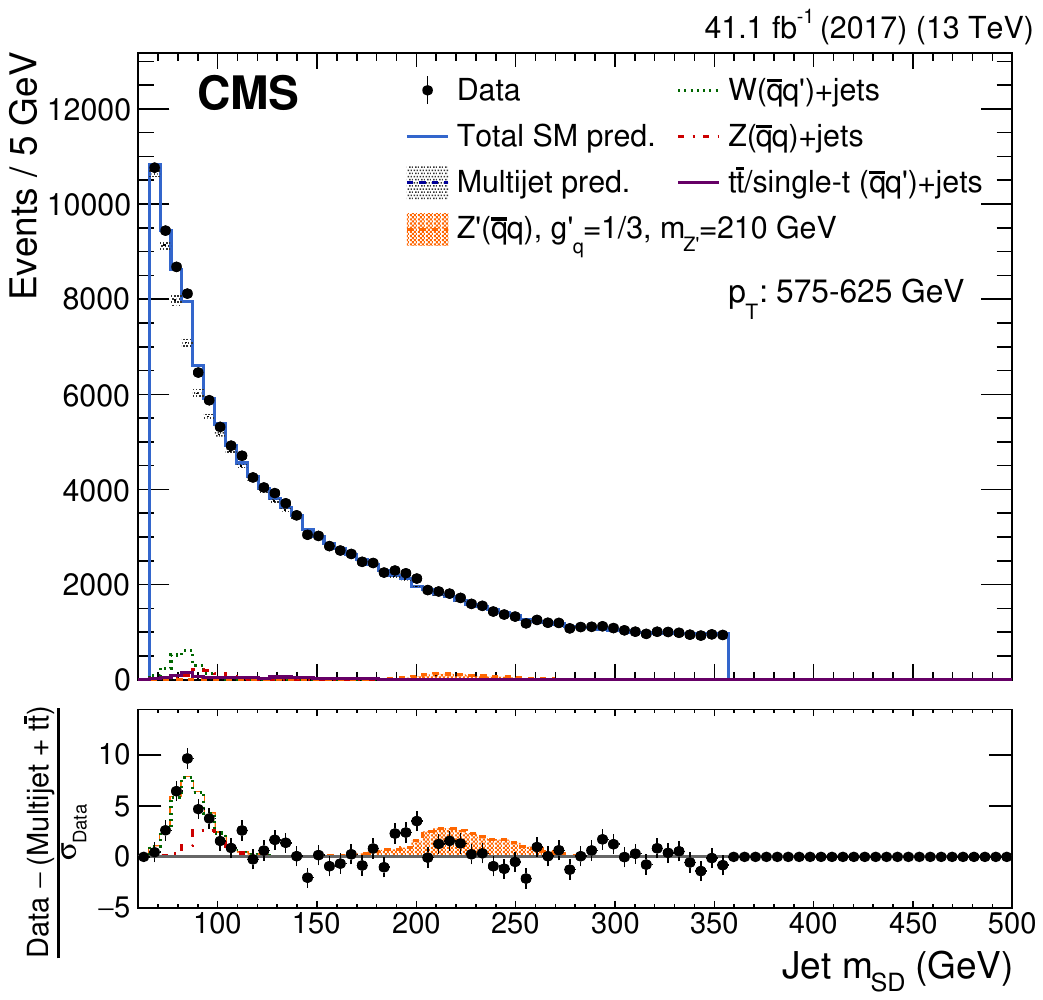}
\caption{Jet \msd distribution in data for CA15 jets for a $\pt$ range of the fit from 575 to 625\GeV, in the search for low-mass vector resonances decaying into quark-antiquark pairs. Data are shown as black points. The QCD multijet background prediction, including uncertainties, is shown by the shaded bands. Smaller contributions from the \PW and \PZ bosons, and top quark background processes are shown as well. A hypothetical \PZpr boson signal with a mass of 210\GeV is also indicated. In the bottom panel, the ratio of the data to its statistical uncertainty, after subtracting the nonresonant backgrounds, is shown. Figure taken from Ref.~\cite{CMS-PAPERS-EXO-18-012}.}
\label{fig:EXO-18-012}
\end{figure}

The QCD multijet background is estimated from data, using a CR consisting of events failing a requirement on $N_2^1$. In simulation, the ``designed decorrelated tagger'' method ensures that the \msd shape in the CR is the same as the one in the SR by construction. The \msd distribution is shown for a representative category in Fig.~\ref{fig:EXO-18-012}.

\cmsParagraph{Search for low-mass resonances decaying into bottom quark-antiquark pairs} An analysis searching for new spin-0 resonances decaying into bottom quark-antiquark pairs, with resonance masses between 50 and 350\GeV is described in Ref.~\cite{CMS-PAPERS-EXO-17-024}.

\begin{figure}
\centering
\includegraphics[width=0.55\textwidth]{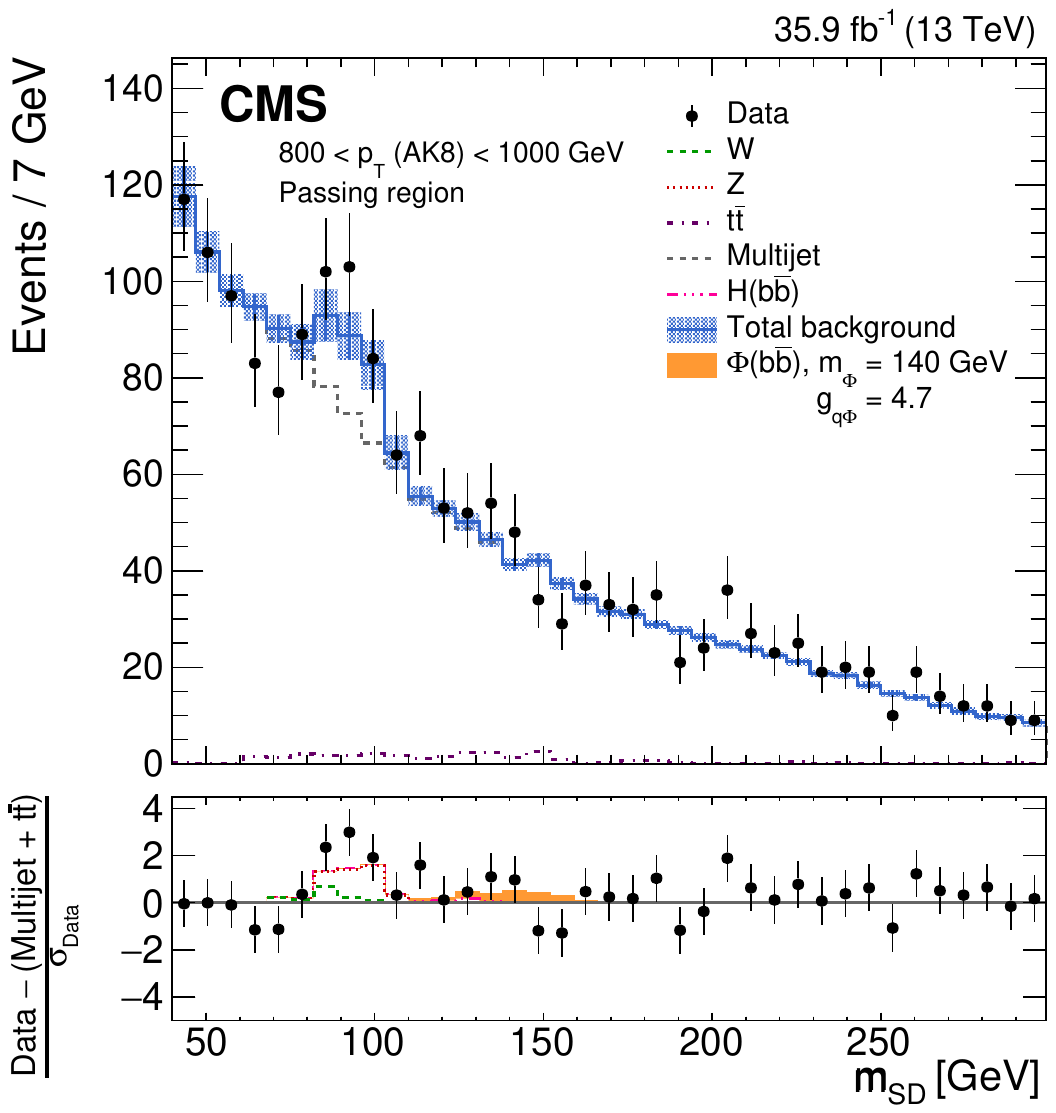}
\caption{The observed and fitted background \msd distributions in the $800 <\pt< 1000\GeV$ category for the AK8 selection in the passing regions, in the search for low-mass resonances decaying into bottom quark-antiquark pairs. The fit is performed under the background-only hypothesis. A hypothetical signal at a mass of 140\GeV is also indicated. The shaded blue band shows the systematic uncertainty in the total background prediction. The bottom panel shows the difference between the data and the nonresonant background prediction, divided by the statistical uncertainty in the data. Figure taken from Ref.~\cite{CMS-PAPERS-EXO-17-024}.}
\label{fig:EXO-17-024}
\end{figure}

The analysis follows the general strategy of Ref.~\cite{CMS-PAPERS-EXO-18-012}, a search for low-mass, boosted dijet resonances, and adapts it for new scalar resonances decaying into $\PQb\PAQb$ pairs, using a data sample corresponding to $\Lint=36\fbinv$, taken during  2016. Resonances are produced with high \pt because of significant ISR, ensuring events pass stringent trigger restrictions set by bandwidth limitations. In such events, the decay products of the resonance are reconstructed as a single large-radius jet with jet substructure consistent with originating from two $\PQb$ quarks. Both AK8 and CA15 jets are considered as candidates, with \pt thresholds of 450 and 500\GeV, respectively. The AK8 algorithm provides better sensitivity at signal masses less than 175\GeV, while the CA15 algorithm provides better sensitivity at higher masses. Jet substructure techniques and dedicated $\PQb$-tagging algorithms are used to distinguish the signal from the QCD background. The signal is identified as a narrow resonance in the \msd spectrum. The main algorithm for distinguishing signal jets from the QCD background, called the ``double-\PQb tagger,'' is a multivariate algorithm based on boosted decision trees, and uses kinematic information from tracks and DVs relative to two leading subjet axes. The $N^1_2$~\cite{Larkoski:2013eya,Moult:2016cvt} variable is also used to further distinguish the two-pronged signal jets from QCD jets. The \msd distribution  is shown for a representative category in Fig.~\ref{fig:EXO-17-024}.

\cmsParagraph{Search for low-mass quark-antiquark resonances produced in association with a photon\label{par:EXO-17-027}} Another strategy to extend dijet searches to small \PZpr boson masses is to focus on events in which the resonance is produced in association with a high-momentum ISR photon. The analyses described previously, Refs.~\cite{CMS-PAPERS-EXO-18-012} and \cite{CMS-PAPERS-EXO-17-024}, probe resonance masses down to about 50\GeV; this bound arises from the offline lower \pt jet threshold of 500\GeV, which causes the lowest-mass resonances to be extremely collimated, as well as directly from HLT selections on the jet mass. Lower masses can be probed by triggering on photons. Specifically, in 2016, the CMS trigger system recorded events containing photons with $\pt>175\GeV$. A search for dijet resonances with masses from 10 to 125\GeV and produced in association with an ISR photon is described in Ref.~\cite{CMS-PAPERS-EXO-17-027}, using data collected in 2016 corresponding to $\Lint=36\fbinv$.

\begin{figure}
\centering
\includegraphics[width=0.55\textwidth]{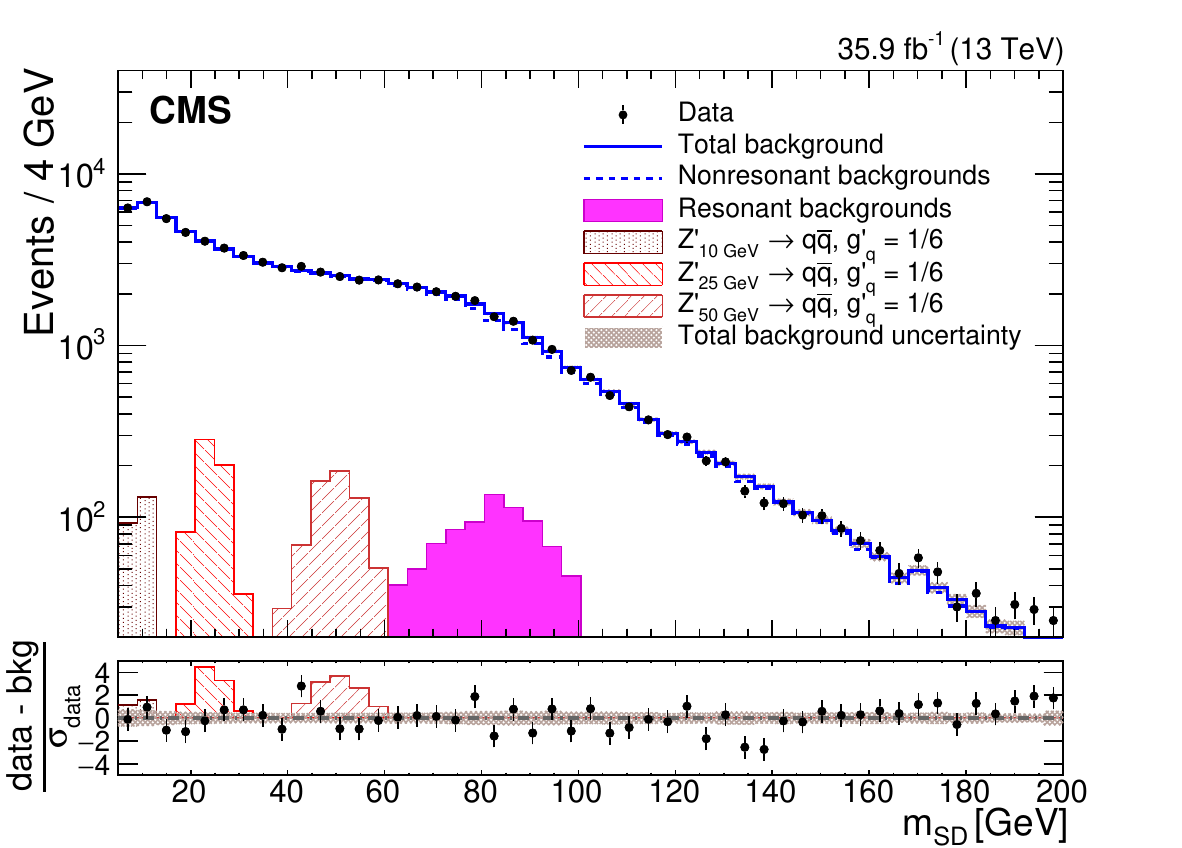}
\caption{The soft drop jet mass distribution of the SR in the search for low-mass quark-antiquark resonances produced in association with a photon, after the main background estimation fit is performed. The nonresonant background is indicated by a dashed line, while the total background composed of the sum of this nonresonant background and the resonant backgrounds is shown by the solid line. Representative signals are plotted for comparison. The bottom panel shows the difference between the data and the final background estimate, divided by the statistical uncertainty of the data in each bin. The shaded region represents the total uncertainty in the background estimate in each bin. Figure taken from Ref.~\cite{CMS-PAPERS-EXO-17-027}.}
\label{fig:EXO-17-027}
\end{figure}

The offline analysis of this dijet resonance search uses events containing a photon with $\pt>200\GeV$. Events with additional photons with $\pt>14\GeV$ or leptons with $\pt>10\GeV$ are discarded to avoid overlap with other searches and to reduce backgrounds from EW sources. The analysis strategy is otherwise similar to Ref.~\cite{CMS-PAPERS-EXO-18-012}, described above. The \PZpr boson is reconstructed as a single AK8 jet and produces a local excess in the \msd spectrum. The main background, coming from photons produced in association with jets from SM processes, is determined using a variation of the ABCD method with additional correction factors to account for the statistical dependencies of the variables. The \msd distribution for the SR is shown in Fig.~\ref{fig:EXO-17-027}.

\cmsParagraph{Search for a prompt dark photon resonance decaying into two muons including data scouting\label{paragraph:EXO-19-018}} Reference~\cite{CMS:2019buh} presents a search for a narrow resonance, in the 11.5 to 200\GeV mass range, decaying into a pair of oppositely charged muons. For masses less than ${\approx} 40\GeV$, a dedicated scouting trigger (as discussed in Section~\ref{subsec:scouting}) with an exceptionally low muon \pt threshold was used. For higher masses, standard triggers were used. The data correspond to $\Lint=97$ and 137\fbinv for the scouting and conventional triggering strategies, respectively. The dimuon mass resolution depends strongly on the pseudorapidity of the muons. Therefore, events are divided into two categories. The barrel category consists of events in which both muons are in the barrel region, and the forward category contains events in which at least one of the two muons is not in the barrel region. 

\begin{figure}
\centering
\includegraphics[width=0.55\textwidth]{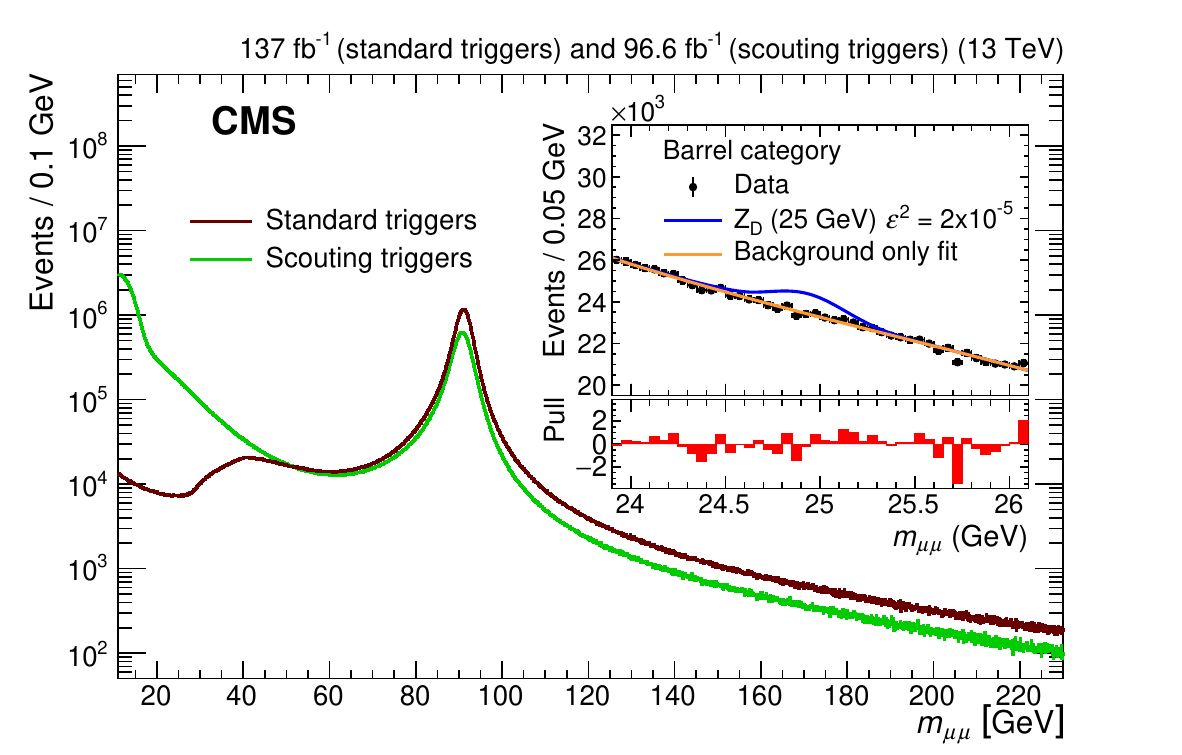}
\caption{The dimuon invariant mass distributions of events selected with the standard muon triggers (brown, darker), and the scouting dimuon triggers (green, lighter), in the search for a prompt dark photon resonance decaying into two muons. Events are required to pass all the selection requirements. The inset shows the data (black points), the signal model (blue line), and the background-only fit (orange line), and it is restricted to events in the barrel category in the mass range 23.9--26.1\GeV. A function describing the background is fit to these data. The bottom panel of the inset shows the bin-by-bin difference between the number of events in data and the prediction from the background fit, divided by the statistical uncertainty. Figure taken from Ref.~\cite{CMS:2019buh}.}
\label{fig:EXO-19-018}
\end{figure}

In the high-mass search performed with the standard triggers, events are required to have at least one well-reconstructed PV and two oppositely charged muons. The muons are required to be isolated and to pass selection requirements based on the quality of their reconstructed tracks. In the search performed using the scouting triggers, events are required to contain two muons of opposite charge that are consistent with originating from the same vertex, with similar requirements on muon isolation and track quality as in the search using standard triggers. The dimuon invariant mass distribution is shown for a representative category in Fig.~\ref{fig:EXO-19-018}.

\cmsParagraph{Search for prompt dimuon resonances with data scouting\label{paragraph:EXO-21-005}} An analysis~\cite{CMS-PAS-EXO-21-005} similar to the one described in Section~\ref{paragraph:EXO-19-018} is performed to search for dimuon resonances with masses below the $\PGUP{1S}$ resonance in the range of 1.1--2.6\GeV and 4.2--7.9\GeV using data collected by the dimuon scouting trigger during 2017--2018 with $\Lint=97\fbinv$. 

\begin{figure}
\centering
\includegraphics[width=0.55\textwidth]{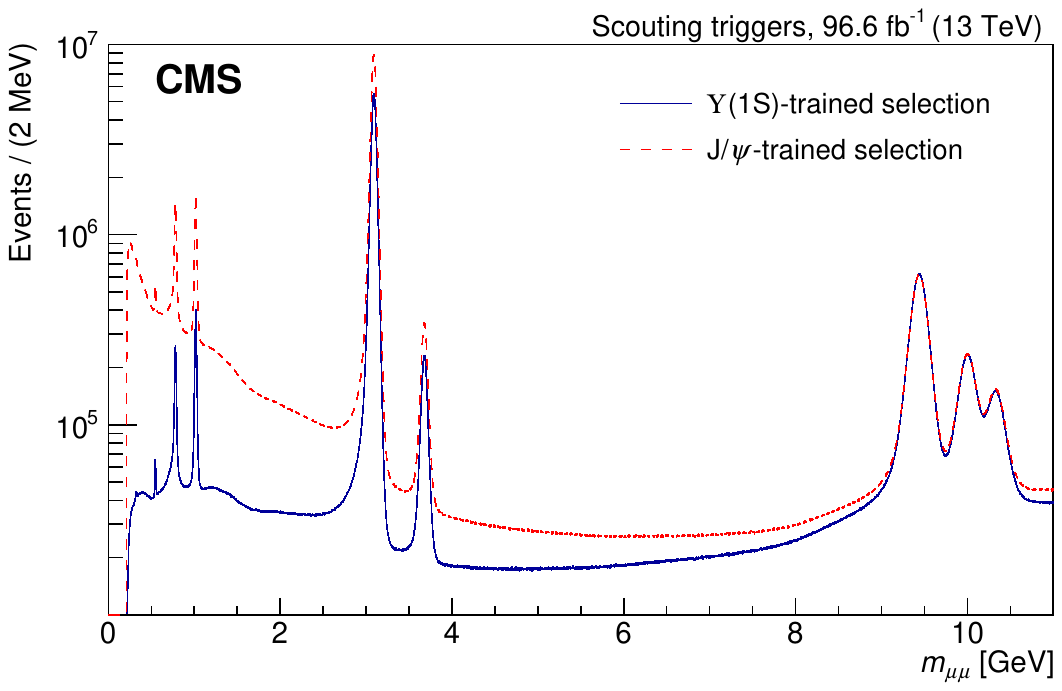}
\caption{The dimuon invariant mass distribution obtained with the muon scouting data collected during 2017--2018 with two sets of selections: the $\PGUP{1S}$-trained muon MVA identification (blue solid line), and the $\PJGy$-trained muon MVA identification (red dashed line). Figure taken from Ref.~\cite{CMS-PAS-EXO-21-005}.}
\label{fig:EXO-21-005}
\end{figure}

The event candidate is required to have at least one PV reconstructed by the HLT system and to contain a pair of oppositely charged muons originating from this vertex. To identify good-quality muon candidates, two multi-variate analysis (MVA) discriminants are used depending on the reconstructed dimuon mass, optimized for the signal kinematic properties in each mass range. The MVA identification utilizes information on the quality of the muon tracks, the relative isolation of the muon, and the vertex associated with the muons. Different vertex displacement criteria with respect to the beam spot are imposed in different mass ranges to account for the increased uncertainty in the PV position from the larger boost of the dimuon system and hence the more collinear tracks for smaller dimuon masses. The dimuon invariant mass distribution with both selections is shown in Fig.~\ref{fig:EXO-21-005}.

\subsubsection{High-mass resonance searches} \label{sec:highMass}

While resonances decaying into leptons have been excluded over a wide mass range and down to small couplings, resonances decaying into quarks are more challenging to detect because of the multijet background at hadronic colliders. Searches for resonances decaying into a quark pair have been performed mainly at high masses (e.g., $m > 1000\GeV$) in the dijet final state, while the low-mass range (e.g., $m < 200\GeV$) has been covered by the search for boosted resonances reconstructed as a single large-radius jet.

Three resonance searches are described in this section. We discuss a search for dijet resonances in events with three jets (which targets more mid-range masses), a search for high-mass dijet resonances, and a search for high-mass dilepton resonances.

The sensitivities of the searches in this section to a range of simplified DM models are shown in Section~\ref{sec:simpDSResults}.        

\cmsParagraph{Search for dijet resonances using events with three jets\label{par:EXO-19-004}} The search presented in Ref.~\cite{EXO-19-004} combines the data scouting technique with the requirement of an additional jet with high \pt to enhance signal sensitivity in the low-mass region. The analysis is performed on part of the data collected in 2016 (corresponding to $\Lint=18.3\fbinv$) when the trigger threshold was particularly low ($\HT > 240\GeV$) in an attempt to extend the search to the lowest mass possible.

\begin{figure}
\centering
\includegraphics[width=0.5\textwidth]{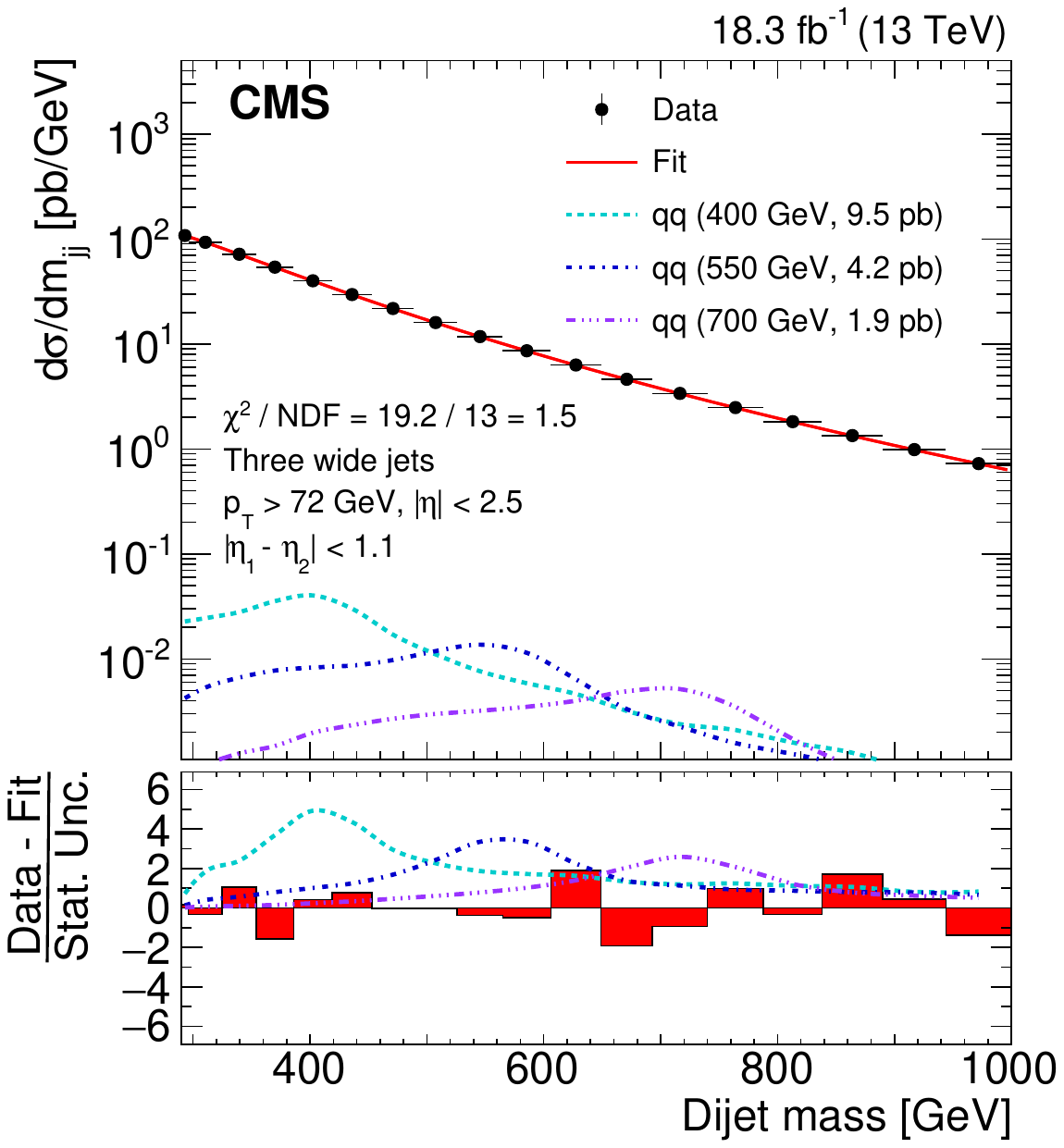}
\caption{Dijet mass spectrum (points) compared to a fitted parameterization of the background (solid curve) in the search for dijet resonances using events with three jets, where the fit is performed in the range $290 <\mjj< 1000\GeV$. The horizontal bars show the widths of each bin in dijet mass. The dashed lines represent the dijet mass distribution from 400, 550, and 700\GeV resonance signals expected to be excluded at 95\% \CL by this analysis. The lower panel shows the difference between the data and the fitted parametrization, divided by the statistical uncertainty of the data. Figure taken from Ref.~\cite{EXO-19-004}.}
\label{fig:EXO-19-004}
\end{figure}

This analysis uses large-radius jets to recover the energy from final-state radiation, improving the dijet mass resolution. A selection on the $\eta$ separation is used to suppress and reduce the QCD multijet background, which is dominated by \PQt-channel production of jets. A bump search is then performed on the dijet mass spectrum, which is shown in Fig.~\ref{fig:EXO-19-004}. 

\begin{figure*}[htb!]
\centering
\includegraphics[width=0.5\linewidth]{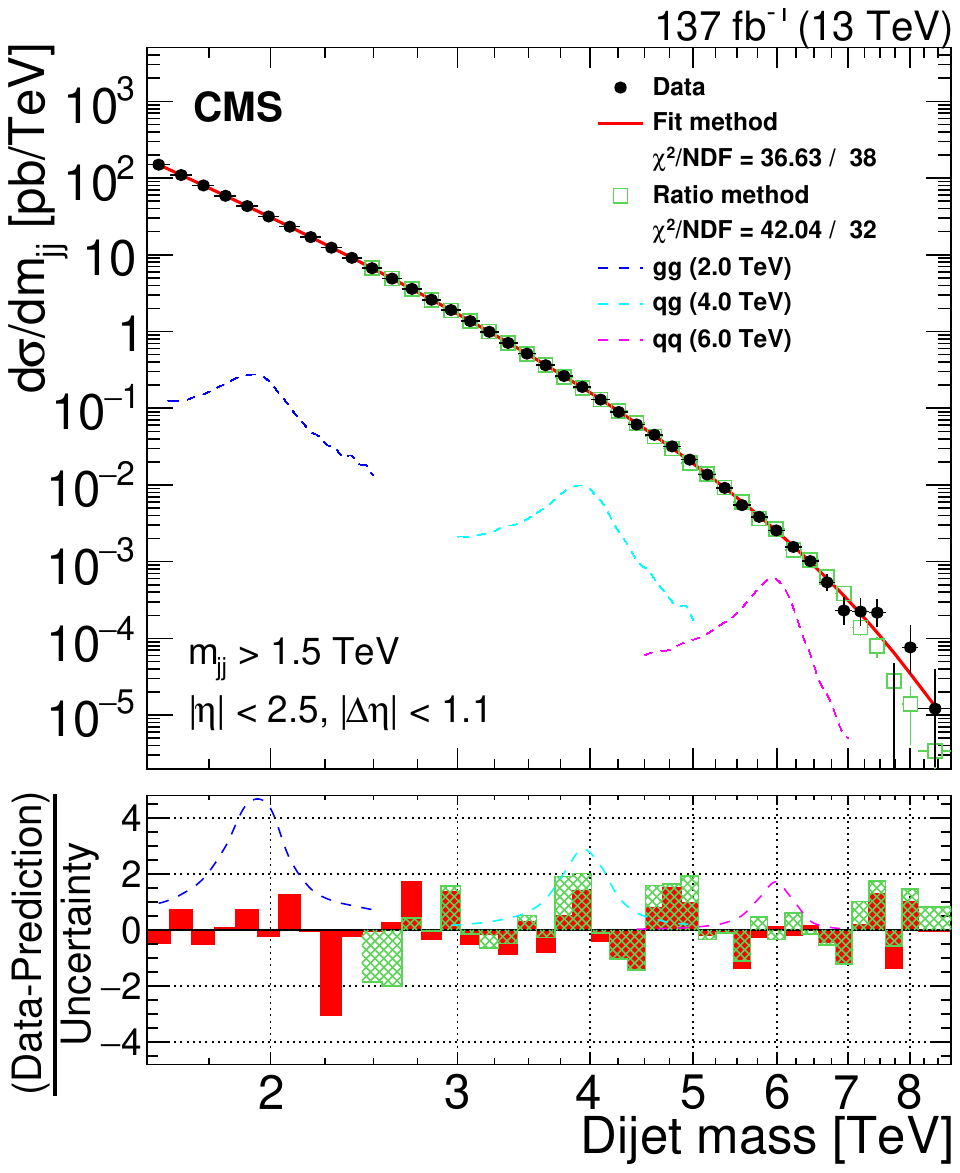}
\caption{Dijet mass spectrum in the SR (points) compared to a fitted parameterization of the background (solid line) and the one obtained from the CR (green squares), in the search for high-mass dijet resonances. The lower panel shows the difference between the data and the fitted parametrization (red, solid), and the data and the prediction obtained from the CR (green, hatched), divided by the statistical uncertainty in the data, which for the ratio method includes the statistical uncertainty in the data in the CR. Examples of predicted signals from narrow gluon-gluon, quark-gluon, and quark-quark resonances are shown (dashed colored lines) with cross sections equal to the observed upper limits at 95\% \CL. Figure taken from Ref.~\cite{CMS:2019gwf}.
\label{fig:EXO-19-012_dijetMass}}
\end{figure*}

\cmsParagraph{Search for high-mass dijet resonances\label{sec:EXO-19-012}} Models in which DM mediators arise from an interaction between quarks and DM produce dijet resonance signatures. The natural width of such mediators increases with the coupling and may vary from narrow to broad, as defined in comparison to the experimental resolution. In Ref.~\cite{CMS:2019gwf}, we describe a largely model-independent search for narrow or broad $s$-channel dijet resonances with masses greater than 1.8\TeV, shown in Fig.~\ref{fig:EXO-19-012_dijetMass}. We use data corresponding to $\Lint=137\fbinv$ collected in Run~2.

Each of the two leading jets is formed into a large-radius jet using an algorithm introduced for previous CMS dijet searches in Ref.~\cite{CMS:2015rkn}. The SR is defined by vetoing events with a large $\eta$ separation between the jets, which maximizes the search sensitivity for isotropic decays of dijet resonances in the presence of QCD dijet background. 

The main background from QCD multijet production is predicted by fitting the $m_{jj}$ distribution with an empirical functional form. For $m_{jj} > 2.4\TeV$, a new background estimation method is introduced, which predicts the background from a CR where the pseudorapidity separation of the two jets, $\abs{\Delta\eta}$, is large. This new background estimation method yields smaller systematic uncertainties.

This strategy, employing the empirical background fit, has also been applied to data collected using scouting in Ref.~\cite{CMS:2018mgb}. The search uses 27\fbinv of data collected during Run 2 and is sensitive to resonance masses from 0.6 to 1.6\TeV.

Mediators with intrinsic widths larger than 50\% have also been probed in CMS dijet events in a dedicated analysis of the dijet angular distributions~\cite{CMS:2018ucw} using a data set corresponding to $\Lint=36\fbinv$ at $\sqrt{s}=13\TeV$. While constraints on $\gq$ from the dijet angular analysis are not competitive with the dijet resonance search, the dijet angular analysis allows to extend the excluded range of widths from 50 to 100\% for mediator masses ${<}4.6\TeV$.

\cmsParagraph{Search for new physics in high-mass dilepton final state\label{sec:EXO-19-019}} The decay of mediator particles into SM particles could be observed through dilepton final states. A search for BSM physics using electron or muon pairs with high invariant mass~\cite{CMS:EXO19019} is sensitive to such mediator particles. Standard reconstruction techniques are used for high-\pt electrons and muons in this search; however, dedicated identification selection criteria are employed to ensure that high efficiency is maintained for both electrons~\cite{CMS:2015xaf} and muons~\cite{CMS:2014lcz}. The $\Pp\Pp$ collision data at $\sqrt{s}=13\TeV$ collected in 2016--2018 are used in the search, corresponding to $\Lint$ up to 140\fbinv. 

\begin{figure}
\centering
\includegraphics[width=0.95\textwidth]{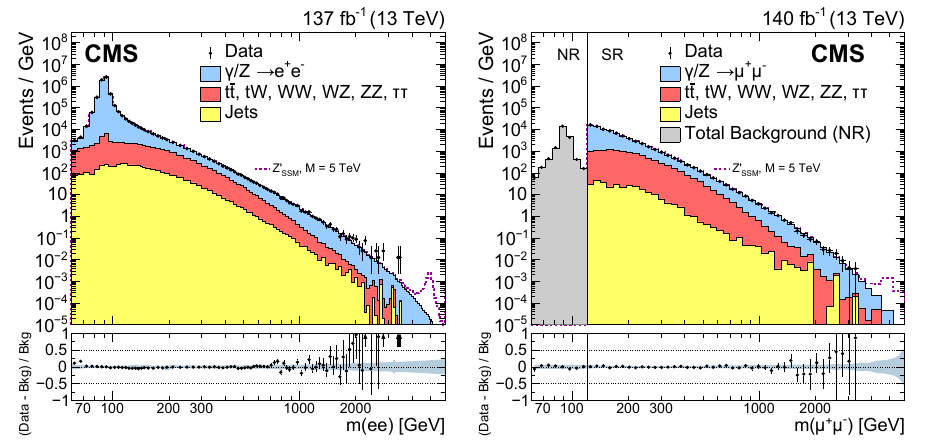}
\caption{The invariant mass distribution of pairs of (\cmsLeft) electrons and (\cmsRight) muons observed in data (black dots with statistical uncertainties) and expected from the SM processes (stacked histograms), in the high-mass dilepton search. For the dimuon channel, a prescaled trigger with a $\pt$ threshold of 27\GeV was used to collect events in the normalization region (NR) with dimuon mass less than 120\GeV. The corresponding offline threshold is 30\GeV. Events in the SR corresponding to masses greater than 120\GeV are collected using an unprescaled single-muon trigger. The bin width gradually increases with mass. The ratios of the data yields after background subtraction to the expected background yields are shown in the lower plots. The blue shaded band represents the combined statistical and systematic uncertainties in the background. Signal contributions expected from simulated resonances are shown. Figures adapted from Ref.~\cite{CMS:EXO19019}.}
\label{fig:EXO-19-019}
\end{figure}

The SM background processes are modeled with simulation (except for leptons produced inside jets or jets misidentified as leptons, which are estimated from CRs in data) and are normalized to the observed data yields in a mass window of 60--120\GeV around the \PZ boson peak, separately for the dielectron and dimuon channels. The search for resonant signatures is performed in a mass window around the assumed resonance mass, whose size depends on the assumed intrinsic decay width of the resonance and the mass-dependent detector resolution. A range of masses and widths is scanned to provide results covering a wide selection of signal models. Unbinned maximum likelihood fits are performed inside the mass windows, allowing the background normalization to be determined from the data. Through setting upper limits on the ratio of the product of the production cross section and the branching fraction of a new narrow dilepton resonance to that of the SM \PZ boson, many experimental and theoretical uncertainties common to both measurements cancel out or are reduced, leaving only uncertainties in the ratio that vary with the dilepton mass to be considered. The dielectron and dimuon invariant mass distributions are shown in Fig.~\ref{fig:EXO-19-019}.

\subsubsection{Other signatures} \label{sec:otherSig}

In this Section, we describe searches for visible and prompt signatures that do not fall into the low- and high-mass resonance categories described in Sections~\ref{sec:lowMass} and \ref{sec:highMass}, respectively. The searches described here include a search for fractionally charged particles, a search for SUEPs, a search for stealth top squarks, a search for ALPs in ultraperipheral PbPb collisions, and a search using the missing-mass technique in CMS and CMS-TOTEM events.

\cmsParagraph{Search for fractionally charged particles (FCPs)} In the search for LLPs carrying a fraction of the electron charge, \ie, $\text{Q}_{\text{FCP}} = \varepsilon e$, where $\varepsilon$ is lower than 1, described in Ref.~\cite{CMS-PAS-EXO-19-006}, we consider a signal generated via DY production using a data set corresponding to $\Lint=138\fbinv$ at $\sqrt{s}=13\TeV$. The experimental signature of an FCP is close to that of a muon, but with a larger mass and a lower charge. Therefore, we require events to contain exactly one or two high-\pt isolated muons.

\begin{figure*}[htb!]
\centering
\includegraphics[width=0.5\linewidth]{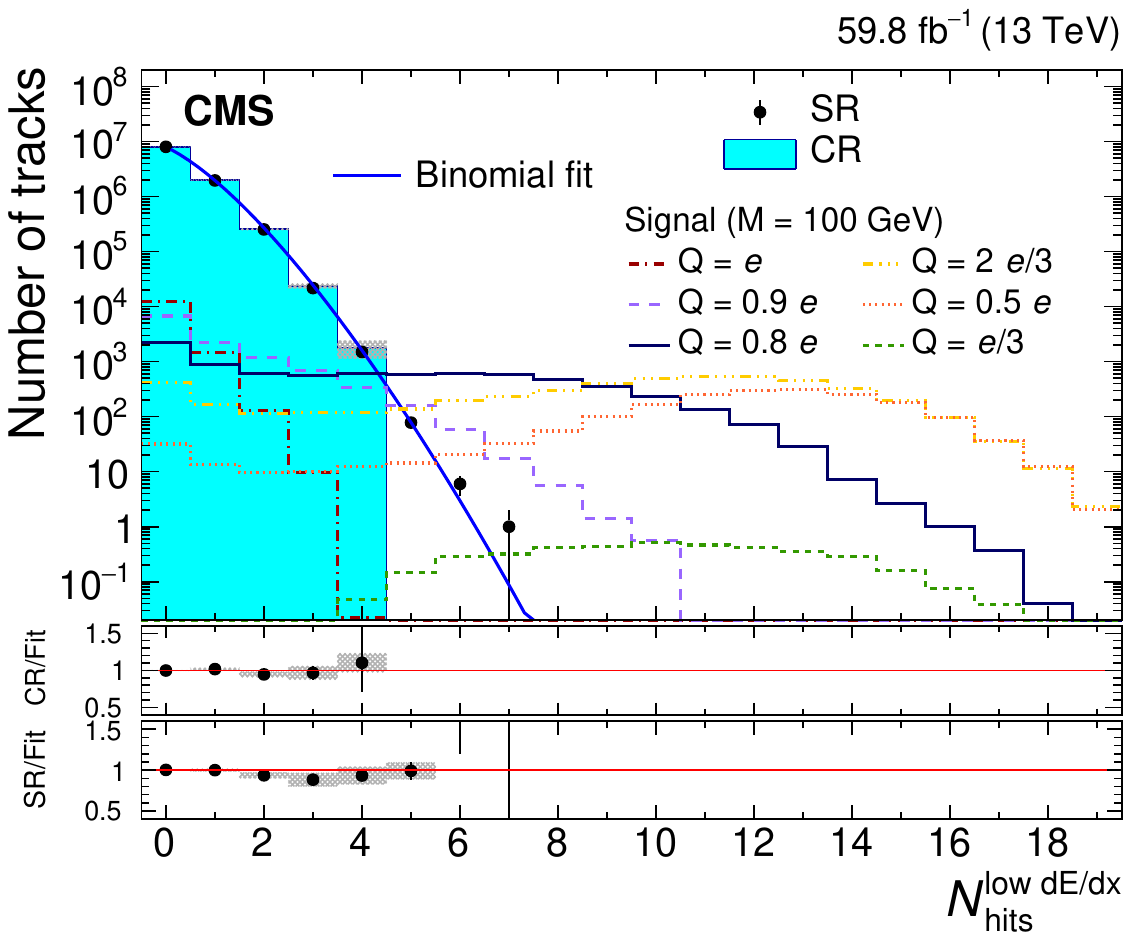}
\caption{Distribution of \nLow in the search and CRs for the early 2018 data set, in the search for fractionally charged particles. The middle (lower) panels show the ratio of the number of tracks observed in the CR (SR) and the fit function. Figure taken from Ref.~\cite{CMS-PAS-EXO-19-006}.\label{fig:fcp}}
\end{figure*}

The analysis strategy relies on the measurement of the ionization loss per unit length (\dEdx) associated with the hits in the modules of the CMS silicon tracker (described in Section~\ref{dEdx}). The energy loss process in silicon is stochastic; the most probable hit \dEdx value for a muon is around 3\MeV/cm. A low-charge particle is expected to deposit lower amounts of energy, systematically across all hits. The scaling goes with the square of the FCP charge, as described by the Bethe-Bloch function. To discriminate signal from background, we build a binomial distribution by asking the following question for each hit on a track: is the \dEdx less than a threshold value? The threshold is adapted layer-by-layer to take into account experimental effects such as radiation damage to a module. The variable \nLow is the total number of hits on a track that pass the requirement, shown in Fig.~\ref{fig:fcp} for 2018 data. It accumulates at small values for charge $e$ particles such as muons and extends to larger values as the charge decreases. We define a CR with the events containing exactly two candidate tracks with an invariant mass between 80 and 100\GeV, \ie, around the \PZ boson mass, and a SR that contains all other events. We fit the \nLow distribution in the CR to estimate our background and compare it to the observation in the SR.

\begin{figure}
\centering
\includegraphics[width=.95\textwidth]{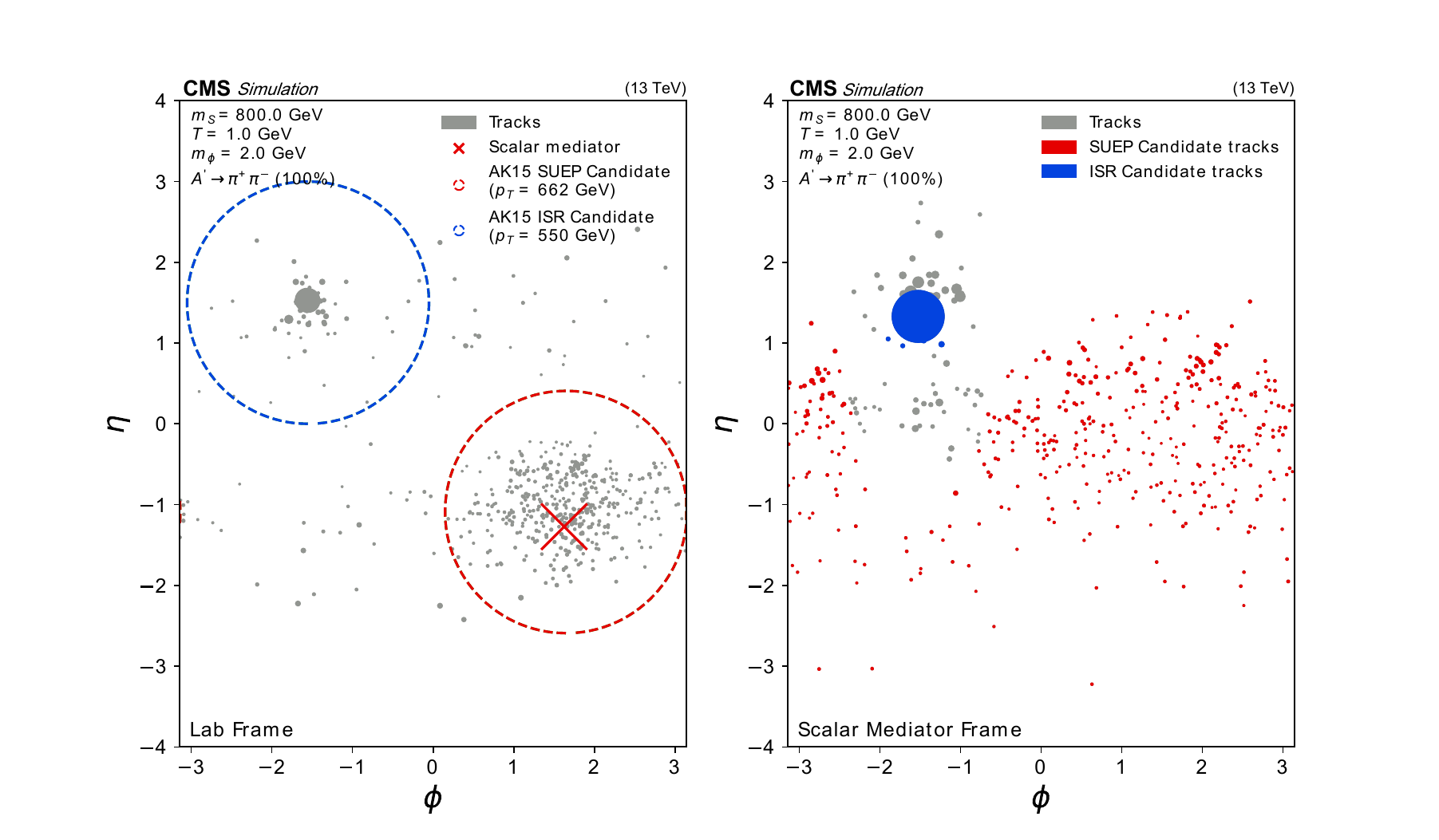}
\caption{An example SUEP event from a representative model with a scalar mediator of mass 800\GeV shown in the lab frame (\cmsLeft) and the generator-level S mediator frame (\cmsRight). The jets are clustered from charged particle tracks associated with the primary vertex using the anti-\kt algorithm with a distance parameter of 1.5. The size of each dot is scaled based on the \pt of the corresponding track.}
\label{fig:EXO-23-002}
\end{figure}

\cmsParagraph{Search for soft unclustered energy patterns\label{sec:SUEPsearch}} A search for SUEPs arising from the decay of a heavy scalar mediator is reported in Ref.~\cite{CMS-PAS-EXO-23-002}. Motivated by HV models with a dark-QCD sector and large 't Hooft coupling, the signature of a SUEP is a high multiplicity of spherically distributed low-momentum charged particles in the final state. The data, which correspond to $\Lint=138\fbinv$, were collected in 2016--2018 using traditional hadronic triggers, which often select events with  high-\pt ISR jets. As a result, boosted topologies are favored in this analysis. The charged particle tracks in the event are clustered into large-radius jets and of the two leading jets, the jet with the larger number of constituent tracks is chosen to be the SUEP candidate. An example signal event is shown in Fig.~\ref{fig:EXO-23-002}.

The primary background in this search comes from QCD multijet events with a large number of tracks. This search utilizes a novel approach to predict the background by estimating the contribution from traditional processes directly from data using the ``extended'' version of the ABCD method described in Section~\ref{sec:abcd}~\cite{Choi:2019mip}. 

The sensitivity of the search is shown in Section~\ref{par:suepSens}.

\cmsParagraph{Search for stealth top squarks\label{sec:signatures_stealth}} As detailed in Section~\ref{sec:theory_stealth}, models of stealth SUSY result in final states where the typical \ptmiss of SUSY searches is replaced with additional visible objects, which are jets in the models considered here. The search described in Ref.~\cite{SUS-19-004} targets pair production of top squarks with decays via the stealth sector through the vector portal (labeled `SYY'), resulting in a final state with two top quarks and six gluons.

\begin{figure}
\centering
\includegraphics[width=.5\textwidth]{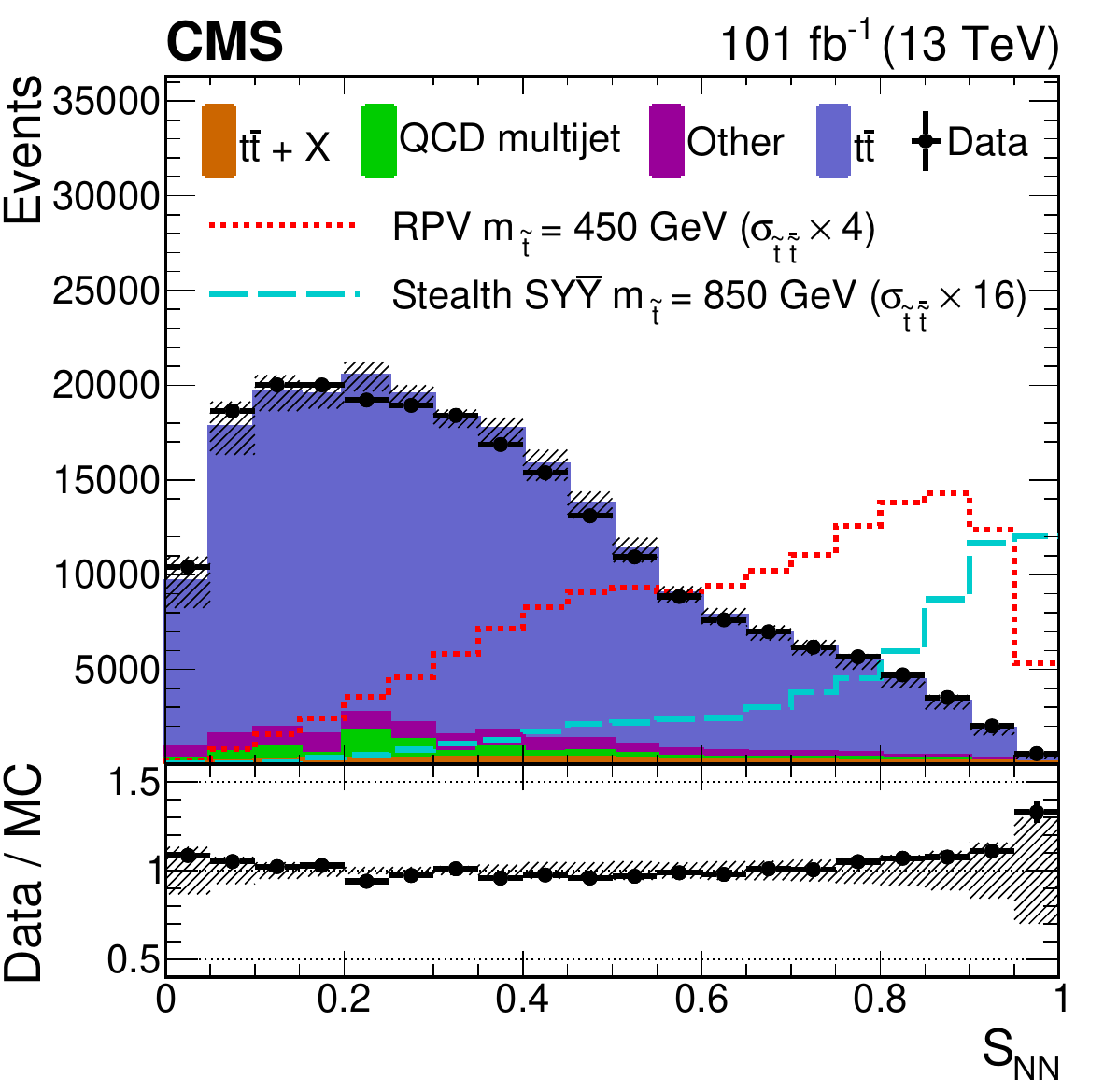}
\caption{The neural network score ($S_{\mathrm{NN}}$) distribution for 2017--2018 shows the data in the SR (black points); simulated background normalized to the number of data events (filled histograms); RPV SUSY signal model with a top squark mass of 450\GeV (red short dashed line); and stealth SYY signal model with a top squark mass of 850\GeV (cyan long dashed line), in the search for stealth top squarks. The band on the total background histogram denotes the dominant systematic uncertainties, as well as the statistical uncertainty for the non-$\ttbar$ components. The lower panel shows the ratio of the number of data events to the number of normalized simulated events with the band representing the difference between the nominal ratio and the ratio obtained when varying the total background by its uncertainty. Figure taken from Ref.~\cite{SUS-19-004}.}
\label{fig:SUS-19-004}
\end{figure}

The search selects events with exactly one electron or muon, at least seven jets, and at least one \PQb-tagged jet, using data corresponding to $\Lint=137\fbinv$, collected in 2016--2018. No requirement is placed on \ptmiss. The signal is distinguished from the background by means of a neural network that uses the jet kinematics as well as overall event shape variables as input features. Crucially, the network is trained to be independent of the jet multiplicity by using the gradient reversal technique~\cite{ganin2014unsupervised}, which is a technique that is used to effectively penalize the network for being able to determine the jet multiplicity for a given event. This enabled the background estimation to be done via a simultaneous fit to the jet multiplicity distribution in four bins of the neural network score. The jet multiplicity is modeled with a recursive fit function based on QCD jet scaling patterns. The distribution of the neural network score for 2017--2018 data and simulation is shown in Fig.~\ref{fig:SUS-19-004}.
The sensitivity of the search is shown in Section~\ref{sec:sensStealth}.

\cmsParagraph{Search for axion-like particles in ultraperipheral PbPb collisions\label{sec:FSQ-16-012}} A search for ALPs (Section~\ref{subsec:ALPs}) that couple to photons has been conducted using PbPb UPCs~\cite{FSQ-16-012}. UPCs are defined as collisions in which the impact parameter is larger than twice the nucleus radius, where passing heavy ions do not break up and are so close that their electromagnetic fields are intense enough to interact as quasi-real photon beams. The PbPb collisions provide an enhancement of a factor given by the atomic number to the power of four for photon-photon scattering processes as compared to $\Pp\Pp$ collisions, since the photon flux scales as the atomic number squared of the emitting ion. The production of a resonant ALP ($\PGg\PGg \to \Pa \to \PGg\PGg$) is expected to modify the rate of the light-by-light scattering process ($\PGg\PGg\to\PGg\PGg$) that shares the same final state. 

\begin{figure}
\centering
\includegraphics[width=.5\textwidth]{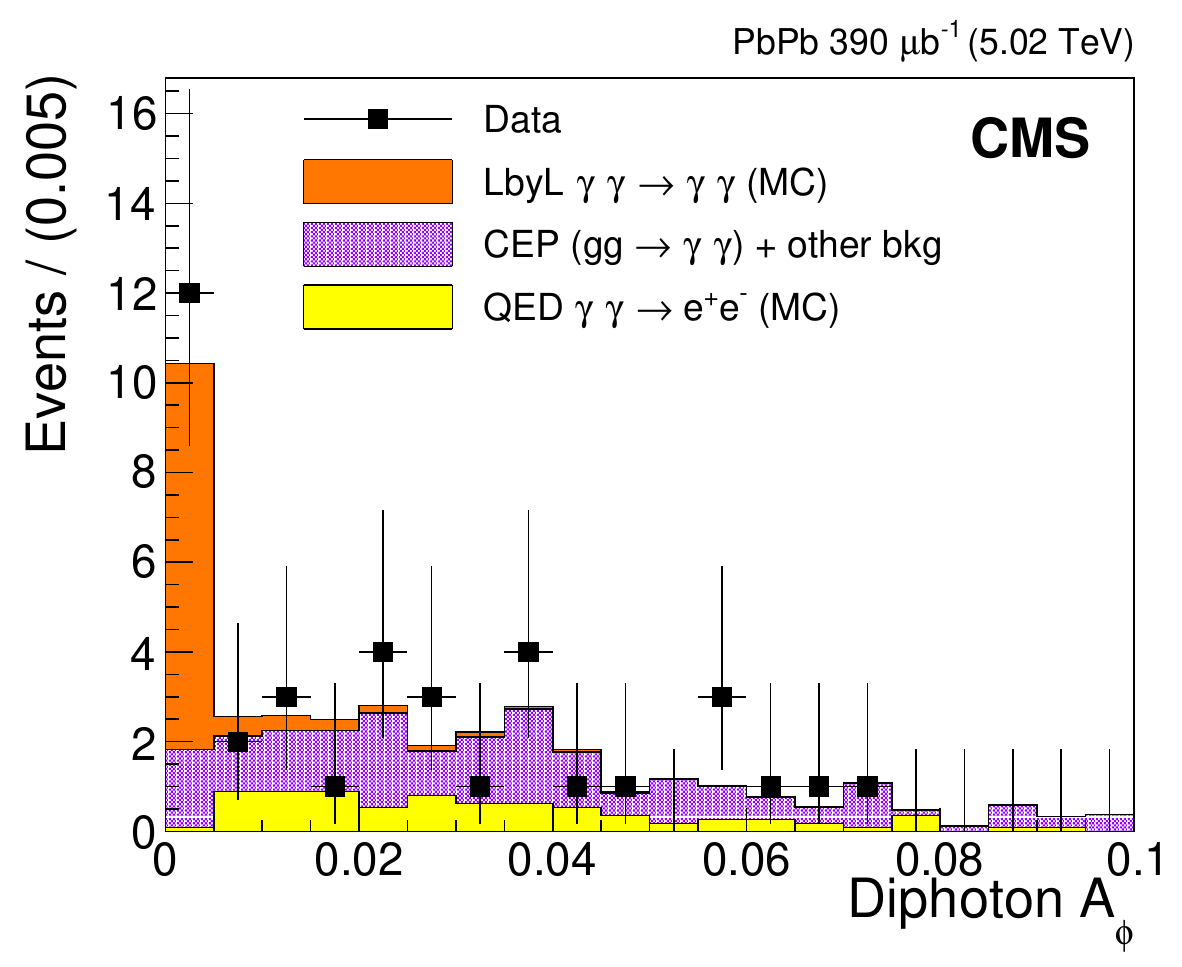}
\caption{Diphoton acoplanarity distribution in the search for axion-like particles in ultraperipheral PbPb collisions, for exclusive events measured in the data after selection criteria (squares), compared to the expected light-by-light scattering signal (orange histogram), quantum electrodynamics $\Pe^{+} \Pe^{-}$ (yellow histogram), and the CEP+other (purple histogram) backgrounds. Signal and quantum electrodynamics $\Pe^{+} \Pe^{-}$ MC samples are scaled according to their theoretical cross sections and integrated luminosity. The error bars around the data points indicate statistical uncertainties. The horizontal bars around the data symbols indicate the bin size. Figure taken from Ref.~\cite{FSQ-16-012}.}
\label{fig:FSQ-16-012}
\end{figure}

Potential backgrounds to ALP production include the major nonresonant light-by-light process, the quantum electrodynamics $\PGg\PGg \to \Pe^{+} \Pe^{-}$ process where both electrons are misidentified as photons, and the CEP $\Pg\Pg \to \PGg\PGg$ where the exclusive diphotons are produced via strong interactions. Events with exactly two photons with $\ET> 2\GeV$ and $\abs{\eta} <2.4$, no extra charged particles, and no calorimeter activity are selected. The nonexclusive diphoton background is eliminated by requiring events to have diphoton acoplanarity ${\mathrm A}_{\phi} < 0.01$ and diphoton transverse momentum $\pt^{\PGg \PGg}< 1\GeV$. The diphoton acoplanarity distribution, before the criterion on this variable is applied, is shown in Fig.~\ref{fig:FSQ-16-012}. The measured diphoton invariant mass distribution is used to search for possible narrow diphoton resonances. The sensitivity of the search is discussed in Section~\ref{sec:ALPportal}.

\cmsParagraph{Search for new physics in central exclusive production using the missing-mass technique with CMS and CMS-TOTEM\label{sec:EXO-19-009}} Studies of CEP processes in high-energy $\Pp\Pp$ collisions provide a unique method to access a class of physics processes, such as new physics via anomalous production of fermions, V bosons (where V is a \PGg, \PW, or \PZ boson), high-\pt jet production, and possibly the production of new resonances or pair production of new particles. The addition of new detectors further extends the coverage and enhances the sensitivity of the LHC experiments thus offering a new opportunity to explore processes and final states previously not covered. The CMS-TOTEM PPS~\cite{CMS:2014sdw} allows the surviving scattered protons during standard running conditions in regular ``high-luminosity'' fills to be measured~\cite{TOTEM:2022vox} (Section~\ref{subsec:PPS}). 

A generic search for a hypothetical massive particle $X$ produced in association with one or more SM particles in CEP processes is performed~\cite{CMS:2023roj}. In particular, the process $\Pp\Pp \to \Pp\Pp+\PZ/\PGg+\PX$ is studied. In the interaction, the two colliding protons survive after exchanging two colorless particles and can be recorded in the PPS. The detection and precise measurement of both forward protons allows a full kinematic reconstruction of the event, including the four-momentum of $X$ measured from the balance between the tagged SM particle(s) and the forward protons. This technique---the ``missing-mass'' technique---allows for searches for BSM particles without assumptions about their decay properties, except that the decay width can be considered narrow enough to produce a resonant mass peak, thus providing a new tool for generic BSM searches. A search for a massive particle produced in association with a \PZ boson or a photon in the final state is considered, using data samples corresponding to $\Lint=37$ and 2.3\fbinv, respectively.

\begin{figure*}[htb!]
\centering
\includegraphics[width=0.75\linewidth]{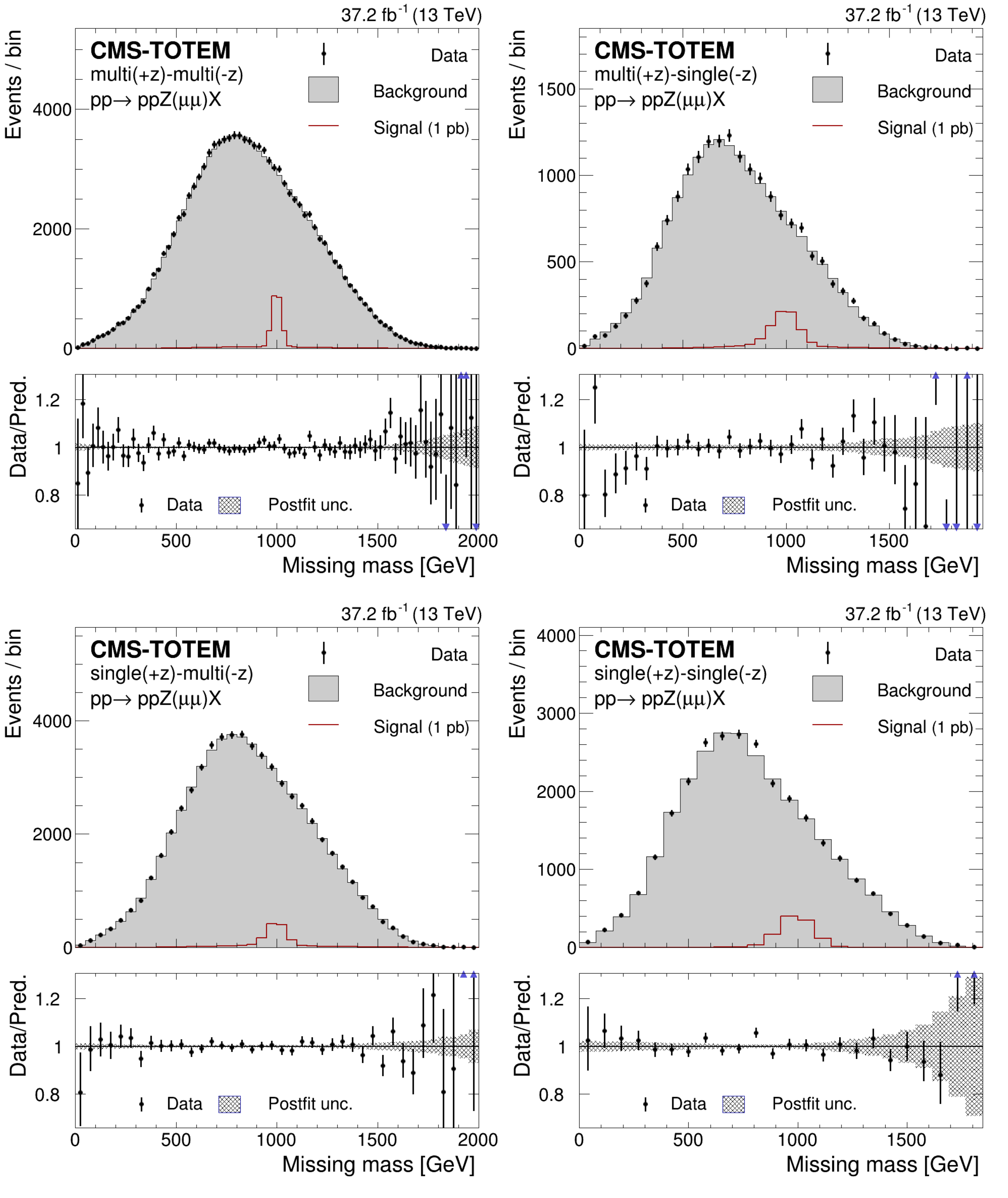}
\caption{Missing-mass distributions in the $\PZ\to\Pgm\Pgm$ final state of the CMS and CMS-TOTEM search using the missing-mass technique. The distributions are shown for protons reconstructed with (from \cmsLeft to \cmsRight) the multi-multi, multi-single, single-multi, and single-single methods, respectively. The background distributions are shown after the fit. The lower panels display the ratio between the data and the background model, with the arrows indicating values lying outside the displayed range. The expectations for a signal with $m_X=1000\GeV$ are superimposed and normalized to 1\unit{pb}. Figure taken from Ref.~\cite{CMS:2023roj}.}
\label{fig:EXO-19-009_missing_mass}
\end{figure*}

The excellent proton momentum reconstruction of PPS allows us to search for missing-mass signatures at high invariant masses with unprecedented resolution. In this high-mass range, EW processes are generally enhanced relative to QCD-induced processes. The main goal is the search for a $\PGg\PGg$-induced exclusive production process in which an unspecified weakly interacting BSM particle with a narrow decay width is produced. No assumption is made on its decay properties. Leptonically decaying \PZ bosons or an isolated photon are selected in the central detector, and the missing mass is constructed from the kinematics of the reconstructed boson in the central detector and the final-state protons in PPS (Fig.~\ref{fig:EXO-19-009_missing_mass}). A hypothetical $X$ resonance is searched for in the mass region between 0.6 and 1.6\TeV.

\subsection{Searches for long-lived particles} \label{sec:signatures_llp}

As mentioned in Sections~\ref{sec:intro} and \ref{sec:LLPreconstruction}, scenarios with LLPs can provide a DM candidate. Here we describe the signatures and searches for LLPs in CMS that provide sensitivity to the DS. We first describe searches for LLPs that decay into displaced leptons in Section~\ref{sec:displacedLeptons}, then searches for LLPs that decay hadronically in Section~\ref{sec:hadronicLLPdecays}, and lastly searches for LLPs and \ptmiss in Section~\ref{sec:LLPsAndMET}. 

\subsubsection{Displaced leptons} \label{sec:displacedLeptons}

Displaced leptons provide a powerful handle to identify LLP decays while maintaining sensitivity to a wide range of models. Events with displaced leptons can provide a striking signature of BSM physics because of the reduced background contribution from SM processes. In this section, we describe several displaced-lepton analyses with distinct signatures. The reconstruction of displaced signatures with the tracker is described in detail in Section~\ref{sec:displacedTracking} and the reconstruction of displaced muons is described in detail in Section~\ref{sec:displacedMuons}.

\begin{figure}[hbtp]
\centering
\includegraphics[width=0.45\textwidth]{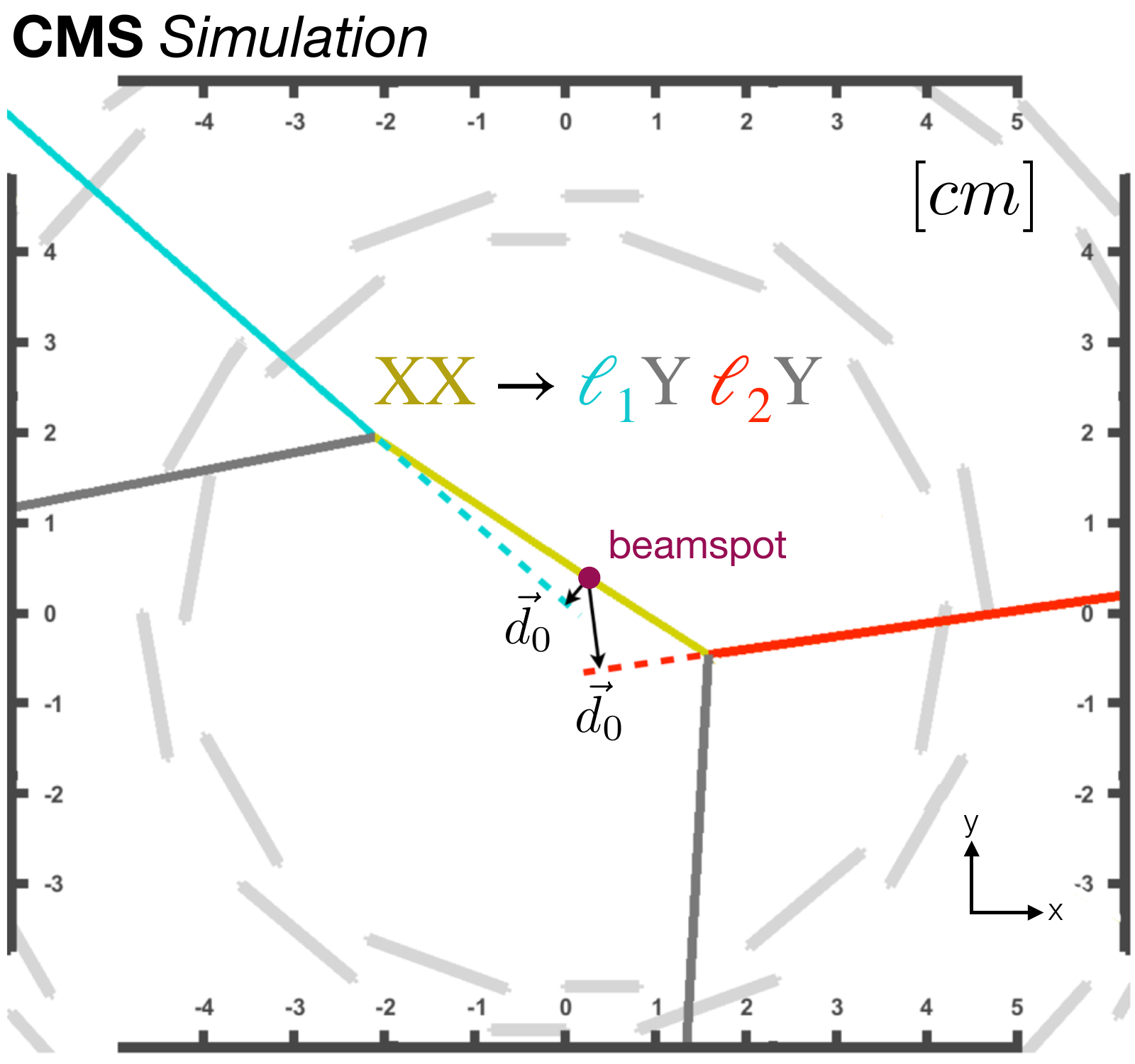}
\caption{
A diagram of a simulated signal event in the search for leptons with large impact parameters, from a transverse view of the interaction point, in the analysis presented in Ref.~\cite{EXO-18-003}. The black arrows indicate the lepton transverse impact parameter vectors.
}
\label{fig:EXO_18_003_signal}
\end{figure}

\cmsParagraph{Search for leptons with large impact parameters in \texorpdfstring{$\Pe\pmb{\Pgm}$, $\Pe\Pe$, and $\pmb{\Pgm}\pmb{\Pgm}$}{e-mu, e-e, and mu-mu} final states\label{sec:EXO-18-003}} The analysis described in Ref.~\cite{EXO-18-003} is carried out on a $\Pp\Pp$ collision data set corresponding to $\Lint=115\fbinv$ at $\sqrt{s}=13\TeV$. This analysis targets the displaced lepton signature by studying events with at least two leptons (any combination of electrons and muons) with transverse impact parameters between 0.01 and 10\cm. Requiring two such leptons with transverse momenta thresholds varying from 35 to 75\GeV, depending on lepton flavor and data-taking year, and relatively little nearby activity is sufficient to reject nearly all SM backgrounds without placing any requirements on the dilepton charge product or flavor combination, constraining other event properties (such as hadronic activity or \ptmiss), or requiring that the leptons form a common vertex. The signature for this search is shown in Fig.~\ref{fig:EXO_18_003_signal}. This approach allows the analysis to be sensitive to effectively any new physics process that involves at least one LLP whose decay includes at least two leptons or two LLPs whose decays each include at least one lepton. This is the only CMS Run~2 search for displaced leptons where the leptons are not required to come from a common DV.

This analysis uses dilepton triggers that do not require the leptons to originate from the collision point. The SM background is dominated by leptons with poorly measured displacement values, and care is taken to reject sources of genuine displaced leptons such as cosmic ray muons, displaced decays of SM mesons, and material interactions. The SM background estimate uses the ABCD method described in Section~\ref{sec:abcd} within 15 orthogonal SRs that differ in lepton flavor, displacement, and momentum, an approach that maximizes the sensitivity to a range of LLP masses and lifetimes. 

The sensitivity of this search to Higgs boson decays to LLPs is discussed in Section~\ref{sec:higgsLLPSens}.

\begin{figure}[hbtp]
\centering
\includegraphics[width=0.5\textwidth]{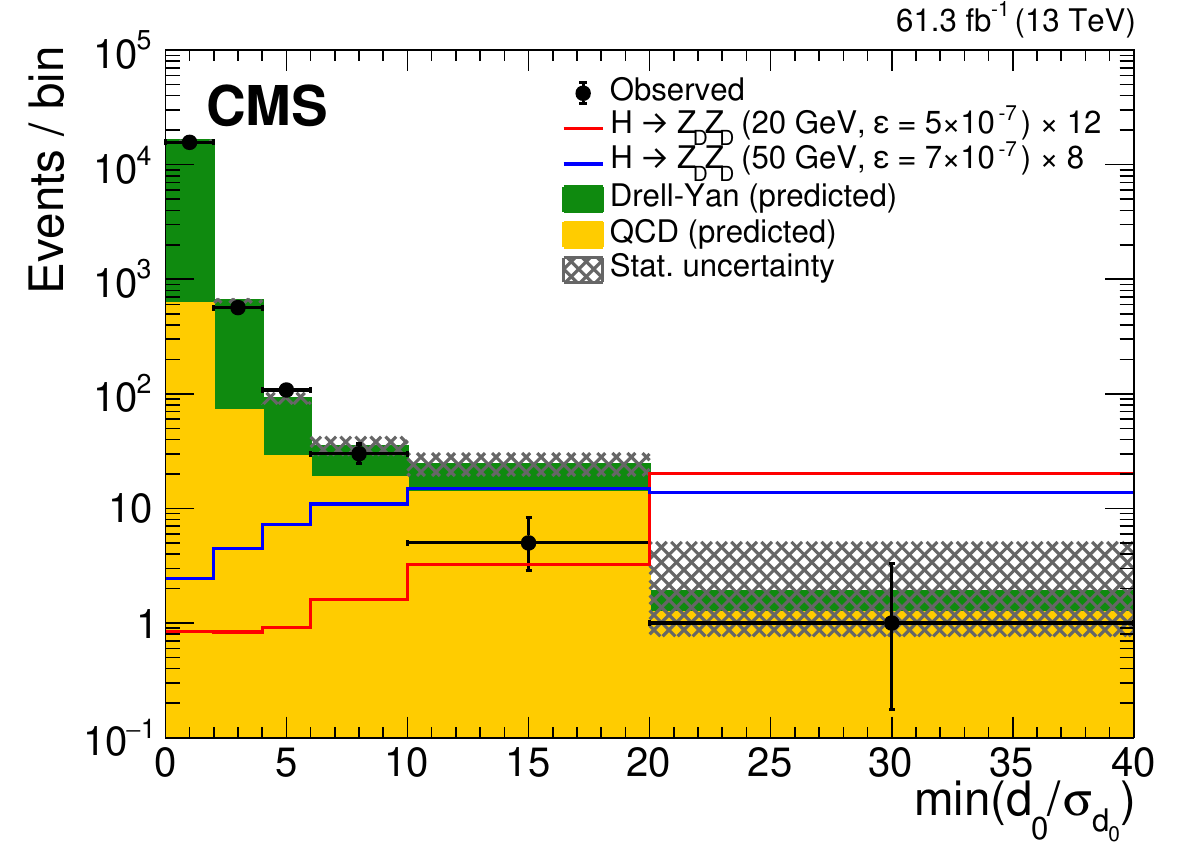}
\caption{
Comparison of the number of events observed in 2018 data with the expected number of background events, as a function of the smaller of the two $d_0$ significance values ($\text{min}(d_0/{\sigma_{d_0}})$) for pairs of global muons that are reconstructed in the tracker and muon system, in the search for muon pairs from a DV. The black points with error bars show the number of observed events; the green and yellow components of the stacked histograms represent the estimated numbers of DY and QCD events, respectively. The last bin includes events in the overflow. The uncertainties in the total expected background (shaded area) are statistical only. Signal contributions expected from simulated decays of exotic Higgs bosons to dark \PZ bosons, with \PZ boson masses of 20 and 50\GeV are shown in red and blue, respectively. Their yields are set to the corresponding combined median expected exclusion limits at 95\% \CL, scaled up as indicated in the legend to improve visibility. Figure adapted from Ref.~\cite{EXO-21-006}.}
\label{fig:EXO_21_006_minD0sig}
\end{figure}

\cmsParagraph{Search for muon pairs from a displaced vertex\label{sec:EXO-21-006}} Reference~\cite{EXO-21-006} presents an inclusive search for an exotic massive LLP decaying into a pair of oppositely charged muons (``dimuon'') originating from a common DV. The DV can be spatially separated from the $\Pp\Pp$ interaction point by a distance ranging from several hundred $\mu$m to several meters. The analysis uses muons produced within the silicon tracker, which can be reconstructed by both the tracker and the muon system, as well as muons produced in the outer tracker layers or beyond, which are reconstructed by only the muon system. The data sample corresponds to $\Lint=98\fbinv$. The minimal set of requirements and loose event selection criteria used in the search allow us to be sensitive to a wide range of LLP models. Figure~\ref{fig:EXO_21_006_minD0sig} shows the distribution of a key discriminating variable, namely, the minimum $d_0$ significance, for globally reconstructed dimuon pairs with 2018 data.

Reference~\cite{CMS-PAS-EXO-23-014} presents a continuation and extension of the search for displaced dimuons produced within and beyond the tracker described in Ref.~\cite{EXO-21-006}. The search is based on data collected during 2022 at $\sqrt{s}=13.6\TeV$, corresponding to $\Lint=36.6\fbinv$ and recorded with an improved set of HLT and level-1 trigger algorithms~\cite{LLPRun3TriggerDPNote}, aimed at increasing the signal efficiency by lowering the $\pt$ thresholds as much as possible without increasing the resulting trigger rate considerably. Overall, the addition of the new trigger algorithms improves the trigger efficiency for LLPs with a mass of a few tens of {\GeVns} and displacement ${\gtrsim}0.1\unit{cm}$ by a factor of 2 to 4, depending on displacement and mass, as compared to Run~2. 

The sensitivity of this search to HAHM scenarios is shown in Section~\ref{sec:darkSUSYHAHM} and the sensitivity to Higgs boson decays to LLPs is shown in Section~\ref{sec:higgsLLPSens}.

\begin{figure}
\centering
\includegraphics[width=.5\textwidth]{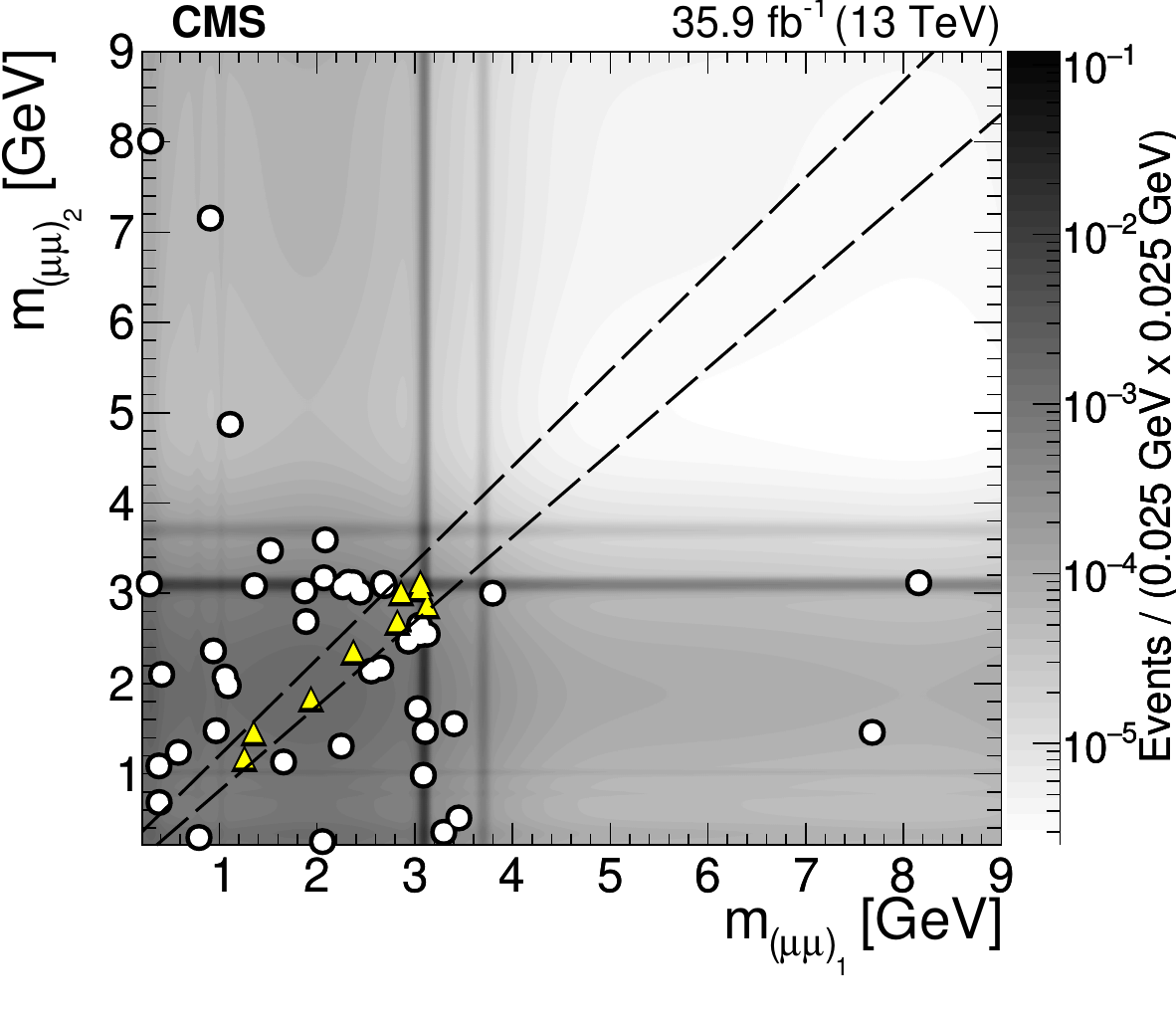}
\caption{Distribution of the invariant masses $m_{(\Pgm\Pgm)1}$ vs. $m_{(\Pgm\Pgm)2}$ of the isolated dimuon systems, in the search for prompt and displaced dimuons in final states with $4\Pgm$+X. Triangles represent data events passing all the selection criteria and falling in the SR $m_{(\Pgm\Pgm)1} \approx m_{(\Pgm\Pgm)2}$ (outlined by dashed lines), and white circles represent data events that pass all selection criteria but fall outside the SR. Figure taken from Ref.~\cite{CMS-HIG-18-003}.}
\label{fig:HIG-18-003}
\end{figure}

\cmsParagraph{Search for prompt and displaced dimuons in final states with \texorpdfstring{$4\pmb{\Pgm}$+X}{4mu+X}\label{paragraph:HIG-18-003}} Reference~\cite{CMS-HIG-18-003} describes another analysis that uses muons to search for evidence of DS particles. In this analysis, we search for the production of two BSM particles per event, selecting pairs of prompt or displaced dimuons reconstructed in the tracker in a data sample with $\Lint=36\fbinv$. Events that can mimic the signal come from pair-production of bottom quarks through QCD processes (QCD \bbbar), double \PJGy production, and EW processes. The 2D distribution of the invariant masses of the isolated dimuon systems is shown in Fig.~\ref{fig:HIG-18-003}.

In the case of the QCD \bbbar background, CRs in data are used to estimate its contribution, while for the \PJGy and EW processes, such as $\PZ\PZ\to 4\PGm$ and $\PZ^{*}/\PGg\to2\PGm$ (where a second \PZ boson is radiated and decays into a muon pair), the backgrounds are estimated with CRs in data and from simulation, respectively. 

The sensitivity of this search to HAHM scenarios is shown in Section~\ref{sec:darkSUSYHAHM}.

\begin{figure}
\centering
\includegraphics[width=.6\textwidth]{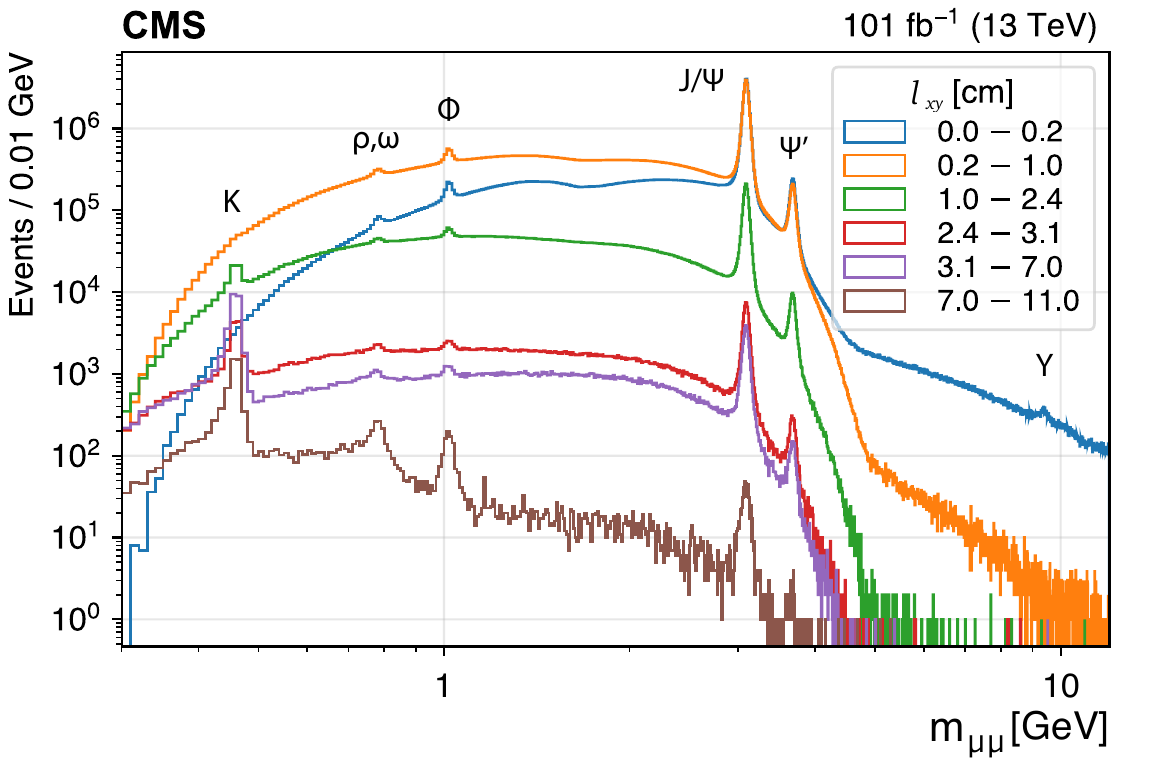}
\caption{The dimuon invariant mass distribution from the search for displaced dimuon resonances with data scouting, shown in bins of $l_{xy}$ as obtained from all selected dimuon events. Figure taken from Ref.~\cite{EXO-23-007}.}
\label{fig:EXO-20-014}
\end{figure}

\cmsParagraph{Search for displaced dimuon resonances with data scouting} \label{paragraph:EXO-20-014}Scouting triggers such as those described in Section~\ref{subsec:scouting} also provide opportunities for DS searches with displaced leptons. A search for narrow, long-lived dimuon resonances~\cite{EXO-20-014} is performed based on data collected during Run~2 in 2017 and 2018, using a dedicated dimuon scouting trigger stream. The selected data correspond to $\Lint=101\fbinv$. The rate of scouting triggers is higher than that of the standard triggers allowing less stringent requirements on the muon \pt. This enables dimuon resonance searches across mass and lifetime ranges that are otherwise inaccessible; in particular, the search described here has sensitivity to masses in the 1--3\GeV range. The scouting trigger algorithms used in this search select events containing muons with $\pt>3\GeV$. The search targets narrow, low-mass, long-lived resonances decaying into a pair of oppositely charged muons, where the lifetime of the LLP is such that the transverse displacement ($l_{xy}$) of its decay vertex is within 11\cm of the PV. The hard cutoff in $l_{xy}$ at 11\cm is a result of the scouting trigger algorithm, which requires muons with hits in at least two layers of the pixel tracker. Muon tracks are used in pairs to form dimuon vertices, considering all possible pairings. These vertices are considered to be candidate DVs, and they may be displaced from the PV or not. The dimuon invariant mass distribution in bins of $l_{xy}$ is shown in Fig.~\ref{fig:EXO-20-014}. The signal is expected to appear as a narrow peak on the dimuon mass continuum, with a resonance width smaller than the experimental mass resolution. Events are required to contain at least one pair of oppositely charged muons associated with a selected DV, and those that contain a single muon pair are then categorized according to transverse displacement and the \pt and isolation of the muon pair. In each category, we define mass windows sliding along the dimuon invariant mass spectrum, and we perform a search for a resonant peak in each mass window. 

The sensitivity of the search to Higgs boson decays to LLPs is discussed in Section~\ref{sec:higgsLLPSens}.

\subsubsection{Hadronic LLP decays} \label{sec:hadronicLLPdecays}

Hadronic decays of LLPs can provide sensitivity to a large variety of DS models. Here we describe several CMS searches that utilize hadronic LLP decays. The decay positions of the LLPs targeted in these searches span a wide range, including decays in the tracker, calorimeters, and even in the muon system.

\cmsParagraph{Search for LLPs decaying into displaced jets\label{parag:disp_jets}} In Ref.~\cite{CMS:2020iwv}, we present a model-independent search for LLPs decaying into jets, with at least one LLP having a decay vertex within the tracker acceptance, which goes up to $\approx$550\mm in the plane transverse to the beam direction. The data sample corresponds to $\Lint= 132\fbinv$. Events were collected with dedicated displaced-jets triggers, which select jets with small numbers of prompt tracks or with displaced tracks. 
With these tracking requirements, the \HT trigger threshold has been lowered from 1000 to 430\GeV, which significantly increases the trigger efficiencies for a large variety of models with LLPs. 

After the trigger selections, we look for all possible pairs of jets in a given event. For each jet pair (dijet), we attempt to reconstruct one DV using the displaced tracks associated with the two jets. 

\begin{figure}
\centering
\includegraphics[width=.55\textwidth]{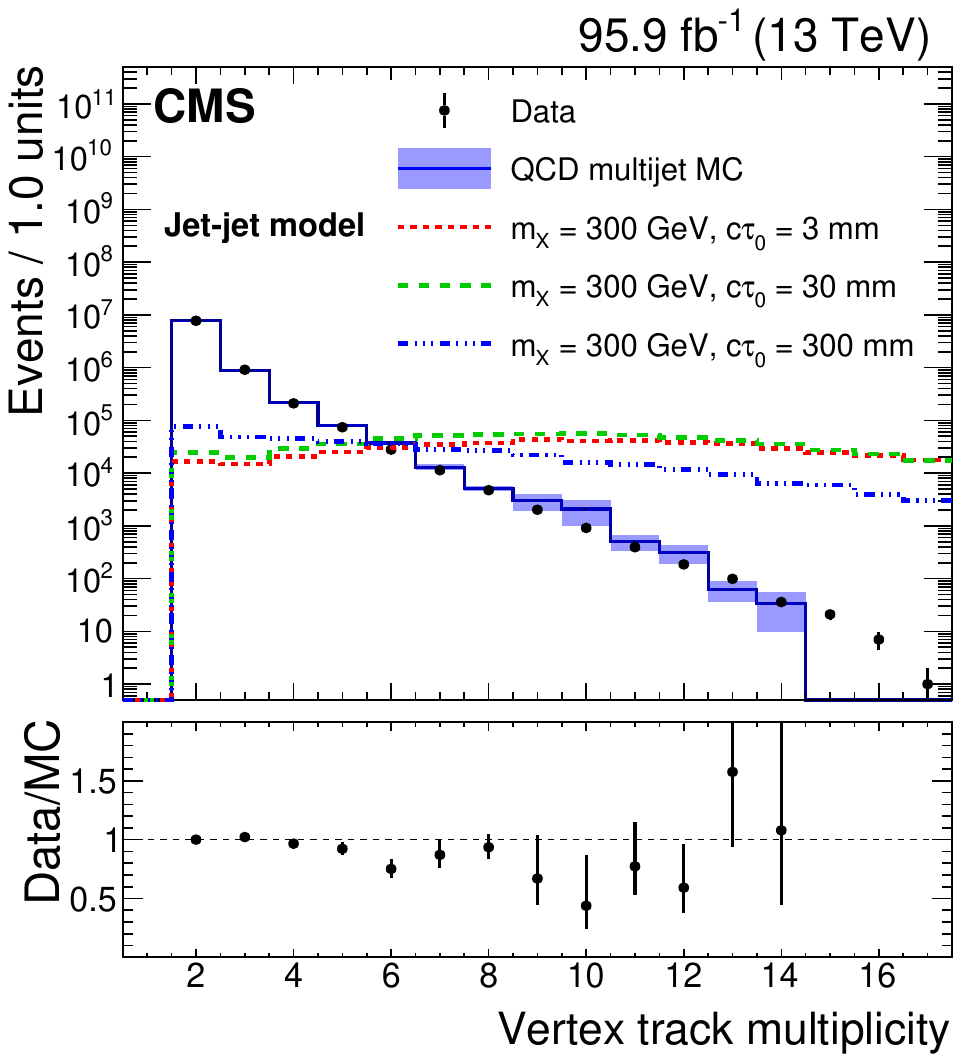}
\caption{Distribution of the vertex track multiplicity, for data, simulated QCD multijet events, and simulated signal events, in the displaced-jets search. For a given event, if there is more than one DV candidate being reconstructed, the one with the largest vertex track multiplicity is chosen. If the track multiplicities are the same, the one with the smallest $\chi^{2}$/ndof is chosen, where ndof is the number of degrees of freedom. The lower panel shows the ratios between the data and the simulated QCD multijet events. The blue shaded error bands and vertical bars represent the statistical uncertainties. Three benchmark signal distributions are shown (dashed lines). For visualization purposes, each signal process is given a cross section that yields 106 events produced in the analyzed data sample. Figure taken from Ref.~\cite{CMS:2020iwv}.}
\label{fig:EXO-19-021}
\end{figure}

The vertex reconstruction is performed using the adaptive vertex fitter described in \ref{sec:displacedTracking}. The properties of the DV, such as the number of tracks and the transverse displacement significance, provide discrimination power to distinguish LLP signatures from SM backgrounds. The distribution of the vertex track multiplicity is shown in Fig.~\ref{fig:EXO-19-021}. The relations among the DV, displaced tracks, and the dijet are also examined to construct more discriminating variables. Using these variables, a multivariate classifier based on a GBDT is developed to further improve the signal-to-background discrimination. The use of displaced jet tagging is described in detail in Section~\ref{sec:jetTagging}.

The sensitivity of the search to Higgs boson decays to LLPs is presented in Section~\ref{sec:higgsLLPSens}. The sensitivities to models containing heavy \PZpr and heavy \Hdark bosons are described in Section~\ref{sec:heavyLLPs}.

\begin{figure}[htbp]
    \centering
    \includegraphics[width=0.5\textwidth]{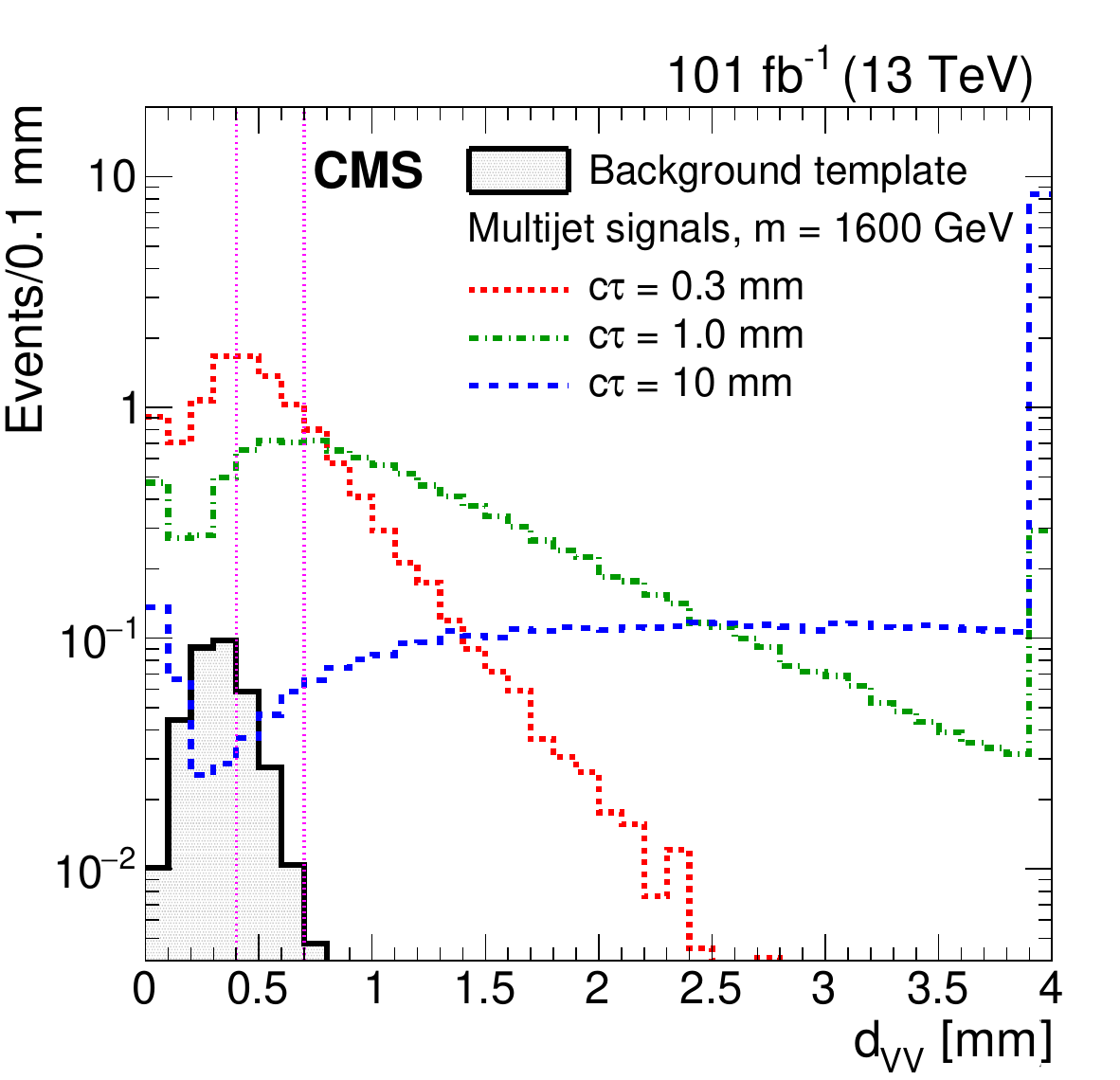}
    \caption{The distribution of distances between vertices in the $x$-$y$ plane, \dVV, for the displaced-vertices search, for three simulated multijet signals each with a mass of 1600\GeV, with the background template distribution overlaid. The production cross section for each signal model is assumed to be the lower limit excluded by Ref.~\cite{EXO-17-018}, corresponding to values of 0.8, 0.25, and 0.15\unit{fb} for the samples with $\ctau = 0.3$, 1.0, and 10\mm, respectively. The last bin includes the overflow events. The two vertical pink dashed lines separate the regions used in the fit. Figure taken from Ref.~\cite{EXO-19-013}.}
    \label{fig:EXO-19-013}
\end{figure}

\cmsParagraph{Search for LLPs decaying to jets with displaced vertices\label{sec:EXO-19-013}} \begin{sloppypar}This inclusive and largely model-independent search for pair-produced LLPs that decay hadronically focuses on LLPs with mean proper decay lengths less than 100~mm~\cite{EXO-19-013}. The reconstruction of DVs is detailed in Section~\ref{sec:displacedTracking}. To perform the search, the LLP decay positions are reconstructed as DVs, which are formed from charged particle tracks using a custom vertex reconstruction algorithm. The search is performed using data corresponding to $\Lint=140\fbinv$ from 2015--2018, collected at $\sqrt{s} = 13\TeV$, and relies on events collected with \HT triggers that require large jet activity. Offline, $\HT>1200\GeV$ is required. After forming the DVs, a series of selection criteria are used to suppress backgrounds. For instance, to eliminate backgrounds originating from material interactions, the DVs are required to be located within the radius of the beam pipe. Several other criteria are additionally used in the search to distinguish signal from background, including requirements on the uncertainty in the beamspot-to-vertex distance, which is crucial for mitigating backgrounds from genuine \PQb quark decay vertices, as well as a requirement that each signal-like vertex be formed from at least five charged particle tracks, to reduce combinatorial backgrounds. The primary search variable is the distance between two signal-like vertices in the $x$-$y$ plane (\dVV), shown in Fig.~\ref{fig:EXO-19-013}, as the LLPs considered are often expected to be produced back-to-back and to each have large $x$-$y$ displacement, while the separation between background vertices tends to be smaller.\end{sloppypar}

The sensitivities of the search to models containing heavy \PZpr and heavy \Hdark bosons are shown in Section~\ref{sec:heavyLLPs}.

\cmsParagraph{Searches for emerging jets\label{subsec:emj_run1}} Emerging jet phenomena may be observable at the LHC detectors when the DS is strongly coupled and the composite dark mesons have a finite lifetime comparable to the detector size, as described in the HV description in Section~\ref{sec:darkqcd}. The signature of an EJ differs from that of an SM jet in that the associated tracks will originate from many vertices, which can appear at various distances from the collision point depending on the dark meson lifetimes. The axis of each vertex within the jet points radially from the collision point. Dark quark production occurs via the decay of a complex scalar mediator \Pbifun, which is charged under both SM QCD and dark QCD. The mediator is produced in pairs at the LHC primarily through gluon-gluon fusion, and it decays into a dark quark and SM quark: $\Pbifun\Pbifun^\dag \to \Pqdark \PAQq\PQq^{\prime}\Paqdark^{\prime}$.

The displacement features from tracks associated with a jet are used to tag the EJ signal. Because there are no dedicated triggers for this signature, an \HT-based trigger is used, as the signal includes multiple hard jets.

The first iteration of the search~\cite{CMS:2018bvr} uses a set of requirements on several jet- and track-based variables to tag the EJs. This search uses a data set corresponding to $\Lint=16\fbinv$, which is approximately half of the 2016 data~\cite{CMS:2018bvr}.

\begin{figure}
\centering
\includegraphics[width=1\textwidth]{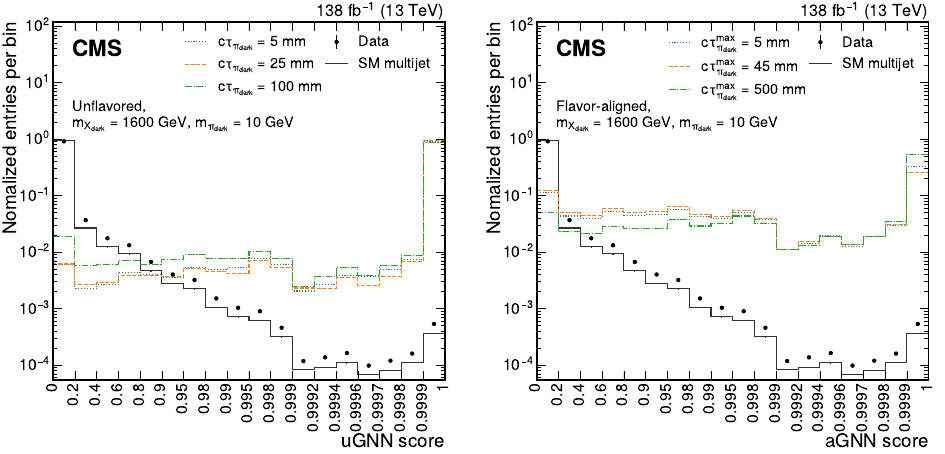}
\caption{Distributions of the GNN output score for the data (points with error bars), SM multijet simulation (dark gray line), and signal simulation (colored lines), for the search for emerging jets. Separate GNNs are trained for the unflavored model (uGNN, \cmsLeft) and the flavor-aligned model (aGNN, \cmsRight). Bins are chosen to correspond to the jet selection criteria applied in the analysis. The sums of the entries are normalized to unity. Figure taken from Ref.~\cite{CMS:2024emj}.}
\label{fig:EXO-22-015}
\end{figure}

The second iteration of the search~\cite{CMS:2024emj} employs both a model-agnostic EJ tagger, similar to the first search, and a more powerful, but model-dependent, graph neural network (GNN) EJ tagger. Distributions of the output score of the GNN are shown in Fig.~\ref{fig:EXO-22-015}. The Run~2 data set corresponding to $\Lint=138\fbinv$ is analyzed. 

The sensitivity of the search is shown in Section~\ref{par:emjsens}.

\cmsParagraph{Search for decays of stopped LLPs\label{parag:stopped_part}} At the LHC, the LLPs could stop inside the detector material if they lose all of their kinetic energy while traversing the detector, which will typically occur for particles with initial velocities $\beta < 0.5$~\cite{Arvanitaki:2005nq}. This energy loss can occur via nuclear processes if they are strongly interacting and/or through ionization if they are charged. The observation of a stopped particle decay signature would not only indicate new physics but also help measure the lifetime of LLPs, giving insights into various BSM scenarios.

If these stopped LLPs have lifetimes longer than tens of nanoseconds, most of their decays would be reconstructed as separate events unrelated to their production~\cite{Graham_stoppedParticlesAtLHC}. Owing to the difficulty of differentiating between the LLP decay products and SM particles from LHC $\Pp\Pp$ collisions, these subsequent decays are most easily identified when there are no proton bunches in the detector. The detector is quiet during these out-of-collision time periods with the exception of rare noncollision backgrounds, such as cosmic rays, beam halo particles, and detector noise. If LLPs come to a stop in the detector, they are most likely to do so in the densest detector materials, which in the CMS detector are the ECAL, the HCAL, and the steel yoke in the muon system. If the stopped LLPs decay in the calorimeters, relatively large energy deposits occurring in the intervals between collisions could be observed. Furthermore, if the stopped LLPs decay into muons, displaced muon tracks out of time with the collisions could be detected. Both signatures require dedicated triggers to select events in between bunch crossings.

\begin{figure}
\centering
\includegraphics[width=0.45\textwidth]{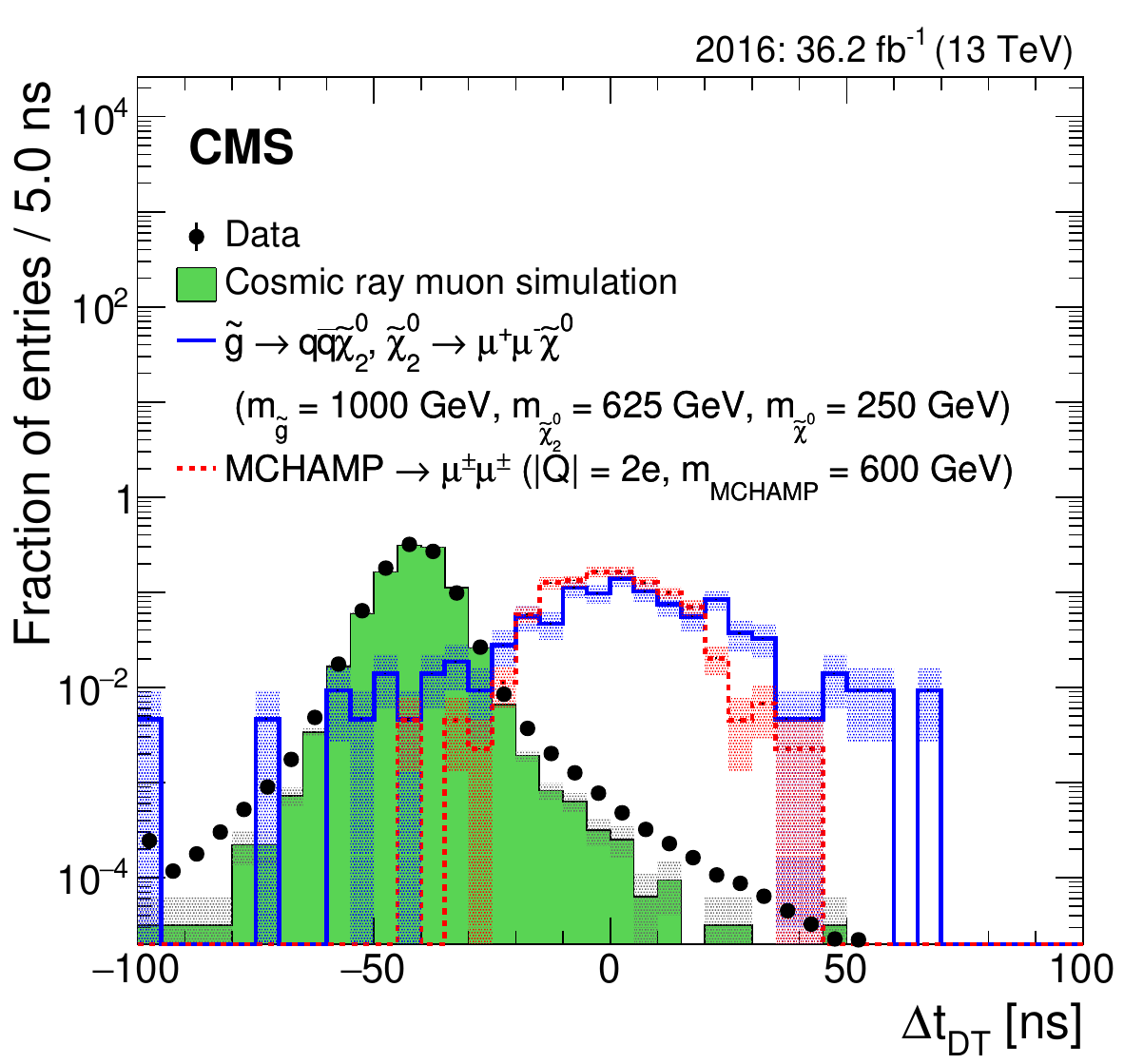}
\caption{The muon timing distribution in the DTs for 2016 data, simulated cosmic ray muon events, and simulated signal events, for the muon channel of the stopped-LLPs search. The gray bands indicate the statistical uncertainty in the simulation. The histograms are normalized to unit area. Figure taken from Ref.~\cite{exo-16-004}.}
\label{fig:EXO-16-004}
\end{figure}

Two searches are performed for stopped LLPs that decay out of time with respect to the presence of proton bunches in the detector~\cite{exo-16-004}. One search targets hadronic decays detected in the calorimeters and the other looks for decays into muon pairs in the muon system. These two search channels are analyzed independently using data collected in 2015 and 2016 with separate dedicated triggers. The triggers select calorimeter deposits or muons during gaps between proton bunches in the LHC beams. The calorimeter (muon) search uses $\sqrt{s}=13\TeV$ data corresponding to $\Lint=38.6\,(39.0)\fbinv$ collected with LHC $\Pp\Pp$ collisions separated by 25\unit{ns} during a search interval totaling 721\,(744) hours. Figure~\ref{fig:EXO-16-004} shows the muon timing distribution used in the muon search.

\subsubsection{Signatures with LLPs and \texorpdfstring{\ptmiss}{missing transverse momentum}} \label{sec:LLPsAndMET}

Some DS models lead to striking signatures with both displaced particles and significant \ptmiss. This \ptmiss can arise from either stable particles, which could be a DM candidate, or from an LLP that escapes the detector before decaying. These signatures often have very low levels of SM backgrounds and can be sensitive to unique DS interpretations. Four of these searches are described below.

\cmsParagraph{Searches for neutral LLPs decaying in the muon system}
\label{sec:msclusters}
        
Reference~\cite{CMS:2021juv} describes the first search at the LHC that uses a muon detector as a sampling calorimeter to identify showers produced by decays of LLPs. The analysis uses a data set corresponding to $\Lint = 137\fbinv$ collected during 2016--2018 with \ptmiss triggers. Based on a unique detector signature, the search is largely model-independent, with sensitivity to a broad range of LLP decay modes and to LLP masses as small as a few {\GeVns}. Decays of LLPs in the muon detectors induce hadronic and electromagnetic showers, giving rise to a high hit multiplicity in localized detector regions. The use of muon detector showers is described in detail in Section~\ref{sec:mdShowers}.

This first search effort used the CSC endcap muon detectors. To identify displaced showers, the CSC hits are clustered to form CSC clusters with a large hit multiplicity, which has a high efficiency of about 80\% for \ddbar and \bbbar decays and 65\% for $\PGt^{+}\PGt^{-}$ decays. A number of selections are applied to suppress SM background clusters from jets that ``punch through'' the calorimeters and make it to the muon system, muons that undergo bremsstrahlung, and decays of SM LLPs, such as the neutral kaon \PKzL.

\begin{figure}[htb!]
    \centering
    \includegraphics[width=0.55\textwidth]{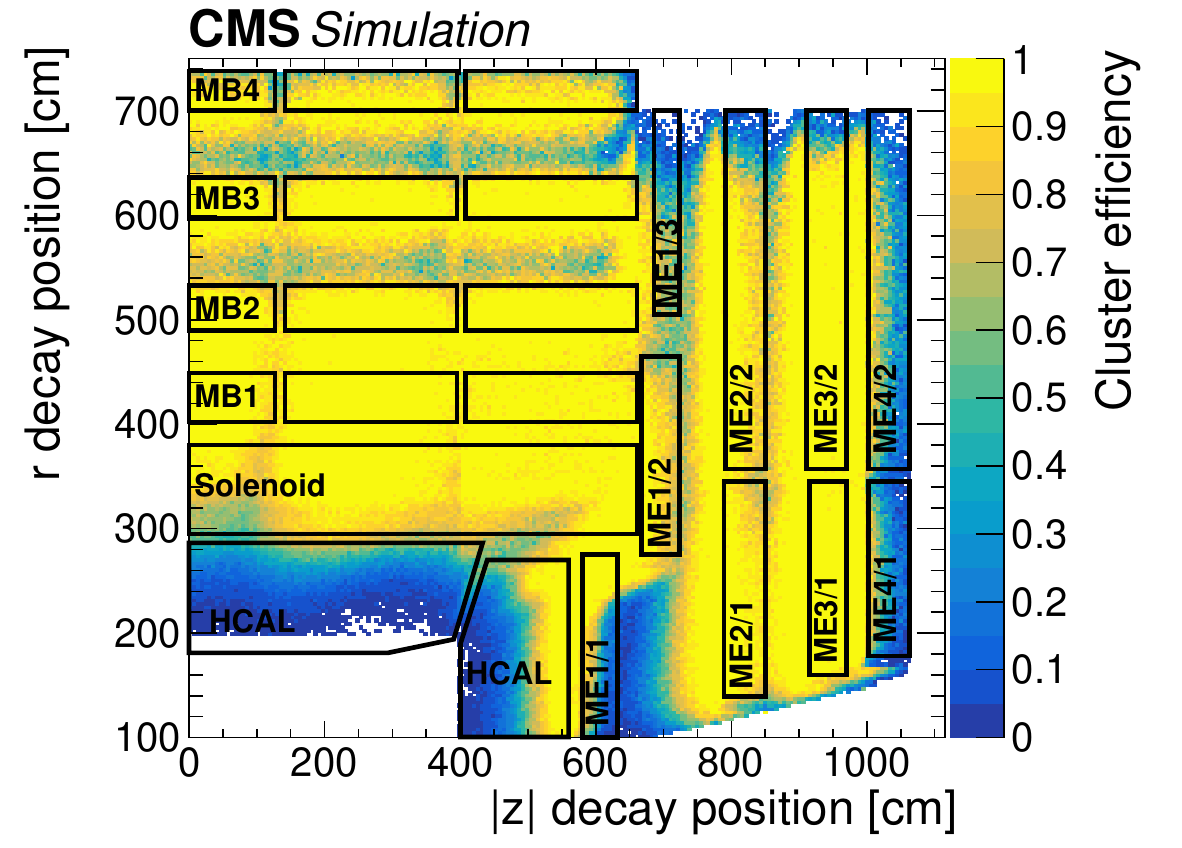}
    \caption{The cluster reconstruction efficiency as a function of the simulated $r$ and $\abs{z}$ decay positions of an LLP with a mass of 40\GeV and a range of $\ctau$ values between 1 and 10\unit{m}, for the search for neutral LLPs decaying in the muon system. Figure taken from Ref.~\cite{CMS:2023arc}.
    }
    \label{fig:msclustereff}
\end{figure}

A second analysis, presented in Ref.~\cite{CMS:2023arc}, is an extension of the muon endcap search described above and in Ref.~\cite{CMS:2021juv}. This second analysis is the first search at the LHC that uses both the barrel and endcap muon detectors as a sampling calorimeter to identify showers produced by decays of LLPs. As in the previous search, the CSC/DT hits are clustered to form muon detector showers with a large hit multiplicity to identify displaced showers in the muon detector. The efficiency for this clustering is shown in Fig.~\ref{fig:msclustereff}.

The sensitivity of the search to EJ signatures is presented in~\ref{par:emjsens}. The sensitivity of the search to Higgs boson decays to LLPs is given in Section~\ref{sec:higgsLLPSens}. The sensitivities to models containing heavy \PZpr and heavy \Hdark bosons are provided in Section~\ref{sec:heavyLLPs}.

\cmsParagraph{Search for inelastic dark matter\label{parag:inelDMEXO20010}} The traditional ``mono-X'' approach can be combined with searches for LLPs to probe new models and new signatures. In this analysis~\cite{CMS-PAS-EXO-20-010}, the final state of interest includes two displaced, nonresonant muons that are produced collinearly with the \ptvecmiss arising from the DM production. The DM and the muons also recoil against an ISR jet. The muons are too soft to be used for triggering, but by requiring the presence of a hard ISR jet in the final state, the use of data recorded with \ptmiss triggers is possible. The data sample corresponds to $\Lint=138\fbinv$. The results are interpreted in the context of an IDM model~\cite{Tucker-Smith:2001myb, Izaguirre:2015zva,Berlin:2018jbm}, described in Section~\ref{sec:iDM}.

\begin{figure}[htb!]
    \centering
    \includegraphics[width=0.55\textwidth]{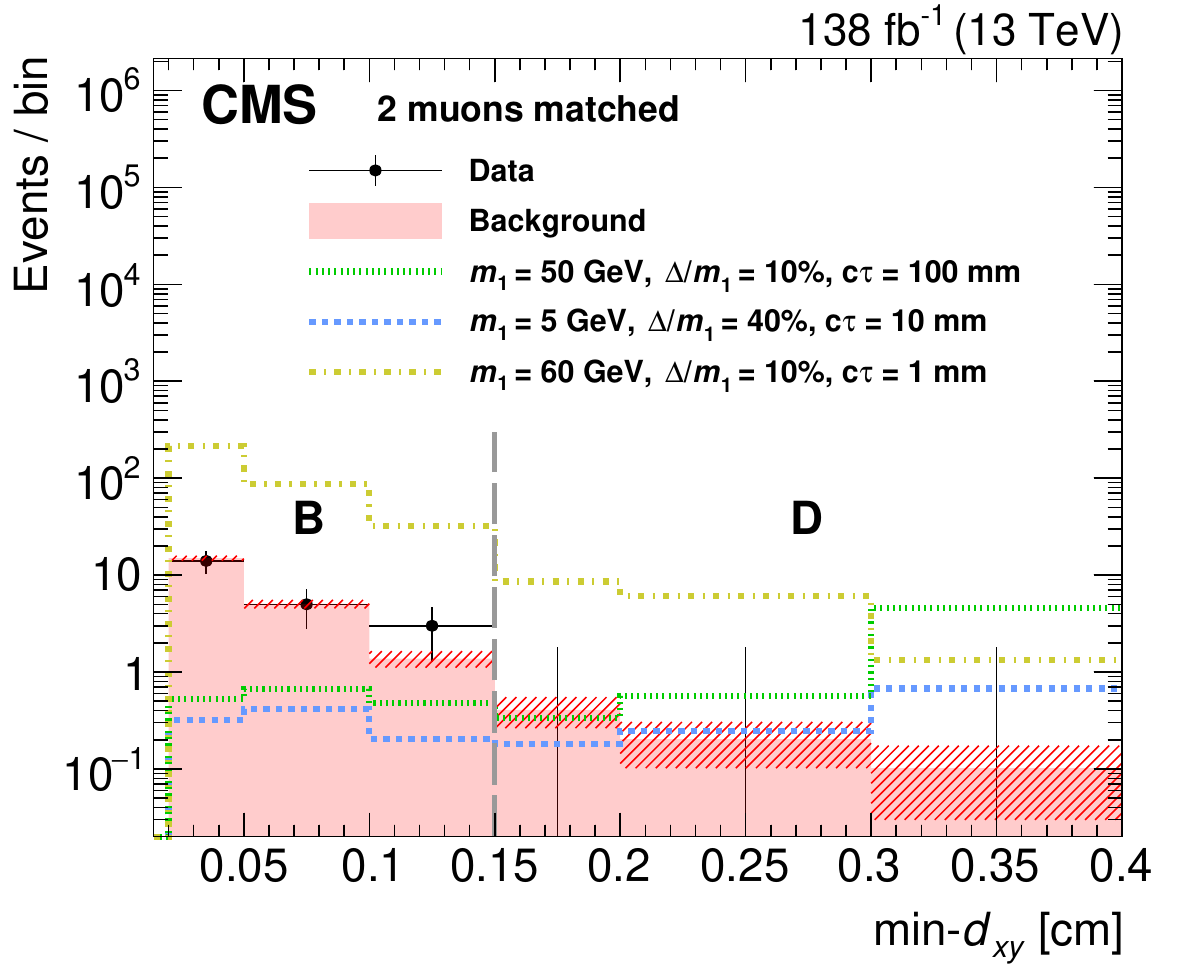}
    \caption{Measured min-$d_{xy}$ distribution in the 2-match category of the IDM search, after requiring the min-$d_{xy}$ muon to pass the isolation requirement $I^{\text{rel}}_{\text{PF}} < 0.25$. Overlaid with a red histogram is the background predicted from the region of the ABCD plane failing the same requirement, as well as three signal benchmark hypotheses (as defined in the legends), assuming $\alpha_{\mathrm{D}} = \alpha_{\mathrm{EM}}$ (the fine-structure constant). The red hatched bands correspond to the background prediction uncertainty. The last bin includes the overflow. Figure taken from Ref.~\cite{CMS-PAS-EXO-20-010}.}
    \label{fig:IDM_mindxy}
\end{figure}

The event selection requires significant \ptmiss and hadronic activity. Two muons reconstructed with the DSA muon reconstruction algorithm~\cite{CMS-DP-2015-015,CMS:2018rym,exo-16-004} are required. As explained in Section~\ref{sec:displacedMuons}, the DSA muon reconstruction algorithm only uses information from the muon spectrometer system, but similar to the approach developed in Ref.~\cite{EXO-21-006}, different categories of events are defined depending on whether the DSA muons can be matched to muons reconstructed using both the tracker and muon spectrometer. The minimum displacement min-$d_{xy}$ distribution is shown in Fig.~\ref{fig:IDM_mindxy} for the most sensitive category.

\begin{figure}[htb!]
    \centering
    \includegraphics[width=0.6\textwidth]{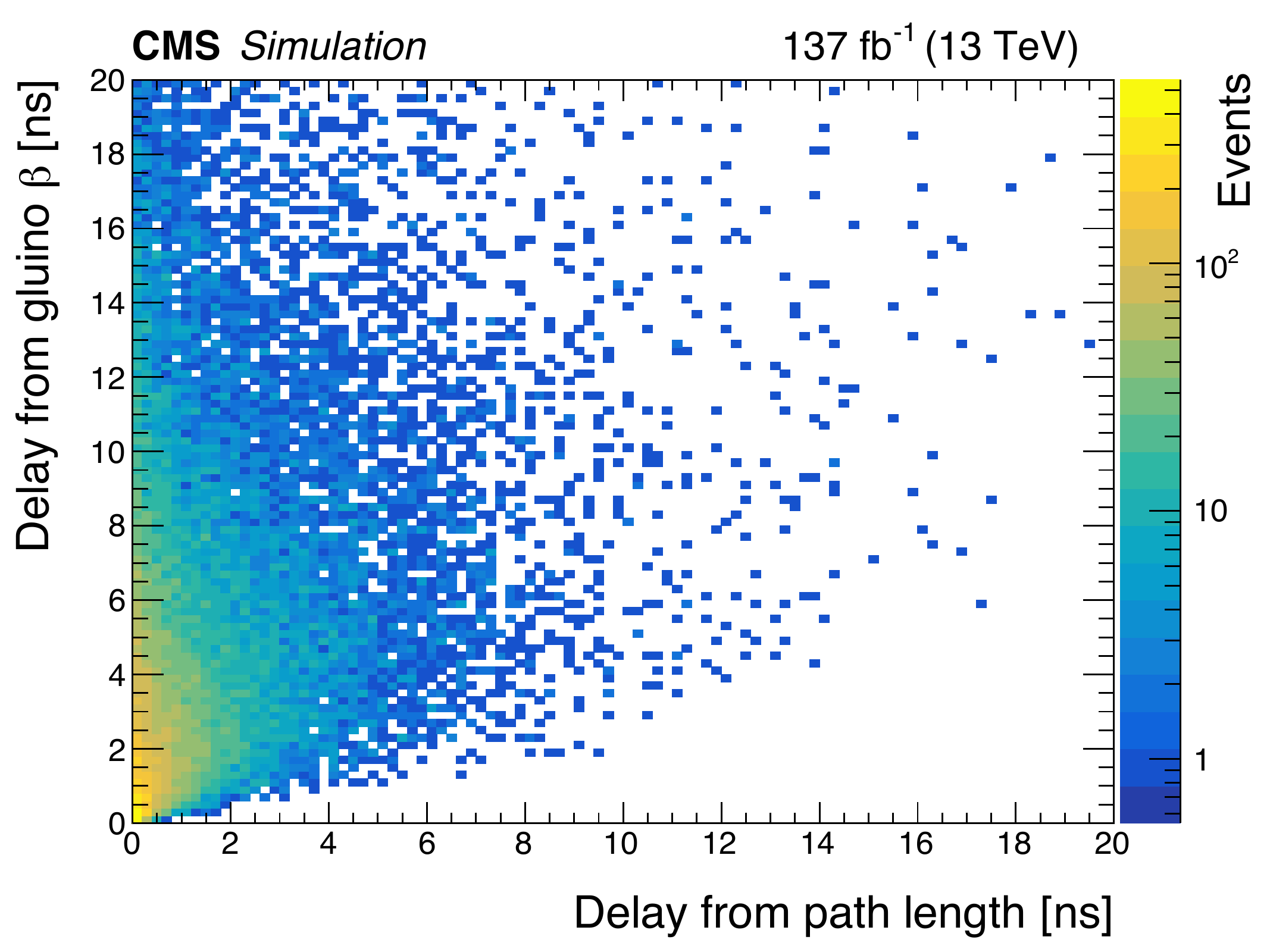}
    \caption{The contributions to the delay of the LLP from the path length and the lower velocity of the parent particle, in the delayed-jets search~\cite{EXO-19-001}. For this model, which features LLPs with proper decay lengths of 10\unit{m} and masses of 3\TeV, the lower velocity dominates the contribution to the delay.
    }
    \label{fig:delaycontributions}
\end{figure}

\cmsParagraph{Search for new physics with delayed jets\label{sec:EXO-19-001}} This search~\cite{EXO-19-001} presents the first use of timing signatures with the ECAL to identify OOT jets from the decays of heavy LLPs~\cite{Chiu:2021sgs}, using a data sample corresponding to $\Lint=137\fbinv$. The use of timing to provide sensitivity to LLPs is discussed in detail in Section~\ref{sec:delayedCalo}. There are two effects that contribute to the time delay of jets from the decay of heavy LLPs relative to deposits from jets originating at the interaction point. First, the total path, composed of the initial LLP trajectory and the subsequent jet trajectories, will be longer, and second, the LLP will move with a lower velocity owing to its high mass, as was shown earlier in Fig.~\ref{fig:timingDiagram}. The two contributions are shown in Fig.~\ref{fig:delaycontributions} for a representative LLP signal model. The ${\approx}0.4\unit{ns}$ timing resolution of the ECAL allows backgrounds to be greatly mitigated. The use of this technique allows signatures with displacements significantly beyond the acceptance of the tracker to be reconstructed. 

This search for heavy BSM LLPs also requires that the events contain significant \ptmiss. The \ptmiss can originate from invisible particles in the final state or from decays occurring beyond the detector acceptance. The \ptmiss is used as a trigger requirement as it allows substantially lower thresholds than \HT triggers. A series of selections is performed to reject backgrounds from both prompt collisions and noncollision processes, such as cosmic ray muons and beam halo. Example selections include using the tracker to veto deposits originating from the interaction point and using the muon systems to reject beam halo and cosmic ray muon deposits. The remaining background components are individually characterized and their residual contributions are predicted using CRs in data.

The sensitivities to models containing a heavy \PZpr boson are discussed in Section~\ref{sec:heavyLLPs}.

\cmsParagraph{Search for LLPs with trackless and out-of-time jets and \texorpdfstring{\ptmiss}{missing transverse momentum}\label{sec:EXO-21-014}} Another search, which uses timing information, targets events with LLP decays into hadronically decaying Higgs or \PZ bosons with \ptmiss~\cite{CMS-PAS-EXO-21-014}. Signal events are characterized by large \ptmiss, either because of the production of particles that do not interact with the detector material, or because of the LLP decaying at a macroscopic distance, outside of the calorimeters, and by the presence of trackless and OOT jets, described in Section~\ref{sec:delayedCalo}. A hadronic LLP decay in the outer regions of the tracker or within the calorimeter volume will result in jets with a low track multiplicity (nearly trackless) and OOT with regard to the LHC collisions. A delay of $\gtrsim 1\unit{ns}$ provides significant separation between signal and background for signal models with LLP masses heavier than 600\GeV. 

\begin{figure}[htb!]
    \centering
    \includegraphics[width=0.5\textwidth]{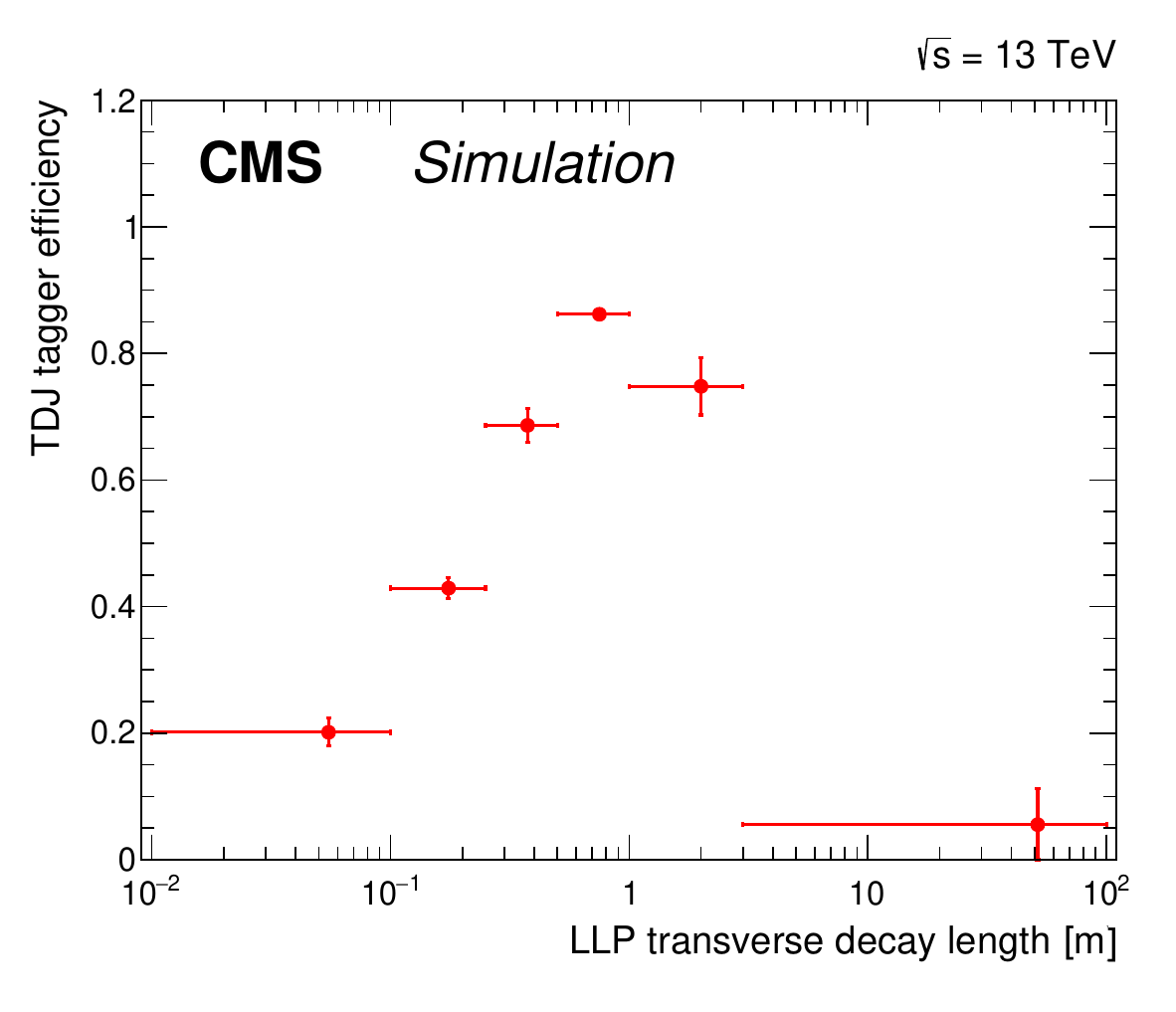}
    \caption{The efficiency of the jet tagger working point used in the trackless and OOT jets and \ptmiss analysis shown as a function of the lab frame LLP transverse decay length. The uncertainties shown account for lifetime dependence and statistical uncertainty. Figure taken from Ref.~\cite{CMS-PAS-EXO-21-014}.
    }
    \label{fig:CMS-PAS-EXO-21-014}
\end{figure}

The search uses \ptmiss as a trigger selection and is performed on a data sample corresponding to $\Lint=138\fbinv$. The jet timing, as well as features of the tracks and the electromagnetic calorimeter crystal hits associated with the jets induced by the LLP decays are the inputs of a DNN that tags trackless and OOT jets. The efficiency of the OOT jet tagger used for this search as a function of LLP transverse decay length is shown in Fig.~\ref{fig:CMS-PAS-EXO-21-014}.

The sensitivities of the search to models containing heavy \PZpr and heavy \Hdark bosons are provided in Section~\ref{sec:heavyLLPs}.

\begin{figure}
\centering
\includegraphics[width=0.5\textwidth]{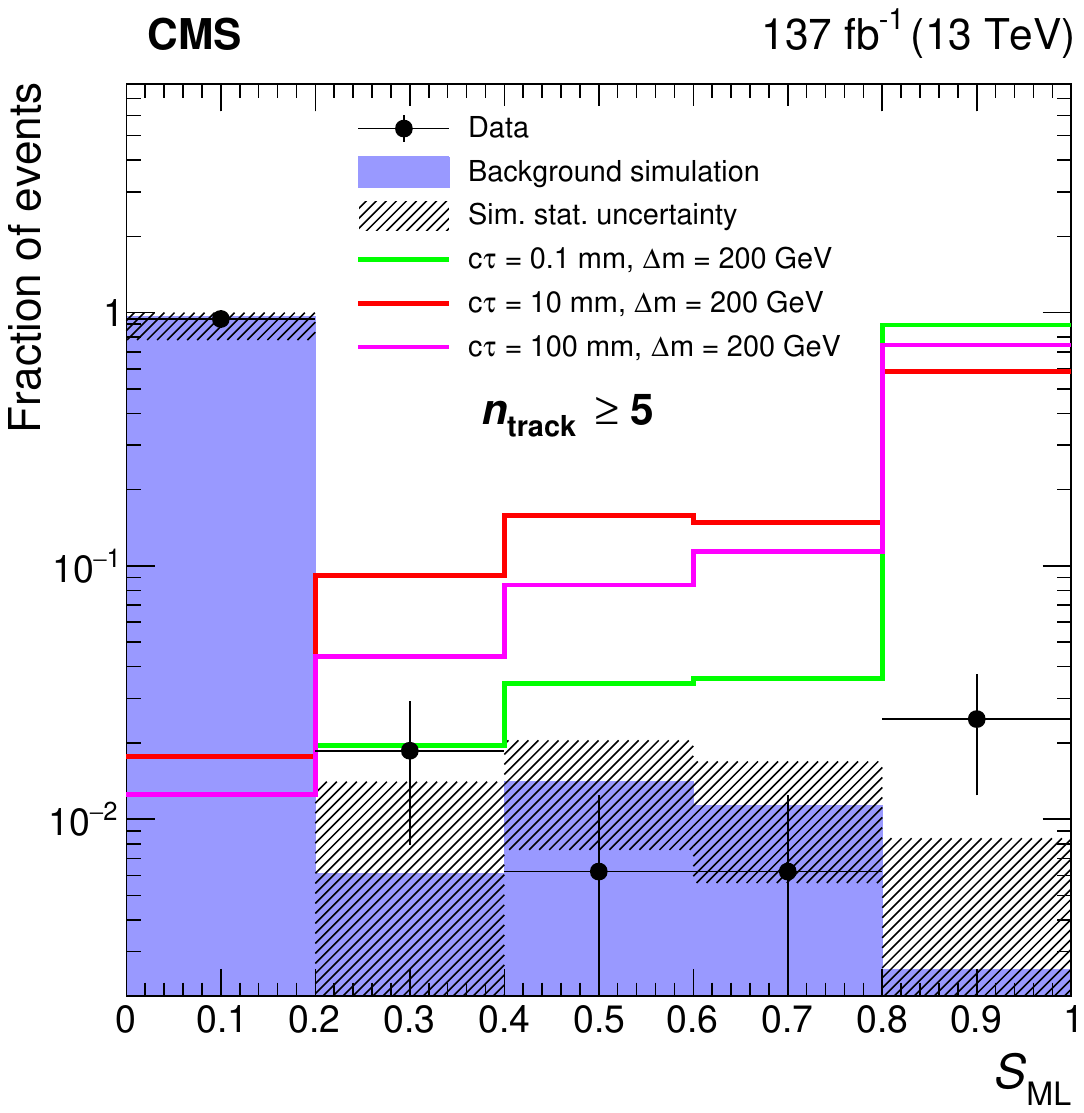}
\caption{Distributions of the output score of the interaction network ($S_{\mathrm{ML}}$) for data, simulated background, and signal, for the displaced vertex plus \ptmiss search. Events with at least five tracks are shown. The distributions are shown for split-SUSY signals with a gluino mass of 2000\GeV and a neutralino mass of 1800\GeV. Different gluino proper decay lengths are shown. All distributions are normalized to unity. Figure taken from Ref.~\cite{CMS:2024trg}.}
\label{fig:EXO-22-020}
\end{figure}

\cmsParagraph{Search for new physics with at least one displaced vertex and \texorpdfstring{\ptmiss}{missing transverse momentum}} This search~\cite{CMS:2024trg} targets LLPs in signatures with at least one DV and \ptmiss using $\Pp\Pp$ collision events taken during 2016--2018 at $\sqrt{s}=13\TeV$. The reconstruction of DVs is detailed in Section~\ref{sec:displacedTracking}. This search expands on Ref.~\cite{EXO-19-013}, which targets a pair of DVs and triggers on \HT. Compared to the search described in Section~\ref{sec:EXO-19-013}, this search aims to target DVs with low \HT (${\gtrsim} 200\GeV$) and a broader range of displacement. In particular, this search targets LLPs with masses between about 1 and 3\TeV and mean proper decay lengths between $0.1$ and 1000\mm, depending on the model. A \ptmiss trigger is used to record events. A customized vertex reconstruction algorithm, which takes displaced tracks and iteratively creates vertices from them, is used to reconstruct DVs. A set of vertex selections is applied to avoid background vertices from material interactions and SM backgrounds originating from decays of particles with nonnegligible lifetimes, such as \PQb hadrons. For LLP events with low \HT, fewer displaced tracks are available to be used for vertex reconstruction, and thus the vertex reconstruction efficiency is smaller. To overcome this difficulty, this search only requires one DV, which improves the search sensitivity to signal events with low \HT and longer LLP lifetime. After the vertex selections, the dominant source of background stems from the accidental crossing of tracks originating from the $\Pp\Pp$ collision, which are fit to a spurious vertex. To further mitigate such background vertices, an interaction network, a machine-learning algorithm based on a GNN, is used as an event classifier. The distribution of the output score of the interaction network is shown in Fig.~\ref{fig:EXO-22-020}.

\section{Summary}
\label{sec:conclusion}

\begin{figure}[htb!]
    \centering
    \includegraphics[width=1.0\textwidth]{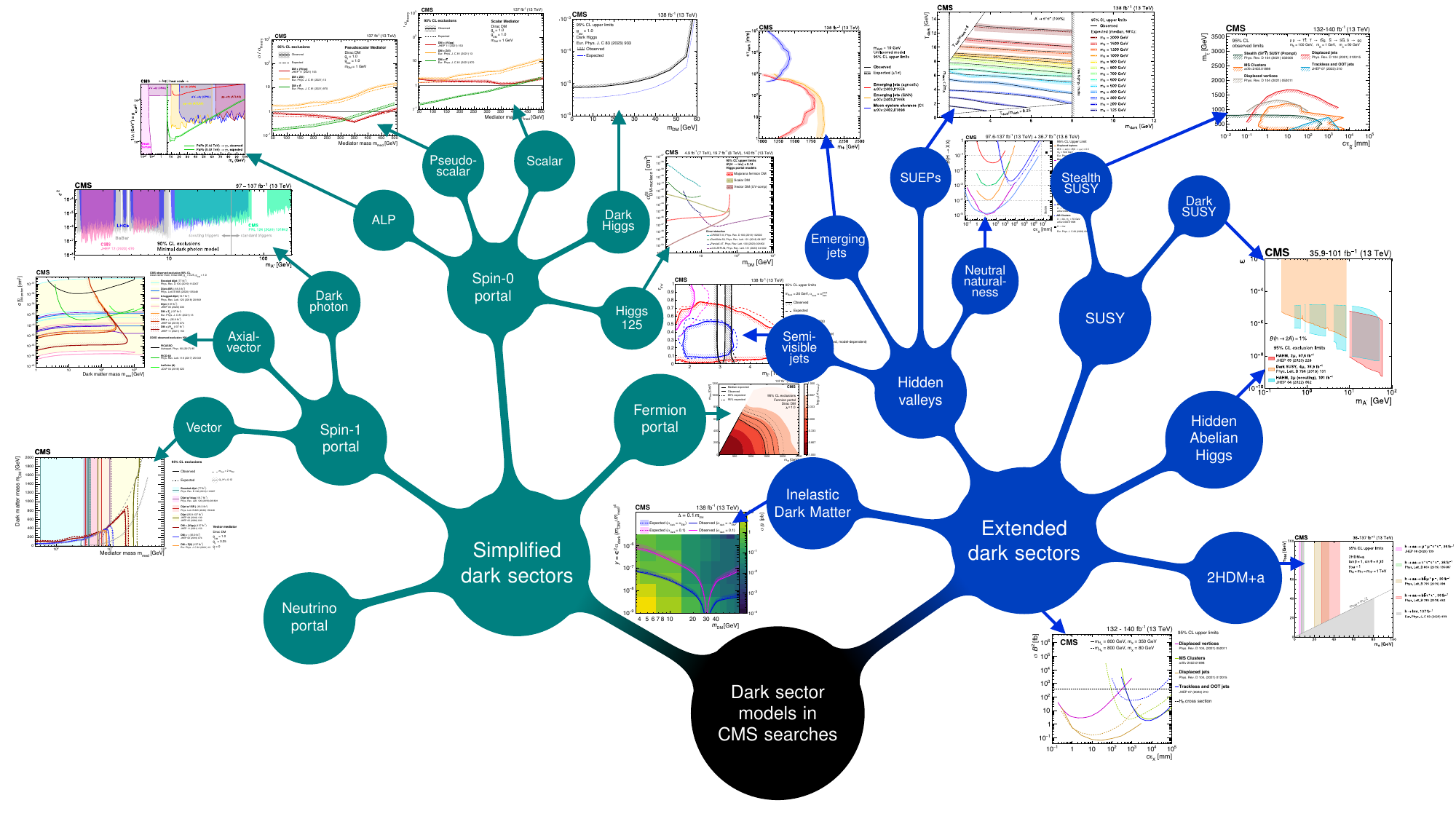}
    \caption{A qualitative depiction of how the results in this Report map onto the models probed in CMS searches for dark sectors.}
    \label{fig:summarySketch}
\end{figure}

A comprehensive review of dark sector (DS) searches with the CMS experiment at the LHC has been presented, using proton-proton and heavy ion collision data collected in Run~2, from 2016 to 2018, or, in some cases, from Run~1 (2011--2012) or Run~3 (2022). These searches have been interpreted in simplified and extended DS models. Figure~\ref{fig:summarySketch} qualitatively illustrates how the results map into this theoretical framework. The broad DS search program spans many different signatures, including those with invisible particles, those with particles promptly decaying into fully visible final states, and those with long-lived particles (LLPs). A number of searches have been newly reinterpreted with DS benchmark scenarios for this Report. In order to perform these searches, several unique techniques of data collection, reconstruction, and analysis were employed, and they are also described in this Report. The broad variety of searches provides sensitivity across a wide range of models and parameter space, and the results represent the most complete set of constraints on DS models obtained by the CMS Collaboration to date.

In particular, this Report has presented the latest constraints from the CMS experiment on a comprehensive set of simplified dark matter models, and it has compared these constraints with those from direct-detection experiments. New interpretations have been shown for extended DS scenarios, including semivisible jets, emerging jets, dark supersymmetry, hidden Abelian Higgs models, and two-Higgs-doublet plus a pseudoscalar models. Several scenarios involving LLPs have been presented, including models with heavy LLPs, stealth supersymmetry, and Higgs boson decays to LLPs.

The searches described in this Report have employed innovative new techniques developed during Run~2 of the LHC to increase sensitivity. First, dedicated triggers have been employed to provide acceptance to previously inaccessible final states, such as triggers that require a displaced jet to reduce the minimum \HT trigger threshold from 1000 to 430\GeV. Second, data scouting has been exploited for displaced muons for the first time, significantly expanding sensitivity to low-mass resonances. Third, entirely new LLP reconstruction techniques have been deployed to significantly expand the sensitivity to displacements beyond the tracker, including the first uses of delayed calorimetry for hadronic final states, the development of muon detector shower reconstruction for both hadronic and electromagnetic final states, and the development of new reconstruction algorithms for highly displaced muons. Fourth, new background predictions and background reduction methods involving advanced machine learning techniques including the use of deep neural networks and graph neutral networks have greatly reduced previously dominant backgrounds and background-related systematic uncertainties. Fifth, new functional forms, as well as new jet substructure and pileup mitigation techniques, have greatly improved searches for resonances and searches that feature large \ptmiss. Together, these developments have greatly extended the ranges of lifetimes, masses, and even types of signatures that can be probed by the CMS detector. This expansion in accessibility can be seen in the sensitivity across up to seven orders of magnitude in signal lifetime, from 0.1 to $10^6$\unit{mm}, for the wide array of displaced signatures shown in Section~\ref{sec:res-hidden-valleys}.

The new signatures targeted by searches in this Report go beyond single displaced objects to emerging jets, complex final states with multiple displaced decays. This new phenomenon appears in many dark quantum chromodynamics models, as do other novel objects including semivisible jets, a mixture of visible and invisible particles, and soft unclustered energy patterns, a spherical distribution of many low-momentum tracks. CMS has now conducted the first dedicated searches for all of these phenomena. In particular, the complementarity between the dedicated semivisible jet search and conventional approaches using invisible or visible final states has been quantified for the first time. The continued expansion of this program to cover a wider range of final states will be facilitated by the new and upcoming developments described below.

Future improvements will further increase the sensitivity of searches for LLPs and other novel final states. For Run~3 of the LHC~\cite{CMS:2023gfb}, a range of new displaced trigger algorithms have been developed and deployed for both the level-1 and the high-level trigger, taking advantage of the experience gained carrying out the searches described in this Report. This includes entirely new delayed calorimetry, using both the electromagnetic and the hadronic calorimeter, and muon detector shower algorithms to increase acceptance for many models by over an order of magnitude~\cite{LLPRun3TriggerDPNote}. New machine-learning-based anomaly detection triggers, using calorimeter or global information, are also being deployed to access final states and kinematic ranges that are not covered by conventional triggers~\cite{DP-2023-079,DP-2023-086}. The performance of data scouting for displaced muons is improved in Run~3 by removing the requirement of a hit in the pixel tracker, thus extending the sensitivity to larger lifetimes. There are also opportunities to employ data parking for several displaced signatures. CMS will continue to seek out opportunities to improve the performance of the full range of prompt and displaced reconstruction algorithms. The hadronic calorimeter timing resolution has been improved from a few ns to around 1\unit{ns}, and the granularity in the readout has been increased, which will enable new searches exploiting this timing and granularity. The sensitivity of many searches will be extended by incorporating new state-of-the-art developments in machine-learning techniques.

The High-Luminosity LHC will include multiple detector upgrades that will substantially enhance the performance of the techniques that have been explored with the current calorimetry, tracking, and trigger capabilities~\cite{Contardo:2015bmq,CMS:2017lum,CERN-LHCC-2017-011,CMS:2017jpq,Hebbeker:2017bix,Zabi:2020gjd,CMS:2667167,CMS-DP-2022-025}. Timing resolution of order $10\unit{ps}$ will be available across multiple upgraded and new subsystems, and there will be a new calorimeter to provide high-granularity energy and position information in the forward region. For the level-1 trigger, tracking will be implemented and sophisticated machine-learning algorithms will be deployed. Finally, increasing the integrated luminosity by an order of magnitude will allow probing rarer processes with smaller cross sections, exceeding existing limits on mediator particles. Together, these upgrades will substantially improve the sensitivity to a wide range of DS models, as shown in several studies of the physics performance at the High-Luminosity LHC~\cite{CMS:2022cju,Dainese:2019rgk,CMS:2022sfl}.

\begin{acknowledgments}
\hyphenation{Bundes-ministerium Forschungs-gemeinschaft Forschungs-zentren Rachada-pisek} We congratulate our colleagues in the CERN accelerator departments for the excellent performance of the LHC and thank the technical and administrative staffs at CERN and at other CMS institutes for their contributions to the success of the CMS effort. In addition, we gratefully acknowledge the computing centers and personnel of the Worldwide LHC Computing Grid and other centers for delivering so effectively the computing infrastructure essential to our analyses. Finally, we acknowledge the enduring support for the construction and operation of the LHC, the CMS detector, and the supporting computing infrastructure provided by the following funding agencies: the Armenian Science Committee, project no. 22rl-037; the Austrian Federal Ministry of Education, Science and Research and the Austrian Science Fund; the Belgian Fonds de la Recherche Scientifique, and Fonds voor Wetenschappelijk Onderzoek; the Brazilian Funding Agencies (CNPq, CAPES, FAPERJ, FAPERGS, and FAPESP); the Bulgarian Ministry of Education and Science, and the Bulgarian National Science Fund; CERN; the Chinese Academy of Sciences, Ministry of Science and Technology, the National Natural Science Foundation of China, and Fundamental Research Funds for the Central Universities; the Ministerio de Ciencia Tecnolog\'ia e Innovaci\'on (MINCIENCIAS), Colombia; the Croatian Ministry of Science, Education and Sport, and the Croatian Science Foundation; the Research and Innovation Foundation, Cyprus; the Secretariat for Higher Education, Science, Technology and Innovation, Ecuador; the Estonian Research Council via PRG780, PRG803, RVTT3 and the Ministry of Education and Research TK202; the Academy of Finland, Finnish Ministry of Education and Culture, and Helsinki Institute of Physics; the Institut National de Physique Nucl\'eaire et de Physique des Particules~/~CNRS, and Commissariat \`a l'\'Energie Atomique et aux \'Energies Alternatives~/~CEA, France; the Shota Rustaveli National Science Foundation, Georgia; the Bundesministerium f\"ur Bildung und Forschung, the Deutsche Forschungsgemeinschaft (DFG), under Germany's Excellence Strategy -- EXC 2121 ``Quantum Universe" -- 390833306, and under project number 400140256 - GRK2497, and Helmholtz-Gemeinschaft Deutscher Forschungszentren, Germany; the General Secretariat for Research and Innovation and the Hellenic Foundation for Research and Innovation (HFRI), Project Number 2288, Greece; the National Research, Development and Innovation Office (NKFIH), Hungary; the Department of Atomic Energy and the Department of Science and Technology, India; the Institute for Studies in Theoretical Physics and Mathematics, Iran; the Science Foundation, Ireland; the Istituto Nazionale di Fisica Nucleare, Italy; the Ministry of Science, ICT and Future Planning, and National Research Foundation (NRF), Republic of Korea; the Ministry of Education and Science of the Republic of Latvia; the Research Council of Lithuania, agreement No.\ VS-19 (LMTLT); the Ministry of Education, and University of Malaya (Malaysia); the Ministry of Science of Montenegro; the Mexican Funding Agencies (BUAP, CINVESTAV, CONACYT, LNS, SEP, and UASLP-FAI); the Ministry of Business, Innovation and Employment, New Zealand; the Pakistan Atomic Energy Commission; the Ministry of Education and Science and the National Science Center, Poland; the Funda\c{c}\~ao para a Ci\^encia e a Tecnologia, grants CERN/FIS-PAR/0025/2019 and CERN/FIS-INS/0032/2019, Portugal; the Ministry of Education, Science and Technological Development of Serbia; MCIN/AEI/10.13039/501100011033, ERDF ``a way of making Europe", Programa Estatal de Fomento de la Investigaci{\'o}n Cient{\'i}fica y T{\'e}cnica de Excelencia Mar\'{\i}a de Maeztu, grant MDM-2017-0765, projects PID2020-113705RB, PID2020-113304RB, PID2020-116262RB and PID2020-113341RB-I00, and Plan de Ciencia, Tecnolog{\'i}a e Innovaci{\'o}n de Asturias, Spain; the Ministry of Science, Technology and Research, Sri Lanka; the Swiss Funding Agencies (ETH Board, ETH Zurich, PSI, SNF, UniZH, Canton Zurich, and SER); the Ministry of Science and Technology, Taipei; the Ministry of Higher Education, Science, Research and Innovation, and the National Science and Technology Development Agency of Thailand; the Scientific and Technical Research Council of Turkey, and Turkish Energy, Nuclear and Mineral Research Agency; the National Academy of Sciences of Ukraine; the Science and Technology Facilities Council, UK; the US Department of Energy, and the US National Science Foundation.

Individuals have received support from the Marie-Curie program and the European Research Council and Horizon 2020 Grant, contract Nos.\ 675440, 724704, 752730, 758316, 765710, 824093, 101115353, 101002207, and COST Action CA16108 (European Union) the Leventis Foundation; the Alfred P.\ Sloan Foundation; the Alexander von Humboldt Foundation; the Belgian Federal Science Policy Office; the Fonds pour la Formation \`a la Recherche dans l'Industrie et dans l'Agriculture (FRIA-Belgium); the Agentschap voor Innovatie door Wetenschap en Technologie (IWT-Belgium); the F.R.S.-FNRS and FWO (Belgium) under the ``Excellence of Science -- EOS" -- be.h project n.\ 30820817; the Beijing Municipal Science \& Technology Commission, No. Z191100007219010; the Ministry of Education, Youth and Sports (MEYS) of the Czech Republic; the Shota Rustaveli National Science Foundation, grant FR-22-985 (Georgia); the Hungarian Academy of Sciences, the New National Excellence Program - \'UNKP, the NKFIH research grants K 131991, K 133046, K 138136, K 143460, K 143477, K 146913, K 146914, K 147048, 2020-2.2.1-ED-2021-00181, and TKP2021-NKTA-64 (Hungary); the Council of Scientific and Industrial Research, India; ICSC -- National Research Center for High Performance Computing, Big Data and Quantum Computing and FAIR -- Future Artificial Intelligence Research, funded by the NextGenerationEU program (Italy); the Latvian Council of Science; the Ministry of Education and Science, project no. 2022/WK/14, and the National Science Center, contracts Opus 2021/41/B/ST2/01369 and 2021/43/B/ST2/01552 (Poland); the Funda\c{c}\~ao para a Ci\^encia e a Tecnologia, grant FCT CEECIND/01334/2018; the National Priorities Research Program by Qatar National Research Fund; the Programa Estatal de Fomento de la Investigaci{\'o}n Cient{\'i}fica y T{\'e}cnica de Excelencia Mar\'{\i}a de Maeztu, grant MDM-2017-0765 and projects PID2020-113705RB, PID2020-113304RB, PID2020-116262RB and PID2020-113341RB-I00, and Programa Severo Ochoa del Principado de Asturias (Spain); the Chulalongkorn Academic into Its 2nd Century Project Advancement Project, and the National Science, Research and Innovation Fund via the Program Management Unit for Human Resources \& Institutional Development, Research and Innovation, grant B37G660013 (Thailand); the Kavli Foundation; the Nvidia Corporation; the SuperMicro Corporation; the Welch Foundation, contract C-1845; and the Weston Havens Foundation (USA).    
\end{acknowledgments}

\bibliography{auto_generated}

\clearpage
\appendix

\section{Glossary of acronyms}
\label{sec:glossary}

\begin{longtable}[l]{ll}
ALP & Axion-like particle\\
AVF & Adaptive vertex fitter\\
BDT & Boosted decision tree\\
BPTX & Beam pickup timing device\\
BSM & Beyond the standard model\\
CA & Cambridge--Aachen\\
CEP & Central exclusive production\\
CHS & Charged-hadron subtraction\\
\CL & Confidence level\\
CMS & Compact Muon Solenoid\\
\CP & Charge conjugation parity\\
CSC & Cathode strip chamber\\
CR & Control region\\
DA & Domain adaptation\\
DM & Dark matter\\
DD & Direct detection\\
DNN & Deep neural network\\
DS & Dark sector\\
DSA & Displaced standalone\\
DT & Drift tube\\
DV & Displaced vertex\\
ECAL & Electromagnetic calorimeter\\
EFT & Effective field theory\\
EJ & Emerging jet\\
EW & Electroweak\\
FCNC & Flavor-changing neutral currents\\
FIP & Feebly interacting particle\\
FSUSY & Folded SUSY\\
GBDT & Gradient-boosted decision tree\\
GMSB & Gauge-mediated SUSY breaking\\
GNN & Graph neural network\\
HAHM & Hidden Abelian Higgs model\\
HCAL & Hadronic calorimeter \\
HI & Heavy ion\\
HLT & High level trigger\\
HNL & Heavy neutral lepton\\
HV & Hidden valley\\
ID & Indirect detection\\
IDM & Inelastic dark matter\\
IP & Impact parameter\\
ISR & Initial-state radiation\\
LHC & Large Hadron Collider\\
LLP & Long-lived particle\\
LO & Leading order\\
MC & Monte Carlo\\
MVA & Multi-variate analysis\\
NLO & Next-to-leading order\\
NNLL & Next-to-next-to-leading logarithm\\
NNLO & Next-to-next-to-leading order\\
OOT & Out of time\\
PDF & Parton distribution function\\
PF & Particle flow\\
PPS & Precision proton spectrometer\\
PU & Pileup \\
PUPPI & Pileup-per-particle identification\\
PV & Primary vertex\\
QCD & Quantum chromodynamics\\
ROC & Receiver operating characteristic\\
RPC & Resistive-plate chamber\\
RPV & $R$-parity violating\\
SD & Spin dependent\\
SI & Spin independent\\
SM & Standard model\\
SR & Signal region\\
SUEP & Soft unclustered energy patterns\\
SUSY & Supersymmetry\\
SVJ & Semivisible jet\\
TF & Transfer factor\\
TH & Twin Higgs (TH)\\
UPC & Ultra-peripheral collision\\
VBF & Vector-boson fusion\\
WIMP & Weakly interacting massive particle\\
2D & Two-dimensional \\
3D & Three-dimensional \\
2HDM & Two-Higgs-doublet model\\
2HDM+a & Two-Higgs-doublet model plus pseudoscalar\\
\end{longtable}

\cleardoublepage \section{The CMS Collaboration \label{app:collab}}\begin{sloppypar}\hyphenpenalty=5000\widowpenalty=500\clubpenalty=5000
\cmsinstitute{Yerevan Physics Institute, Yerevan, Armenia}
{\tolerance=6000
A.~Hayrapetyan, A.~Tumasyan\cmsAuthorMark{1}\cmsorcid{0009-0000-0684-6742}
\par}
\cmsinstitute{Institut f\"{u}r Hochenergiephysik, Vienna, Austria}
{\tolerance=6000
W.~Adam\cmsorcid{0000-0001-9099-4341}, J.W.~Andrejkovic, T.~Bergauer\cmsorcid{0000-0002-5786-0293}, S.~Chatterjee\cmsorcid{0000-0003-2660-0349}, K.~Damanakis\cmsorcid{0000-0001-5389-2872}, M.~Dragicevic\cmsorcid{0000-0003-1967-6783}, P.S.~Hussain\cmsorcid{0000-0002-4825-5278}, M.~Jeitler\cmsAuthorMark{2}\cmsorcid{0000-0002-5141-9560}, N.~Krammer\cmsorcid{0000-0002-0548-0985}, A.~Li\cmsorcid{0000-0002-4547-116X}, D.~Liko\cmsorcid{0000-0002-3380-473X}, I.~Mikulec\cmsorcid{0000-0003-0385-2746}, J.~Schieck\cmsAuthorMark{2}\cmsorcid{0000-0002-1058-8093}, R.~Sch\"{o}fbeck\cmsorcid{0000-0002-2332-8784}, D.~Schwarz\cmsorcid{0000-0002-3821-7331}, M.~Sonawane\cmsorcid{0000-0003-0510-7010}, S.~Templ\cmsorcid{0000-0003-3137-5692}, W.~Waltenberger\cmsorcid{0000-0002-6215-7228}, C.-E.~Wulz\cmsAuthorMark{2}\cmsorcid{0000-0001-9226-5812}
% ...
\par}
\vskip\cmsinstskip
\dag:~Deceased\\
$^{1}$Also at Yerevan State University, Yerevan, Armenia\\
$^{2}$Also at TU Wien, Vienna, Austria\\
% ...
\end{sloppypar}
\end{document}